\documentclass[12pt]{JHEP} 
\usepackage{epsfig}
\newcommand\fverb{\setbox\pippobox=\hbox\bgroup\verb}
\newcommand\fverbdo{\egroup\medskip\noindent%
			\fbox{\unhbox\pippobox}\ }
\newcommand\fverbit{\egroup\item[\fbox{\unhbox\pippobox}]}
\newbox\pippobox

\catcode `@=11 \@addtoreset{equation}{section} \catcode `@=12

\newcommand{\be}{\begin{equation}}
\newcommand{\ee}{\end{equation}}
\newcommand{\bea}{\begin{eqnarray}}
\newcommand{\eea}{\end{eqnarray}}
\newcommand{\beas}{\begin{eqnarray*}}
\newcommand{\eeas}{\end{eqnarray*}}
\newcommand{\eq}[1]{(\ref{#1})}
\newcommand{\del}{\partial}
\newcommand{\nl}{\hfill\break}
\newcommand{\gst}{g_s}
\newcommand{\nn}{\nonumber}

\newcommand{\dd}{D1-D5 }
\font\cmsss = cmss14
\def\H{\hbox{\cmsss H}}
\def\GR{general relativity}
\def\half{\frac12}
\def\c#1{{\hat{#1}}}
\def\ttt{{\tt t}}
\def\fig#1{Fig \ref{#1}}
\def\gap#1{\vspace{#1 ex}}
\def\ads{\ensuremath{\hbox{AdS}_3}}
\def\cft{CFT$_2$}
\def\ni{\noindent}
\def\mysec#1#2#3{\gap#1\noindent\underbar{\it #3}\gap#2}
\def\gravity{\cite{MTW,chp1:Wald,feyn-gravity,wein-gravity,hawking-ellis}}
\def\singularity{\cite{Horowitz:1995ta,Johnson:1995bf,Johnson:1999qt,
Gubser:2000nd,Natsuume:2001ba}}

\title{Microscopic Formulation of Black Holes in String Theory}
\author{by
Justin R.~David$^{a}$, Gautam Mandal$^{b}$ and Spenta R.~Wadia$^b$\\
$^a$ {\it Department of physics, \\
University of Californina, Santa Barbara, CA 93106, USA.}\\
\vspace*{1ex}

$^b$ {\it Tata Institute of Fundamental Research,\\ 
Homi
Bhabha Road, Mumbai 400 005, India.}
\\
E-mail: \email{justin@vulcan.physics.ucsb.edu,
mandal@theory.tifr.res.in, wadia@theory.tifr.res.in}}

\preprint{\hepth{0203048}}

\abstract{In this Report we review the microscopic formulation of the five
dimensional black hole of type IIB string theory in terms of the D1-D5
brane system.  The emphasis here is more on the brane dynamics than on
supergravity solutions.  We show how the low energy brane dynamics,
combined with crucial inputs from AdS/CFT correspondence, leads to a
derivation of black hole thermodynamics and the rate of Hawking
radiation. Our approach requires a detailed exposition of the gauge
theory and conformal field theory of the D1-D5 system. We also discuss
some applications of the AdS/CFT correspondence in the context of
black hole formation in three dimensions by thermal transition and by
collision of point particles.
}

\begin{document}

\newpage
\section{Introduction}

\subsection{Quantum theory and general relativity\label{qm-gr}} 
Quantum theory and the general theory of relativity form the basis of
modern physics. However, these two theories seem to be fundamentally
incompatible.  Quantizing general relativity leads to a number of
basic problems:

\begin{enumerate}
\item 
Ultraviolet divergences render general relativity ill-defined as a
quantum theory (see, e.g. S. Weinberg in \cite{Hawking:ig}). This
specifically means that if we perform a perturbation expansion around
flat Minkowski space-time (which is a good first approximation to our
world) then to subtract infinities from the divergent diagrams we have
to add an infinite number of counterterms to the Einstein-Hilbert
action with coefficients that are proportional to appropriate powers
of the ultraviolet cutoff.

There is good reason to believe that string theory
\cite{Green:sp,Polchinski:rq} solves this ultraviolet problem because
the extended nature of string interactions have an inherent
ultraviolet cutoff given by the fundamental string length
$\sqrt{\alpha'}$.  Furthermore, for length scales much larger than the
string length the Einstein-Hilbert action emerges
\cite{Yoneya:jg,Scherk:1974mc} as a low energy effective action from
string theory, with Newton's constant (for type II strings in ten
dimensions) given by,
\be
G_N^{(10)}= 8 \pi^6 g_s^2 \alpha'^4,
\label{gn-ten}
\ee
where $g_s$ is the string coupling. 

\item There are many singular classical solutions of general
relativity (for standard textbooks on classical general relativity,
see, e.g., \gravity), including the Schwarzs\-child black hole and the
Big Bang model of cosmology.  Black holes and their higher dimensional
analogues (black branes) also appear as solutions of low energy string
theory. A quantum theory of gravity must (a) present an understanding
regarding which of these singular geometries can arise from a well
defined quantum mechanics in an appropriate limit, and (b) formulate
such a quantum mechanics where possible.  String theory has been able
to ``resolve'' a class of singularities in this way, but a complete
understanding of the issue of singularities is still lacking (see
\singularity\ for a partial list of related papers;
\cite{Natsuume:2001ba} contains a review and a comprehensive list of
references).
 
\item While the above problems are related to the high energy (short
distance) behaviour of \GR, there exists another basic problem when we
quantize matter fields in the presence of a black hole, which does not
ostensibly depend on high energy processes.  This problem is called
the information puzzle (\cite{Hawk1,Hawk2}, for early reviews see, e.g.,
\cite{Hawking:ig,'tHooft:ij,Giddings:1995gd}). In the following we 
shall explain the issue in some detail and subsequently summarize the
attempts within string theory to resolve the puzzle in a certain class
of black holes.

Besides being a long-lasting problem of \GR, this is an important
problem for string theory for the following reason. String theory has
been proposed as a theory that describes all elementary particles and
their interactions.  Presently the theory is not in the stage of
development where it can provide quantitative predictions in particle
physics. However if string theory can resolve some logical problem
that arises in the applications of standard quantum field theory to
general relativity, then it is a step forward for string theory.

\item Finally, any quantum theory of gravity should lead to
an understanding of the problem of the cosmological constant
(for reviews see, e.g. \cite{Carroll:2000fy},\cite{Weinberg:2000yb},
\cite{Witten:2000zk}).

\end{enumerate}

\subsection{Black holes and the information puzzle}

Let us briefly review some general properties of black holes
\gravity. Black holes are objects which result as end points of
gravitational collapse of matter. For an object of mass greater than
roughly three solar masses (see, e.g. \cite{Celotti:1999tg}), the
gravitational force overcomes all other forces and the matter
generically collapses into a black hole (in some exceptional cases a
naked singularity might result). This would suggest that to specify a
black hole it is necessary to give in detail the initial conditions of
the collapse. As we will see below a black hole is completely
specified by a few parameters only.

To introduce various concepts related to black holes we will discuss
two examples of black holes. First, let us consider the Schwarzschild
black hole in $3+1$ dimensions. It is a time independent, spherically
symmetric solution of Einstein gravity without matter. The metric is
given by
\be
ds^2= -\left(1 - \frac{2 G_N  M}{r}\right) dt^2 + 
\left(1 - \frac{2 G_N  M}{r}\right)^{-1} dr^2 + r^2 d\Omega^2 
\label{sch-bh}
\ee
where $G_N$ is the Newton's constant\footnote{We will use the notation
$G_N^{(D)}$ or $G_N^D$ for Newton's constant in $D$ spacetime
dimensions, reserving the default $G_N$ for $D=4$.} and  
$d\Omega^2 = \left( d\theta^2 + \theta^2 d\phi^2 \right)$ is the
metric on $S^2$.  We have chosen
units so that the velocity of light is $c=1$.  The surface $r=2 G_N M$
is called the event horizon. It is a coordinate singularity
($g_{rr}=\infty$) but not a curvature singularity (e.g., Ricci scalar
is finite here). Light-like geodesics and time-like geodesics starting
``inside'' the event horizon ($r<2 G_N M$) end up at $r=0$ (the
curvature singularity) in a finite proper time. This means that
classically the black hole is truly black, it cannot emit
anything. Note that the solution is completely specified by only one
parameter $M$, which coincides with the ADM mass (see 
\eq{mass}) of the black hole.

Next we consider the Reissner-N\"{o}rdstrom (RN) black hole. 
It is a time independent, spherically symmetric solution of Einstein
gravity coupled to the electromagnetic field. The solution is given
by the following backgrounds:
\bea
ds^2&=& -\left( 1-\frac{2G_N M}{r} + \frac{G_NQ^2}{r^2} \right) dt^2
+ \left( 1-\frac{2G_N M}{r} + \frac{G_NQ^2}{r^2} \right)^{-1} dr^2 +
r^2 d\Omega^2 ,\nn \\
A_0 &=&\frac{Q}{r}, \quad A_i=0, \; i=1,2,3.
\label{rn-bh}
\eea
where $A_0$ is the time component of the vector potential. This
solution carries a charge $Q$ (the Schwarzschild black hole,
Eqn. \eq{sch-bh}, corresponds to the special case
$Q=0$). There are two coordinate singularities
($g_{rr}=\infty$) at $r= r_+$ (outer horizon) and $r= r_-$ (inner
horizon)
\be
\label{def-rpm}
r_\pm  = G_N M \pm \sqrt{(G_N M)^2 -G_N Q^2}
\ee
The event horizon coincides with the outer horizon $r=r_+$. When $M=
 |Q|/\sqrt{G_N}$, $r_+$ coincides with  $r_-$.
Such a black hole is called  an extremal black hole. 
Note that the black hole \eq{rn-bh} is completely specified by its
mass $M$ and the charge $Q$. 

The \dd\ black hole, which will be the main subject of this report, is
similar to the RN black hole; it is charged under Ramond-Ramond fields
in type IIB string theory and has the same causal structure,
equivalently the same Penrose diagram (see Section
\ref{penrose-diagram}), as the RN black hole.

The point $r=0$ is a curvature singularity for both the Schwarzschild
solution \eq{sch-bh} and the Reissner-N\"{o}rdstrom solution
\eq{rn-bh}. The singularity is ``clothed'' by the event horizon
in the physical range of parameters ($M>0$ for Schwarzschild,
$M>|Q|/\sqrt{G_N}$ for RN). The ranges of mass: $M<0$ for
Schwarzschild, $M<|Q|/\sqrt{G_N}$ for RN, are unphysical because these
result in a naked singularity (see, e.g. \cite{Horowitz:1995ta}).  The
limiting case for the RN black hole, $M= |Q|/\sqrt{G_N}$, is called an
extremal black hole, as already mentioned above.

In general, collapsing matter results in black holes which are
completely specified by the mass $M$, the $U(1)$ charges $Q_i$ and the
angular momentum $J$. This is called the no hair theorem (see,
e.g.,\cite{Chrusciel:sn,Bekenstein:1996pn,Carter:1997im,
Marolf:1999cn} for a review). Whatever other information (for example,
multipole moments) present decays exponentially fast with a
characteristic time $ \tau = r_h/c$, during the collapse. ($r_h$ is
the radius of the horizon and $c$ is the speed of light). Thus, all
detailed information carried by the collapsing matter is completely
lost.

The other beautiful result of classical general relativity is the so
called area law. It says that \cite{hawking-ellis} the area of the
horizon of a black hole cannot decrease with time, and if two (or
more) black holes (of areas $A_1, A_2, \ldots$) merge to form a single
black hole, the area $A$ of its horizon will satisfy
\be
\label{area-thm}
A \ge A_1 + A_2 + \ldots
\ee
This general result is easy to verify for the Schwarzschild and
Reissner-N\"{o}rdstrom black holes that we have discussed above. 

Bekenstein argued that unless a black hole had entropy, an infalling
hot body would lead to a violation of the second law of thermodynamics
because the entropy of the hot body will be lost once it is absorbed
by the black hole, thus causing entropy to decrease. He postulated
that the entropy of a black hole is proportional to the area of the
event horizon, the constant of proportionality being universal for all
black holes. The area law now have natural thermodynamic
interpretations and the macroscopic observables allowed by the no hair
theorems play the role of macroscopic thermodynamic variables. In the
classical theory, however, there is no notion of the absolute entropy
of a system. Such a concept, as we shall see now, requires quantum
theory.

In his pioneering work in the seventies, Hawking \cite{Hawk1}
quantized matter fields in the background geometry of a black hole. He
found by that the Schwarzschild black hole is not truly black. A
semi-classical calculation showed that it emits radiation with the
spectrum of a black body at a temperature T given by
\be
\label{sch-temp}
T_H= \frac{\hbar}{8\pi G_N M}
\ee
The quantum nature of this effect is clearly evident from the fact
that the temperature is proportional to $\hbar$. For the 
Reissner-N\"{o}rdstrom black hole the temperature of 
Hawking radiation is 
\be
\label{rn-temp}
T_H= \frac{(r_+ - r_-)\hbar}{4\pi r_+^2}
\ee
where $r_\pm$ are defined in \eq{def-rpm} as functions of
$Q,M$. 
A brief derivation of the temperature formulae, using the Euclidean
approach \cite{Gibbons:1976ue} is presented in Appendix
\ref{gibbons-hawking}.
One notes that the extremal Reissner-N\"{o}rdstrom black hole
has $T_H=0$ and  does not Hawking radiate.

In general the Hawking temperature turns out to be a function of mass,
the charge(s) and the angular momentum alone. Thus even semi-classical
effects do not provide further information of the black hole. The
works in the seventies culminated in the following laws of black holes
\cite{Bekenstein:ur,Bekenstein:ax,Bardeen:gs,Hawking:de}
which are analogous to the laws of thermodynamics.\\

\begin{enumerate}
\item 
First Law: Two neighbouring equilibrium states of a black
hole, of mass $M$, charges $Q_i$ and angular momentum $J$, 
are related by
\be
dM= T_H d\left[\frac{A}{4G_N\hbar}\right] + \Phi_i dQ_i + \Omega dJ
\label{1:first-law}
\ee
where $A$ is the area of the event horizon, $\Phi$ the electric surface
potential and $\Omega$ the angular velocity. For the special case of the
Reissner-N\"{o}rdstrom black hole the first law reduces to
\be
dM= T_H d\left[\frac{A}{4G_N\hbar}\right] + \Phi dQ.
\label{1:rn-first-law}
\ee
where $\Phi= Q/r_+$. This is explicitly verified in
appendix \ref{gibbons-hawking}, Eqn. \eq{a1:first-law}. 

\item  Second Law: Black holes have entropy $S$ given by
\bea
S &=& \frac{A}{4G_N\hbar}= \frac14 \frac{A}{A_{Pl}}
\label{bek-hawk}\\
A_{Pl} &=& \frac{G_N\hbar}{c^3}
\label{a-pl}
\eea
where we have reinstated in the last equation the speed of light
$c$. The generalized second law says that the sum of the entropy of
the black hole and the surroundings never decreases (this is a
generalized form of \eq{area-thm}).
\end{enumerate}

Formula \eq{bek-hawk}, called the Bekenstein-Hawking entropy formula,
is very important because it provides a counting of the effective
number of degrees of freedom of a black hole which any theory of
quantum gravity must reproduce. We note that $A_{Pl}$ in \eq{a-pl} is
a basic unit of area (Planck unit), involving all three fundamental
constants ($A_{Pl}= 2.61 \times 10^{-66} \hbox{cm}^2$ in four
dimensions). The Bekenstein-Hawking formula simply states that the
entropy of a black hole is a quarter of its horizon area measured in
units of $A_{\rm Pl}$.

The entropy of the Schwarzschild black hole, according to
\eq{bek-hawk}  is
\be
S= 4\pi G_N M^2
\ee
while that of the RN black hole is
\be
S= \frac{\pi r_+^2}{G_N \hbar}
\ee
The Hawking radiation as calculated in semi-classical general
relativity is a mixed state.  It turns out to be difficult to
calculate the correlations between the infalling matter and outgoing Hawking
particles in the standard framework of general relativity. Such a
calculation would require a good quantum theory of gravity where
controlled approximations are possible \cite{'tHooft:1996tq}.

If we accept the semi-classical result that black holes emit radiation
that is \underbar{exactly} thermal then it leads to the information
puzzle (\cite{Hawk1,Hawk2,Hawking:ig,'tHooft:ij,Giddings:1995gd}):

Initially the matter that formed the black hole is in a pure quantum
mechanical state. Here in principle we know all the quantum mechanical
correlations between the degrees of freedom of the system.  In case
the black hole evaporates completely then the final state of the
system is purely thermal and hence it is a mixed state. This evolution
of a pure state to a mixed state is in conflict with the standard laws
of quantum mechanics which involve unitary time evolution which sends
pure states to pure states.

Hence it would appear that we have to modify quantum mechanics, as was
advocated by Hawking \cite{Hawk2}. However, in the following
we shall argue that if we replace the paradigm of quantum field theory
by that of string theory (Section \ref{framework}), we are able to
retain quantum mechanics and resolve the information puzzle (for a
certain class of black holes) by discovering the microscopic degrees
of freedom of the black hole.

It is well worth pointing out that the existence of black holes in
nature (for which there is mounting evidence \cite{Celotti:1999tg})
compels us to resolve the conundrums that black holes present. One can
perhaps take recourse to the fact that for a black hole whose mass is
a few solar masses the Hawking temperature is very tiny ($\sim
10^{-8}$ degree Kelvin), and not of any observable consequence.
However the logical problem that we have described above cannot be
wished away and its resolution makes a definitive case for the string
paradigm as a correct framework for fundamental physics as opposed to
standard local quantum field theory.

This assertion implicitly assumes that in string theory there exists a
controlled calculational scheme to calculate the properties of black
holes.  Fortunately there does exist a class of black holes in type
IIB string theory compactified on a 4-manifold ($T^4$ or $K_3$) which
has sufficient supersymmetry to enable a precise calculation of low
energy processes of this class of black holes. This aspect is the main
focus of this review.

\subsection{The string theory framework for black holes}\label{framework}

We now briefly describe the conceptual framework of black hole
thermodynamics in string theory. A black hole of a given mass $M$,
charges $Q_i$ and angular momentum $J$ is defined by a density 
matrix (see also \eq{chp4:density-matrix}):
\be
\label{density}
\rho  =  \frac{1}{\Omega }\sum_{i\in {\mathcal S}}
|i\rangle \langle i|
\ee
where $|i\rangle $ is a microstate which can be any of a set
${\mathcal S}$ of states ({\em microcanonical ensemble}) all of which
are characterized by the above mentioned mass, charges and angular
momentum.

A definition like \eq{density} is of course standard in quantum
statistical mechanics, where a system with a large number of degrees
of freedom is described by a density matrix to derive a thermodynamic
description. Using \eq{density} thermodynamic quantities like the
temperature, entropy and the rates of Hawking radiation can be derived
for a black hole in string theory (see Section 8 for details). In
particular the Bekenstein-Hawking formula is derived from Boltzmann's
law:
\be
\label{Boltz}
S=\ln \Omega
\ee
where $\Omega$ is the number of microstates of the system. We are
using units so that the Boltzmann constant is unity.

Given this we can calculate formulas of black hole thermodynamics just
like we calculate the thermodynamic properties of macroscopic objects
using standard methods of statistical mechanics. Here the quantum
correlations that existed in the initial state of the system are in
principle all present and are only erased by our procedure of defining
the black hole state in terms of a density matrix. In this way one can
account for not only the entropy of the system which is a counting
problem but also the rate of Hawking radiation which depends on
interactions.

Let us recall the treatment of radiation coming from a star, or a lump
of hot coal. The `thermal' description of the radiation coming is the
result of averaging over a large number of quantum states of the coal.
In principle by making detailed measurements on the wave function of
the emitted radiation we can infer the precise quantum state of the
emitting body. For black holes the reasoning is similar.

Hence in the string theory formulation the black hole can exist as a
pure state: one among the highly degenerate set of states that are
characterized by a small number of parameters. Let us also note that
in Hawking's semi-classical analysis, which uses quantum field theory
in a given black-hole space-time, there is no possibility of a
microscopic construction of the black hole wave functions.  To repeat,
in string theory Hawking radiation is \underbar{not} thermal and in
principle we can reconstruct the initial state of the system from the
final state, which therefore resolves the information paradox.

We shall see that in the case of the five-dimensional black hole of
type IIB string theory it is possible to construct a precise
microscopic description of black hole thermodynamics. We now summarize
the four basic ingredients we need to describe and calculate
Hawking radiation  for near extremal black holes
which have low Hawking temperature:
\begin{enumerate}
\item The microscopic constituents of the black hole. In the case of the 
5-dimensional black hole of type IIB string theory the microscopic
modeling is in terms of a system of \dd branes wrapped on $S^{1}\times
M_{4}$, where $M_{4}$ is a 4-dim. compact manifold, which can be
either $T^{4}$ or $K_{3}$. In this report we will mainly focus on
$T^{4}$.
\item The spectrum of the low energy degrees of freedom of the bound state of 
the \dd  system. These are derived at weak coupling
and we need to know if the spectrum survives at strong coupling.
\item The coupling of the low energy degrees of freedom to supergravity modes.
\item The description of the black hole as a density matrix. This 
implies expressions for decay and absorption probabilities which are
related to S-matrix elements between initial and final states of the
black hole.
\end{enumerate}

To understand the microscopic calculation of Hawking rate in a
nutshell, consider a black hole of mass $M$ and charge $Q$, described
by a microcanonical ensemble ${\mathcal S}$. Consider the process of
absorption of some particles by the black hole which changes the mass
and charge to $M', Q'$ (corresponding to another microcanonical
ensemble, say ${\mathcal S}')$. Let $\Omega$ ($\Omega'$) be the total
number of states in ${\mathcal S}$ (resp. ${\mathcal S}'$). The
absorption probability from a state 
$|i \rangle\in {\mathcal S}$ to a state
$|f \rangle \in {\mathcal S}'$ is given by
\be
\label{probabs}
P_{\rm abs}(i\rightarrow f)=\frac{1}{\Omega}\sum_{\it i,f}
|\langle f|S|i \rangle |^{2} 
\ee 
where, by definition, we sum over the final states and
average over the initial states.
Similarly, the decay probability from a state $|i>\in {\mathcal S}'$ 
to a state $|f>\in {\mathcal S}$ is given by
\be
\label{probdecay}
P_{\rm decay}(i\rightarrow f)=\frac{1}{\Omega'}\sum_{\it i,f}|
\langle f|S|i \rangle |^{2}
\ee 
Point (3) above allows us to calculate the
matrix element $\langle f|S|i \rangle $ in string theory, thus leading to
a calculation of absorption cross-section and Hawking rate.
We will elaborate on this in  detail in Section 8. 

One of the important issues in this subject is that the string theory
ingredients (points (1) and (2) above) are usually known in the 
case when the effective open string coupling is small. In this case
the Schwarzschild radius $R_{sch}$ of the black hole is smaller than
the string length $l_s$ and we have a controlled definition of a
string state.  As the coupling is increased we go over to the
supergravity description where $R_{sch} \gg l_s$ and we have a black
hole. Now it is an issue of dynamics whether the spectrum of the
theory undergoes a drastic change or not, determining whether the
description of states in weak coupling, which enabled a thermodynamic
description, remains valid or not. In the model we will consider we
will see that the description of the weak coupling effective
Lagrangian goes over to strong coupling because of supersymmetry. It
is an outstanding challenge to understand this problem when the weak
coupling theory has little or no supersymmetry \cite{susskind,
polhoro,Horowitz:1998jc,venezia}. We will briefly touch upon this
issue in Section 11.

\subsection{Plan of this Report\label{sec:plan}}

The plan of the rest of this Report is as follows.

In Section 2, we discuss in some detail the construction of the
five-dimensional black hole solution in type IIB supergravity. We
provide a brief background of how to construct BPS solutions in
M-theory from first principles by explicitly solving Killing spinor
equations; we focus on the example of the M2 brane and its
intersections. We then construct the D1-D5 black string solution by
dualizing M2$\perp$M2, and the extremal (BPS) D1-D5 black hole by
dualizing M2$\perp$M2$\perp$M2. We then describe an algorithm of how
to generate non-BPS solutions from BPS ones; we provide a motivation
for this algorithm in Appendix B. We compute the Bekenstein-Hawking
entropy and the Hawking temperature of these black holes. We next
discuss how to generate non-zero $B_{\rm NS}$ in the D1-D5 system by
exact dualities in supergravity. In the absence of the $B_{\rm NS}$
vev, the D1-D5 system is marginally bound and can fragment into
clusters of smaller D1-D5 systems. We show, by constructing the exact
supergravity solution with non-zero $B_{\rm NS}$, that the BPS mass
formula becomes non-linear in the masses of the D1 and D5 branes and
that a binding energy is generated. The picture of the bound state
thus obtained is compared with gauge theory and conformal field theory
in Sections 4 and 7. Next, we describe the
appearance of AdS$_3$ and BTZ black holes as the ``near-horizon
limit'' of the D1-D5 string and the D1-D5 black hole, respectively;
this discussion is used as background material (together with Appendix
C) for the discussions of AdS/CFT in Sections 6 and 11.
In the context of the near-horizon physics we also discuss
a connection between the pure five-brane system and the
two-dimensional black hole.

In Section 3, we describe the semiclassical derivation of absorption
cross-section (also called graybody factor) and the rate of Hawking
radiation for the near-extremal five-dimensional black hole
constructed in Section 2. We describe various fluctuations around the
black hole solutions. We specifically focus on (a) minimal scalars
which couple only to the five-dimensional Einstein-frame metric, and
(b) fixed scalars which couple also to other background fields like
the Ramond-Ramond fluxes. We derive case (a) in more detail in which
we show that the computation of the graybody factor and the Hawking
rate essentially depends only on the near-horizon limit. Both cases
(a) and (b) are compared with the D1-D5 conformal field theory
calculations in Section 8 where we find exact agreement with the
results of Section 3. The agreement involves new first principles
calculations in the D-brane conformal field theory which involve
significant conceptual departures from the phenomenological approach
adopted in the early calculations (see
\cite{Dha-Man-Wad96,Das-Mat96,Mal-Str96} for the original papers and
\cite{Aharony:1999ti} for a review). We mention the
new points in detail in Section 8 (esp. Secs \ref{sec:blow-up},
\ref{chp3:sec-fixed}). The  results in Section 3 and Section 8 have
a priori different regions of validity; a comparison therefore involves
an extrapolation which is made possible by non-renormalization
theorems discussed in Section 9.

In Section 4, we discuss the gauge theory for the \dd system. This
system is closely related to the five-dimensional black hole
solution. The \dd system, as discussed in Section 2, consists of $Q_1$
D1 branes and $Q_5$ D5 branes wrapped on a $T^4$ in type IIB
theory. It is a black string in six dimensions. The five-dimensional
black hole solution is obtained by wrapping the black string on a
circle and introducing Kaluza-Klein momenta along this direction. For
the purpose of understanding Hawking radiation from the
five-dimensional black hole it is sufficient to study the low energy
effective theory of the \dd system. The low energy theory of the \dd
system is a $1+1$ dimensional supersymmetric gauge theory with gauge
group $U(Q_1) \times U(Q_5)$. It has 8 supersymmetries. The matter
content of this theory consists of hypermultiplets transforming as
bi-fundamentals of $U(Q_1) \times U(Q_5)$. We review in detail the
field content of this theory and its symmetries. The bound state of
the D1 and D5 branes is described by the Higgs branch of this
theory. We show the existence of this bound state when the
Fayet-Iliopoulos terms are non-zero. The Higgs branch of the \dd
system flows in the infrared to a certain ${\mathcal{N}} = (4,4)$
superconformal field theory (SCFT). A more detailed understanding of
this conformal field theory can be obtained by thinking of the bound
D1 branes as solitonic strings of the D5 brane theory. From this point
of view the target space of the SCFT is the moduli space of $Q_1$
instantons of a $U(Q_5)$ gauge theory on $T^4$. This moduli space is
known to be a resolution of the orbifold
$(\tilde{T})^{Q_1Q_5}/S(Q_1Q_5)$ which we denote by
${\mathcal{M}}$. $S(Q_1Q_5)$ stands for the symmetric permutation
group of $Q_1Q_5$ elements. The torus $\tilde{T}^4$ can be distinct
from the compactification torus $T^4$. We review the evidence for this
result. The mere fact that the low energy theory is a (super)conformal
field theory with a known central charge, is used in Section
\ref{sec:quick} to provide a short-hand derivation of the
Bekenstein-Hawking entropy and Hawking temperature.  The complete
derivation, including that of the graybody factor and Hawking rate, is
however possible only in Section 8 after we have discussed in detail
the precise SCFT in Section 5,the couplings to the bulk fields in
Section 6  and the location of the SCFT used in the moduli space
of the ${\mathcal M} =(4,4)$ SCFT on ${\mathcal M}$. It turns out the
supergravity calculation and the SCFT are at different points in the
moduli space and the reason they agree is due to the
non-renormalization theorems discussed in section 9.

In Section 5 we formulate the SCFT on ${\mathcal M}$, we will discuss
in detail a realization of this orbifold SCFT as a free field theory
with identifications. The symmetries of this SCFT including a new
$SO(4)$ algebra will also be discussed. As the SCFT relevant for the
\dd system is a resolution of the SCFT on the orbifold ${\mathcal M}$,
we construct all the marginal operators of this SCFT including
operators which correspond to blowing up modes.  We classify the
marginal operators according to the short multiplets of the global
part of the ${\mathcal N} =(4,4)$ superalgebra and the new $SO(4)$
algebra. We then explicitly construct all the short multiplets of the
SCFT on ${\mathcal M}$ and classify them according to the global
subgroup of the ${\mathcal N} =(4,4)$ superconformal algebra.
These short multiplets are shown to be in one to one correspondence
with the spectrum of short-multiplets of supergravity in the near
horizon geometry of the \dd system in section 6. 

Now that we have discussed the microscopic degrees of freedom in
section 6 we take the next step towards the microscopic understanding
of Hawking radiation. We find the precise coupling of the fields of
the supergravity to the microscopic SCFT.  This is given by a specific
SCFT operator ${\mathcal O}(z, \bar{z})$, which couples to the
supergravity field $\phi$ in the form of an interaction.
\be
\label{introbbcoup}
S_{\rm int} = \mu \int d^2 z \phi (z, \bar{z}) {\mathcal O} (z, \bar z)
\ee
where $\mu$ is the strength of the coupling. As the ${\mathcal N} =
(4,4)$ SCFT on ${\mathcal M}$ is an ``effective'' theory of the \dd
system, it is difficult to fix the coupling of this theory to the
supergravity fields using {\em ab-initio} methods . We fix the
operator in the SCFT corresponding to the supergravity field using the
method of symmetries. The near horizon limit of the \dd system
exhibits enhanced symmetries. This is a special case of the AdS/CFT
correspondence. As we have seen in section 2 the near horizon geometry
of the \dd system reduces to that of $AdS_3\times S^3\times T^4$. The
AdS/CFT correspondence \cite{Mal97} states that string theory on
$AdS_3\times S^3\times T^4$ is dual to the $1+1$ dimensional SCFT
describing the Higgs branch of the \dd system on $T^4$. For large
$Q_1Q_5$ the radius of $S^3$ is large in string units. The
Kaluza-Klein modes on $T^4$ is of the order of string length. Thus
type IIB string theory on $AdS_3\times S^3\times T^4$ passes over to 6
dimensional $(2,2)$ supergravity on $AdS_3\times S^3$. We will work in
the supergravity limit. The evidence for this correspondence comes
from symmetries. The isometries of $AdS_3$ correspond to the global
part of the Virasoro group of the SCFT. The R-symmetry of the SCFT is
identified with the isometry of the $S^3$.  The number of
supersymmetries of the bulk get enhanced to 16 from 8.  These
correspond to the global supersymmetries of the ${\mathcal N} =(4,4)$
superalgebra of the SCFT. Thus the global part of the superalgebra of
the SCFT is identified with the $AdS_3\times S^3$ supergroup,
$SU(1,1|2)\times SU(1,1|2)$. Therefore a viable strategy to fix the
coupling is to classify both the bulk fields and the SCFT operators
according to the symmetries. The question then would be if this
procedure can fix the SCFT operator required for analysing Hawking
radiation. We will review the classification of the entire set of
Kaluza-Klein modes of the six-dimensional supergravity on $S^3$ as
short multiplets of $SU(1,1|2) \times SU(1,1|2)$
\cite{chp1:Deboer}. We use symmetries, including the new global
$SO(4)$ algebra, to identify the marginal operators constructed in
section 5 with their corresponding supergravity fields. The enables us
to identify the operators corresponding to the minimal scalars. We
also identify the quantum numbers of the fixed scalars and the
intermediate scalars.

From section 5 and section 6 we see that there is a one to one
correspondence between the moduli of the ${\mathcal N}= (4,4)$ SCFT on
${\mathcal M}$ and the moduli of supergravity. In section 7 we address
the question where in this moduli space is the free field orbifold
SCFT, which we use in our calculations, with respect to the
supergravity solution. Let us first examine the supergravity side.  In
section 2 we constructed the supergravity solution with the self-dual
Neveu-Schwarz $B$ field turned on. The self dual NS B-field is a
moduli of the \dd background.  The mass formula for these solutions
was nonlinear and had a binding energy for the break up of the \dd
system into constituent D1 and D5 branes. This is unlike the case when
no moduli is turned on. The mass formula in that case is linear and
the system is marginally stable with respect to decay into constituent
branes.  We see in section 7 that the effective theory from
supergravity governing the decay of the \dd system is a Liouville
theory. This is derived by using probe branes near the boundary of
$AdS_3$. These branes are called long strings.  We then derive this
Liouville theory from the \dd gauge theory. As we have seen in section
4, absence of a bound state implies that the Fayet-Iliopoulos and
$\theta$ terms have to be set to zero . In section 7 we see that the
effective theory near the origin of the Higgs branch ({\em i.e.} when
these terms are small but non-zero) is the same Liouville theory as
seen in supergravity. This Liouville theory is strongly coupled and
singular at the origin of the Higgs branch. The free orbifold SCFT on
${\mathcal M}$, on the other hand, is regular and finite. Thus we
conclude that the free orbifold theory {\em must correspond}, (a) in
the gauge theory to a point where the Fayet-Iliopoulos and/or $\theta$
terms do not vanish, equivalently (b) in supergravity when some of the
moduli are turned on which lead to non-zero binding energy, as
mentioned above. More specifically we will see that the orbifold
theory corresponds to a point where the $\theta$ term in the gauge
theory of the \dd system is turned on. Thus the simple supergravity
solution of Sections 2.3 and 2.4 is at a different point in moduli
space compared to the SCFT. In section 9 we describe why calculations
done in the free orbifold SCFT can be valid at the point in moduli
space which corresponds to the simple supergravity solution.

In Section 8, we derive the thermodynamics of the five-dimensional
black hole from the microscopic viewpoint (SCFT).  We use the free
field orbifold ${\mathcal N}=(4,4)$ SCFT on ${\mathcal M}$ as the
microscopic theory.  The \dd black hole is identified as an excited
state with definite left and right conformal weights over the Ramond
sector of this theory. Using Cardy's formula we get the asymptotic
state counting for a conformal field theory in terms of its central
charge and the level number. Use of the Cardy formula in the SCFT for
the black hole state shows that the entropy evaluated from SCFT
precisely matches with the entropy of \dd black hole. We then discuss
Hawking radiation of minimal scalars from the \dd black hole. In order
to do this we have to fix the strength of the coupling $\mu$ of the
minimal scalar to its boundary operator in \eq{introbbcoup}.  We
determine this by matching the two point functions evaluated in the
SCFT and supergravity. It is only here we use the more quantitative
version of the AdS/CFT correspondence.  Once the coupling, and hence
$S_{int}$ is determined, Hawking radiation/absorption cross-section
from the microscopic SCFT can be derived as a purely quantum
scattering process.  We formulate the absorption cross-section
calculation from the SCFT as an evaluation of the thermal Green's
function of the operators ${\mathcal O}$ in \eq{introbbcoup}. The
absorption cross-section evaluated from SCFT agrees with the
semi-classical calculation evaluated in supergravity in section 3.  In
Sections \ref{sec:blow-up} and \ref{chp3:sec-fixed} we point out the
differences between the present SCFT results and those in the early
works (reviewed in \cite{Aharony:1999ti}) which, adopted a
phenomenological approach for the D-brane degrees of freedom as well
as the coupling to supergravity fields. The phenomenological method
had a discrepancy for 4 minimal scalars (which correspond to blow-up
modes in the SCFT) and fixed scalars from the semi-classical results
of Section 3. In fact using the phenomenological Dirac-Born-Infeld
action for the D-brane degrees of there is no accounting at all for
the minimal scalars corresponding to the blow-up modes.  We show that
the first principles calculation presented in Section 8 remove these
discrepancies. We also outline the absorption cross-section
calculation for the intermediate scalars.

Though the ${\mathcal N}= (4,4)$ SCFT and the near horizon moduli of
supergravity agree, we have seen in section 7 that the free field
orbifold theory is at a different point in moduli space compared to
the supergravity solution.  In section 9 we see why calculations done
using the free orbifold theory are valid and agree with that of
supergravity.  We show how the spectrum of short multiplets, the
entropy and the calculation of Hawking radiation are independent of
moduli both on the SCFT and supergravity side due to
non-renormalization theorems. In discussing non-renormalization
theorems for Hawking radiation we first examine the dilute gas
approximations and the low energy approximation made in the
semi-classical calculation in section 3 and show how they appear in
the SCFT.

In section 10 we go beyond the supergravity approximation in the study
of the near horizon geometry, $AdS_3\times S^3\times T^4$ and study
string theory in this curved background to explore the AdS/CFT
correspondence.  Since string theory in Ramond-Ramond backgrounds are
difficult to study it is convenient to study the near horizon geometry
of the S-dual to the \dd system, the NS1-NS5 brane system. We review
the spectrum of strings in this background and show that the long
strings found in section 7 play an important role in the completion of
the spectrum. We then review the formulation of string theory on
Euclidean $AdS_3$ and derive the world sheet algebra of the long
string. It is seen, as expected from S-duality, that the world sheet
theory is a lioville theory which coincides with that of the single
long D1-string near near the boundary of $AdS_3$ in section 7. We then
review the elvaluation of the partition function of a gas of strings
in thermal $AdS_3$. From this partition function it can be verified
that long strings are present in the spectrum as expected from the
analysis of the spectrum with the Lorentzian signature metric.

In Section 11, we discuss some applications of AdS/CFT correspondence
(introduced in Section 6). In the first part we discuss the thermal
phase transition (Hawking-Page) in AdS$_3$ and how to understand it in
terms of the dual CFT picture. We show that the Euclidean partition of
asymptotically \ads\ spaces can be evaluated in the leading
semiclassical approximation as a sum over an SL(2,Z) family of
saddle-point configurations, of which two members are \ads and the BTZ
black hole. We discuss the issue of boundary conditions of this
partition function (see also Appendix C) and relate it to the boundary
conditions of the CFT partition function.  We calculate the BTZ
partition function in \ads\ supergravity as well as in the boundary
SCFT and show that they agree. We show that the supergravity partition
function is dominated at low energies by \ads, and high energies by
the Euclidean BTZ black hole; from the CFT viewpoint this phenomenon
gets related to the fact that at low temperatures the NS sector
dominates and high temperatures the Ramond sector dominates. The
modular invariance of the boundary SCFT points gets related to a
similar modular invariance in the three-dimensional supergravity. We
next describe some new spaces which are also asymptotically AdS$_3$;
these are conical spaces which are created by static point
particles. We discuss various ways of understanding them as solutions
of string theory, the simplest being the case of $Z_N$ cones which are
described as orbifolds. We briefly discuss the CFT duals.  We end the
discussion with conical spaces created by moving point particles and
describe the scenario in which two such particles with sufficient
energy to form a BTZ black hole.  The dual CFT is left as an open
problem.

In Section 12 we conclude with a discussion of some open problems.

\newpage

\section{Construction of  classical solutions}

The aim of this section will be to construct the classical solution
representing the five-dimensional black hole in \cite{Cal-Mal96}.
Rather than presenting the solution and showing that it solves the low
energy equations of type II superstring, we will describe some aspects
of the art of solution-building.  There are many excellent reviews of
this area (see, for example, \cite{Ste98,Obe-Pio98,Gau97,Mal96,
Cve-review,Youm:1997hw, David:1999dx,Mandal:2000rp, Wadia:2000wy}, 
other general reviews on black holes in
string/M theories include \cite{Peet,Skenderis,Duff}), so we shall be
brief.  The method of construction of various classical solutions, we
will see, will throw light on the microscopic configurations
corresponding to these solutions.

Two widely used methods for construction of classical
solutions are

\noindent (a) the method of harmonic superposition\nl
(b) $O(d,d)$ transformations
 
\noindent
We will mainly concentrate on the first one below. For 
a more detailed account including the second method,
see, e.g., \cite{Youm:1997hw}.

As is well-known by now, classical solutions of type II string
theories can be obtained from those of M-theory
\cite{Tow-Mtheory,Wit-Mtheory} through suitable compactification and
dualities. We will accordingly start with classical solutions of
M-theory, or alternatively, of 11-dimensional supergravity. 

We should note two important points:  

\noindent (a)For classical supergravity description of these solutions to be
valid, we need the curvature to be small (in the scale of the
11-dimensional Planck length $l_{11}$ for solutions of M-theory, or of
the string length $l_s$ for string theories)\nl (b) Since various
superstring theories are defined (through perturbation theory) only in
the (respective) weakly coupled regimes, in order to meaningfully talk
about classical solutions of various string theories we need the
string coupling also to be small.

For the RR charged type II solutions (charge Q) that we will describe
below, both the above conditions can be met if $Q \gg 1/\gst \gg 1$
(that is, $\gst Q \gg 1, \gst \ll 1$).

\subsection{ Classical solutions of M-theory}

The massless modes of M-theory are those of 11-dimensional
supergravity: the metric $G_{MN}$, the gravitino $\psi_M$ and a
three-form $A^{(3)}_{MNP}, M=0,1,\ldots,10$.

The (bosonic part of the) classical action is
\be
\label{2.1}
S_{11} = \frac{1}{2 \kappa_{11}^2}\int d^{11}x
[ \sqrt{-G} (R - {1\over 48} (dA^{(3)})^2) - 
\frac{1}{6} A^{(3)} \wedge dA^{(3)} \wedge dA^{(3)}] 
\ee
There are two basic classical solutions of this Lagrangian, the M2
and M5 branes, whose intersections account for 
most  stable supersymmetric solutions of M-theory
\cite{Pap-Tow96,Tse96,Gau97}.

\subsubsection{The 2-brane solution of M-theory: M2}

It will suffice for us to describe only the M2-brane solution
\cite{M2-ref} in some detail.

\noindent 
Statement of the problem: we want to find (a) a relativistic 2-brane
solution of \eq{2.1} (say stretching along $x^{1,2}$) with (b) 
some number of unbroken supersymmetries.

Condition (a) implies that the solution must have a $SO(2,1)_{0,1,2}
\times SO(8)_{3,4,\ldots,10}$ symmetry, together with
translational symmetries along $x^{0,1,2}$
\footnote{The subscripts denote
which directions are acted on by the $SO$ groups. We denote
spacetime coordinates by $x^M, M=0,1,\ldots,10$.}.
This uniquely leads to
\bea
\label{2.2}
ds_{11}^2 &=& e^{2A_1(r)}  dx^\mu dx_\mu + e^{2A_2(r)} dx^m dx_m
\nn\\
A^{(3)} &=& e^{A_3(r)} dt\wedge dx^1 \wedge dx^2 
\eea
where $\mu=0,1,2$ denote directions parallel to the world-volume
and $m=3,\ldots,9,10$ denote the transverse directions. $r^2
\equiv x^mx_m$.
 
Condition (b) implies that there should exist a non-empty set of
supersymmetry transformations $\epsilon$ preserving the solution
\eq{2.2}; in particular the gravitino variation
\bea
\label{2.3}
\delta_\epsilon \psi_M &=& D_M \epsilon 
+ \frac{1}{288} (\Gamma_M^{NPQR} - 8 \delta_M^N \Gamma^{PQR})
F_{NPQR} \epsilon =0,\; \hbox{where}\nn\\
D_M \epsilon &\equiv & 
\left(\del_M + {1\over4} \omega^{BC}_M\Gamma_{BC} \right)\epsilon
\eea
must vanish for some $\epsilon$'s.

It is straightforward to see that Eqn. \eq{2.3} vanishes for $M=\mu$
(world volume directions) if
\bea
\label{2.4}
\del_\mu \epsilon &=& 0, \nn\\
A_3 &=& 3 A_1 \nn\\ 
{\rm and}\;\Gamma^{\c{0}\c{1}\c{2}}\epsilon &=& \epsilon
\eea
where the caret ~~$\c{}$~~ denotes local Lorenz indices. 
(Flipping the sign of $A^{(3)}$ would correspond to $-\epsilon$ on the
right hand side of the last line of \eq{2.4}: this would correspond
to an anti-brane solution in our convention.)

The $M=m$ (transverse) components of \eq{2.3} give rise to 
the further conditions
\bea
\label{2.6}
A_1 &=& - 2 A_2 \nn\\
\epsilon &=& e^{A_3/6} \epsilon_0
\eea

\mysec02{Harmonic equation}

The equations \eq{2.4} and \eq{2.6} fix the
three functions $A_i$ in \eq{2.2} in terms of
just one function, say $A_3$. It is easy to
determine it by looking at the equation of motion
of the three-form potential:
\be
\label{2.7}
\del_M(\sqrt{-g} F^{MNPQ}) + {1\over 2.(4!)^2}
\epsilon^{NPQABCDEFGH}F_{ABCD} F_{EFGH}=0
\ee
The second term is clearly zero for our ansatz
\eq{2.2} for $A^{(3)}$. The first term, evaluated for
$(P,Q,R)=(0,1,2)$ gives 
to
\be
\del_m\del_m(e^{-A_3})=0
\ee
Thus, the full M2 solution is given by
\bea
\label{2.8}
ds_{11}^2 &=& H^{1/3}[ H^{-1}  dx^\mu dx_\mu +  dx^m dx_m]
\nn\\
A^{(3)} &=&   H^{-1} dt\wedge dx^1 \wedge dx^2
\eea
where $H=H(r)$ satisfies the harmonic equation
in the transverse coordinates
\be
\label{2.9}
\del_m \del_m H = 0
\ee
The simplest solution for $H$,  in an  asymptotically flat space,
is given by 
\be
\label{harmonic-function}
H= 1 + k/r^6
\ee 
Clearly, {\em multi-centred solutions} are also allowed:
\be
\label{multi-centred}
H= 1 + \sum_i {k_i\over | \vec{x} - \vec{x_i}|^6}
\ee
where $\vec x$ denotes the transverse directions $x^m$.

We note that, the constant, $1$, in \eq{harmonic-function} is
essentially an integration constant. Clearly, it can also be zero;
such choices have led to M/string theory solutions involving AdS
spaces.  The point of this remark is to emphasize that the
near-horizon geometry ($r\to 0$), important in the context of AdS/CFT
correspondence \cite{Mal97,Wit98-ads,Gub-Kle-Pol98}, in which $H=
k/r^6$, corresponds to a complete solution in its own right without
the appendage of the asymptotically flat regions. We will return
to the AdS/CFT correspondence many times in this as well as later
Sections.

\mysec11{ADM mass}

The integration constant $k$ 
in \eq{harmonic-function} affects the asymptotic fall-off of the
metric as well as of the field strength, and is related
to the ADM mass (per unit area of the 2-brane) $M$ and to the 
gauge charge (per unit area) $q$.
Using the definitions\footnote{We follow
the normalizations in \cite{Ste98} which 
differ from, e.g. \cite{Mal96}.}
\bea
\label{mass}
M &=& \int_{S^7} d^7\Sigma^m (\del^n h_{mn} - \del_m h),
\\
\label{charge}
q &=& \int_{S^7} d^7\Sigma^m F_{m012},
\\
\label{mass-charge}
\!\!\!\!\!\!\!\!\!\! \hbox{we get~~} 
M &=& 6k \Omega_7 = q
\eea
Here $S^7$ represents the sphere at $r^2=x^mx_m=\infty$,%
\footnote{The total
ADM mass, which diverges, includes 
integrals over $x^{1,2}$ as well; we ignore
them here since we are interested in the mass per unit area.
Similar remarks apply to the charge.} 
$h_{MN} \equiv g_{MN} - \eta_{MN} $, $h \equiv \sum_{M=1}^{10} 
h_{MM}$, and $\Omega_n \equiv 2 \pi^{(n+1)/2}/\Gamma({n+1\over 2})$
is the volume of the unit sphere $S^n$.

\mysec11{BPS nature}

The mass-charge equality in the last equation \eq{mass-charge} is
characteristic of a ``BPS solution''. We provide a very brief 
introduction below. The 11-dimensional supersymmetry algebra 
\cite{Cre-Jul-Sch78} is
\be
\label{susy-algebra}
\{Q, Q\} = C(\Gamma^M P_M + \Gamma^{MN} U_{MN} +
\Gamma^{MNPQR}V_{MNPQR}),
\ee
where $C$ is the charge conjugation matrix and $P,U$ and $V$
are various central terms. When \eq{susy-algebra}
is  evaluated \cite{Azc-Gau-Izq-Tow89} for the above M2 solution, we get 
\be
\label{s-a-m2}
\frac{1}{V_2} \{Q_\alpha, Q_\beta \}= (C \Gamma^\c0)_{\alpha\beta}
M +  (C\Gamma^{\c1\c2})_{\alpha\beta} Q,  
\ee
where we have used the notation
\be 
P_\c0 = V_2 M, \quad  U_{\c1\c2} = V_2 q,
\ee
$V_2$ being the spatial volume of the 2-brane (assumed
compactified on a large $T^2$). 

Now, the positivity of the $Q^2$ operator implies that $ M \ge q $
where the inequality is saturated when the right hand side of
\eq{s-a-m2} has a zero eigenvector. For our solution \eq{2.8}, we see
from \eq{2.4} that the unbroken supersymmetry transformation parameter
satisfies
\be
\label{unbroken-susy}
(1 - \Gamma^{\c0\c1\c2})\epsilon =0
\ee
This clearly leads to $M=q$.  This is a typical example of how
classical solutions with (partially) unbroken supersymmetries satisfy
the extremality condition mass= charge.

We note here that the remaining half of the supersymmetry
transformations, the complement of the ones in \eq{unbroken-susy}, are
non-linearly realized in the M2 geometry and can be regarded as
spontaneously broken supersymmetries. Interestingly, the supersymmetry
variations under these transformations vanish in the near-horizon
limit which has the geometry \cite{Gib-Tow93} $ AdS_4 \times S^7. $ As
a result the broken supersymmetry transformations reemerge as
unbroken, leading to an enhancement of the number of supersymmetry
charges $16\to 32$ in the near-horizon limit, a fact that plays a
crucial role in the AdS/CFT correspondence.

\mysec11{Identification as a ``Black brane''} 

The M2-brane itself has a black hole geometry.  If we compactify the
directions $1,2$ on a 2-torus, we have a \underbar{black hole}
solution in the remaining nine extended dimensions.  The compactified
solution is constructed by placing the multiple centres $\vec{x_i}$ in
\eq{multi-centred} at the sites of a lattice defining the 2-torus. The
horizon is situated at $r=0$.  The detailed geometry has been
discussed in many places, {\it e.g} in \cite{Ste98}. Since our main
object of interest is the five-dimensional black hole, and we will use
the M2-brane as essentially a building block for that solution, we
defer the geometrical discussion till we discuss the latter.

Without compactification too, the above solution is ``black'', but it
has extensions in 1,2 directions and is called a black 2-brane.

\subsubsection{Intersecting M2-branes}

We will now use the above solution as a building block to
construct more complicated solutions corresponding to intersecting
branes.

\mysec11{M2 $\perp$ M2}

We consider first two orthogonal M2 branes, along $x^{1,2}$
and $x^{3,4}$ respectively. The geometry
of the solution corresponds to 
a spacetime symmetry consisting
of rotations  $SO(2)_{1,2}\times SO(2)_{3,4} \times
SO(6)_{5,6,7,8,9,10}$ plus Killing vectors $(\del_t,
\del_1,\ldots, \del_4)$. This leads to
\bea
\label{m2-m2-ansatz}
ds_{11}^2 &=& e^{2A_1} (-dt^2) + e^{2A_2} (dx_1^2 + dx_2^2)
+ e^{2A_3} (dx_3^2 + dx_4^2) + e^{2A_4} dx_i dx_i \nonumber\\
A^{(3)} &=& e^{A_5} dt\wedge dx^1 \wedge dx^2 + 
e^{A_6} dt\wedge dx^3 \wedge dx^4
\eea

Now, as before, the desire to have a BPS solution leads to
existence of unbroken supersymmetry, or
$\delta_\epsilon \psi_M=0$. This now yields \underbar{four} 
different type of equations, depending on whether
the index $M$ is $0, \{1,2\},\{3,4\}$ or the rest. 
These express the six functions above
in terms of two independent functions $H_1, H_2$.
These functions turn out to harmonic  in the
common transverse directions when one imposes
closure of SUSY algebra or equation of motion. The solution
ultimately is
\bea
\label{m2-m2-sol}
ds_{11}^2 &=& (H_1 H_2)^{1/3}
[ -{dt^2\over H_1 H_2 } + 
{dx_1^2 + dx_2^2\over H_1} + {dx_3^2 + dx_4^2\over H_2} 
+ dx_i dx_i]
\nn\\
A^{(3)} &=& {1\over H_1} dt\wedge dx^1 \wedge dx^2 +  
{1\over H_2} dt\wedge dx^3 \wedge dx^4
\eea
The above is an example of ``harmonic superposition of branes''.
(see, e.g., \cite{Tse96}).

\gap{1}

\noindent\underbar{\it Delocalized nature of the solution}

\gap{1}

We note that the ansatz above \eq{m2-m2-ansatz}, as well as the
solution \eq{m2-m2-sol}, represent a ``delocalized solution''.  A
localized $M2 \perp M2$ intersection would destroy translational
symmetries along the spatial world-sheet of both the 2-branes. The
subject of localized intersection is interesting in its own right
(see, e.g. \cite{Sur-Mar98} which is especially relevant to the \dd
system), although we do not have space to discuss them here. The
delocalization here involves ``smearing'' the first M2 solution along
the directions $x^3, x^4$ (by using a continuous superposition in
\eq{multi-centred}, see e.g. \cite{Gau97}), and ``smearing'' the
second M2 solution along $x^1, x^2$.

\mysec22{M2 $\perp$ M2 $\perp$ M2}

Extending the above method, we get the following supergravity solution
for three orthogonal M2-branes, extending respectively along
$x^{1,2}, x^{3,4}$ and $x^{5,6}$:
\bea
\label{triple-m2}
ds_{11}^2 &=& (H_1H_2H_3)^{1/3}
[(H_1 H_2 H_3)^{-1} (-dt^2) + 
H_1^{-1} (dx_1^2 + dx_2^2) 
\nn\\
&+& H_2^{-1}  (dx_3^2 + dx_4^2) +
H_3^{-1}  (dx_5^2 + dx_6^2)  +  dx_i dx_i]
\nn\\
A^{(3)} &=& H_1^{-1} dt\wedge dx^1 \wedge dx^2 
+  
H_2^{-1} dt\wedge dx^3 \wedge dx^4
\nn\\
&~~& +
H_3^{-1} dt\wedge dx^5 \wedge dx^6
\eea
\subsection{The 6D black string solution of IIB on $T^4$
\label{sec:6d-blackstring}}

In the following we will construct solutions of type II string
theories using the above M-theory solutions by using various duality
relations which we will describe as we go along. For an early account
of black p-brane solutions in string theory, see \cite{Hor-Str91}.

We apply the transformation $ T_{567} R_{10}$   to
the $M2 \perp M2$ solution \eq{m2-m2-sol}:

\gap1

\begin{tabular}{l  c  l c l  }
M-theory  & ${\buildrel R_{10} \over \rightarrow}$
        & IIA & ${\buildrel T_{567} \over \rightarrow}$
        & IIB               \\
        &           &           &         &                      \\
M2 (8,9)   &        & D2 ( 8,9) &         & D5 ( 5,6,7,8,9)      \\
M2 (6,7)   &        & D2  (6,7) &         & D1 ( 5)              \\
\end{tabular}

\gap1

The first transformation $R_{10}$ denotes the reduction from M-theory
to type IIA. To do this, one first needs to compactify the $M2 \perp
M2$ solution along $x^{10}$ (by using the multi-centred harmonic
functions, with centres separated by a distance $2\pi R_{10}$ along
$x^{10}$). Essentially, at transverse distances large compared to
$R_{10}$, this amounts to replacement of the harmonic function $1/r^4$
by $1/r^3$ and a suitable modification of the integration constant to
reflect the appropriate quantization conditions.  At this stage, one
still has 11-dimensional fields. To get to IIA fields, we use the
reduction formula
\bea
ds_{11}^2 &=& \exp[-2\phi/3] ds_{10}^2 + \exp[4\phi/3] (dx^{10}
        + C^{(1)}_\mu dx^\mu )^2
\nn\\
A &=& B \wedge dx^{10} + C^{(3)}
\eea
It is instructive to verify at this stage that the classical D2
solutions do come out of the M2-brane after these transformations. We
use the notation $C^{(n)}$ for the $n$-Form Ramond-Ramond (RR)
potentials in type II theories.

The second transformation $T_{567}$ involves a sequence of T-dualities
(for a recent account of T-duality transformations involving RR
fields, see \cite{Has99}). We denote by $T_m$ T-duality along the
direction $x^m$.  $T_{567}$ denotes $T_5 T_6 T_7$.

The final transformation, not explicitly written in the above table,
is to wrap $x^{6,7,8,9}$ on $T^4$. We will
denote the volume of the $T^4$ by
\be
V_{T^4} \equiv \alpha^{\prime 2} (2\pi)^4 \tilde v
\label{vol-t4}
\ee
Assuming the number of the two orthogonal sets of M2-branes to be
$Q_5, Q_1$ respectively, the final result is: $Q_5$ strings from
wrapping D5 on $T^4$ and $Q_1$ D-strings. This is the
\dd system in IIB supergravity
\footnote{Strictly speaking, we should wrap the D1-D5 string
on a large circle to avoid the Gregory-Laflamme
instability \cite{Gregory:1993vy}.}, characterized by the
following solution:
\bea
\label{6d-black-string}
ds_{10}^2 &=& f_1^{-1/2}f_5^{-1/2} (-dt^2 + dx_5^2) 
 +  f_1^{1/2}f_5^{1/2} dx_i dx^i  +
                f_1^{1/2}f_5^{-1/2} dx_a dx^a
\nn\\
C^{(2)}_{05} &=& -\half (f_1^{-1} -1)  \nn\\
F^{(3)}_{ijk} &=& \epsilon_{ijkl} \del_l f_5, \; F^{(3)}= dC^{(2)} 
\nn\\
e^{-2\phi} &=& f_5 f_1^{-1}
\nn\\
f_{1,5} &=& (1 + r^2_{1,5}/r^2)  
\eea
Here $C^{(2)}$ is the 2-form RR gauge potential
of type IIB string theory. The parameters $r_1, r_5$ are
defined in terms of $Q_1, Q_5$, see Eqns. \eq{quantization}.

\gap{1}

\ni\underbar{\it Spacetime Symmetry}

\gap1

The spacetime symmetry ${\mathcal S}$ of the above 
solution is:
\be
\label{space-time-symmetry}
{\mathcal S}= SO(1,1) \times  SO(4)_E \times 
\hbox{$`SO(4)_I$'}
\ee
where $SO(1,1)$ refers to directions $0,5$, $SO(4)_E$
to directions $1,2,3,4$ ($E$ for external)  $`SO(4)_I$' to
directions $6,7,8,9$ ($I$ stands for internal; the
quotes signify that the symmetry is broken by wrapping
the directions on a four-torus although for low energies
compared to the inverse radii it remains a symmetry
of the supergravity solution).

\gap1

\ni\underbar{\it Supersymmetry}

\gap1

The unbroken supersymmetry can be read off either
by recalling those of the M-theory solution and following the dualities 
or by solving the Killing spinor equations (analogous to  \eq{2.3}). 
The result is: 
\bea 
\label{supersymmetry}
\Gamma^{056789} \epsilon_L &=& \epsilon_R
\nn\\
\Gamma^{05} \epsilon_L &=& \epsilon_R
\eea
The first line corresponds to the unbroken supersymmetry appropriate for
the D5-brane (extending in 5,6,7,8,9 directions). The second line
refers to the D1-brane.  
(The superscripts in $\Gamma^{ab..}$ denote local
Lorenz indices like in \eq{2.3}, although we have dropped 
the carets.)

To solve Eqn. \eq{supersymmetry}, we recast \eq{supersymmetry}
as 
\[
\Gamma^{6789}\epsilon_L = \epsilon_L,\;
\epsilon_R= \Gamma^{05} \epsilon_L
\]
Since $\epsilon_L$ has 16 independent real
components to begin with, the first equation
cuts it down by a half, thus leaving 8 real
components. These, by virtue of the
second equation, completely determine
$\epsilon_R$, thus leaving no further degrees
of freedom.

Thus, there are eight unbroken real supersymmetries. Recall that type
II string theory has 32 supersymmetries (i.e., $\epsilon_L,
\epsilon_R$ each has 16 independent real components). In the
near-horizon limit, eight of the broken supersymmetries
reemerge as unbroken.

\subsection{The extremal 5D black hole solution}\label{extreme}

Let us now compactify $x^5$ along a circle of radius $R_5$ and wrap
the above solution along $x^5$ to get a spherically
symmetric object in five dimensions.
Let us also ``add'' gravitational waves (denoted $W$) moving to the
``left'' along $x^5$. This gives us the BPS version
\cite{Str-Vaf96,Cal-Mal96} of the five-dimensional black hole. Adding
such a wave can be achieved either (a) by applying the
Garfinkle-Vachaspati transformation \cite{Gar-Vac92} to the black
string solution \eq{6d-black-string} and wrapping it on $S^1$, or (b)
augmenting the $M2\perp M2$ solution \eq{m2-m2-sol} by a third,
transverse, set of M2-branes along $x^{5,10}$ (cf. \eq{triple-m2}),
and passing it through the same sequence of transformations
$T_{567}R_{10}$ as before (see the Table in Section 
\ref{sec:6d-blackstring}), with
the result that the third set of M2 branes becomes a gravitational
wave (= momentum mode) along the $x^5$-circle:
\gap1

\begin{tabular}{l  c  l c l  }
M-theory  & ${\buildrel R_{10} \over \rightarrow}$
        & IIA & ${\buildrel T_{567} \over \rightarrow}$
        & IIB               \\
        &           &           &         &                      \\
M2 (8,9)   &        & D2 ( 8,9) &         & D5 ( 5,6,7,8,9)      \\
M2 (6,7)   &        & D2  (6,7) &         & D1 ( 5)              \\
M2 ( 5,10) &        & NS1 ( 5 ) &         & W  ( 5)   \\
\end{tabular}\label{duality-table-big}

\gap1

\noindent 
The last transformation essentially reflects the fact that T-duality
changes winding modes to momentum modes. ($W$ denotes a gravitational
\underbar{w}ave and not a \underbar{w}inding mode.)

The final configuration corresponds to D5 branes along $x^{5,6,7,8,9}$
and D1 branes along $x^5$, with a non-zero amount of (left-moving)
momentum. If the number of the three sets of M2 branes are $Q_1,Q_5$
and $N$ respectively, then these will correspond to the numbers of
D1-, D5-branes and the quantized left-moving momentum respectively.

The solution for the extremal five-dimensional \dd\ black hole
is thus given by
\bea
\label{extremal-5d}
ds_{10}^2 &=& f_1^{-1/2}f_5^{-1/2} (-du dv + (f_n-1) du^2) 
\nn\\
        &~~& +  f_1^{1/2}f_5^{1/2} dx_i dx^i  +
                f_1^{1/2}f_5^{-1/2} dx_a dx^a
\nn\\
C^{(2)}_{05} &=& -\half (f_1^{-1} -1)  \nn\\
F^{(3)}_{ijk} &=& \epsilon_{ijkl} \del_l f_5 \nn\\
e^{-2\phi} &=& f_5 f_1^{-1}
\nn\\
f_{1,5,n} &=& (1 + (\frac{r_{1,5,n}}{r})^2)  
\eea
The parameters $r_1, r_5, r_n$ are defined in terms
of $Q_1, Q_5, N$ respectively, see Eqn. 
\eq{quantization}.

\gap1

\ni\underbar{\it Symmetries}

\gap1

Curling up $x^5$ and adding momentum along it reduces the spacetime
symmetry and supersymmetry of the solution \eq{extremal-5d}, compared
to \eq{space-time-symmetry},\eq{supersymmetry}. Thus the spacetime
symmetry is $SO(4)_E \times \hbox{`{SO(4)$_I$}'}$ while the number of
supersymmetries is reduced to four due to an additional condition on
the Killing spinor: $\Gamma^{05} \epsilon_{L,R} = \epsilon_{L,R}$.
 
\gap1

\ni\underbar{\it Charge quantization}

\gap1

The parameters $r_{1,5,n}^2$ in
\eq{extremal-5d} are related to the
integer-quantized charges $Q_{1,5}$ and momentum $N$ by
\bea 
\label{quantization}
r_1^2 &=&  c_1 Q_1, \; c_1 = 
{4G^5_N R_5\over \pi \alpha' g_s}= \frac{g_s \alpha'}{\tilde v}
\nn\\
r_5^2 &=&  c_5 Q_5, \; c_5 = g_s \alpha'
\nn\\
r_n^2 &=&  c_n N, \; c_n = {4G_N^5 \over \pi R_5}=\frac{g_s^2
\alpha^{\prime2}}{\tilde v R_5^2}
\eea
where 
\be
G_N^5 = \frac{G_N^{10}}{ (2\pi R_5 V_{T^4})}=\frac{\pi g_s^2
\alpha^{\prime2}}{4 \tilde v R_5^2}
\label{gn-5}
\ee
In the above we have used \eq{gn-ten}, \eq{vol-t4}.
For a detailed discussion of quantization conditions like
\eq{quantization}, see, e.g. \cite{Bre-Lu-Pop-Ste97,Mal96,Ste98}. Here
$G_N^d$ denotes the $d$-dimensional Newton's constant. $V_{T^4}$ is
the volume of the four-torus in the directions $x^{6,7,8,9}$, while
$R_5$ is the radius of the circle along $x^5$.

\gap1

\ni\underbar{\it Explicit five-dimensional form}

The background \eq{extremal-5d} is written in a ten-dimensional
form. It is easy to derive the five-dimensional metric and other
fields. For the five-dimensional Einstein metric we use the reduction
formula 
\be
ds_{10}^2 = e^{2\chi} dx_a dx^a + 
e^{2\psi}(dx_5 + A_\mu dx^\mu)^2
+ e^{-(8\chi + 2\psi+\phi)/3} ds_5^2
\label{maharana}
\ee
(the first two exponential factors are simply the definitions of the
scalars $\chi, \psi$; the factor in front of $ds_5^2$ can be found
easily by demanding that $ds_5^2$ is the five-dimensional Einstein
metric). Here $\mu=1,2,3,4$.

Using \eq{maharana}, the five-dimensional Einstein metric is given by
\bea
\label{extremal-5d-red}
ds_5^2 &=& - f^{-2/3}(r) dt^2 + f^{1/3}(r)(dr^2 + 
r^2 d\Omega_3^2)
\nn\\ 
f(r) &=&  f_1(r) f_5(r) f_n(r) 
\eea
where $f_{1,5,n}(r)$ are defined in \eq{extremal-5d}.

\gap1
\ni\underbar{\it Area and Entropy}
\gap1

The above metric has a horizon at $r=0$, which has a finite
radius $R_h$ and a finite area $A_h$, given by
\be
A_h = 2 \pi^2 R_h^3, \, 
R_h = \left( r_1 r_5 r_n \right)^{1/3}  
\label{extremal-area}
\ee
The Bekenstein-Hawking entropy \eq{bek-hawk},  is given by
\be
S= 2\pi^2 \frac{\sqrt{c_1c_5c_n}}{4 G_N^5}\sqrt{Q_1 Q_5 N}
= 2\pi\sqrt{Q_1 Q_5 N}
\label{extremal-entropy}
\ee
where in the second step we have used 
\be
\label{2:curious}
\sqrt{c_1c_5c_n}= \frac{4 G_N^5}{\pi}
\ee
which follows easily from \eq{quantization}. The fact
that all ``moduli'' like the coupling and radii disappear from 
\eq{extremal-entropy},  and that the
entropy is ultimately given only in terms of quantized charges, is
remarkable.

We defer a  discussion of the geometry of this solution
till the next section \ref{non-extreme} where we describe the
non-extremal version.

\subsection{Non-extremal five-dimensional black hole}\label{non-extreme}

We have explained above how to construct from first principles the BPS
(hence extremal) version of the 5D black hole solution. We will now
present an algorithm (without proof and specialized to intersections
of M2) of how to generalize these constructions to their non-extremal
(nonsupersymmetric) versions \cite{Cve-Tse96}. A heuristic
motivation  for this algorithm is presented in 
appendix B.

\gap1

\noindent\underbar{Rule 1}: 
In the  transverse part of the metric (including time) make
the following substitution:
\bea 
\label{rule1}
dt^2 &\to& h(r) dt^2,\quad  dx^i dx_i \to h^{-1}(r) dr^2 + r^2 
d\Omega_{d-1}^2
\nn\\ 
h(r) &=& 1 - \mu/r^{d-2}\nn
\eea
with the harmonic function now defined as
\be 
H(r) = 1 + \tilde Q/r^{d-2} 
\ee
where $\tilde Q$ is a combination
of the non-extremality parameter $\mu$ and some ``boost''
angle $\delta$:  
\be 
\label{boost}
\tilde Q = \mu \sinh^2 \delta
\ee
(for multicentred solutions, $ \tilde Q_i = \mu \sinh^2 \delta_i$.)

\gap1

\noindent\underbar{Rule 2}: 

In the expression for $F_4= dA$, make the substitution
\bea
\label{rule-2}
H &\to& \tilde H(r)= 1+ \frac{\bar Q}{r^{d-2} + \tilde Q
- \bar Q} = \left( 
1- \frac{\bar Q}{r^{d-2}}H^{-1}
\right)^{-1},
\nn\\ 
\bar Q &=& \mu \sinh \delta \cosh \delta 
\eea

\gap1

\noindent Applying this rule to the $M2\perp M2 \perp M2$ case
\eq{triple-m2}, we get
\bea
\label{non-extremal-triple-m2}
ds_{11}^2 &=& (H_1 H_2 H_3)^{-1/3} [ - H_1 H_2 H_3 h dt^2 +
H_1 (dy_1^2 + dy_2^2)  
\nn\\
&+& H_2 (dy_3^2 + dy_4^2) +  H_3
 (dy_5^2 + dy_6^2) + h^{-1} dr^2 
\nn\\ 
&+& r^2 d\Omega_{d-1}^2]
\eea
The rest of the story is similar to the BPS case described
in the previous subsection. Namely, we apply the duality
transformation $T_{567}R_{10}$ as in the Table
in Section \ref{extreme}: by first reducing the
M-theory solution \eq{non-extremal-triple-m2} to IIA
and then T-dualizing to IIB, and finally wrapping it on
$T^4 \times S^1$.

Under the reduction from M-theory to type IIA in ten dimensions,
we get
\bea
\label{non-extreme-2a}
e^{-2\phi} &=& f_1 f_5^{-1}
\nn\\
ds_{10}^2 &=& f_1^{-1/2} f_5^{-1/2}[-dt^2 + dx_5^2 
\nn\\
&+& (1-h)(\cosh \alpha_n dt + \sinh \alpha_n dx_5)^2]
\nn\\
&+&  f_1^{1/2} f_5^{1/2}
(\frac{dr^2}{h} + r^2 d\Omega_3^2) +  F_1^{1/2}F_5^{-1/2} dx_a dx^a 
\nn\\ 
h &=& 1 - r_0^2/r^2
\eea
where $a=6,..,9$, $(r, \Omega_3)$ are polar coordinates for
$x^{1,2,3,4}$. $f_1,f_5$ (also $f_n$ in
\eq{non-extremal-5d}) are defined as in
\eq{extremal-5d}, except that the
parameters $r_1,r_5$ (also $r_n$) are no more defined by
\eq{quantization}, but by their
non-extremal counterparts \eq{non-extremal-5d},\eq{charges}. The
parameter $r_0^2$ is the same as the non-extremality parameter $\mu$
of \eq{rule1}, while $\alpha_{1,5,n}$ are related to the boost angle
of \eq{boost}.

This is still a IIA solution. In order to get the IIB version, we have
to apply the sequence $T_{567}$. We omit the details here which are
fairly straightforward. At the end, after we further use the
Kaluza-Klein reduction \eq{maharana} we get the following
five-dimensional Einstein metric \cite{Hor-Mal-Str96}:
\bea
\label{non-extremal-5d}
ds_5^2 &=& - h f^{-2/3} dt^2 + f^{1/3}(\frac{dr^2}{h} + 
r^2 d\Omega_3^2)
\nn\\
f &=& f_1 f_5 f_n = (1 + r_1^2/r^2)(1 + r_5^2/r^2)(1 + r_n^2/r^2) 
\nn\\
r_{1,5,n}^2 &=& r_0^2 \sinh^2 \alpha_{1,5,n}
\eea
There are six independent parameters of the metric: $\alpha_{1,5,n}$,
$r_0, R_5,\tilde v \equiv V_{T^4}/(2\pi l_s)^4$ ($l_s =
\sqrt{\alpha'}$). The boost angles and the non-extremality parameters
are related to the three charges and the mass $M$ as follows:
($F^{(3)}\equiv dC^{(2)}$)
\bea
\label{charges}
Q_1 &=&\frac{V}{4\pi^2g}\int e^{\phi} *F^{(3)} =  
\frac{r_0^2\sinh 2\alpha_1}{2 c_1}
\nn\\
Q_5 &=&\frac{1}{4\pi^2 \gst}\int F^{(3)}= 
\frac{r_0^2\sinh 2\alpha_5}{2 c_5}
\nn\\
N  &=& \frac{r_0^2\sinh 2\alpha_n}{2 c_n}
\nn\\
M &=& \frac{R_5 \tilde v r_0^2}{2 \left(\alpha'\right)^{2} \gst^2}
(\cosh 2\alpha_1 + \cosh 2\alpha_5+ \cosh 2\alpha_n)
\eea
There is another very interesting representation of the above-mentioned
six parameters in terms of what looks like brane-, antibrane-numbers
and left-, right-moving momenta:
\be
N_{1,\bar 1} = \frac{r_0^2 e^{\pm 2\alpha_1}}{4 c_1}, \,
N_{5,\bar 5} = \frac{r_0^2 e^{\pm 2\alpha_5}}{4 c_5}, \,
N_{L,R} = \frac{r_0^2 e^{\pm 2\alpha_n}}{4 c_n}
\label{def-nl-nr}
\ee
The coefficients $c_1,c_5, c_n$ are as in \eq{quantization}.
Clearly 
\be
N_1 - N_{\bar 1}= Q_1, N_5 - N_{\bar 5}= Q_5, 
N_L - N_R= N.
\label{2:anti}
\ee
The extremal limit corresponds to taking $r_0\to 0, \alpha_{1,5,n} \to
\infty$ keeping the charges $Q_{1,5}, N$ finite. We comment on the
brane-antibrane interpretation in Section \ref{2:entropy}.

\subsubsection{Geometry}\label{penrose-diagram}

It is easy to see that the above solution is a five-dimensional
black hole, with horizon at $r=r_0$. The horizon has a finite
area $A_h$, given by
\bea
\label{area}
A_h &=& 2\pi^2 r_0^3 \cosh\alpha_1 \cosh\alpha_5\cosh\alpha_n
\nn\\  
&=& 8 \pi G_N^{5} (\sqrt N_1
+\sqrt N_{\bar 1})(\sqrt N_5
+\sqrt N_{\bar 5})(\sqrt N_L+ \sqrt N_R)
\eea
Here we have used \eq{2:curious} and \eq{def-nl-nr}.

\begin{figure}[htb]
   \vspace{0.5cm}
\centerline{
   {\epsfxsize=5.5cm
   \epsfysize=5cm
   \epsffile{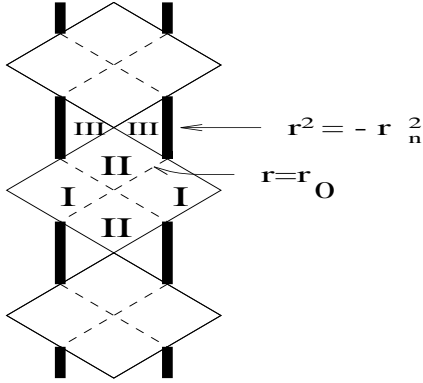}}
}
\caption{Carter-Penrose diagram for the non-extremal 5D black hole}
\label{f-penrose}
\end{figure}

The fact that the horizon has a finite area indicates that
the singularity lies ``inside'' $r=r_0$. It is not
at $r=0$, however, which corresponds to the inner horizon
(where light-cones ``flip'' the second time as one
travels in). To locate the singularity one needs
to use other coordinate patches which extend the manifold
further ``inside''. The singularity is time-like
and the Carter-Penrose diagram (\fig{f-penrose}) %
is similar to that of the non-extremal Reissner-Nordstrom metric
(see \eq{rn-bh},\eq{def-rpm}). The inner and outer
horizons (cf. \eq{def-rpm}) in the present case are 
$r_-=0, r_+=r_0$.

\subsubsection{Hawking temperature and 
Bekenstein-Hawking entropy}\label{2:entropy}

By using the formula \eq{a1:temp} we get the following Hawking
temperature ($\hbar=1$) (see Appendix \ref{gibbons-hawking} for details)
\be
\frac1{T_H}=  2\pi r_0 \cosh(\alpha_1)
\cosh(\alpha_5)\cosh(\alpha_n)
\label{2:temp}
\ee
We will compare this with the CFT result for the temperature 
\eq{4:hawking} in Section \ref{sec:gauge} (see
also Section 8).

By using the formula $S= A_h/4 G_N^5$ (cf.
\eq{bek-hawk}) and  \eq{area}, we get 
\be
\label{non-extremal-entropy}
S_{\rm BH} = 2\pi (\sqrt N_1
+\sqrt N_{\bar 1})(\sqrt N_5
+\sqrt N_{\bar 5})(\sqrt N_L+ \sqrt N_R)
\ee
Of course, the extremal entropy \eq{extremal-entropy}
corresponds to the special case $N_R=  N_{\bar 1}=
N_{\bar 5}=0$ (use \eq{2:anti}). A
somewhat more general  case is when $ N_{\bar 1}=
\sqrt N_{\bar 5}=0, N_R \not= 0$; the entropy
in that case is given by
\be
\label{non-extremal-entropy-special}
S_{\rm BH} = 2\pi \sqrt{Q_1 Q_5}\left( \sqrt{N_L} +
\sqrt{N_R} \right)
\ee
The entropy formulae \eq{extremal-entropy} and
\eq{non-extremal-entropy} are U-duality invariant, in the following
sense.  Consider an $S(3)$ subgroup of the U-duality group of type IIB
on $T^5$, which permutes the three charges $Q_1, Q_5$ and $N$. Such an
$S(3)$ is generated by\\ (a) $T_{6789}$ which sends $Q_1\to Q_5,
Q_5\to Q_1, N\to N$, and \\ (b) $T_{9876}ST_{65}$ which sends $Q_1 \to
Q_5, Q_5\to N, N\to Q_1$.\\ The entropy formula \eq{extremal-entropy}
remains invariant under these permutations. Since the ``anti''-objects
are also permuted among each other by these U-duality transformations,
we can say that the entropy formula \eq{non-extremal-entropy} is also
U-duality invariant.

\subsubsection{Comments on brane-antibrane and other non-BPS solutions
\label{sec:ddbar}}

It should be noted that the ``brane-antibrane'' representation of the
above non-extremal black hole is only {\em suggestive}
at the moment. The subject of supergravity representation of 
brane-antibrane and other non-BPS systems is very much open; for
a partial list of papers see \cite{Sen:1997pr,Eyras:2000ig,
Lozano:2000dv,Brax:2000cf,Bertolini:2000jy,Bain:2000sa,
Intriligator:2000pk,Liang:2001sp,Emparan:2001bb,
Alexander:2001ks,Alberghi:2001fy}.

\subsection{Supergravity Solution with Non-zero vev of $B_{NS}$}

Our discussion so far has been devoted to supergravity solutions in
which the values of all the moduli fields were set to zero. Such
solutions have the characteristic that the mass of the \dd
system is a sum of the charges that characterize the system. Such
bound states are marginal, without any binding energy, and can
fragment into clusters of \dd branes. The corresponding CFT has
singularities.  In order to obtain a stable bound state and a
non-singular CFT we have to turn on certain moduli fields. We will
consider the case when $B_{NS}$ is non-zero.

The construction of the supergravity solution that corresponds to a
$\frac{1}{4}$ BPS configuration, with a non-zero $B_{NS}$ was
presented in \cite{russo,DMWY}.  $B_{NS}$ has non-zero components only
along the directions $6,7,8,9$ of the internal torus.  {}From the view
point of open string theory this is then a non-commutative torus.

Here we will summarize the result. The solution contains, besides D1
and D5 brane charges, D3 brane charges that are induced by the
$B_{NS}$. For simplicity we consider only non-zero values for $B_{79}$
and $B_{68}$. The asymptotic values are given by $B^{(\infty)}_{79} =
b_{79}$ and $B^{(\infty)}_{68} = b_{68}$. It is important that at
least 2 components of the $B_{NS}$ are non-zero, in order to be able
to discuss the self-dual and anti-self-dual components.

Below we present the full solution which can be derived by a solution
generating technique. Details can be found in \cite{DMWY}.
\bea
\label{full-solution}
ds^2 &=& (f_1f_5)^{-1/2} (-dt^2 + (dx^5)^2) + (f_1f_5)^{1/2} (dr^2 +
r^2 d \Omega^2_3) \nonumber \\ 
&& + (f_1f_5)^{1/2} \left\{ Z^{-1}_\varphi ( (dx^6)^2 + (dx^8)^2) +
Z^{-1}_\psi ((dx^7)^2 + (dx^9)^2) \right\} ,\nn \\ 
e^{2\phi} &=& f_1 f_5 / Z_\varphi Z_\psi , \nn\\ 
B^{(2)}_{NS} &=& (Z^{-1}_\varphi \sin\varphi \cos\varphi (f_1-f_5) +
b_{68})
dx^6 \wedge dx^8 \nonumber \\ 
&& + (Z^{-1}_\psi \sin\psi \cos\psi (f_1-f_5) + b_{79}) dx^7 \wedge
dx^9 , \nn\\ 
F^{(3)} &=& \cos\varphi \cos\psi \tilde K^{(3)} + \sin\varphi \sin\psi
K^{(3)} , \nn\\
F^{(5)} &=& Z^{-1}_\varphi (-f_5 \cos\varphi \sin\psi K^{(3)} + f_1
\cos\psi
\sin\varphi \tilde K^{(3)}) \wedge dx^6 \wedge dx^8 \nonumber \\ 
&&+ Z^{-1}_\psi (-f_5 \cos\psi \sin\varphi K^{(3)} + f_1 \cos\varphi
\sin\psi \tilde K^{(3)}) \wedge dx^7 \wedge dx^9 ,\nn \\ 
Z_{\varphi, \psi} &=& 1 + {\mu_{\varphi,\psi} \over 2} \left({\alpha'
\over r^2}\right) ,
\nn\\
\mu_\varphi &=& \mu_1 \sin^2\varphi + \mu_5
\cos^2 \varphi ,\  \mu_\psi =
\mu_1 \sin^2\psi + \mu_5
\cos^2 \psi.
\eea
Here $b_{68}$ and $b_{79}$ are arbitrary constants which we have added
at the end by a T-duality transformation that
shifts the NS B-field by a constant. Note that for $\varphi = \psi =
0$ and $b_{68} = b_{79} = 0$, the above solution reduces to the
known solution for \dd system without B-field.

The above solution depends upon 4 parameters $\mu_1$, $\mu_5$, and the
angles $\phi $ and $\psi $, and in general represents a system of D1,
D5 and D3 branes.  Since we are seeking a solution that has no source
D3 branes we require that the D3 brane charges are only induced by the
presence of the non-zero $B_{NS}$. This leads to certain conditions on
the solutions which we do not derive here, but whose physical
implication we analyze. We discuss both the asymptotically flat and 
near horizon geometry.

\subsubsection{Asymptotically Flat Geometry}

In this case the induced D3 brane charges along the $(5,7,9)$ and  
$(5,6,8)$ directions are 
\be
\label{induced-d3}
Q_3 = B^{(\infty)}_{79} Q_5 , \ \ \ Q'_3 = B^{(\infty)}_{68} Q_5 ,
\ee
where
$B^{(\infty)}_{79} = b_{79} , \ \ \ B^{(\infty)}_{68} = b_{68}.$
There is an induced contribution to the D1 brane charge. The charge
$Q_{1s}$ of the source D1 branes is
\be
\label{induced-d1}
Q_{1s} = Q_1 - b_{68} \ b_{79} \ Q_5 .
\ee
while the D5 brane charge remains unaffected by the moduli.

\gap1

\noindent\underbar{\it Mass}
\gap1

Let us now study the mass formula as a function of the charges and the moduli.
The mass corresponding to the
${1\over 4}$ BPS solution
\cite{Obe-Pio98}, which coincides with the ADM mass, is given
in terms of the appropriate charges by
\be
\label{10}
M^2 = (Q_1 + Q_5)^2 + (Q_3 - Q'_3)^2
\ee
This can in turn be expressed in terms of $Q_{1s}$, $Q_5$ and $b_{68}, b_{79}$
\be
\label{mass-flat-mod}
M^2 = (Q_{1s}+b_{68}b_{79}Q_5 + Q_5)^2 + Q_5^2 (b_{68} - b_{79})^2
\ee
We must consider the mass as a function of the moduli, holding
$Q_{1s}$ and $Q_5$ fixed. We see that for non-zero moduli we have a
true bound state that turns marginal when the moduli are set to
zero. To locate the values of the moduli which minimize the mass, we
extremize the mass with respect to the moduli. The extremal values of
the moduli are
\be
\label{12}
b_{68} = - b_{79} = \pm \sqrt{Q_{1s}/Q_5 -1}  ,
\ee
This says that the $B_{NS}$ moduli are self-dual, 
in the asymptotically flat metric.
The mass at the critical point of the true bound state is then given by
\be
\label{13}
M^2 = 4 Q_{1s} Q_5 
\ee
\subsubsection{Near Horizon Geometry}

In this case, absence of D3-brane sources is ensured if we set
\be
\label{no-d3-nh}
Q^{(h)}_3 = B^{(h)}_{79} Q_5 , \ \ \ Q^{(h)'}_3 = B^{(h)}_{68} Q_5 ,
\ee
where
\bea
\label{14a}
B^{(h)}_{68} &=& {\mu_1 - \mu_5 \over \mu_\varphi} \sin\varphi
\cos\varphi + b_{68} , \\ [2mm]
\label{14b}
B^{(h)}_{79} &=& {\mu_1 - \mu_5 \over \mu_\psi} \sin\psi
\cos\psi + b_{79} ,
\eea
are the horizon values of the two nonzero components of the
B-field. 
Moreover, we see can that
in this case
\be
\label{15}
{B^{(h)}_{68}\over \mu_\psi} = - {B^{(h)}_{79} \over \mu_\varphi} ,
\ee
which is the self-duality condition on the B-field in the near horizon
geometry. We also note that the volume of $T^4$ at the horizon is given
by
\be
\label{16}
V^{(h)}_{T^4} = {\mu_1\mu_5 \over \mu_\varphi \mu_\psi} = {Q^{(h)}_{1s}
\over Q_5} .
\ee

The D1-brane charge that arises from
source D1-branes in this case is given by
\be
\label{17}
Q^{(h)}_{1s} = Q^{(h)}_1 - B^{(h)}_{68} B^{(h)}_{79} Q_5 .
\ee
One can show that
\be
\label{18}
Q^{(h)}_{1s} = Q_{1s}
\ee
where $Q_{1s}$ is given by (\ref{17}). Thus we see that not only do the
parameters $b_{68}$ and $b_{79}$ have the same values here as in the
asymptotically flat case, even the source D1-branes are identical,
despite the total D1-brane charges being very different in the two
cases.

\gap1

\noindent\underbar{\it Mass}
\gap1

The ${1\over 4}$ BPS mass formula in terms of the various charge
densities in this case is
\be
\label{19}
\left({M^{(h)} \over V^{(h)}_{T^4}}\right)^2 = \left({Q^{(h)}_1 \over
V^{(h)}_{T^4}} + Q_5\right)^2 + \left({Q^{(h)}_3\over
\sqrt{g_{77}g_{99}}} - {Q^{(h)'}_3 \over \sqrt{g_{66} g_{88}}}\right)^2
\ee
Using (\ref{no-d3-nh})-(\ref{18})
it can be easily seen that
\be
\label{20}
\left(M^{(h)}\right)^2 =  V^{(h)}_{T^4} \left(4Q_{1s}Q_5\right).
\ee
Apart from the extra factor of the $T^4$ volume in the near horizon
geometry, this is exactly the same as \eq{13}. The extra volume factor
correctly takes into account the difference in the 6-dimensional
Newton's constant between the asymptotically flat and near horizon
geometries because of the difference in the $T^4$ volume in the two
cases. We have already seen that the B-field is automatically
self-dual in the near horizon geometry and that the volume of $T^4$
satisfies the condition given by \eq{16} and \eq{17}. We now see that
the mass of the bound state is already at the fixed point value. Thus
the solution we have here provides an explicit demonstration of the
attractor mechanism \cite{Ferrara}.

The significance of this solution is that it is the description of a
stable bound state in the near horizon geometry. As we shall discuss
later this situation corresponds to a non-singular dual CFT.

\subsection{Near-horizon limit 
and $AdS_3\times S^3$}\label{near-horizon-ads} 

In this section we will exhibit the form of the classical solution in
the near horizon limit of Maldacena\cite{Mal97}.  This subsection,
together with Appendix C, will be used as background
for discussions of AdS/CFT correspondence  in Sections 6 and
11.

The basic idea of the near horizon limit is that, near the horizon of
a black hole or a black brane, the energies of particles as seen by
the asymptotic observer get red-shifted:
\be
E_{\infty}=\sqrt{g_{00}}E
\ee
For the metric \eq{6d-black-string} the red-shift factor is
\be
\sqrt{g_{00}} = (f_1f_5)^{-1/4}
\ee
Clearly as $r\rightarrow \infty$ the red shift factor is
unity. However near the horizon we get the equation
\be
E_{\infty} = \frac{r}{R}E
\ee
where $R$, Eqn. \eq{radius-ads}, is the length scale that
characterizes the geometry. In the near-horizon region, defined by
\be
\label{near-horizon-def}
r \ll R
\ee
we see that the energy observed by the asymptotic observer goes to
zero for finite values of E. This means that near the horizon (defined
by \eq{near-horizon-def}) an excitation of arbitrary energy looks
massless.  For massless modes this means that they have almost
infinitely long wavelengths and for massive modes they appear as long
wavelength massless excitations.  If one examines the potential energy
of a particle in the above geometry then in the near horizon limit the
potential barrier becomes very high so that the modes near the horizon
cannot get out. In the exact limit of $Q_1$ and $Q_5$ going to
infinity the horizon degrees of freedom become exactly massless and
decouple from the bulk degrees of freedom.  As we shall see later it
is in this limit that the bulk string theory is dual to a SCFT which
also exhibits massless behavior in the infrared.

\subsubsection{The three-dimensional anti de Sitter space or AdS$_3$} 

We now apply these ideas to the metric of the \dd\ black string with
the KK charge $N=0$, namely \eq{6d-black-string}.

In the region \eq{near-horizon-def} the metric and other backgrounds
are still given by \eq{6d-black-string}, except that the harmonmic
functions change to
\be
f_1 = \frac{16 \pi ^4 g_s \alpha '^3 Q_1}{V_4 r^2}, \;\;\;\; f_5=
\frac{ g_s \alpha' Q_5 }{r^2},\;
\ee
Here $r^2 = x_1^2 + x_2^2 + x_3^2 + x_4^2$ denotes the 
distance measured in the transverse direction to all the D-branes.
The above metric differs from \eq{6d-black-string} only
in that the harmonic functions do not have the ``$1$'' term
any more (see remarks after \eq{multi-centred}).

A more precise scaling limit of the geometry is given by
 \cite{chp3:MalStr98}
\bea
\label{prescal}
\alpha' \rightarrow 0, \;&\;&\; \frac{r}{\alpha '} 
\equiv U=  \mbox{fixed} \\  \nonumber
v \equiv \frac{V_4}{16\pi^4\alpha^{\prime 2}} = \mbox{fixed}, \;&\;&\;
g_6 = \frac{g_s}{\sqrt{v}} = \mbox{fixed}
\eea
In this limit the metric in (\ref{6d-black-string}) becomes
\be
\label{near-horizon}
ds^2 = \alpha' \Big(ds_3^2 + ds^2[S^3] + ds^2[T^4]\Big)
\ee
where
\be
\label{2:ads}
ds_3^2 = \Big[ \frac{U^2}{l^2}(-dx_0^2 + dx_5^2)  
+ l^2 \frac{dU^2}{U^2} \Big],
\ee
represents three-dimensional anti-de Sitter space
AdS$_3$ (see Appendix C, Eq.
\eq{ads-poincare}) and
\bea
\label{s3-t4}
ds^2[S^3] &=& l^2 d\Omega_3^2, \nn \\
ds^2[T^4] &=&  \sqrt{\frac{Q_1}{vQ_5}}(dx_6^2+ \ldots dx_9^2), 
\eea
represent a three-sphere and a four-torus.  Thus the near horizon
geometry is that of $AdS_3\times S^3\times T^4$.  Our notation for
coordinates here is as follows: $AdS_3: (x_0, x_5, r)$,
$S^3:(\Omega_3=(\chi, \theta, \phi))$, $T^4:(x_6, x_7, x_8, x_9)$. $r,
\chi,\theta,\phi$ are spherical polar coordinates for the directions
$x_1,x_2,x_3,x_4$. The length scale $l$ is
the dimensionless radius of $S^3$ and the anti-de Sitter space:
\be
l= \frac{R}{\sqrt{\alpha'}},\; 
R = \sqrt{\alpha'} (g_6^2 Q_1Q_5)^{1/4}.
\label{radius-ads}
\ee
Note that the effective string coupling in the near horizon limit is
given by
\[
g_{eff}=g_6\sqrt {Q_1/Q_5}
\]  
The formulas for the blackhole entropy and temperature, which depend
only on the near horizon properties of the geometry, do not change in
the near horizon limit.

In Section \ref{near-horizon-symmetry} we will discuss in detail the
symmetries of the near-horizon geometry \eq{near-horizon}. The
spacetime symmetries as well as supersymmetries get enhanced
compared to \eq{space-time-symmetry}, \eq{supersymmetry}.

\subsubsection{The BTZ black hole}\label{near-horizon-btz}

The above discussion was about the near-horizon limit 
of the six dimensional black string. 
We now turn to the near-horizon limit of the five-dimensional
black hole \eq{non-extremal-5d}. The near
horizon scaling limit is given by \cite{Sfetsos:1998xs,chp3:MalStr98}
\be
\label{chp4:Maldalim}
\alpha^{\prime}\rightarrow 0, \;\;\;  r\rightarrow 0,\;\;\;
r_0\rightarrow 0
\ee
with
\bea
 U\equiv \frac{r}{\alpha^{\prime}} =\mbox{fixed} \;&\;&\; U_0\equiv
 \frac{r_0}{\alpha^{\prime}} =\mbox{fixed} \\ \nonumber v\equiv
 \frac{V_4}{16\pi^4\alpha^{\prime 2}} =\mbox{fixed} \;&\;&\; g_6=
 \frac{g_s}{\sqrt{v}} =\mbox{fixed} \;\;\; R_5 = \mbox{fixed} 
\eea
In this limit the metric of the \dd\ black hole
(cf. Eqn. \eq{non-extreme-2a}) reduces to the following
\be
\label{chp4:Maldametric}
ds^2 = \alpha' \Big( ds_3^2 + ds^2[S^3] + ds^2[T^4] \Big)
\ee
where the metric on the 3-sphere and 4-torus are
still given by \eq{s3-t4}, whereas
\be
ds_3^2 = \frac{\alpha^{\prime} U^2}{l^2}(-dx_0^2 + dx_5^2)  
+ \frac{\alpha^{\prime} U_0^2}{l^2} (\cosh \sigma dt + \sinh\sigma
dx_5^2)^2 +  \frac{\alpha^{\prime} l^2}{U^2- U_0^2} dU^2
\label{2:btz}
\ee
represents now the BTZ geometry, as we will show below. Thus the
near-horizon geometry is BTZ$\times T^4\times S^3$.  Our co-ordinate
definitions here are as follows: $t, \phi, \tilde{r}$ refer to BTZ
co-ordinates, $\Omega_3$ stands for the $S^3$ and $x_6, x_7, x_8, x_9$
stand for the co-ordinates of $T^4$.  To identify $ds_3^2$ with the
BTZ metric (i.e., the black hole in three-dimensional anti de-Sitter
space discovered by \cite{chp1:Btz}), we make the coordinate
redefinitions given below \cite{Sfetsos:1998xs,chp3:MalStr98}
\bea
\tilde{r}^2&=& (U^2 + U_0^2\sinh^2\sigma)\frac{R_5^2}{l^2} \\ \nonumber
r_+&=& \frac{R_5 U_0\cosh\sigma}{l} \\   \nonumber
r_-&=& \frac{R_5 U_0\sinh\sigma}{l} \\  \nonumber
\phi= \frac{x_5}{R_5}, \;&\;&\; t = \frac{lx^0}{R_5}
\eea
The metric \eq{2:btz} in these new coordinates is given by
\be
\label{chp4:btz}
ds_3^2= -\frac{\alpha^{\prime}(\tilde{r}^2-r_+^2)
(\tilde{r}^2-r_-^2)}
{l^2\tilde{r}^2} dt^2
+ \frac{\alpha^{\prime} \tilde{r}^2l^2}{(\tilde{r}^2-r_+^2)
(\tilde{r}^2-r_-^2)} dr^2
+\alpha^{\prime}\tilde{r}^2\left( d\phi + \frac{r_+r_-}
{\tilde{r}^2l}dt\right)^2 
\ee
In this form, the metric coincides with that of a BTZ black hole
(cf. Eqs. \eq{a3:btz},\eq{a3:lapse}), with mass $M$ 
and angular momentum $J$ given by (cf. \eq{a3:rpm})
\be
\label{chp4:parameters}
M = \frac{r_{+}^2 + r_{-}^2}{l^2}, \quad J = \frac{\sqrt{\alpha'}2r_+
r_-}{l}
\ee
The mass $M$ and the angular momentum $J$ for the BTZ black hole are
related to the parameters of the D1-D5 black hole by
\be
\label{chp4:parmetersd1-d5}
\frac{M}{2} = L_0 + \bar{L}_0 = \frac{N_L + N_R}{Q_1Q_5},
\quad
\frac{J}{2\sqrt{\alpha '} l} = 
L_0 -\bar{L}_0 = \frac{N_L - N_R}{Q_1Q_5}
\ee
where $N_L$, $N_R$ are defined in \eq{def-nl-nr} and $L_0$,
$\bar{L}_0$ are the levels of the SCFT.  The extremal limit is given
by $r_{+}=r_{-}$. From \eq{chp4:parameters} and
\eq{chp4:parmetersd1-d5} we see that in the extremal limit $N_R=0$ as
expected for the \dd\ black hole.

It is important to mention the global properties of 
the metric \eq{chp4:btz}, especially in relation to those
of the AdS$_3$ metric in \eq{near-horizon} above. Let us
consider the simplest BTZ solution first, namely with
$r_+=r_-=0$.  Substituting these values
in \eq{chp4:btz} we find the metric is given by
\be
\label{chp4:zero-btz}
\frac{ds_3^2}{\alpha^{\prime}} = -\frac{\tilde{r}^2}{l^2}dx_0^2 +
\frac{l^2}{\tilde{r}^2}dr^2 + \tilde{r}^2d\phi^2
\ee
By comparison with \eq{near-horizon} one can see that 
this metric is locally $AdS_3$ except for the global identification 
$\phi\equiv \phi + 2\pi$. This of course reflects the
fact that the $r_+=r_-=0$ BTZ solution corresponds to the
near-horizon limit of the \dd
string (with $N_L= N_R=0$) wrapped on a circle along $
x^5\equiv x^5 + 2\pi R_5$. 

This periodic identification has two important implications:

(a) Firstly, that the zero-mass BTZ black hole is a quotient of the
AdS$_3$ space by a discrete isometry. Indeed, as has been shown in
\cite{chp3:MalStr98} the global property of the more general
near-horizon solution \eq{chp4:btz} also corresponds to an appropriate
quotient of AdS$_3$ by a discrete isometry, consistent with the
expected global properties of BTZ black holes
\cite{chp1:Btz}.


(b) Secondly, the difference between the geometry of the zero mass
$BTZ$ black hole and that of the AdS$_3$ (although identical locally)
leads to an important difference in the boundary conditions for the
fermions. For the case of AdS$_3$ the fermions are anti-periodic in
$\phi$ and for the zero mass BTZ black hole they are periodic in
$\phi$. One can easily see that the constant time slice of the metric
in \eq{2:ads} has the topology of a disk. This forces the fermions to
be anti-periodic in $\phi$ for AdS$_3$. For the case of the zero mass
BTZ black hole in \eq{chp4:zero-btz}, the constant time slice has a
singularity at $\tilde{r}=0$. Therefore the fermions can be both
periodic or anti-periodic.  An analysis of the Killing spinors in the
background of the zero mass BTZ black hole shows that the fermions in
fact have to be periodic \cite{chp4:CouHen}.

\subsubsection{The two-dimensional black hole\label{sec:2dbh}}

In this subsubsection we will discuss the near-horizon geometry of a
related (pure 5-brane) system and its connection to the two
dimensional black hole
\cite{Mandal:1991tz, Witten:1991yr,Elitzur:1991cb}.

Consider the supergravity solution of the non-extremal black hole in
type IIB string theory \eq{non-extremal-5d} with $Q_1=0$ and $\sigma
=0$. The \dd supergravity solution then reduces to the non-extremal
D5-brane. We will now show that there is a near horizon region where
the geometry can be approximated to that of two dimensional black hole
\cite{Mandal:1991tz, Witten:1991yr,Elitzur:1991cb}.  The ten
dimensional geometry with $Q_1=0$ is given by
\bea
\label{non-extremal-five}
e^{-2\phi} &=& \frac{1}{g_s^2} \left( 
1+ \frac{r_5^2}{r^2} \right) \\ \nonumber
H_{\theta\phi\psi} &=& Q_5\alpha'  \\ \nonumber
ds^2 &=& \left( 1+ \frac{g_s Q_5 \alpha'}{r^2} \right)^{-\frac{1}{2}}
\left(-(1- \frac{r_0^2}{r^2}) dt^2 + dx_5^2 + \cdots dx_9^2 
\right) \\ \nonumber
&+& \left(1+ \frac{g_sQ_5 \alpha'}{r^2} \right)
\left( \frac{dr^2}{1-\frac{r_0}{r^2}}dr^2 + r^2 d\Omega_3^2 \right)
\eea
To obtain the near horizon geometry
We use the IMSY limit \cite{Itzhaki:1998dd} 
for the case of D5 branes. This is
given by
\bea
U = \frac{r}{\alpha'} = \mbox{fixed}, &\quad 
U_0 = \frac{r_0}{\alpha'} = \mbox{fixed},  \\ \nonumber
g_{\rm{YM}}^2 = (2\pi)^3 g_s \alpha' = \mbox{fixed} &\quad \mbox{with  }
\alpha'\rightarrow 0
\eea
Here $g_{\rm{YM}}$ is the Yang-Mills coupling on the D5-brane.
This geometry is nonconformal and the dilaton depends on the scale
$U$.
For $\sqrt{Q_5} \ll g_{\rm{YM}} U$, the string coupling and the
curvature in string units is large, and therefore the valid discription
of the background is obtained by performing an S-duality. 
The solution reduces to the near horizon geometry of
non-extremal NS 5-branes.
The metric and the dilaton after S-duality is given by
\bea
\label{s-dual-fivebrane}
e^{2\phi} &=& \frac{(2\pi)^3 Q_5}{g_{\rm{YM}}^2 U^2} \\ \nonumber
ds^2 &=& 
-\left(1- \frac{U_0^2}{U^2}\right)dt^2 + 
(dx_5^2 + \cdots dx_9^2) + g_s\alpha' Q_5 \left(
\frac{dU^2}{U^2(1-\frac{U_0}{U})} +
d\Omega_3^2\right)
\eea
Here we have scaled the metric by $g_s$ so that the 10 dimenional
Newton's constant is invariant.
To see this is the metric of the (2d black hole)$\times S^3 \times
R^5$ we change coordinates by substituting $U=U_0\cosh
\gamma$. Then we get
\bea
\label{twodbh}
e^{2\phi} &=& \frac{ (2\pi)^3 Q_5}{g_{\rm{YM}}^2 U_0^2} 
\frac{1}{\cosh^2 \gamma} \\ \nonumber
ds^2 &=& -\tanh^2\gamma dt^2 + g_s\alpha' Q_5(d\gamma^2 + d\Omega_3^2)
+ ( dx_5^6 + \cdots dx_9^2)
\eea
Now it is easily seen that the geometry reduces to that of (2d black
hole)$\times S^3\times R^5$. This near horizon limit of extremal NS
5-branes was obtained in \cite{Maldacena:1997cg}.

\newpage

\section{Semi-classical derivation of Hawking radiation}

We described in some detail the construction of the \dd black hole,
\eq{extremal-5d},\eq{non-extremal-5d}, in the last section. 
We will now address the issue of absorption and Hawking radiation by
this black hole. Both absorption and Hawking radiation involve
interesting questions, as we remarked in the introduction.  For
instance, classically the black hole only absorbs and does not
emit. One of our goals will be to ultimately interpret this in the
microscopic model, explaining thereby a crucial aspect of the event
horizon. Secondly, the semiclassical treatment of Hawking radiation
leads to the information puzzle, and we would like to see how standard
scattering processes described in terms of the microscopic model gives
rise to such a radiation within a unitary quantum theory.

Before we proceed to the microscopic description, however, we will
devote the present section to briefly review the (semi)classical
calculations of absorption/emission of particles (in the type IIB
spectrum) by the \dd black hole \eq{non-extremal-5d}.

The absorption cross-section and emission rate of a particular field
depend on how the field propagates and backscatters from the geometry
of the black hole. We will look at the equation of propagation of
scalar fluctuations.

We begin by writing down the IIB Lagrangian
\cite{Mah-Sch93,Bergshoeff:1995as,Cal-Gub-Kle96} 
compactified on $T^5$ (of which
the \dd black hole \eq{non-extremal-5d}, \eq{charges} is a solution):
\bea
\label{IIBlag}
\!\!\!\!\!\!\!\!\!\!
S_5 &=& \frac{1}{2\kappa_5^2} \int d^5 x \sqrt{-g} 
\Big[ R - \frac{4}{3}
(\del_\mu \phi_5)^2  
- \frac{1}{4} G^{ab} G^{cd}(\del_\mu G_{ac}\del^\mu G_{bd}
+ e^{2 \phi_5} \sqrt{G} \del_\mu C_{ac}\del^\mu C_{bd}) 
\nn \\
&& - \frac{e^{-4\phi_5/3}}{4} G_{ab}F^{a}_{\mu\nu}
F^{b \mu\nu} - \frac{e^{2\phi_5/3}}{4}
\sqrt{G}  G^{ab} H_{\mu\nu a}H_b^{\mu\nu}
-  \frac{e^{(4\phi_5/ 3)}}{12} \sqrt {G}  
H^2_{\mu\nu\lambda}\Big]
\nn\\
\eea
Notation: $a,b,\ldots=5,\ldots,9$ denote the directions along $T^5$,
while $\mu,\nu,\ldots=0,\ldots,4$ denote the non-compact
directions. We have included in the above Lagrangian only
the following ten-dimensional fields: 
\begin{itemize}
\item
the ten dimensional dilaton $\phi$,
\item
the ten-dimensional string-frame metric $ds^2$ ~~ \footnote{
related to the Einstein frame metric $ds^2_E$ as 
$ds^2= \exp[\phi/2]ds^2_E$} written as
\be
\label{3:kk}
ds^2 = g_{\mu\nu}dx^\mu dx^\nu + G_{ab}
\left( dy^a + A^{a}_\mu dx^\mu \right)
\left( dy^b + A^{b}_\nu dx^\nu \right)
\ee
which identifies $A^{a}_\mu$ as the KK vector fields,
\item
and the RR 2-form field $C^{(2)}$ written as
\[
C^{(2)} = 
C_{\mu\nu} dx^\mu \wedge dx^\nu + C_{ab} dx^a\wedge  dx^b
\]
\end{itemize}
The various fields appearing in \eq{IIBlag} are
defined in terms of the above fields, as follows
\begin{itemize}
\item
the five-dimensional dilaton
$ \phi_5 = \phi_{10} - (1/4)\ln [{\rm det}_{ab} G_{ab}],  $
\item
the KK field strengths $F^{a} = dA^{a},$
\item
and the $H$-fields given by \cite{Mah-Sch93}
\beas
H_{\mu\nu a} &=& {\tt F}_{a\mu\nu} - 
C_{ab}  F^{b}_{\mu\nu}
\nn\\
H_{\mu\nu \lambda} &=& 
\del_\mu{\tt B}_{\nu\lambda} -\frac12 A_{\mu}^{a}
{\tt F}_{a\nu\lambda} - \frac12 {\tt A}_{a\mu}
F^a_{\nu\lambda} + {\rm cyc.perm.}
\eeas
where
\[
{\tt A}_{a\mu}=
C_{\mu a} + C_{ab} A_{\mu}^{b}, \; {\tt F}_a = d{\tt A}_{a},
\;
{\tt B}_{\mu\nu} =C_{\mu\nu}+ 
 A_{[\mu}^{a}{\tt A}_{\nu] a} - A_\mu^{a}C_{a b}A_\nu^{b}
\]
\end{itemize}
The five-dimensional Newton's constant
$16\pi (G^5_N)^2 \equiv 2 \kappa_5^2$ is defined as
in \eq{gn-5}.

We will now simplify the Lagrangian even further, by assuming that (a)
of the KK-gauge fields only $A^5_\mu$ is non-zero and is of the
``electric'' type (b) $C_{ab}=0$. This is a consistent truncation, and
\dd black hole \eq{non-extremal-5d} is a solution of the
truncated system.

We will consider below two separate sets of scalar fluctuations:
\begin{enumerate}
\item This set of fluctuations $h_{ab}, a\not= b, \; 
a,b=6,7,8,9 $ are defined by
\be
\label{def-h}
G_{ab}= f_1^{1/2}f_5^{-1/2} \left(\delta_{ab} + h_{ab}
\right),\;  a,b=6,7,8,9
\ee
Recall that $\langle G_{ab} \rangle= 
f_1^{1/2}f_5^{-1/2} \delta_{ab}$ represents the background value
(cf. \eq{non-extreme-2a}).
\item This set of fluctuations are defined by 
\begin{itemize}
\item
$e^{2\nu} \equiv G_{66}$ 
(assumed equal to $ G_{77}= G_{88}= G_{99}),$
\item
${\tt \phi}\equiv  \phi_5 + \frac12 \nu_5,$
where $e^{2\nu_5} = G_{55}$,
\item
$\lambda \equiv \frac34 \nu_5 - \frac12 \phi_5.$
\end{itemize}
\end{enumerate}

For the fluctuations in Case 1, the action \eq{IIBlag} reduces to the
following action 
\be
S = -\frac1{8\kappa_5^2} \int d^5x \sqrt{-g} \del_\mu h_{ab}\del^\mu
h_{ab}
\label{3:action-minimal}
\ee
whereas for the three fluctuations in Case 2,
the action \eq{IIBlag} reduces to
\bea
S &&= 
\frac{1}{2\kappa_5^2} \int d^5x \sqrt{-g}
\Big[R - (\del_\mu {\tt \phi})^2
\nn \\
 &&-   \frac43 (\del_\mu \lambda)^2 - 4 (\del_\mu \nu)^2
- \frac14 e^{\frac83\lambda}(F^5_{\mu\nu})^2 -
\frac14 e^{-4/3\lambda + 4 \nu}{\tt F}_{5,\mu\nu}^2 
-\frac1{12} e^{4/3\lambda  + 4\nu} H_{\mu\nu\lambda}^2
\Big]
\label{3:action-fixed}
\eea
The background values of the 5-D Einstein metric  $g_{\mu\nu}$
and the other fields are to be read off from the \dd \
black hole solution \eq{non-extremal-5d}, \eq{charges}.

Note that $h_{ab},a\not= b$ and ${\tt \phi}$ couple only to the
$g_{\mu\nu}$ (see Eqn.  \eq{non-extremal-5d}); because of this
property they are called ``minimal scalars''. On the other hand, $\nu,
\lambda$ couple to the dilaton and the RR fields as well; these are
called ``fixed scalars'' because their value at the horizon cannot be
arbitrarily chosen but are fixed by the charges $Q_1, Q_5, N$.

In the following we will first calculate the
absorption crosssection for minimal scalars, and
later briefly mention the case of fixed scalars.

\subsection{Minimal scalar}

For the semiclassical absorption/emission \cite{Dha-Man-Wad96,
Das-Mat96,Mal-Str96}, all we need is the equation for propagation of
the fluctuation $h_{ab}$ (or $\phi$) 
on the black hole metric $g_{\mu\nu}$. We will denote the
minimal scalar fluctuation generically by the symbol
$\varphi$; since it couples only to the five-dimensional
Einstein metric \eq{3:action-minimal}, the equation of motion is
\[
D_\mu \del^\mu \varphi =0
\]
For the five-dimensional black hole metric \eq{non-extremal-5d}
the above  equation becomes for the s-wave mode:
\be
[\frac{h}{r^3}\frac{d}{dr} (h r^3 \frac{d}{dr}) + f w^2]R_w(r) =0
\label{R-eqn}
\ee
where $ \varphi  = R_w(r) \exp[-iwt]. $

\begin{figure}[!ht]
\begin{center}
\leavevmode
\epsfbox{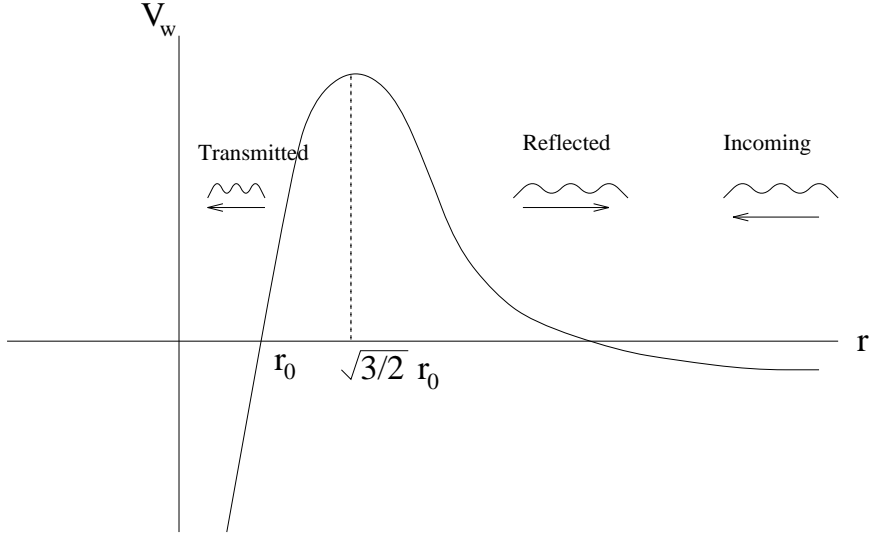}
\end{center}
\caption{Potential for minimal scalar}
\label{f-potential}
\end{figure}

The idea behind the absorption calculation is very simple. 
In terms of $\psi = r^{3/2} R$ the above equation becomes 
\be
[-\frac{d^2}{dr^2_*}  + V_w(r_*)]\psi = 0
\ee
where
\be
V_w (r_*) = -w^2 f + 
\frac{3}{4 r^2}(1 + 2 r_0^2/r^2 - 3 r_0^4/r^4)
\ee
The shape of the potential is given by (\fig{f-potential}).
Absorption is caused by the tunnelling of an incoming
wave into the ``pit of the potential''.

\gap{1}

\ni\underbar{\it Near and Far solutions:}

\gap{1}

It is not possible to solve the wave equation exactly.  However, we
can devise near and far zones where the potential simplifies enough to
admit known solutions.  If the zones have an overlap region then
matching the near and far wave-functions and their radial derivatives
will provide the solution for our purpose. In the following we will
closely follow \cite{Mal-Str96}.  We will work in the following range
of frequency and parameters
\bea
r_0, r_n &&\ll  r_1,r_5
\nn\\
wr_5 &&\ll 1
\label{dilute-gas}
\eea

The far and near solutions will be matched at an intermediate point
$r_m$ such that
\be
r_0, r_n \ll r_m \ll r_1, r_5, \qquad wr_1 \ll r_m/r_1
\ee
The existence of such an intermediate point $r_m$
is guaranteed by \eq{dilute-gas}.

\mysec11{Far zone ($r \ge r_m$):} 

\ni Here the potential $V_w$ becomes (in terms of $\rho=wr$)
\be
V_w(\rho)= -w^2(1 - \frac{3}{4 \rho^2})
\ee
This gives a Bessel equation, so that
\bea
\psi &=& \alpha F(\rho) + \beta G(\rho)
\nn\\
F(\rho) &=& \sqrt{\pi \rho/2}J_1(\rho),\quad 
G(\rho) = \sqrt{\pi \rho/2}N_1(\rho)
\eea
Using $R= r^{-3/2} \psi$ and the asymptotic forms
of the above Bessel functions, we find the
following asymptotic form for $R$:
\be
R= \frac{1}{r^{3/2}}\Big[\frac{e^{iwr}}2 \Big(\alpha
e^{-i3\pi/4} - \beta e^{-i\pi/4} \Big)
+ \frac{e^{-iwr}}2 \Big(\alpha
e^{i3\pi/4} - \beta e^{i\pi/4} \Big) \Big]
\label{far}
\ee

\mysec11{Near zone ($r \le r_m$):}

\ni Here we have
\be
\frac{h}{r^3}\frac{d}{dr} (h r^3 \frac{d}{dr}R) 
+ [\frac{(wr_nr_1r_5)^2}{r^6}
+ \frac{w^2r_1^2 r_5^2}{r^4} ]R_w(r) =0
\ee
which is a Hypergeometric equation, with solution \cite{Mal-Str96}
\bea
\label{3:near}
R &=& A \tilde F + B \tilde G 
\nn\\
\tilde F &=& z^{-i(a+b)/2} F(-ia, -ib, 1-ia -ib,z)
\nn\\
\tilde G &=& z^{i(a+b)/2} F(-ia, -ib, 1-ia -ib,z)
\nn\\
z &=& (1- r_0^2/r^2), \; a= w/(4 \pi T_R),\; b=w/(4 \pi T_L)
\eea
where we have introduced two parameters $T_L, T_R$, given by
\be
T_{L,R} = \frac{r_0}{2\pi r_1 r_5}e^{\pm \alpha_n}=
\frac{1}{2\pi r_0 \sinh(\alpha_1)\sinh(\alpha_5)\exp[\mp \alpha_n]}
\label{3:temp}
\ee
These, as will see in Section \ref{sec:cft}, play the role of
`left-' and `right'-moving temperatures (cf. Eqn. \eq{4:temp}). 
In the second step we have used the definition of the
gravitational lengths $r_1, r_5$ in \eq{non-extremal-5d}.

We now impose on the ``near solution'' \eq{3:near} the condition that
the wave at the horizon should not have any outcoming component: it
should be purely ingoing (no ``white hole''). This gives $B=0$.

\gap1

\ni\underbar{\it Matching}

\gap1

We now match $R$ and $dR/dr$ between the near and far regions at some
point $r_m$ in the overlap region.
This gives
\be
\sqrt{\pi/2} w^{3/2} \alpha/2 = A
\frac{\Gamma(1-ia-ib)}{\Gamma(1-ib)\Gamma(1-ia)}, \quad
\beta/\alpha \ll  1
\label{beta-0}
\ee

\gap1
\ni\underbar{\it Fluxes}
\gap1

The equation \eq{R-eqn} for $R$ implies $\frac{d}{dr} {\mathcal F}=0$,
where
\be
{\mathcal F}(r) = \frac{1}{2i}[ R^* hr^3 dR/dr - {\rm c.c.}]
\ee
In order to find out what fraction of the flux gets
absorbed at the horizon, we compute the ratio
\be
R_1 = {\mathcal F}(r_0)/{\mathcal F}^{in}(\infty)= r_0^2 
\frac{a+b}{w|e_1|^2} w^3 \pi/2
\label{flux-ratio}
\ee
where the superscript ``in'' indicates the flux calculated
from the ``ingoing'' part of the wave at infinity.

\subsubsection{Absorption Cross-section:}

In order to define absorption cross-section in the standard
way, one has to consider plane waves and not $s$-waves.
It is easy to derive that
\be
e^{-iwz} = (4\pi/w^3) e^{-iwr}Z_{000} + {\rm other\;partial\;waves}
\ee
where we use the notation $Z_{lm_1m_2}$ for the $S_3$ analogs of the
spherical harmonics $Y_{lm}$. From this and the standard definition of
absorption cross-section we get
\be
\sigma_{abs} = (4\pi/w^3) R_1 
\nn
\ee
which evaluates to \cite{Mal-Str96}
\be
\label{class-abs}
\sigma_{abs}= 2 \pi^2 r_1^2 r_5^2 \frac{\pi w}{2} 
\frac{\exp(w/T_H)-1}{(\exp(w/2T_R)-1) (\exp(w/2T_L) -1)}
\ee
where $T_{L,R}$ is given by \eq{3:temp}, and $T_H$,
to be identified below with the Hawking temperature, is given by
the harmonic mean 
\be
\frac1{T_H} = \frac12 \left(\frac1{T_L}+ \frac1{T_R} \right)
= 2\pi r_0 \sinh(\alpha_1) \sinh (\alpha_5) \cosh(\alpha_n)
\label{3:harmonic}
\ee
Note that in the regime \eq{dilute-gas}, the Hawking temperature
agrees with \eq{2:temp}. We will make this
comparison in Section \ref{sec:quick} where we will
also compare the
temperatures \eq{3:temp},\eq{3:harmonic} with the values
obtained from the D1-D5 CFT (see also Section 8).
  
In the $w\to 0$ limit, one gets \cite{Dha-Man-Wad96}
\be
\sigma_{abs} = A_h
\ee
where $A_h$ denotes the area of the event horizon.

\subsubsection{Hawking radiation:}

The semiclassical calculation of Hawking radiation is performed
through the standard route of finding Bogoliubov coefficients
representing mixing of negative and positive frequency modes due to
evolution from ``in'' to ``out'' vacua, defined with respect to
Minkowski observers existing in the asymptotically flat regions at
$t=-\infty$ and $t=+\infty$ respectively \cite{Hawk1} (see,
e.g. \cite{'tHooft:ij,Birrell:1982ix} for more leisurely derivations).

The rate of radiation is given by
\be
\label{class-decay}
\Gamma_H = \sigma_{abs} (e^{w/T_H}-1)^{-1} 
\frac{d^4k}{(2 \pi)^4}
\ee
As we remarked above, the Hawking temperature, given by Eqns.
\eq{3:harmonic}, agrees with the temperature \eq{2:temp} in the region
\eq{dilute-gas} (see Section \ref{sec:quick}).
 
We will see in Section 8 how $\Gamma_H$ and $\sigma_{abs}$ are
reproduced in the D-brane picture.

\subsubsection{Importance of near-horizon physics}

\ni It is interesting to note two points for later use:

(a) The ``near zone'' described above is
simply the near-horizon region as in the AdS/CFT context (see Sections
6 and 8),

(b)  With the inequality $\beta \ll \alpha$ in
\eq{beta-0} the
solution \eq{far} in the Far zone simply becomes
\be
R= \frac{\alpha}{\sqrt 2} e^{iwr}
\label{far-matched}
\ee
which is just a free incoming wave, with flux ${\mathcal
F}^{in}(\infty)= |\alpha|^2$.  As we saw (Eqn. \eq{flux-ratio}), the
parameter $\alpha$ also disappears from the ultimate calculation
because of the division by the flux at infinity.

Thus, at the end of the day, it is only the near-horizon geometry,
together with the mere existence of the asymptotically flat region,
which ultimately determines the absorption cross-section
\eq{class-abs} and the Hawking flux \eq{class-decay}.

\subsection{Fixed scalars}

The graybody factor for the fixed scalars ${\tt phi}, \lambda$
\cite{Cal-Gub-Kle96, Klebanov:1996gy,Krasnitz:1997gn,
Taylor-Robinson:1997kx,Lee:1998xz} (also reviewed
in \cite{Gubser:ex}) follows from a similar, but more involved,
analysis of the (coupled) equations of motion of these two fields
which follow from the action \eq{3:action-fixed}. These were solved
for general $Q_1, Q_5, N$ in \cite{Krasnitz:1997gn}.  The method for
computing the absorption cross-section and the Hawking rate is similar
to those employed for the minimal scalars. As we noted above, the
important ingredient in the semiclassical calculation is the
near-horizon equation of motion; this turns out be Eq.
\eq{fixscalar}. Using this, we arrive at the following result for the
absorption cross-section for fixed scalars (for $w,T_R
\ll T_L$)
\be
\sigma_{abs}= \frac14 A_h (wr_n)^2
\left(1 + \frac{4\pi^2 T_H^2}{w^2} \right)
\label{3:abs-fixed}
\ee
where the temperatures $T_L, T_R, T_H$ are defined in
\eq{3:temp},\eq{3:harmonic}, and the area $A_h$ of the horizon is
defined in \eq{area}. 

As we will see in Section 8, understanding the absorption and emission
of fixed scalars from D-brane models is a subtle
problem\cite{Krasnitz:1997gn}.  Resolution of this problem requires
\cite{chp1:DavManWad1} a new insight from AdS/CFT 
correspondence about coupling of D-branes to supergravity.

\newpage

\section{The microscopic modeling 
of black hole and gauge theory of the \dd system}\label{sec:gauge}

In section \ref{sec:6d-blackstring} 
we discussed the supergrativy solution of the 
\dd black string solution. The solution  with $N=0$, then it
it consists of $Q_1$ D1-branes and $Q_5$ D5-branes. 
The realization that solitons carrying Ramond-Ramond charges can be
represented at weak string coupling 
by open strings with Dirichlet boundary conditions 
\cite{Pol-tasi96} allows the
formulation of the microscopic theory for the \dd system.  We will be
interested in only low energy degrees of freedom of the \dd system,
and thus we ignore all the massive string modes.  There are two ways
to proceed in the study of the massless modes, and we shall discuss
both of them. The first method is a description in terms of a
2-dimensional gauge theory 
of the D-branes and the second method involves identifying
D1 branes with instantons of a 4 dimensional gauge theory.  The latter
description is more accurate and is valid for instantons of all sizes.
The 2-dimensional gauge theory description is valid near the point in 
the moduli space of instantons when the instantons have shrunk to zero
size \cite{Douglas:1997vu}. 
We will discuss this more approximate description first and detail the
domain of validity of this description.

\subsection{The \dd System and the ${\mathcal N}=4$, $U(Q_1) 
\times U(Q_5)$ gauge theory in 2-dimensions}

Consider type IIB string theory with five coordinates, say $x^5\cdots
x^9$, compactified on $S^1\times T^4$. The microscopic model for the
solution \eq{extremal-5d} 
with $N=0$ consists $Q_1$ D1-branes and $Q_5$ D5-branes
\cite {Str-Vaf96,Cal-Mal96}. The D1-branes are along the $x^5$
coordinate compactified to a circle $S^1$ of radius $R_5 \equiv R$,
while the D5-branes are parallel to $x^5$ and $x^6,\cdots,x^9$
compactified on a torus $T^4$ of volume $V_{T^4}\equiv V_4$. The
charge $N$ is related to the momenta of the excitations of this system
along $S^1$. We will work in the following region of parameter space:
\bea
&& V_4 \sim O({\alpha'}^2) \nn\\
&& 
R \gg l_s \equiv \sqrt{\alpha'}
\label{4:scales}
\eea
Let us briefly discuss the implications of the above region
in parameter space.
The size of the torus $T^4$ 
is of the order of string scale the masses of the 
winding and momentum
modes of the strings are of order of $1/l_s$. This implies that for
energies $E\ll 1/l_s$ we can neglect these modes. On the other hand
the  radius of the $S^1$ is much larger than string scale. Therefore
for $E\ll 1/l_s$ the winding modes can be neglected but one has to
retain all the momentum modes. Effectively we can then treat the
circle as non-compact. 
We now discuss the symmetries preserved by this configuration of
D-branes.
The $SO(1,9)$ symmetry of
10 dimensions is broken down to $SO(1,1)\times SO(4)_E \times
SO(4)_I$. The $SO(4)_E$ stands for rotations of the $6,7,8,9$
directions. As the $6,7,8,9$ directions are compactified on the torus
$T^4$, the $SO(4)_I$ symmetry is also broken. But we can still use the
$SO(4)_I$ algebra to classify states and organize fields. This 
configuration of D-branes preserves 8 supersymmetries out of the 32
supersymmetries of type IIB theory. From the fact that we are
retaining only momentum modes along the $x^5$ the 
low energy effective action for the collective modes of this D-brane
configuration is $1+1$ dimensional. 

More precisely,
we shall see that the low-energy dynamics of this D-brane system is
described by a $U(Q_1)\times U(Q_5)$ gauge theory in two dimensions
with ${\mathcal N}=4$ supersymmetry \cite {Mal96,Has-Wad97b}. 
The gauge theory
will be assumed to be in the Higgs phase because we are interested in
the bound state where the branes are not separated from each other in
the transverse direction. In order to really achieve this and prevent
branes from splitting off we will turn on the Fayet-Iliopoulos
parameters. 
We shall show in section \ref{sugramod}  that
in supergravity these parameters correspond to the vev of the
Neveu-Schwarz $B_{NS}$. In principle we can also turn on the $\theta $
term in the gauge theory. This corresponds to a vev of a certain
linear combination of the RR $0$-form and $4$-form.

\begin{figure}[t]
   \vspace{0.5cm}
\centerline{
   {\epsfxsize=10cm
   \epsfysize=8cm
   \epsffile{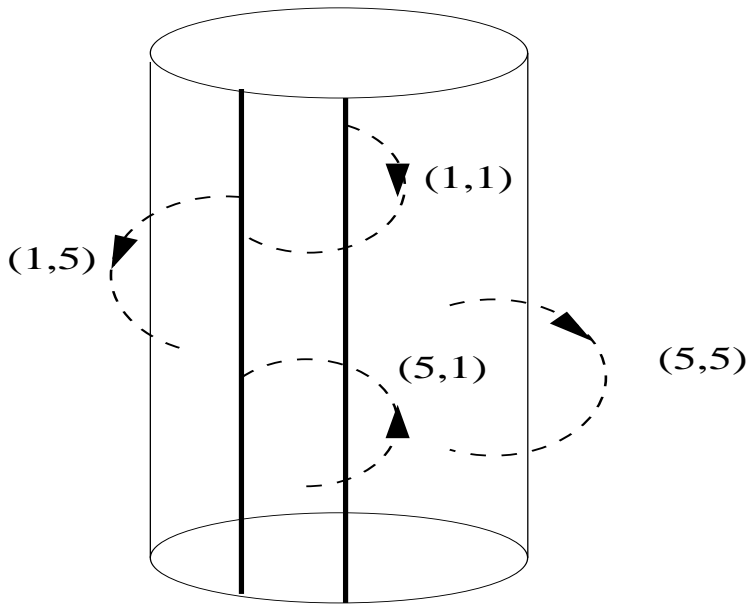}}
}
\caption{Open strings in the D1-D5 system}
\label{f-d1d5}
\end{figure}

The elementary excitations of the D-brane system 
(see Figure \ref{f-d1d5}) 
correspond to open
strings with two ends attached to the branes and there are three
classes of such strings: the (1,1), (5,5) and (1,5) strings. The
associated fields fall into vector multiplets and hypermultiplets,
using the terminology of ${\mathcal N}=2, D=4$ supersymmetry. 

\gap1
\noindent\underbar{\it $(1,1)$ strings} 
\gap1

\ni
The part of the spectrum coming from (1,1) strings is
simply the dimensional reduction, to $1+1$ dimensions (the
$(t,x^5)$-space), of the ${\mathcal N}=1, U(Q_1)$ gauge theory in $9+1$
dimensions \cite{Pol-tasi96}.  

The bosonic fields of this theory can be organized into the vector
multiplet and the hypermultiplet of ${\mathcal N}=2$ theory in
four-dimensions as
\bea
\mbox{Vector multiplet:} \; A_0^{(1)}, A_5^{(1)}, Y_m^{(1)}, m=1,2,3,4
\\   \nonumber
\mbox{Hypermultiplet:} \; Y_i^{(1)}, i=6,7,8,9
\eea
The $A_0^{(1)}, A_5^{(1)}$ are the $U(Q_1)$ gauge fields in the
non-compact directions. The $Y_m^{(1)}$'s and $Y_i^{(1)}$'s are gauge
fields in the compact directions of the ${\mathcal N}=1$ super
Yang-Mills in ten-dimensions. They are hermitian $Q_1\times Q_1$
matrices transforming as adjoints of $U(Q_1)$. The hypermultiplet of
${\mathcal N}=2$ supersymmetry are doublets of the $SU(2)_R$ symmetry
of the theory. The adjoint matrices $Y_i^{(1)}$'s can be arranged as
doublets under $SU(2)_R$ as
\be
N^{(1)} = 
\left(
\begin{array}{c}
N_1^{(1)} \\ \nonumber
N_2^{(1)\dagger}
\end{array}
\right)
=\left(
\begin{array}{c}
Y_9^{(1)} + i 
Y_8^{(1)}  \\
Y_7^{(1)} - i 
Y_6^{(1)} 
\end{array}
\right)
\ee

\gap1

\noindent\underbar{\it $(5,5)$ strings} 

\gap1
\ni
The field content of these massless open strings is similar to the the
$(1,1)$ strings except for the fact that the gauge group is $U(Q_5)$
instead of $U(Q_1)$.  Normally one would have expected the gauge
theory of the $(5,5)$ strings to be a dimensional reduction of
${\mathcal N}=1$ $U(Q_5)$ super Yang-Mills to $5+1$ dimensions. In the
region \eq{4:scales} where we are working, we can ignore the
Kaluza-Klein modes on $T^4$, effectively leading to a theory in $1+1$
dimensions. The vector multiplets and the hypermultiplets are given by
\bea
\mbox{Vector multiplet:} \; A_0^{(5)}, A_5^{(5)}, Y_m^{(5)} \, m=1,2,3,4
\\   \nonumber
\mbox{Hypermultiplet:} \; Y_i^{(5)} \, i=6,7,8,9
\eea
The $A_0^{(5)}, A_5^{(5)}$ are the $U(Q_5)$ gauge fields in the
non-compact directions. The $Y_m^{(5)}$'s and $Y_i^{(5)}$'s are gauge
fields in the compact directions of the ${\mathcal N}=1$ super
Yang-Mills in ten-dimensions. They are hermitian $Q_5\times Q_5$
matrices transforming as adjoints of $U(Q_5)$. The hypermultiplets
$Y_i^{(5)}$'s can be arranged as doublets under $SU(2)_R$ as
\be
N^{(5)} = 
\left(
\begin{array}{c}
N_1^{(5)} \\ \nonumber
N_2^{(5)\dagger}
\end{array}
\right)
=
\left(
\begin{array}{c}
Y_9^{(5)} + i 
Y_8^{(5)}  \\
Y_7^{(5)} - i 
Y_6^{(5)} 
\end{array}
\right)
\ee
Since $x^m$ are compact, the (1,1) strings can also have winding modes
around the $T^4$. These are, however, massive states in the
$(1+1)$-dimensional theory and can be ignored. This is because their
masses are proportional to $1/\sqrt{\alpha '}$ (see
\eq{4:scales}), which can be neglected for energies $E\ll 1/l_s$. 
Similarly, the part of the spectrum coming from (5,5)
strings is the dimensional reduction, to $5+1$ dimensions, of the
${\mathcal N}=1, U(Q_5)$ gauge theory in $9+1$ dimensions. In this
case, the gauge field components $A^{(5)}_m$ ($m=6,7,8,9$) also have a
dependence on $x^m$. Momentum modes corresponding to this dependence
are neglected because the size of the 4-torus is of the order of the
string scale $\sqrt{\alpha '}$. The neglect of the winding modes of
the $(1,1)$ strings and the KK modes of the $(5,5)$ strings is
consistent with T-duality.  A set of four T-duality transformations
along $x^m$ interchanges D1- and D5-branes and also converts the
momentum modes of the (5,5) strings along $T^4$ into winding modes of
(1,1) strings around the dual torus \cite{WT}. Since these winding
modes have been ignored, a T-duality covariant formulation requires
that we should also ignore the associated momentum modes.

\gap1

\noindent\underbar{\it $(1,5)$ and $(5,1)$ strings} 

\gap1
\ni
The field content obtained so far is that of ${\mathcal N}=2, \,
U(Q_1)\times U(Q_5)$ gauge theory, in 1+5 dimensions, reduced to 1+1
dimensions on $T^4$.

The $SO(4)_I\sim SU(2)_L\times SU(2)_R$ rotations on the tangent space
of the torus act on the components of the adjoint hypermultiplets
$X^{(1,5)}_i$ as an $R$-symmetry. To this set of fields we have to add
the fields from the (1,5) sector that are constrained to live in 1+1
dimensions by the ND boundary conditions. These strings have their
ends fixed on different types of D-branes and, therefore, the
corresponding fields transform in the fundamental representation of
both $U(Q_1)$ and $U(Q_5)$.  The ND boundary conditions have the
important consequence that the (1,5) sector fields form a
hypermultiplet which is chiral w.r.t. $SO(4)_I$.  The chirality
projection is due to the GSO projection. Hence the $R$-symmetry group
is $SU(2)_R$.
\be
\chi = 
\left(
\begin{array}{c}
A \\
B^{\dagger}
\end{array}
\right)
\ee
A few comments are in order:
\begin{enumerate}
\item The inclusion of these
fields, coming from the (1,5) and (5,1) strings, breaks the
supersymmetry by half, to the equivalent of ${\mathcal N}=1$ in $D=6$,
and the final theory only has $SU(2)_R$ $R$-symmetry.
\item The fermionic superpartners of these hypermultiplets which 
arise from the Ramond sector of the massless
excitations of $(1,5)$ and $(5,1)$ strings carry spinorial indices
under $SO(4)_E$ and they are singlets under $SO(4)_I$.
\item The $U(1) \times \overline{U(1)}$ subgroup is important. 
One combination involving the sum of the $U(1)$'s 
leaves the hypermultiplet invariant. 
$(A_{a'a}, B_{a'a})$ have charges
$(+1, -1)$ under the relative $U(1)$.
\item $\chi$ is a chiral spinor of $SO(4)_I$ with convention 
$\Gamma_{6789} \ \chi =  \chi$ .
\item 
Since we are describing the Higgs phase in which all the branes sit on
top of each other we have $Y_i^{(1,5)}=0$.
\item
There are two coupling constants in the gauge theory, the
coupling constant of the D1-brane gauge theory 
$g_1^2 = g_s\/(2\pi \alpha')$ and
the coupling constant of the D5-brane gauge theory 
$g_5^2 = g_s\/(2\pi \alpha' \tilde{v}$ Here $\tilde{v}$ is related to
the volume of $T^4$ by $V_{T^4} = \alpha^{\prime 2} (2\pi)^4\tilde{v}$.
Since we are interested in low energies $E\ll 1/l_s$ the 
gauge theory is strongly coupled. 
\item In the above discussion from the geometry of the configuration, 
the fields $X_i^{(1,5)}$ along the
torus directions and the fields $\chi$ are compact, since they
parmetrize positions along the compact directions.
However, 
it is not consistent with gauge invariance 
to take hypermultiplets of the 
${\mathcal N} =2$ mulitplet to be compact 
\footnote{
This can be seen as
follows.
The D-term equations in the ${\mathcal N} =2$ theory admit a symmetry
under the complexified gauge symmetry $GL(C, Q_1) \times GL(C, Q_5)$,
(see for instance in \cite{witt-higgs-br}). This symmetry involves an
arbitary scaling.}.
Therefore the hypermultiplets are non-compact. 
Since we are interested in energies $E\ll 1/l_s$ the expectation values
of the hypers (which have units of energy )
$X_i^{(1,5)}, \chi \ll 1/l_s$. Thus, the $(1,1), (5,5) (1,5)$ strings 
do not probe the entire domain of $T^4$. Therefore even though
geometrically we are on $T^4$ it is  consistent with the fact that the
hypers are non-compact as we are intersted in $E\ll 1/l_s$. 
\item
The above discussion of domain of validity of the gauge theory ties in
nicely with the fact that the gauge theory is an approximate
description while the instanton moduli space description is a more
global description. The gauge theory is valid when the hypers get
small expectation values. The hyper-multiplet of the gauge theory
corresponds to the scales of the instanton (via the ADHM construction
for instantons on $R^4$). 
Thus the gauge theory is
valid when the instantons have shrunk to almost zero size
\cite{Douglas:1997vu}.
\end{enumerate}
In summary, the gauge theory of the \dd  system is a $1+1$ dimensional
$(4,4)$ supersymmetric gauge theory with gauge group $U(Q_1)\times
U(Q_5)$. The matter content of this theory consists of hypermultiplets
$Y^{(1)}$'s, $Y^{(5)}$'s transforming as adjoints of $U(Q_1)$ and
$U(Q_5)$ respectively. It also has the hypermultiplets $\chi$'s which
transform as bi-fundamentals of $U(Q_1)\times \overline{U(Q_5)}$.

\subsection{The Potential Terms}

The Lagrangian of the above gauge theory can be worked out from the 
dimensional reduction of $d=6$,\, ${\mathcal N}=1$ gauge theory.
The potential energy density of the vector and hyper multiplets is a
sum of 4 positive terms (in this section for convenience of notation 
we have defined $Y_i^{(1)}=Y_i, Y_i^{(5)}=X_i, 
Y_m^{(1)}=Y_m, Y_m^{(5)}=X_m$) \cite{Mal96, Has-Wad97b} : 
\bea
\label{v}
V &=& V_1 + V_2 + V_3 + V_4 \\ [2mm]
\label{v.1}
V_1 &=& - {1\over 4g^2_1} \sum_{m,n} 
\mbox{Tr}_{U(Q_1)} [Y_m, Y_n]^2 - {1\over
4g^2_5} \sum_{m,n} \mbox{Tr}_{U(Q_5)} [X_m, X_n]^2 \\ [2mm]
\label{v.2}
V_2 &=& - {1\over 2g^2_1} \sum_{i,m} 
\mbox{Tr}_{U(Q_1)} [Y_i, Y_m]^2 - {1\over
2g^2_5} \sum_{i,m} \mbox{Tr}_{U(Q_5)}[X_i, X_m]^2 \\ [2mm]
\label{v.3}
V_3 &=& {1\over 4} \sum_m \mbox{Tr}_{U(Q_1)} (\chi X_m - Y_m \chi) (X_m
\chi^\dagger -
\chi^\dagger Y_m) \\ [2mm]
V_4 &=& {1\over 4} 
\mbox{Tr}_{U(Q_1)} (\chi i \Gamma^T_{ij} \chi^+ + i [Y_i,
Y_j]^+ - \zeta^+_{ij} {1\!\!\!1 \over Q_1})^2 \nonumber \\ [2mm]
\label{v.4}
&& + {1\over 4} \mbox{Tr}_{U(Q_5)} (\chi^+ i \Gamma_{ij} \chi + i [X_i,
X_j]^+ - \zeta^+_{ij} {1\!\!\!1 \over Q_5})^2
\eea
The potential energy $V_4$ comes from a combination of $F$ and $D$
terms of the higher dimensional gauge theory. $\Gamma_{ij} = {i\over 2}
[\Gamma_i, \Gamma_j]$ are spinor rotation matrices. The notation
$a^+_{ij}$ denotes the self-dual part of the anti-symmetric tensor
$a_{ij}$.

In $V_4$ we have included the Fayet-Iliopoulos (FI) terms
$\zeta^+_{ij}$, which form a triplet under $SU(2)_R$. Their inclusion
is consistent with ${\mathcal N}=4$ SUSY. The FI terms can be
identified with the self dual part of $B_{ij}$, the anti-symmetry
tensor of the NS sector of the closed string theory
\cite{Sei-Wit99}. This identification at this stage rests on the fact
that (i) $\zeta^+_{ij}$ and $B^+_{ij}$ have identical transformation
properties under $SU(4)_I$ and (ii) at the origin of the Higgs branch
where $\chi = X = Y = 0$, $V_4 \sim \zeta^+_{ij}
\zeta^+_{ij}$. This signals a tachyonic mode from the view point of
string perturbation theory
\cite{Sei-Wit99}. The tachyon mass is easily computed and
this implies the relation $\zeta^+_{ij} \zeta^+_{ij} \sim B^+_{ij}
B^+_{ij}$. These issues are discussed further
in Section 7.

\subsection{D-Flatness Equations and the Moduli Space}

The supersymmetric ground state (semi-classical) is characterized by
the 2-sets of D-flatness equations which are obtained by setting $V_4
= 0$. They are best written in terms of the $SU(2)_R$ doublet fields
$N^{(1)}$ and $N^{(5)}$ :
\bea
N^{(1)} &=& \pmatrix{N^{(1)}_1 \cr N^{(1)\dagger}_2} 
= \pmatrix{Y_9 + i Y_8
\cr Y_7 + i Y_6} \nonumber \\ [2mm]
\label{3.7}
N^{(5)} &=& \pmatrix{N^{(5)}_1 \cr N^{(5)\dagger}_2} = 
\pmatrix{X_9 + iX_8 \cr X_7 + iX_9}
\eea

We also define $\zeta=\zeta^{+}_{69}$ and $\zeta_{c}=
\zeta^{+}_{67}+ i\zeta^{+}_{68}$.
With these definitions the 2 sets of D-flatness conditions become:
\bea
\label{D-terms}
(AA^{\dagger} - B^{\dagger}B)_{a'b'} + 
[N^{(1)}_1, N^{(1)\dagger}_1]_{a'b'} -
[N^{(1)}_2,
N^{(1)\dagger}_2]_{a'b'} = {\zeta \over Q_1} \delta_{a'b'} \\ \nonumber
(AB)_{a'b'} + [N^{(1)}_1, N^{(1)\dagger}_2]_{a'b'} = {\zeta_{c} \over
Q_1} \delta_{a'b'} \\ \nonumber
(A^{\dagger}A - BB^{\dagger})_{ab} + 
[N^{(5)}_1, N^{(5)\dagger}_1]_{ab} - [N^{(5)}_2,
N^{(5)\dagger}_2]_{ab} = {\zeta \over Q_5} \delta_{ab} \\ \nonumber
(A^+B^+)_{ab} + [N^{(5)}_1, N^{(5)\dagger}_2]_{ab} = {\zeta_{c} \over
Q_5} \delta_{ab} \nonumber
\eea
Here $a', b'$ runs from $1\cdots Q_1$ and $a,b$ runs from $1\cdots Q_5$.
The hypermultiplet moduli space is a solution of the above equations
modulo the gauge group $U(Q_1) \times U(Q_5)$. A detailed discussion
of the procedure was given in \cite {Has-Wad97b,DMWY}. 
Below we summarize the main points.

If we take the trace parts of (\ref{D-terms})
we get the {\em same} set of 3
equations as the D-flatness equations for  a $U(1)$ theory
with $Q_1 Q_5$ hypermultiplets, with $U(1)$ charge assignment $(+1,
-1)$ for $(A_{a'b}, B^T_{a'b})$. Thus,
\bea
\label{Tcpn1}
\sum_{a'b} (A_{a'b} A^\ast_{a'b} - B^T_{a'b} B^{T\ast}_{a'b}) &=& \zeta
\\
\label{Tcpn2}
\sum_{a'b} A_{a'b} B^T_{a'b} &=& \zeta_c
\eea
For a given point on the surface defined by (\ref
{Tcpn1}),(\ref{Tcpn2}) the traceless parts of (\ref{D-terms}) lead to
$3Q^2_1 + 3Q^2_5 - 6$ constraints among $4Q^2_1 + 4Q^2_5 - 8$ degrees
of freedom corresponding to the traceless parts of the adjoint
hypermultiplets $N^{(1)}$ and $N^{(5)}$. Using $Q^2_1 + Q^2_5 - 2$
gauge conditions corresponding to $SU(Q_1) \times SU(Q_5)$ we have
$(3Q^2_1 + 3Q^2_5 - 6) + (Q^2_1 + Q^2_5 - 2) = 4Q^2_1 + 4Q^2_5 - 8$
conditions for the $(4Q^2_1 + 4Q^2_5 -8)$ degrees of freedom in the
traceless parts of $N^{(1)}$ and $N^{(5)}$. The 8 degrees of freedom
corresponding to $\mbox{Tr} X_i$ and 
$\mbox{Tr}Y_i$, $i = 6,7,8,9$ correspond to the
centre-of-mass of the D5 and D1 branes respectively.

\subsection{The Bound State in the Higgs Phase}

Having discussed the moduli space that characterizes the SUSY ground
state we can discuss the fluctuations of the transverse vector
multiplet scalars $X_m$ and $Y_m$, $m = 1,2,3,4$. In the Higgs phase
since $\langle X_m \rangle = \langle Y_m \rangle = 0$ and $\chi =
\overline \chi$ lies on the surface defined by 
(\ref{Tcpn1}),(\ref{Tcpn2}).  The relevant action of fluctuations in
the path integral is,
\bea
S = \sum_m \int  && dt dx_5 (\mbox{Tr}_{U(Q_5)} \partial_\alpha X_m
\partial^\alpha X_m + 
\mbox{Tr}_{U(Q_1)} \partial_\alpha Y_m \partial^\alpha
Y_m)
\\ \nonumber
+ \int && dt dx_5 (V_2 + V_3)
\eea

We restrict the discussion to the case when $Q_5=1$ and $Q_1$ is
arbitrary.  In this case the matrix $X_m$ is a real number which we
denote by $x_m$.  $\chi$ is a complex column vector with components
$(A_{a'},B_{a'})$. Since we are looking at the
fluctuations of the $Y_m$ only to quadratic order in the path
integral, the integrals over the different $Y_m$ decouple from each
other and we can treat each of them separately. Let us discuss the
fluctuation $Y_1$ and set $(Y_1)_{a'b'}=\delta_{a'b'} y_{1{a'}}$. Then
the potential $V_3$, (\ref{v.3}) becomes
\bea
V_3= \sum_{a'} (|A_{a'}|^2 + |B_{a'}|^2 )(y_{1a'} -x_1)^2
\eea
We will prove that $|A_{a'}|^2 + |B_{a'}|^2$ can never vanish if the
FI terms are non-zero.  In order to do this let us analyze the second 
D-term equation (\ref{D-terms})
\bea
\label{Tcpn-3}
A_{a'}B_{b'}+ [N^{(1)}_1,N^{(1)\dagger}_2]_{a'b'} = {\zeta_c \over Q_1}
\delta_{a'b'}
\eea
We can use the complex gauge group $GL(C,Q_1)$ to diagonalize the
complex matrix $N^{(1)}_1$ \cite {witt-higgs-br}. Then, (\ref{Tcpn-3})
becomes
\bea
A_{a'}B_{b'}+ (n_{a'} - n_{b'})(N^{(1)\dagger}_2)_{a'b'} ={\zeta_c \over
Q_1}
\delta_{a'b'}
\eea
For $a' \neq b'$, this determines the non-diagonal components of
$N^{(1)}_2$
\bea
(N^{(1)\dagger}_2)_{a'b'} = - { A_{a'}B_{b'} \over n_{a'} - n_{b'} }
\eea
For $a=b$, we get the equations
\bea
A_{a'}B_{a'}={\zeta_c \over Q_1} ,
\eea
which imply that
\bea
|A_{a'}| |B_{a'}| = { |\zeta_c| \over Q_1}
\eea
with the consequence that $|A_{a'}|$ and $|B_{a'}|$ are non-zero for
all $a' = 1,..,Q_1$. This implies that $(|A_{a'}|^2 + |B_{a'}|^2) >
0)$, and hence the fluctuation $(y_{1a'} - x_1)$ is massive. If we
change variables $y_{1a'}
\rightarrow y_{1a'} +x_1$, then $x_1$ is the only flat direction. This
corresponds to the global translation of the 5-brane in the $x_1$
direction.

A similar analysis can be done for all the remaining directions
$m=2,3,4$ with identical conclusions. This shows that a non-zero FI
term implies a true bound state of the $Q_5=1$, $Q_1=N$ system.  If
$FI=0$, then there is no such guarantee and the system can easily
fragment, due to the presence of flat directions in $(Y_m)_{a'b'}$.

What the above result says is that when the FI parameters are non-zero
the zero mode of the fields $(Y_m)_{a'b'}$ is massive. If we regard
the zero mode as a collective coordinate then the Hamiltonian of the
zero mode has a quadratic potential which agrees with the near horizon
limit of the Liouville potential derived in
\cite {Sei-Wit99,DMWY}.

The general case with an arbitrary number of $Q_1$ and $Q_5$ branes
seems significantly harder to prove and is an open question, 
but the result is very plausible
on physical grounds. If the potential for a single test $D1$ brane is
attractive, it is hard to imagine any change in this fact if there are
2 test $D1$ branes, because the $D1$ branes by themselves can form a
bound state.

\subsection{The Conformally Invariant Limit of the Gauge Theory}

In the previous section we showed that the \dd system leads to
a bound state in the Higgs phase. The next question is about
the low energy collective excitations of the bound state.
They are described by  
the sigma model  corresponding to
the hyper-multiplet moduli defined by the equations (\ref{D-terms}).
The Lagrangian (bosonic part) is
\bea
\label{sigmamodel}
S = \sum_m \int && dt dx_5 (tr_{U(Q_5)} \partial_\alpha X_i
\partial^\alpha X_i + tr_{U(Q_1)} \partial_\alpha Y_i \partial^\alpha
Y_i) \\ \nonumber
+ \int && dt dx_5 (\partial_\alpha \chi \partial^\alpha \chi^{\dagger})
\eea
This is a difficult non-linear system, with ${\mathcal N}=4$
SUSY. Since we are interested in the low energy dynamics we may
as well ask
whether there is a SCFT fixed point. This SCFT fixed point is relevant
in the study of the near horizon geometry (\ref{2:ads}). 
Such a SCFT must have (4,4)
supersymmetry (16 real supersymmetries) with a central charge
$c=6(Q_1Q_5 +1)$.  Now note that the equations
(\ref{Tcpn1}),(\ref{Tcpn2}) describe a hyper-Kahler manifold and hence
the sigma model defined on it is a SCFT with (4,4) SUSY. We can then
consider the part of the action involving the $X_i$ and $Y_i$ which
are solved in terms of the $\chi $ as giving a deformation of the
SCFT. Now this deformation clearly breaks the superconformal symmetry.

The sigma model action at the conformally invariant point is
\be
\int dt dx_5\sum_{a'b} (\del_\alpha A_{a'b} 
\del_\alpha A^\ast_{a'b} +
\del_\alpha B^T_{a'b}\del_\alpha B^{T\ast}_{a'b})
\ee
The sigma model fields are constrained to be on the surface defined by
(\ref{Tcpn1}),(\ref{Tcpn2}). Further after appropriate gauge fixing
the residual gauge invariance inherited from the gauge theory is the
Weyl group $S(Q_1)\times S(Q_5)$ \cite{Has-Wad97b}.  The Weyl
invariance can be used to construct gauge invariant strings of various
lengths. If $Q_1$ and $Q_5$ are relatively prime it is indeed possible
to prove the existence of a single winding string with minimum unit of
momentum given by $\frac{1}{Q_1Q_5}$. This is associated with the
longest cyclic subgroup of $S(Q_1)\times S(Q_5)$. Cyclic subgroups of
shorter length cycles lead to strings with minimum momentum
$\frac{1}{l_1l_5}$, where $l_1$ and $l_5$ are the lengths of the
cycles \cite{Has-Wad97b}. In a different way of describing these
degrees of freedom we shall see in the next sections that strings of
various lengths are associated with chiral primary operators of the
conformal field theory on the moduli space of instantons on a 4-torus.

\subsection{Quick derivation of entropy and temperatures from CFT
\label{sec:quick}}

We pause in this section to show that certain deductions about
thermodynamic properties can be made just by the knowledge of the
central charge ($c= 6 Q_1 Q_5$) and the level of the Virasoro algebra
of the unitary superconformal field theory mentioned above. This
information is sufficient to calculate the number of microstates (a
more detailed and complete derivation is given in Section 8).

To find the microstates of the \dd black hole we look for states with
$L_0 =N_L$ and $\bar{L}_0 = N_R$.  The asymptotic number of distinct
states of this SCFT is given by Cardy's formula \cite{Cardy:ie}
\bea
\Omega &&= \Omega_L \Omega_R
\nn\\
\Omega_{L,R} &&= \exp[2\pi \sqrt{c N_{L,R}/6}]
=\exp [2\pi  \sqrt{Q_1Q_5 N_{L,R}}]
\label{cardy}
\eea
{}From the Boltzmann formula one obtains
\be
\label{microentropy}
S= \ln \Omega = 2\pi( \sqrt{Q_1Q_5 N_L}+\sqrt{Q_1Q_5 N_R})
\ee
This \underbar{exactly} reproduces the Bekenstein-Hawking entropy
\eq{non-extremal-entropy-special}. The exact
agreement is surprising since for arbitrary $N_R, N_L \not= 0$ 
the states being considered  are far from being supersymmetric.

The quantity $\Omega_L(N_L)$ defines (the number of states in) a
microcanonical ensemble, in which we fix the energy of the left-movers
to be
\[
E_{L}= \frac{N_L}{R_5}
\]
The equivalent canonical ensemble is defined by
\bea
Z_L && \equiv {\rm Tr}\ \exp[-\beta_L E_L] 
= \sum_{N_L=0}^{\infty}\Omega_L(N_L) \exp[-\beta_L N_L/R_5]
\nn\\
&& = \sum_{N_L=0}^{\infty}
 \exp[2\pi \sqrt{Q_1Q_5 N_L} -\beta_L N_L/R_5]
\eea
For large enough temperature $T_L
(=1/\beta_L)$, the sum in the second line is
dominated by a saddle point value that occurs at
\be
\frac{\beta_L}{R_5}= \pi \sqrt{\frac{Q_1 Q_5}{N_L}}
\ee
This determines the temperature of the left-movers; a
similar reasoning works for the right movers. The two
temperatures are given by
\be
\label{4:temp}
T_L=\frac{1}{\beta_L}=\frac1{\pi R_5}\sqrt{\frac{N_L}{Q_1Q_5}}
\;\;\; \mbox{and} \;\;\;
T_R=\frac{1}{\beta_R}=\frac1{\pi R_5}\sqrt{\frac{N_R}{Q_1Q_5}}
\ee
The temperature $T_H=1/\beta_H$ of the full system is conjugate
to the total energy $E=E_L +E_R$, and is  given by
\footnote{To derive the ``harmonic'' rule in \eq{4:harmonic}
(cf. also \eq{3:harmonic}), define the full partition function as
\[
Z \equiv  Z_L Z_R = {\rm Tr}\ \exp[-
(\beta_L E_L + \beta_R E_R) ]
\]
which we can rewrite as
\[
Z= {\rm Tr}\ \exp[-\frac12(\beta_L + \beta_R)\left( E_L + E_R \right)
-\frac12(\beta_L - \beta_R)\left( E_L - E_R \right)]
\]
Performing the sum over $E_L - E_R$ (equivalently, over $N_L - N_R$),
and ignoring the unimportant multiplicative constant we get
\[ Z= {\rm Tr}\ \exp[-\beta_H (E_L + E_R)]
\]
where
\[
\beta_H =\frac12(\beta_L + \beta_R)
\]
}
\bea
\beta_H &&= \frac12 \left( \beta_L + \beta_R \right)
\ \Rightarrow \frac1{T_H}= \frac12 (\frac1{T_L}+ \frac1{T_R})
\label{4:harmonic} \\
T_H &&= 1/\beta_H = \frac{2\sqrt{N_L N_R}}{\pi R_5\sqrt{Q_1Q_5}
\left(\sqrt {N_L} + \sqrt{ N_R}\right)}
\label{4:hawking}
\eea 
\mysec11{Comparison with Supergravity}

We have so far encountered three expressions for
temperature, in Eq. \eq{2:temp} in Section 2,
in Eqs. \eq{3:temp},\eq{3:harmonic} in Section 3,
and the expressions \eq{4:temp},\eq{4:hawking} above.
It will be interesting to compare them.

Note that the dilute gas regime \eq{dilute-gas}
implies (see Eqs \eq{non-extremal-5d})
\be
\sinh \alpha_1, \sinh \alpha_5 \gg 1, \sinh \alpha_n
\ee
In this region (see \eq{charges},\eq{def-nl-nr})
\[
Q_{1,5} \approx N_{1,5} \gg N_{\bar 1, \bar 5}
\] 
This gives
\[
r_0 \cosh \alpha_{1,5} \approx \frac{r_0}2 \exp[\alpha_{1,5}] 
\approx  \sqrt{c_{1,5} Q_{1,5}}
\]
Eq. \eq{def-nl-nr} gives 
\[
r_0 e^\alpha_n = 2 \sqrt{c_n N_L}, \;
r_0 e^{-\alpha_n} = 2 \sqrt{c_n N_R}
\]
Using these expressions (and also \eq{2:curious}) it is
easy to see that \eq{2:temp}, \eq{3:harmonic} both
reduce to \eq{4:hawking}. Also \eq{3:temp} reduce
to \eq{4:temp}.

\subsection{D1 branes as solitonic strings of the D5 gauge theory}

\label{chp4:instanton}

In the previous subsection we found that the Higgs branch of the gauge
theory of the \dd system flows in the infrared to ${\mathcal N}
=(4,4)$ SCFT on a target space $\widetilde{\mathcal M}$ 
with central charge
$6Q_1Q_5$ (we have excluded the centre of mass degrees of
freedom). For black hole processes like Hawking radiation it is
important to have a better handle on the target space ${\mathcal
M}$. In this section we review the arguments which show that the
target space $\widetilde{\mathcal M}$ is a resolution of the orbifold
$T^4 \times (\tilde{T}^4)^{Q_1 Q_5}/S(Q_1Q_5)$. 
(We use the notation
$\tilde{T}^4$ to distinguish it from the compactification torus
$T^4$.) This discussion gives a succinct description of the bound
state in the Higgs phase.

The $Q_1$ D1-branes can be thought of as $Q_1$ instantons in the $5+1$
dimensional $U(Q_5)$ super Yang-Mills theory of the $Q_5$ D5-branes
\cite{chp2:Vafa1,chp2:Douglas}. To see this note that the DBI action of the 
D5-branes have a coupling
\be
\int d^6x \;\; C^{(2)}  \wedge \mbox{Tr} [F^{(5)} \wedge F^{(5)}]
\ee
The non-trivial gauge configurations which are independent of $x^0,
x^5$ and have zero values of $A_0^{(5)}$ and $A_5^{(5)}$ but non-zero
values of $\mbox{Tr} [F^{(5)} \wedge F^{(5)}]$ act as sources of the
Ramond-Ramond two-form $C^{(2)}_{05}$. If these gauge field
configurations have to preserve half the supersymmetries of the
D5-brane action they should be self dual. Thus they are instanton
solutions of four-dimensional Euclidean Yang-Mills of the $6,7,8,9$
directions.

Additional evidence for this comes from the fact that the integral
property of $\mbox{Tr} [F^{(5)}\wedge F^{(5)}]$ corresponds to the
quantization of the D1-branes charge. The action for a $Q_1$ instanton
solution is $Q_1/g_{YM}^2$. This agrees with the tension of $Q_1$
D1-branes, namely $Q_1/g_s$. If one is dealing with non-compact
D5-branes and D1-branes it is seen that the D-flatness conditions of
the D1-brane theory is identical to the ADHM construction of $Q_1$
instantons of $U(Q_5)$ gauge theory \cite{chp2:Douglas}.  In fact, the
last two equations in \eq{D-terms} are the relevant ADHM equations in
this case \cite{Witten:1995tz}.

{}From the discussion in the preceding paragraphs we conclude that,
excluding Wilson lines of the  $U(Q_5)$ gauge theory, 
$\widetilde{\mathcal M}$  can be thought of as the
moduli space of $Q_1$ instantons of a $U(Q_5)$ gauge theory on
$T^4$. This moduli space is known to be the Hilbert scheme of $Q_1Q_5$
points on $\tilde{T}^4$ \cite{chp2:Dijkgraaf}. $\tilde{T}^4$ can be
different from the compactification torus $T^4$.  This is a smooth
resolution of the singular orbifold
$(\tilde{T}^4)^{Q_1Q_5}/S(Q_1Q_5)$. We will provide physically
motivated evidence for the fact that the moduli space of $Q_1$
instantons of a $U(Q_5)$ gauge theory on $T^4$ is a smooth resolution
of the orbifold $(\tilde{T}^4)^{Q_1Q_5}/S(Q_1Q_5)$ using string
dualities. The evidence is topological and it comes from realizing
that the cohomology of $\widetilde{\mathcal M}$ is the degeneracy of the ground
states of the \dd gauge theory. We can calculate this degeneracy in
two ways. One is by explicitly counting the cohomology of
$(\tilde{T}^4)^{Q_1 Q_5}/S(Q_1 Q_5)$ \cite{chp2:Vafa2}.  The second
method is to use string dualities as discussed below. Both these
methods give identical answers. Thus at least at the level of
cohomology we are able to verify that the moduli space of $Q_1$
instantons of a  $U(Q_5)$ gauge theory on $T^4$ is smooth resolution
of $(\tilde{T}^4)^{Q_1 Q_5}/S(Q_1 Q_5)$.

Consider type IIB string theory compactified on $S^1\times T^4$ with a
fundamental string having $Q_5$ units of winding along $x^6$ and $Q_1$
units of momentum along $x^6$. On performing the sequence of dualities
$ST_{6789}ST_{56}$ we can map the fundamental string to the \dd 
system (we can de-compactify the $x^5$ direction finally) with $Q_1$
D1-branes along $x^5$ and $Q_5$ D5-branes along $x^5, x^6, x^7, x^8,
x^9$ \cite{chp2:Sen}. Therefore using this U-duality sequence the BPS
states of this fundamental string (that is, states with either purely
left moving or right moving oscillators) maps to ground states of the
\dd system. The number of ground states of the \dd system is given
by the dimension of the cohomology of $\widetilde{\mathcal M}$. From the
perturbative string degeneracy counting the generating function of BPS
states with left moving oscillator number $N_L$ is given by
\be
\sum_{N_L=o}^\infty d(N_L)q^{N_L} = 256\times \prod_{n=1}^\infty
\left(\frac{1+ q^n}{1-q^n} \right)^8
\ee
where $d(N_L)$ refers to the degeneracy of states with left moving
oscillator number $N_L$. The \dd system is U-dual to the perturbative
string with $N_L=Q_1Q_5$.  

Explicit counting of the cohomology of
$(\tilde{T}^4)^{Q_1Q_5}/S(Q_1Q_5)$ gives
$d(Q_1Q_5)/256$. The factor $256$ comes from quantization of the
center of mass coordinate along $1,2,3,4$ directions and the $6,7,8,9$
directions. The center of mass coordinate is represented by the
$U(1)$ of $U(Q_1)\times U(Q_5)$. Therefore the low energy
theory of the bound \dd system is a SCFT on the target space
\be
R^4\times T^4\times (\tilde{T^4})^{Q_1 Q_5}/S(Q_1 Q_5) 
\ee
It is also useful to interpret the moduli represented by the free $T^4$
form the gauge theory of $Q_5$ D5-branes. As this theory is on a torus
$T^4$ it admits Wilson-lines along all the cyles of the $T^4$. The
free torus in the above equation stand for Wilson-lines in the theory
of the D5-branes. From the \dd gauge theory point of view the free 
$T^4$ belongs to the Higgs branch as it stands for the center of mass
coordinate on $T^4$ parametrized by 
$\rm{Tr}_{U(Q_5)} (Y^{(5)}_i) + \rm{Tr}_{U(Q_1)} (Y^{(1)}_i$) while
$R^4$ belongs to the Coulomb branch. It is parameterized by
$\rm{Tr}_{U(Q_5)} (Y^{(5)}_m) + \rm{Tr}_{U(Q_1)} (Y^{(1)}_m$). 
Thus the Higgs branch of the \dd gauge theory flows in the
infrared to a ${\mathcal N} =(4,4)$ SCFT on $T^4 
\times (\tilde{T^4})^{Q_1 Q_5}/S(Q_1 Q_5)$. 
The SCFT on $T^4$ is free while the symmetric product contains
interesting dynamics.  From now on we will denote the
symmetric product orbifold $(\tilde{T}^4)^{Q_1Q_5}/S(Q_1Q_5)$ by
${\mathcal M }$.

\newpage

\def\re#1{{\bf #1}}
\section{The SCFT on the orbifold ${\mathcal M}$}
\label{sec:cft}

As we have seen in the last section the Higgs branch of the \dd system
flows in the infrared to a product of a 
${\mathcal N} = (4,4)$ SCFT on a resolution of the orbifold
${\mathcal M}$ with a  free theory on $T^4$.
As the SCFT on $T^4$ is free and decoupled we 
will first focus on the symmetric product.
We will formulate the SCFT on ${\mathcal M}$ 
as a free field theory with identifications and
discuss its symmetries. In particular we find a  new $SO(4)$ algebra
which is useful in the classification of states.
We then construct the operators which correspond to the moduli of the
SCFT including the 
four operators which correspond to the
resolution of the orbifold. Finally we
explicitly construct 
the chiral primaries of the ${\mathcal
N}=(4,4)$ SCFT on the symmetric product orbifold ${\mathcal M}$. 

The ${\mathcal N}= (4,4)$ SCFT on ${\mathcal M}$ 
is described by the free Lagrangian
\be
\label{free}
S = \frac{1}{2} \int d^2 z\; \left[\del
x^i_A \bar\del x_{i,A} + 
\psi_A^i(z) \bar\del \psi^i_A(z) + 
\widetilde\psi^i_A(\bar z) \del \widetilde \psi^i_A(\bar z) 
 \right]
\ee
Here $i$ runs over the $\widetilde{ T^4}$ coordinates
1,2,3,4 and $A=1,2,\ldots,Q_1Q_5$ labels various copies
of the four-torus. The symmetric group $S(Q_1Q_5)$
acts by permuting the copy indices. It introduces various twisted
sectors which we will discuss later. The free field realization of
this SCFT has ${\mathcal N}=(4,4)$ superconformal symmetry. 
To set up our notations and conventions we review the
${\mathcal N}=4$ superconformal algebra.

\subsection{The ${\mathcal N}=4$ superconformal algebra}
%

The algebra is generated by the stress energy
tensor, four supersymmetry currents, and a local $SU(2)$ $R$ symmetry
current. The operator product expansions(OPE) of the algebra 
with central charge $c$ are given by (See for example \cite{chp2:Yu}.)
\bea
\label{chp2:scft-algebra}
T(z)T(w) &=& \frac{\del T(w)}{z-w} + \frac{2 T(w)}{(z-w)^2} +
\frac{c}{2 (z-w)^4},  \\  \nonumber
G^a(z)G^{b\dagger }(w) &=& 
\frac{2 T(w)\delta_{ab}}{z-w} + \frac{2 \bar{\sigma}^i_{ab} \del J^i}
{z-w} + \frac{ 4 \bar{\sigma}^i_{ab} J^i}{(z-w)^2} + 
\frac{2c\delta_{ab}}{3(z-w)^3}, \\
\nonumber
J^i(z) J^j(w) &=& \frac{i\epsilon^{ijk} J^k}{z-w} + \frac{c}{12 (z-w)^2}
, \\  \nonumber
T(z)G^a(w) &=& \frac{\del G^a (w)}{z-w} + \frac{3 G^a (z)}{2 (z-w)^2},
\\   \nonumber
T(z) G^{a\dagger }(w) &=& \frac{\del G^{a\dagger} (w)}{z-w} + 
\frac{3 G^{a\dagger} (z)}{2 (z-w)^2}, \\   \nonumber
T(z) J^i(w) &=& \frac{\del J^i (w)}{z-w} + \frac{J^i}{(z-w)^2}, \\
\nonumber
J^i(z) G^a (w) &=& \frac{G^b(z) (\sigma^i)^{ba}}{2 (z-w)}, \\
\nonumber
J^i(z) G^{a\dagger}(w) &=& -\frac{(\sigma^i)^{ab} G^{b\dagger
}(w)}{2(z-w)} 
\eea
Here $T(z)$ is the stress energy tensor, $G^a(z), G^{b\dagger}(z)$ the
$SU(2)$ doublet of supersymmetry generators and $J^i(z)$ the $SU(2)$
$R$ symmetry current. The $\sigma$'s stand for Pauli matrices and the
$\bar{\sigma}$'s stand for the complex conjugates of Pauli matrices.
In the free field realization described below, the above holomorphic
currents occur together with their antiholomorphic counterparts, which
we will denote by $\widetilde J(\bar z), \widetilde G(\bar z)$ and
$\widetilde T(\bar z)$. In particular, the R-parity group will be
denoted by $SU(2)_R \times \widetilde{SU(2)}_R$.

\subsection{Free field realization of ${\mathcal N}=(4,4)$ SCFT 
on the orbifold ${\mathcal M}$}

A free field realization of the ${\mathcal N}=4$ superconformal
algebra with $c=6Q_1Q_5$ can be constructed out of $Q_1Q_5$ copies of
four real fermions and bosons. The generators are given by
\bea
T(z) &=& \del X_A (z)\del X^\dagger_A(z) 
+ \frac{1}{2}\Psi_A (z)\del \Psi^{\dagger}_A (z) 
- \frac{1}{2}\del\Psi_A (z) \Psi^{\dagger}_A (z) 
\\   \nonumber
G^a(z) &=&
\left(
\begin{array}{c}
G^1(z)\\
G^2(z)
\end{array}
\right) =
\sqrt{2} \left(
\begin{array}{c}
\Psi^1_A (z) \\
\Psi^2_A (z) \end{array}  \right)  \del X^2_A (z)  +
\sqrt{2} \left( \begin{array}{c}
-\Psi^{2\dagger}_A (z) \\
\Psi^{1\dagger}_A (z)  \end{array}  \right)  \del X^1_A (z) 
\\    \nonumber
J^i_R(z) &=& \frac{1}{2} \Psi_A(z)\sigma^i\Psi^\dagger_A (z)\\
\nonumber
\eea
We will use the following notation for the zero mode
of the R-parity current:
\be
J^i_R = \frac{1}{2}\int\frac{dz}{2\pi i} 
\Psi_A(z)\sigma^i\Psi^\dagger_A(z)
\ee
In the above the summation over $A$ which runs from $1$ to $Q_1Q_5$ is
implied.
The bosons $X$ and the fermions $\Psi$ are 
\bea
\label{defn}
X_A(z) &=& (X^1_A(z), X^2_A(z)) = \sqrt{1/2} (x^1_A(z) + i x^2_A(z),
x^3_A(z) + i x^4_A(z)),   \\  \nonumber
\Psi_A (z) &=& (\Psi^1_A(z), \Psi^2_A(z)) = \sqrt{1/2} (\psi^1_A(z) +
i\psi^2_A(z), \psi^3_A(z) + i\psi^4_A(z)) \\   \nonumber
X_A^\dagger (z) &=& 
\left(
\begin{array}{c}
X_A^{1\dagger} (z) \\
X_A^{2\dagger} (z)
\end{array}
\right)   = \sqrt{\frac{1}{2}}
\left(
\begin{array}{c}
x^1_A(z)-ix^2_A (z)\\
x^3_A(z)-ix^4_A(z)
\end{array}
\right) \\   \nonumber
\Psi_A^\dagger (z) &=&
\left(
\begin{array}{c}
\Psi_A^{1\dagger} (z) \\
\Psi^{2_A \dagger} (z)
\end{array}
\right)     =\sqrt{\frac{1}{2}} 
\left(  
\begin{array}{c}
\psi^1_A(z) - i\psi^2_A(z)\\
\psi^3_A(z) - i\psi^4_A(z)
\end{array}
\right)
\eea

\subsection{The $SO(4)$ algebra}

In addition to the local $R$ symmetry the free field realization of
the ${\mathcal N}=4$ superconformal algebra has additional global
symmetries which can be used to classify the states. There are $2$
global $SU(2)$ symmetries which correspond to the $SO(4)$ rotations of
the $4$ bosons $x^i$. The corresponding charges are given by 
\bea
I_1^i &=& 
\frac{1}{4}\int\frac{dz}{2\pi i} X_A \sigma^i \del X_A^\dagger 
-\frac{1}{4}\int\frac{dz}{2\pi i} \del X_A \sigma^i X_A^\dagger 
+ \frac{1}{2}\int\frac{dz}{2\pi i}
\Phi_A \sigma^i \Phi_A^\dagger  \\  \nonumber
I_2^i &=& 
\frac{1}{4} 
\int\frac{dz}{2\pi i}{\mathcal X}_A\sigma^i\del{\mathcal X}_A^\dagger
-\frac{1}{4} 
\int\frac{dz}{2\pi i}\del{\mathcal X}_A\sigma^i{\mathcal X}_A^\dagger
\eea
Here 
\bea
{\mathcal X }_A = (X^1_A, -X^{2\dagger}_A) \;&\;&\;\;
{\mathcal X}^\dagger  =
\left(
\begin{array}{c}
X^{1\dagger}_A \\
-X^2_A 
\end{array}  \right) \nonumber \\
\Phi_A = (\Psi^1_A, \Psi^{2\dagger}_A ) \;&\;&\;\;
\Phi_A^\dagger =
\left(
\begin{array}{c}
\Psi^{1\dagger}_A  \\
\Psi^2_A 
\end{array} \right).
\eea
These charges are generators of $SU(2)\times SU(2)$ algebra:
\bea
[I_1^i, I_1^j] = i\epsilon^{ijk} I_1^k \;&\;&\;\;
[I_2^i, I_2^j] = i\epsilon^{ijk} I_2^k 
\\  \nonumber
[I_1^i, I_2^j] &=&0
\eea
The commutation relation of these new global charges with the various
local charges are given below
\bea
\label{so4-on-g}
[I_1^i, G^a(z)] =0 \;&\;&\;\;
[I_1^i, G^{a\dagger}(z)] =0 \\ \nonumber 
[I_1^i, T(z)] =0 \;&\;&\;\;
[I_1^i, J(z)]=0 \\ \nonumber 
[I_2^i, {\mathcal G}^a(z)] = 
\frac{1}{2}{\mathcal G}^{b} (z)\sigma^i_{ba} \;&\;&\;\;
[I_2^i, {\mathcal G}^{a\dagger} (z) ]
= - \frac{1}{2}\sigma^i_{ab}{\mathcal G}^{b\dagger}(z) \\ \nonumber
[I_2^i, T(z)] =0  \;&\;&\;\;
[I_2^i, J(z)]=0
\eea
where
\bea
{\mathcal G} = ( G^1, G^{2\dagger}) \;&\;&\;\;
{\mathcal G}^\dagger=
\left(
\begin{array}{c}
G^{1\dagger} \\
G^2
\end{array}
\right)
\eea

The following commutations relation show that 
the bosons transform as $({\bf 2}, {\bf 2})$ under $SU(2)_{I_1}\times
SU(2)_{I_2}$
\bea
\label{boson}
[I_1^i, X^a_A] = \frac{1}{2} X^b_A\sigma^i_{ba} \;&\;&\;\;
[I_1^i, X^{a\dagger}_A] = -\frac{1}{2}\sigma^i_{ab}X^{b\dagger}_A 
\\   \nonumber
[I_2^i, {\mathcal X}^a_A ] =
\frac{1}{2} {\mathcal X}^b_A \sigma^i_{ba}  \;&\;&\;\;
[I_2^i, {\mathcal X}^{a\dagger}_A] =
-\frac{1}{2}\sigma^i_{ab}{\mathcal X}^{b\dagger}_A 
\eea
The fermions transform as $({\bf 2},{\bf 1})$ under $SU(2)_{I_1}\times 
SU(2)_{I_2}$ as can be seen from the commutations relations
given below.
\bea
[I_1^i, \Phi^a_A] = \frac{1}{2}\Phi^b_A \sigma^i_{ba} \;&\;&\;\;
[I_1^i, \Phi^{a\dagger}_A] =
-\frac{1}{2}\sigma^i_{ab}\Phi^{b\dagger}_A  \\   \nonumber
[I_2^i, \Psi^a] =0  \;&\;&\;\;
[I_2^i, \bar{\Psi}^a]=0
\eea
We are interested in  studying the states of the ${\mathcal N}=(4,4)$ SCFT
on ${\mathcal M}$. The classification of the states and their symmetry
properties can be analyzed by studying the states of a free field
realization of a ${\mathcal N}=(4,4)$ SCFT on $R^{4Q_1Q_5}/S(Q_1Q_5)$. 
This
is realized by considering the holomorphic and the anti-holomorphic
${\mathcal N}=4$ superconformal algebra with $c=\bar{c}=6Q_1Q_5$
constructed out of $Q_1Q_5$ copies of four real fermions and bosons. 
So we have an anti-holomorphic component for each field, generator and
charges discussed above. These are labelled by the same symbols used
for the holomorphic components but distinguished by a tilde.

The charges $I_1, I_2$ constructed
above  generate $SO(4)$
transformations only on  the {\em holomorphic} bosons $X_A(z)$. 
Similarly, we can construct charges
$\widetilde{I_1}, \widetilde{I_2}$ which generate $SO(4)$
transformations only on  the {\em antiholomorphic} 
bosons $\widetilde{X_A}(\bar z)$. 
Normally one would expect these
charges to give rise to a global $SO(4)_{hol}
\times SO(4)_{antihol}$ symmetry. However, 
the kinetic term of the bosons in the
free field realization is not invariant under independent holomorphic
and antiholomorphic $SO(4)$ rotations. It is
easy to see, for example by using the Noether
procedure, that there is a residual $SO(4)$ symmetry
generated by the charges 
\bea
J_I= I_1  + \widetilde{I}_1 \;\;&\;&\;
\widetilde{J}_I = I_2 + \widetilde{I}_2
\eea
We will denote this symmetry as $SO(4)_I =
SU(2)_I\times \widetilde{SU(2)}_I$, where the
$SU(2)$ factors are generated by $J_I,
\widetilde{J}_I$. These charges satisfy the
property that (a) they 
correspond to $SO(4)$ transformations of the
bosons $X_A(z,
\bar z)= X_A(z) 
+ \widetilde{X_A}(\bar z)$ and (b) they fall into representations of the
${\mathcal{N}}=(4,4)$ algebra (as can 
be proved by using the commutation relations \eq{boson}
of the $I$'s). The bosons $X(z,\bar{z})$
transform as $(\bf 2 , \bf 2)$ under $SU(2)_I\times \widetilde{SU(2)}_I$. 

\subsection{The supergroup $SU(1,1|2)$}

The global part of the ${\mathcal N}=4$ superconformal algebra forms the
supergroup $SU(1,1|2)$. 
Let $L_{\pm,0} ,J^{(1),(2),(3)}_R$  be
the global charges of the currents
$T(z)$ and $J^{(i)}_R(z)$ and  $G^a_{1/2,-1/2} $  the
global charges of the supersymmetry currents $G^a(z)$ 
in the Neveu-Schwarz sector. From the OPE's \eq{chp2:scft-algebra}
we obtain the following commutation relations for the global charges.
\bea
[L_0, L_{\pm}] = \mp L_{\pm} \;\;&\;&\;\; [L_{1} , L_{-1}] = 2L_{0} 
\\ \nonumber
\{ G^a_{1/2} , G^{b\dagger}_{-1/2} \} &=& 2\delta^{ab}L_0 + 2
\sigma^i_{ab} J^{(i)}_{R} \\  \nonumber
\{ G^a_{-1/2} , G^{b\dagger}_{1/2} \} &=& 2\delta^{ab}L_0 - 2
\sigma^i_{ab} J^{(i)}_{R} \\  \nonumber
[J^{(i)}_R, J^{(j)}_R] &=& i\epsilon^{ijk}J^{(k)}_R \\ \nonumber
[L_0, G^a_{\pm 1/2}] = \mp\frac{1}{2} G^a_{\pm 1/2} \;\;&\;&\;\;[L_0, G^{a\dagger}_{\pm 1/2}] = \mp\frac{1}{2} G^{a\dagger}_{\pm 1/2} 
\\ \nonumber
[L_+ , G^a_{1/2}] = 0 \;\;&\;&\;\; [L_- , G^a_{-1/2}] = 0 \\ \nonumber
[L_- , G^a_{1/2}] = -G^a_{-1/2} \;\;&\;&\;\; [L_+ , G^a_{-1/2}] = G^a_{1/2} \\ \nonumber
[L_+ , G^{a\dagger}_{1/2}] = 0 \;\;&\;&\;\; [L_- , G^{a\dagger}_{1/2}] = 0 \\ \nonumber
[L_- , G^{a\dagger}_{1/2}] = -G^a_{-1/2} \;\;&\;&\;\; [L_+ , G^{a\dagger}_{-1/2}] = G^a_{1/2} \\ \nonumber
[J^{(i)}_R, G^a_{\pm 1/2} ] = \frac{1}{2} G^{b}_{\pm 1/2} (\sigma^i)^{ba}
\;\;&\;&\;\; [J^{(i)}_R, G^{a\dagger}_{\pm 1/2} ] = -\frac{1}{2}
 (\sigma^i)^{ba}G^{b\dagger}_{\pm 1/2} \\  \nonumber
\eea
The above commutation relations form the algebra of the
supergroup $SU(1,1|2)$. The global part of the ${\mathcal N}= (4,4)$
superconformal algebra form the super group $SU(1,1|2)\times
SU(1,1|2)$. 

\subsection{Short multiplets of $SU(1,1|2)$}
\label{chp2:short-multiplets}

The representations of the supergroup $SU(1,1|2)$
are classified according to the conformal weight 
and $SU(2)_R$ quantum number. The highest weight states
$ |\mbox{hw}\rangle = 
|h,{\bf j}_R,j_R^3 =j_R \rangle $ satisfy the following
properties
\bea
L_1 |\mbox{hw}\rangle = 0 &\;\;\; 
L_0 |\mbox{hw}\rangle = h|\mbox{hw}\rangle \\ \nonumber
J^{(+)}_{R}|\mbox{hw}\rangle =0  &\;\;\; 
J_R^{(3)}|\mbox{hw}\rangle = j_R|\mbox{hw}\rangle\\  \nonumber
G_{1/2}^a|\mbox{hw}\rangle =0 &\;\;\; G_{1/2}^{a\dagger}
|\mbox{hw}\rangle =0
\eea
where $J^+_R = J^{(1)}_R + i J^{(2)}_R$.
Highest weight states which satisfy 
$
G^{2\dagger }_{-1/2}|\mbox{hw}\rangle =0 ,\;\;\;
G^1_{-1/2}|\mbox{hw}\rangle =0
$
are chiral primaries. They satisfy $h=j$. We will denote 
these states as $|\mbox{hw}\rangle _{S}$. Short multiplets are
generated from the chiral primaries through the action of the raising
operators $J_{-}, G^{1\dagger }_{-1/2}$ and $G^2_{-1/2}$. The structure
of the short multiplet is given below
\bea
\label{short}
\begin{array}{cccc}
\mbox{States} & j & L_0 & \mbox{Degeneracy} \\
|\mbox{hw}\rangle_{S} & h & h& 2h+1 \\
G^{1\dagger }_{-1/2}|\mbox{hw}\rangle_{S}, 
G^2_{-1/2}|\mbox{hw}\rangle_{S} 
&\;\;\;\;\;h-1/2& \;\;\;\;\;h+1/2 \;\;\;\;\;& 
2h + 2h = 4h \\
G^{1\dagger }_{-1/2} G^2_{-1/2}|\mbox{hw}\rangle_{S} & h-1& h+1& 2h-1
\end{array}
\eea
The short multiplets of the supergroup $SU(1,1|2)\times SU(1,1|2)$ are
obtained by the tensor product of the above multiplet. We denote the
short multiplet of  $SU(1,1|2)\times SU(1,1|2)$ as
$(\bf{2h +1}, \bf{2h'+1})_S$. These stand for the degeneracy of the
bottom component, the top row in  \eq{short}. The top component of
the short multiplet are the states belonging to  the last row in
\eq{short}. The short multiplet $(\bf{2}, \bf{2})_S$ is special, it
terminates at the middle row of \eq{short}. For this case, the top
component is the middle row. These states have $h=\bar{h}=1$
and transform as $(\bf{1}, \bf{1})$ of $SU(2)_R\times
\widetilde{SU(2)}_R$. There are $4$ such states for each $(\bf{2},
\bf{2})_S$.

\subsection{The resolutions of the symmetric product} 
\label{chp2:res}

The Higgs branch of the \dd system at low energies 
apart from the SCFT on the free torus $T^4$ is a SCFT on a
resolution of the orbifold ${\mathcal M}$. 
So it is important for us
to understand the operators corresponding to the moduli and the
resolution of the orbifold ${\mathcal M}$. 
To this end we  construct all the marginal operators of the ${\mathcal
N}=(4,4)$ SCFT on the symmetric product orbifold 
${\mathcal M}$. We will
find the four operators which correspond to resolution of the orbifold
singularity. 

\subsubsection{The untwisted sector}
Let us first focus on the operators constructed from the
untwisted sector. The operators of lowest conformal weight 
are
\bea 
\label{chiral}
\Psi^1_A(z) \widetilde{\Psi}^1_A(\bar{z})  \;&\;&\;  
\Psi^1_A(z)\widetilde{\Psi}^{2\dagger}_A(\bar z) \\   \nonumber
\Psi^{2\dagger}_A(z)\widetilde{\Psi}^1_A(\bar z )  \;&\;&\;
\Psi^{2\dagger}_A(z)\widetilde{\Psi}^{2\dagger}_A(\bar z) 
\eea
where summation over $A$ is implied. These four operators have conformal
dimension $(h, \bar{h})=(1/2, 1/2)$ and
$(j_R^3, \widetilde{j}_R^3)= (1/2, 1/2)$ under
the R-symmetry  $SU(2)_R\times \widetilde{SU(2)}_R$.
Since $(h, \bar{h})=(j_R^3, \widetilde{j}_R^3)$,
 these operators are chiral primaries and
have non-singular operator product expansions (OPE) with the
supersymmetry currents 
$G^1(z), G^{2\dagger}(z), \widetilde{G}^1(\bar z),
\widetilde{G}^{2\dagger}(\bar z)$. 
These properties indicate that they belong to the bottom component of
the short multiplet $(\bf 2, {\bf 2})_S $ 
\footnote{We restrict our
operators to be single trace operators. This excludes operators of the
type $\sum_{A=1}^{Q_1Q_5} \Psi_A^1(z)
\sum_{B=1}^{Q_1Q_5} \tilde{\Psi}_B^{2\dagger}(\bar{z})$. 
Multi-trace
operators in the AdS/CFT correspondence has been discussed in 
\cite{Aharony:2001dp,Witten:2001ua}.}.
Each of the four chiral
primaries gives rise to four top components of the short multiplet
$(\bf 2, {\bf 2})_S$. They are given by the leading pole ($(z-w)^{-1}
(\bar z - \bar w)^{-1}$) in the OPE's
\bea
\label{OPE}
G^2(z)\widetilde{G}^2(\bar z){\mathcal P} (w, \bar w)
\; &\;&\; 
G^2(z)\widetilde{G}^{1\dagger}(\bar z){\mathcal P} (w, \bar w)
\\ \nonumber
G^{1\dagger}(z) \widetilde{G}^2(\bar z){\mathcal P}(w, \bar w) \; &\;& \;
G^{1\dagger}(z)\widetilde{G}^{1\dagger} 
(\bar z){\mathcal P} (w, \bar w)
\eea
where ${\mathcal P}$ stands for any of the four chiral primaries in
\eq{chiral}. From the superconformal algebra it is easily seen that
the top components constructed above have weights $(1,1)$ and
transform as $(\bf 1, {\bf 1} )$ under $SU(2)_R\times
\widetilde{SU(2)}_R$.  The OPE's \eq{OPE}\ can be easily evaluated. We
find that the $16$ top components of the $4 (\bf 2, {\bf 2})_S$ short
multiplets are $\del x_A^i \bar{\del} x_A^j$.

We classify the above operators belonging to the top component
according to representations of (a) the $SO(4)_I$ rotational symmetry
of the $\widetilde{T}^4$, (The four torus $\tilde{T}^4$ breaks this
symmetry but we assume the target space is $R^4$ for the
classification of states)
(b) $R$ symmetry of the SCFT and (c) the
conformal weights. As all of these operators belong to the top
component of $(\bf 2, \bf 2 )_{\bf S}$ the only property which
distinguishes them is the representation under $SO(4)_I$. 
The quantum
numbers of these operators under the various symmetries are
\bea
\label{untwist-operator}
\begin{array}{lccc}
\mbox{Operator}&SU(2)_I\times \widetilde{SU(2)}_I&
SU(2)_R\times\widetilde{SU(2)}_R& (h, \bar{h}) \\
\del x^{ \{ i }_A(z) \bar{\del}x^{ j\} }_A (\bar z) -
\frac{1}{4}\delta^{ij}
\del x^k_A(z) \bar{\del}x^k_A (\bar z) &(\bf 3, \bf 3) & (\bf 1,\bf 1) 
& (1, 1) \\  
\del x^{[i}_A(z) \bar{\del}x^{j]}_A (\bar z) & (\bf 3, \bf 1) +
(\bf 1, \bf 3) & (\bf 1, \bf 1)& (1,1) \\ 
\del x^i_A(z) \bar{\del}x^i_A (\bar z) &(\bf 1, \bf 1)& (\bf 1, \bf 1)
& (1,1)
\end{array}
\eea
Therefore we have $16$ marginal operators from the untwisted sector.
As these are top components they can be added to the free SCFT as
perturbations without violating the ${\mathcal N}=(4,4)$ supersymmetry.

\subsubsection{$Z_2$ twists.}
\label{chp2:z-2twists}

We now construct the marginal operators from the various twisted
sectors of the orbifold SCFT.  The twist fields of the SCFT on the
orbifold ${\mathcal M}$ are labeled by the conjugacy classes of the
symmetric group $S(Q_1 Q_5)$ 
\cite{chp2:VafWit,Dixon:1986jc,chp2:DijMooVerVer}. The conjugacy
classes consist of cyclic groups of various lengths. The various
conjugacy classes and the multiplicity in which they occur in
$S(Q_1Q_5)$ can be found from the solutions of the equation
\be
\sum nN_n = Q_1 Q_5
\ee
where $n$ is the length of the cycle and $N_n$ is the multiplicity of
the cycle. Consider the simplest nontrivial conjugacy class which is
given by $N_1 = Q_1 Q_5 -2, N_2 = 1$ and the rest of $N_n =0$.  
A representative element of this class is 
\be
\label{group-element}
(X_1\rightarrow X_2, \; X_2\rightarrow X_1), 
\; X_3\rightarrow X_3 , \; \ldots ,
\; X_{Q_1Q_5} \rightarrow X_{Q_1Q_5}
\ee
Here the $X_A$'s are related to the $x_A$'s appearing in 
the action \eq{free} by \eq{defn}.

To exhibit the singularity of this group action we go over to the
following new coordinates
\be
X_{cm} = X_1+ X_2 \;\; \mbox{and}\;\; \phi = X_1-X_2
\ee
Under the group action \eq{group-element}\ 
$X_{cm}$ is invariant and $\phi
\rightarrow -\phi$. Thus the singularity is {\em locally} of the type
$R^4/Z_2$. The bosonic twist operators for this orbifold singularity
are given by following OPE's
\cite{chp2:DixFriMarShe}
\bea
\del \phi^1 (z) \sigma^1(w, \bar{w} ) = 
\frac{ \tau^1(w, \bar w ) }{ (z-w)^{1/2} }  \; &\;& \;
\del {\phi}^{1\dagger} (z) \sigma^1(w, \bar{w} ) = 
\frac{ \tau'^1(w, \bar w ) }{ (z-w)^{1/2} }  \\   \nonumber
\del \phi^2 (z) \sigma^2(w, \bar{w} ) =
\frac{ \tau^2(w, \bar w ) }{ (z-w)^{1/2} }  \; &\;& \;
\del {\phi}^{2\dagger} (z) \sigma^2(w, \bar{w} ) =
\frac{ \tau'^2(w, \bar w ) }{ (z-w)^{1/2} }  \\  \nonumber
\bar{\del} \widetilde{\phi}^1 (\bar{z}) \sigma^1(w, \bar{w} ) = 
\frac{ \widetilde\tau'^1(w, \bar w ) }{ (\bar z-\bar w)^{1/2} }  \; &\;&
\;
\bar{\del} \widetilde{\phi}^{1\dagger} (\bar z) \sigma^1(w, \bar{w} ) = 
\frac{ \widetilde\tau^1(w \bar w ) }{ (\bar z-\bar w)^{1/2} } 
\\   \nonumber
\bar{\del} \widetilde{\phi}^2 (\bar z) \sigma^2(w, \bar{w} ) =
\frac{ \widetilde\tau'^2(w, \bar w ) }{ (\bar z-\bar w)^{1/2} }  \; &\;&
\;
\bar{\del} \widetilde{\phi}^{2\dagger} (\bar z ) \sigma^2(w, \bar{w} ) =
\frac{ \widetilde\tau^2(w, \bar w ) }{ (\bar z-\bar w)^{1/2} }  
\eea
The $\tau$'s are excited twist operators.
The fermionic twists are constructed from bosonized currents defined
by
\bea
\chi^1(z) = e^{iH^1(z)} \; &\;& \; \chi^{1\dagger}(z) = e^{-iH^1(z)} \\
\nonumber
\chi^2(z) = e^{iH^2(z)} \; &\;& \; \chi^{2\dagger}(z) = e^{-iH^2(z)} \\
\nonumber
\eea
Where the $\chi$'s, defined as
$\Psi_1 - \Psi_2$, are the superpartners of the bosons $\phi$.

{}From the above we construct the supersymmetric twist fields which act
both on fermions and bosons as follows:
\bea
\label{defSigma}
\Sigma^{(\frac{1}{2}, \, \frac{1}{2})}_{(12)} = \sigma^1(z,\bar z)
\sigma^2 (z,\bar z)
e^{iH^1(z)/2} e^{-iH^2(z)/2} 
e^{i\widetilde{H}^1(\bar z )/2} e^{-i\widetilde{H}^2(\bar z )/2}
\\  \nonumber
\Sigma^{(\frac{1}{2},\,  -\frac{1}{2})}_{(12)} = \sigma^1(z,\bar z)
\sigma^2 (z,\bar z)
e^{iH^1(z)/2} e^{-iH^2(z)/2} 
e^{- i\widetilde{H}^1(\bar z )/2} e^{i\widetilde{H}^2(\bar z )/2} 
\\  \nonumber
\Sigma^{(-\frac{1}{2}, \,  \frac{1}{2})}_{(12)} = \sigma^1(z,\bar z)
\sigma^2 (z,\bar z)
e^{-iH^1(z)/2} e^{+iH^2(z)/2} 
e^{i\widetilde{H}^1(\bar z )/2} e^{-i\widetilde{H}^2(\bar z )/2} 
\\   \nonumber
\Sigma^{(-\frac{1}{2}, \,  -\frac{1}{2})}_{(12)} = 
\sigma^1(z,\bar z) \sigma^2 (z,\bar z)
e^{-iH^1(z)/2} e^{+iH^2(z)/2} 
e^{-i\widetilde{H}^1(\bar z )/2} e^{+i\widetilde{H}^2(\bar z )/2} 
\\  \nonumber
\eea
The subscript $(12)$ refers to the fact that these twist operators were
constructed for the representative
group element \eq{group-element}\ which exchanges the $1$ and
$2$ labels of the coordinates of $\widetilde{T}^4$. 
The superscript stands
for the $(j^3_R, \widetilde{j}^3_R)$ quantum numbers.
The twist operators for the orbifold ${\mathcal M}$ belonging to the
conjugacy class under consideration is obtained by summing over these
$Z_2$ twist operators for all representative elements of this class.
\be
\Sigma^{(\frac{1}{2}, \,  \frac{1}{2})} =
\sum_{i=1}^{Q_1Q_5} \sum_{j=1, j\neq i}^{Q_1Q_5}
 \Sigma^{(\frac{1}{2}, \,  \frac{1}{2})}_{(ij)}
\ee
We can define the rest of the twist operators for the orbifold in a
similar manner. The conformal dimensions of these operators are
$(1/2,1/2)$. They transform as $(\bf 2 , \bf 2)$ under the $SU(2)_R
\times \widetilde{SU(2)}_R$ symmetry of the SCFT. They belong to the
bottom component of the short multiplet $(\bf 2, \bf 2)_S$. The
operator $\Sigma^{(\frac{1}{2}, \, \frac{1}{2})}$ is a chiral primary.
As before the $4$ top components of this short multiplet,
which we denote by
\bea
T^{(\frac{1}{2}, \, \frac{1}{2})}, \;\; T^{(\frac{1}{2}, \,
-\frac{1}{2})} \\ \nonumber
T^{(-\frac{1}{2}, \, \frac{1}{2})}, \;\;
T^{(-\frac{1}{2}, \, -\frac{1}{2})} 
\eea 
are given  by the leading pole in the following OPE's respectively
\bea 
\label{raising}
G^2(z)\widetilde{G}^2(\bar z)\Sigma^
{(\frac{1}{2}, \,  \frac{1}{2})} (w, \bar w), \;\;
G^2(z) \widetilde{G}^{1\dagger}(\bar z)\Sigma^{(\frac{1}{2}, \,  
\frac{1}{2})} (w, \bar w),  \\  \nonumber
 G^{1\dagger}(z) \widetilde{G}^2(\bar z)
 \Sigma^{(\frac{1}{2}, \, \frac{1}{2})} (w, \bar w), \;\;
G^{1\dagger}(z)\widetilde{G}^{1\dagger}
(\bar z)\Sigma^{(\frac{1}{2}, \,\frac{1}{2})} (w, \bar w)
\eea
These are the $4$ blow up modes of the $R^4/Z_2$ singularity
\cite{chp2:CveDix}\ and they have conformal weight $(1,1)$%
\footnote{Relevance of $Z_2$ twist operators
to the marginal deformations of the SCFT has earlier
been discussed in \cite{chp2:HasWad1,chp2:DijVerVer}.}. 
They transform as
$(\bf 1 , \bf
1)$ under the $SU(2)_R \times \widetilde{SU(2)_R}$.  As before, since
these are top components of the short multiplet $(\bf 2 , \bf 2)_S$
they can be added to the free SCFT as perturbations without violating
the ${\mathcal N} = (4,4)$ supersymmetry of the SCFT. 
The various quantum numbers of
these operators are listed below.
\bea
\begin{array}{lccc}
\label{twist-operator}
\mbox{Operator} & (j^3, \widetilde{j}^3)_I & 
\;\;\;SU(2)_R\times \widetilde{SU(2)}_R \;\;\;&  (h, \bar{h}) \\
{\mathcal T}^1_{(1)}=T^{(\frac{1}{2}, \,  \frac{1}{2})} & 
( 0, 1) & (\bf 1, \bf 1) & (1,1) \\
{\mathcal T}^1_{(0)}= T^{(\frac{1}{2}, 
\,  -\frac{1}{2})}+ T^{(-\frac{1}{2}, \,  \frac{1}{2})} &  
(0,0) & (\bf 1, \bf 1) & (1,1) \\
{\mathcal T}^1_{(-1)}= T^{(-\frac{1}{2}, \,  -\frac{1}{2})} 
& (0,-1) & (\bf 1, \bf 1) & (1,1) \\
{\mathcal T}^0= T^{(-\frac{1}{2}, \, -\frac{1}{2})} - T^{(-\frac{1}{2}, 
\,  -\frac{1}{2})}
& (0, 0) & (\bf 1, \bf 1) & (1,1) 
\end{array}  
\eea
The first three operators of the above table can be organized as a
$(\bf 1, \bf 3)$ under $SU(2)_I\times\widetilde{SU(2)}_I$. We will
denote these $3$ operators as ${\mathcal T}^1$. The last
operator transforms as a scalar $(\bf 1, \bf 1)$ under 
$SU(2)_I\times\widetilde{SU(2)}_I$ and is denoted by ${\mathcal T}^0$. 
The simplest way of figuring out the $(j^3, \widetilde{j}^3)_I$
quantum numbers in the above
table is to note that
(a) the $\Sigma$-operators
of \eq{defSigma} are singlets under $SU(2)_I\times\widetilde{SU(2)}_I$,
as can be verified by computing the action on them of
the operators $I_1, I_2$ and $\widetilde{I}_1,
\widetilde{I}_2$, (b) the ${\mathcal T}$-operators are
obtained from $\Sigma$'s by the action of the supersymmetry
currents as in \eq{raising} and (c) the
quantum numbers of the supersymmetry
currents under $I_1, I_2$ and $\widetilde{I}_1,
\widetilde{I}_2$ are given by \eq{so4-on-g}. 

\subsubsection{Higher twists}
\label{chp2:high-twist}

We now show that the twist operators corresponding to any other
conjugacy class of $S(Q_1 Q_5)$ are irrelevant. Consider the class
with $N_1= Q_1Q_5-3, N_3 =1$ and the rest of $N_n=0$. A representative
element of this class is 
\be
\label{z3-element}
(X_1\rightarrow X_2, X_2 \rightarrow X_3, X_3 \rightarrow X_1), 
\; X_4\rightarrow X _4, \ldots ,
\; X_{Q_1Q_5}\rightarrow X_{Q_1Q_5}. 
\ee
To make the action of this group element transparent we diagonalize
the group action as follows. 
\bea
\left(
\begin{array}{c}
\phi_1 \\ \phi_2 \\ \phi_3
\end{array} \right)
=
\left(
\begin{array}{ccc}
1 & 1 & 1 \\
1 & \omega & \omega^2 \\
1 & \omega^2 & \omega^4 
\end{array}
\right)
\left(
\begin{array}{c}
X_1 \\ X_2 \\  X_3
\end{array}
\right)
\eea
where $\omega = \exp(2\pi i/3)$.
These new coordinates are identified under the group action 
\eq{z3-element} $\phi_1
\rightarrow \phi_1$, $\phi_2 \rightarrow \omega^2 \phi_2$  and
$\phi_3 \rightarrow \omega \phi_3$. 
These identifications are locally characteristic of the
orbifold
\be
R^4\times R^4/\omega \times R^4/\omega^2
\ee
The dimension of the supersymmetric twist operator which twists the
coordinates by a phase $e^{2\pi i k /N}$ in $2$ complex dimensions is
$h(k,N)= k/N$\cite{chp2:DixFriMarShe}. The twist operator which implements the
action of the group element \eq{z3-element} combines the
supersymmetric twist operators acting on $\phi_2$ and $\phi_3$
and therefore has total dimension  
\be
h =h(1,3) + h(2,3) = 
1/3+ 2/3 =1 
\ee 
It is
the superpartners of these which could be candidates for the blow up
modes. However, these have weight $3/2$, These operators are therefore
irrelevant.

For the class $N_1= Q_1Q_5 -k$ , $N_k =k$ and the rest of $N_n =0$,
the total dimension of the twist operator is 
\be
h = \sum_{i=1}^{k-1} h(i,k) = (k-1)/2
\ee 
Its superpartner has dimension $k/2$. Now it is easy to see that all
conjugacy classes other than the exchange of $2$ elements give
rise to irrelevant twist operators. Thus the orbifold ${\mathcal M}$ is
resolved by the $4$ blow up modes corresponding to the conjugacy class
represented by \eq{group-element}. We have thus
identified the $20$ marginal operators of the ${\mathcal N}=(4,4)$ SCFT on
$\widetilde{T}^4$. They are all top components of the $5(\bf 2 ,\bf
2)_S$ short multiplets. The $5(\bf 2, \bf 2)_S$ have $20$ 
operators of conformal dimensions $(h, \bar{h})=(1/2, 1/2)$. These are
relevant operators for the SCFT. It would be interesting to
investigate the role of these relevant operators. As they are chiral
primaries they would break only 
half of the supersymmetries of the SCFT and
therfore the renormalization group flow induced by these operators
would persumably be tractable for study.

\subsection{The chiral primaries of the ${\mathcal N}=(4,4)$ SCFT on
${\mathcal
M}$ }
\label{chp2:chprimary}

In this section we will explicitly construct all the chiral primaries 
corresponding to single particle states 
of the SCFT on the orbifold ${\mathcal M}$. 
For this purpose we will have to 
construct the
twist operator corresponding to the conjugacy class $N_1 = Q_1Q_5 -k,
N_k =k$ and the rest of $N_n=0$. 

\subsubsection{The k-cycle twist operator}
\label{chp2:k-cycle}
We will extend the method of construction of 
the 2-cycle twist operator of
Section \ref{chp2:z-2twists} to the construction of the k-cycle twist
operator. Consider the conjugacy class given by $N_1 =Q_1 Q_5 -k, N_k
=k$ and the rest of $N_n=0$. A representative element of this class is
the following group action 
\be
\label{chp2:g-action}
(X_1 \rightarrow X_2 , \ldots , X_k\rightarrow X_1 ), X_{k+1}
\rightarrow
X_{k+1}, \ldots ,
X_{Q_1 Q_5} \rightarrow X_{Q_1 Q_5} .
\ee
We can diagonalize the group action as follows
\be
\left(
\begin{array}{c}
\phi_k \\
\phi_{k-1} \\
\phi_{k-2} \\
\vdots \\
\phi_1
\end{array}
\right)
=
\left(
\begin{array}{ccccc}
1 & 1 & 1 & \ldots & 1 \\ 
1 & \omega & \omega^2 & \ldots & \omega^{k-1} \\ 
1 & \omega^2 & \omega^4 & \ldots & \omega^{2(k-1)} \\ 
\vdots &\vdots & \vdots & \ldots & \vdots \\
1 & \omega^{k-1} & \omega^{(k-1)2} & \ldots & \omega^{(k-1)(k-1)}
\end{array}
\right)
\left(
\begin{array}{c}
X_1 \\
X_2 \\
X_3 \\
\vdots \\
X_k
\end{array}
\right)
\ee
where $\omega = e^{2\pi i /k}$. These new coordinates are identified
under the group action \eq{chp2:g-action} as
\be
\phi_1 \rightarrow \omega \phi_1 ,\;\; \phi_2 \rightarrow \omega^{2} \phi_2 ,
\;\; \phi_3 \rightarrow \omega^3 \phi_3 , \;\; \ldots ,\;\; \phi_{k-1}
\rightarrow \omega^{k-1} \phi_{k-1},\;\;
\phi_k \rightarrow \omega^k \phi_k
\ee
These identifications are locally characteristic of the orbifold
\be
R^4\times R^4/\omega \times R^4/\omega^2 \times \ldots \times 
R^4/\omega^{k-1}
\ee
The coordinate $\phi_m$ is twisted by the phase $\omega^{m}$ ( $m$ runs
from $1\ldots k $).
The bosonic twist operators corresponding to this twist are defined by
the following OPE's
\bea
\label{chp2:OPEs}
\del \phi^1_m (z) \sigma^1_m(w, \bar{w} ) = 
\frac{ \tau^1_m(w, \bar w ) }{ (z-w)^{1-m/k} }  \; &\;& \;
\del {\phi}^{1\dagger}_m (z) \sigma^1_m(w, \bar{w} ) = 
\frac{ \tau'^1_m(w, \bar w ) }{ (z-w)^{m/k} }  \\   \nonumber
\del \phi^2_m (z) \sigma^2_m(w, \bar{w} ) =
\frac{ \tau^2_m(w, \bar w ) }{ (z-w)^{1-m/k} }  \; &\;& \;
\del {\phi}^{2\dagger}_m (z) \sigma^2_m(w, \bar{w} ) =
\frac{ \tau'^2_m(w, \bar w ) }{ (z-w)^{m/k} }  \\  \nonumber
\bar{\del} \widetilde{\phi}^1_m (\bar{z}) \sigma^1_m(w, \bar{w} ) = 
\frac{ \widetilde\tau'^1_m(w, \bar w ) }{ (\bar z-\bar w)^{m/k} } 
\; &\;&
\;
\bar{\del} \widetilde{\phi}^{1\dagger}_m 
(\bar z) \sigma^1_m(w, \bar{w} ) = 
\frac{ \widetilde\tau^1_m(w ,\bar w ) }{ (\bar z-\bar w)^{1-m/k} } 
\\   \nonumber
\bar{\del} \widetilde{\phi}^2_m (\bar z) \sigma^2_m(w, \bar{w} ) =
\frac{ \widetilde\tau'^2_m(w, \bar w ) }{ (\bar z-\bar w)^{m/k} } 
\; &\;& \;
\bar{\del} \widetilde{\phi}^{2\dagger}_m (\bar z ) 
\sigma^2_m(w, \bar{w} ) =
\frac{ \widetilde\tau^2_m(w, \bar w ) }{ (\bar z-\bar w)^{1-m/k} }  
\eea
As in Section \ref{chp2:z-2twists} $\tau$'s are excited twist
operators. The fermionic twists are constructed from bosonized currents
defined by
\bea
\chi^1_m(z) = e^{iH^1_m(z)} \; &\;& \; 
\chi^{1\dagger}_m(z) = e^{-iH^1_m(z)} \\
\nonumber
\chi^2_m(z) = e^{iH^2_m(z)} \; &\;& \; 
\chi^{2\dagger}_m(z) = e^{-iH^2_m(z)} \\
\nonumber
\eea
Where the $\chi_m$'s are 
the superpartners of the bosons $\phi_m$'s. The twist operators
corresponding to the fermions $\chi_m$'s are given by 
$e^{\pm i mH_m/k}$.

We now assemble all these operators to construct the k-cycle twist
operator which is a chiral primary. 
The k-cycle twist operator is given by
\be
\Sigma_{(12\ldots k)}^{(k-1)/2} =
\prod_{m=1}^{k-1} \left[
\sigma^1_m (z, \bar{z})
\sigma^2_m (z, \bar{z})
e^{i m H_m^1(z)/k}
e^{-i m H_m^2(z)/k}
e^{i m \tilde{H}_m^1(\bar{z})/k}
e^{-i m \tilde{H}_m^2(\bar{z})/k}
\right]
\ee
The subscript $(12\ldots k)$ refers to the fact that these twist
operators were constructed for the representative group element 
\eq{chp2:g-action} which cyclically permutes the $1,\ldots , k$ labels
of the coordinates of $\tilde{T}^4$. The superscript $(k-1)/2$ stands
for the conformal dimension of this operator. As we saw in Section
\ref{chp2:high-twist} 
the conformal dimension of the twist operator for the conjugacy
class $N_1 = Q_1Q_5 -k, N_k =k$ and the rest of $N_n=0$ is
$(h, \bar{h}) = ((k-1)/2, (k-1)/2)$. 
The twist operator for the conjugacy
class under consideration is obtained by summing over the k-cycle
twist operators for all representative element of these class.
\be
\Sigma ^{(k-1)/2} (z, \bar{z} ) = \sum_{ \{ i_i, \ldots , i_k \} }
\Sigma_{i_1i_2\ldots i_k} (z, \bar{z} )
\ee
where the sum runs over all  $k$-tuples $\{ i_i \ldots , i_k \}$ such
that $i_i\neq i_2 \neq \ldots \neq i_k$. $i_m$ take values from $1$ to
$Q_1Q_5$. The operator $\Sigma^{(k-1)/2}$ is a chiral primary with
conformal dimension 
$(h, \bar{h}) = ((k-1)/2, (k-1)/2)$ and $(j_R^3, \tilde{j}_R^3)
=((k-1)/2, (k-1)/2)$. As the largest cycle is of length $Q_1Q_5$, the
maximal dimension of the k-cycle twist operator is 
$((Q_1Q_5-1)/2, (Q_1Q_5-1)/2)$. It belongs to the bottom component of the 
short multiplet $(\bf{k}, \bf{k})_S$.
The other components of the short multiplet $(\bf{k}, \bf{k})_S$ 
corresponding to the k-cycle twists can be generated by the
action of supersymmetry currents and the R-symmetry currents of the
${\mathcal N}=(4,4)$ theory on ${\mathcal M}$.

\subsubsection{The complete set of chiral primaries}

We have seen is Section \ref{chp2:res} there are five chiral
primaries corresponding to the short multiplet $5(\bf{2},
\bf{2})_S$. In this section we will construct the
complete set of chiral primaries from single particle states of the
SCFT on ${\mathcal M}$. It is known that the chiral primaries 
with weight $(h, \bar{h})$ of a
${\mathcal N}=(4,4)$ superconformal field theory on a manifold $K$
correspond to the elements of the cohomology ${\mathcal H}_{2h\,
2\bar{h}}(K)$ \cite{chp2:Wit-susy}. 
The chiral primaries are formed by the product of
the chiral primaries corresponding to the cohomology of the diagonal
$\tilde{T}^4$ denoted by $B^4$ 
(the sum of all copies of $\tilde{T}^4$)
and the various k-cycle chiral primaries constructed in
Section \ref{chp2:k-cycle}. We will list the chiral primaries below

\gap1
\ni\underbar{\it Chiral primaries with  $h-\bar{h} =0$}
\gap1

All the k-cycle chiral primaries have $h-\bar{h}=0$. To construct
chiral primaries with $h-\bar{h} =0$ we need the four chiral
primaries which 
correspond to the cohomology ${\mathcal H}_{11}(B^4)$  
with weight $(1/2,1/2)$. They are given in \eq{chiral}.
Using this we can construct the following chiral primaries
\bea 
\Sigma^{(k-1)/2}(z, \bar{z})
\Psi^1_A(z) \widetilde{\Psi}^1_A(\bar{z})  \;&\;&\;  
\Sigma^{(k-1)/2}(z, \bar{z})
\Psi^1_A(z)\widetilde{\Psi}^{2\dagger}_A(\bar z) \\   \nonumber
\Sigma^{(k-1)/2}(z, \bar{z})
\Psi^{2\dagger}_A(z)\widetilde{\Psi}^1_A(\bar z )  \;&\;&\;
\Sigma^{(k-1)/2}(z, \bar{z})
\Psi^{2\dagger}_A(z)\widetilde{\Psi}^{2\dagger}_A(\bar z) 
\eea
where summation over $A$ is implied. These four operators have 
conformal dimension $(k/2, k/2)$. There is one more chiral primary
corresponding to the cohomology ${\mathcal H}_{22}(B^4)$ for which $h-\bar{h}=0$. 
It is given by 
\be
\Psi^1_A(z) \Psi^{2\dagger}_A(z)
\widetilde{\Psi}^1_A(\bar{z})  \widetilde{\Psi}^{2\dagger}_A(\bar z) 
\ee
where summation over all indices of $A$ is implied. 
This chiral primary corresponds to the top form of 
$B^4$. The cohomology ${\mathcal H}_{00}(B^4)$ gives rise to a 
chiral primaries of conformal dimension $(k/2,k/2)$. 
It is given by
\be
\Sigma^{(k-2)/2}(z, \bar{z})
\Psi^1_A(z) \Psi^{2\dagger}_A(z)
\widetilde{\Psi}^1_A(\bar{z})  \widetilde{\Psi}^{2\dagger}_A(\bar z) 
\ee
From the equation above we see that these chiral primaries exist only
of $k\geq 2$. Finally we have the chiral primary
$
\Sigma^{(k)/2}(z, \bar{z})
$
of conformal dimension $(k/2,k/2)$. Thus for $k\geq 2$ 
and $k\leq Q_1Q_5 -1$ there are $6$
chiral primaries of dimension $(k/2,k/2)$ 

The complete list of chiral primaries with $(h, \bar{h})$ 
with $h -\bar{h} =0$ 
corresponding
to single particle states are given by
\be
\begin{array}{lc}
(h, \bar{h}) & \mbox{Degeneracy} \\
   &  \\ 
(1/2, 1/2) &  5 \\
(1, 1)    &   6 \\
(3/2,3/2)     &   6 \\
\vdots    & \vdots \\
((Q_1Q_5 -1)/2 , (Q_1Q_5 -1)/2 ) & 6 \\
( (Q_1Q_5)2, (Q_1Q_1)/2 )  & 5 \\
( (Q_1Q_5 + 1)/2, (Q_1Q_5 + 1)/2 ) & 1 
\end{array}
\ee
In the above table we have ignored the vacuum with weight $(h,
\bar{h})=(0,0)$.

\gap1
\noindent\underbar{\it Chiral primaries with  $h-\bar{h} =1/2$ }
\gap1

The chiral primaries of $B^4$ which correspond to
the elements of the cohomology ${\mathcal H}_{10}(B^4)$ are given by
\be
\sum_{A=1}^{Q_1Q_5}
\Psi^1_A(z) \;\;\; \mbox{and} \;\;\;
\sum_{A=1}^{Q_1Q_5}
\Psi^{2\dagger}_A(z)
\ee
We can construct chiral primaries with weight $( (k+1)/2 , k/2) )$ by
taking the product of the above chiral primaries with the twist
operator $\Sigma^{k/2}(z, \bar{z})$. These give the following
chiral primaries 
\be
\Sigma^{k/2}(z, \bar{z}) 
\sum_{A=1}^{Q_1Q_5}
\Psi^1_A(z) \;\;\; \mbox{and} \;\;\;
\Sigma^{k/2}(z, \bar{z}) 
\sum_{A=1}^{Q_1Q_5}
\Psi^{2\dagger}_A(z)
\ee
The chiral primary of the diagonal $B^4$ which 
correspond to the
elements of the cohomology ${\mathcal H}_{21}(B^4)$  are 
\be
\Psi^1_A(z) \Psi^{2\dagger}_A(z)
\widetilde{\Psi}^1_A(\bar{z})  \;\; \mbox{and} \;\;
\Psi^1_A(z) \Psi^{2\dagger}_A(z)
\widetilde{\Psi}^{2\dagger}_A(\bar z) 
\ee
Here summation over all the three indices of $A$ is implied. 
From these the  one can construct chiral 
primaries with weight $((k+1)/2, k/2)$ are follows
\be
\Sigma^{(k-1)/2}(z, \bar{z}) 
\Psi^1_A(z) \Psi^{2\dagger}_A(z)
\widetilde{\Psi}^1_A(\bar{z})  \;\; \mbox{and} \;\;
\Sigma^{(k-1)/2}(z, \bar{z}) 
\Psi^1_A(z) \Psi^{2\dagger}_A(z)
\widetilde{\Psi}^{2\dagger}_A(\bar z) 
\ee
Therefore there are 4 chiral primaries with weight $((k+1)/2, k/2)$
for $1\leq k\leq (Q_1Q_5 -1)$ and 2 chiral primaries with weight
$((Q_1Q_5+1)/2, Q_1Q_5/2)$. There are also 2 chiral primaries with
weight $(1/2, 0)$.

\gap1
\noindent\underbar{\it Chiral primaries with  $\bar{h}-h =1/2$ }
\gap1

The procedure for constructing these chiral primaries are identical to
the  case $h -\bar{h} =1/2$. 
The four chiral primaries with weight $(k/2, (k+1)/2)$ are given by

\bea
 \Sigma^{k/2}(z, \bar{z}) 
\sum_{A=1}^{Q_1Q_5}\widetilde{\Psi}^1_A(\bar{z}) \;&\;&\;
\Sigma^{k/2}(z, \bar{z}) 
\sum_{A=1}^{Q_1Q_5}\widetilde{\Psi}^{2\dagger}_A(\bar z) \\ \nonumber
\Sigma^{(k-1)/2}(z, \bar{z}) 
\Psi^1_A(z) 
\widetilde{\Psi}^1_A(\bar{z})  \widetilde{\Psi}^{2\dagger}_A(\bar z) 
\;&\;&\;
\Sigma^{(k-1)/2}(z, \bar{z}) 
 \Psi^{2\dagger}_A(z)
\widetilde{\Psi}^1_A(\bar{z})  \widetilde{\Psi}^{2\dagger}_A(\bar z) 
\eea
 There are 4 chiral primaries with weight $(k/2, (k+1)/2)$
for $1\leq k\leq (Q_1Q_5 -1)$  2 chiral primaries with weight
$(Q_1Q_5/2, (Q_1Q_5+1)/2)$ and 2 chiral primaries with weight $(0,
1/2)$.

\gap1
\noindent\underbar{\it Chiral primaries with  $h-\bar{h} =1$}
\gap1

As in the previous cases let us first look at the
chiral primaries corresponding to the cohomology element 
${\mathcal H}_{20}(B^4)$. There is only one element which is given by
\be
\Psi^1_A(z) \Psi^{2\dagger}_A(z)
\ee
where summation over $A$ is implied.  
There is a single  chiral 
primary with weight $((k+2)/2, k/2)$  constructed out of the above
chiral primary is  
\be
\Sigma^{k/2}(z, \bar{z}) 
\Psi^1_A(z) \Psi^{2\dagger}_A(z)
\ee
Thus there is a one chiral primary with weight $((k+2)/2, k/2)$  
for $0\leq k\leq (Q_1Q_5-1)$.
The operator product expansion of two chiral primaries will give rise
to other chiral primaries consistent with conservation laws.
There are known to form a ring. It will be interesting to understand
the structrure of this ring.

\gap1
\noindent\underbar{\it Chiral primaries with  $\bar{h}-h =1$}

The construction of these is parallel to the case 
for $h-\bar{h}=1$. The single
chiral primary with weight $(k/2, (k+2)/2)$ for 
$0\leq k\leq (Q_1Q_5-1)$ is given by 
\be
\Sigma^{k/2}(z, \bar{z}) 
\widetilde{\Psi}^1_A(\bar{z})  \widetilde{\Psi}^{2\dagger}_A(\bar z) 
\ee

\gap1
There are no chiral primaries with $h-\bar{h}>1$ or $\bar{h} -h>1$.
From the construction of the chiral primaries we see that such chiral
primaries can exist only if there is an element in ${\mathcal
H}_{r0}(B^2)$ or ${\mathcal H}_{0r}(B^2)$ with $r>1$. As the homology
groups of $B^4$ is identical to that of a four torus we know that such
elements do not exist. 

\subsection{Short 
multiplets of ${\mathcal N}= (4,4)$ SCFT on ${\mathcal M}$}
\label{chp2:sec-shortmultiplets}

Using the results of Section \ref{chp2:chprimary}
we will write the complete
set of short multiplets of single particle states of
the ${\mathcal N}=(4,4)$ SCFT onf ${\mathcal M}$. 
In Section 6 we will compare this set of short multiplets with that
obtained from supergravity. We will see in Section 6.1 
that supergravity is a good approximation in string theory only when
$Q_1\rightarrow \infty, Q_5\rightarrow \infty$. Therefore we write
down the list of short multiplets for 
$(\tilde{T}^4)^{(\infty)}/S(\infty)$. 
Basically this means that the list
of chiral primaries of the previous Section \ref{chp2:chprimary} 
does not terminate. 

We have seen that each chiral primary  of weight $(h, h')$ gives rise
gives rise to the short multiplet 
$({\bf 2h} +{\bf 1} , {\bf 2h^{'}} +{\bf 1}))_S$. Therefore the results of
Section \ref{chp2:chprimary} indicate that the list of
shormultiplets corresponding to the single particle states of
${\mathcal N}=(4,4)$ SCFT on 
$(\tilde{T}^4)^{(\infty)}/S(\infty)$. 
is given by
\bea
\label{chp2:eq.short}
&5 (\re 2 , \re 2 )_S + 6 \oplus_{\re m \geq \re 3 } (\re m , \re m
)_S \\ \nonumber 
&2(\re 1, \re 2)_S + 2(\re 2, \re 1)_S + (\re 1, \re 3)_S + 
(\re 3, \re 1)_S \\ \nonumber
&\oplus_{\re m \geq \re 2 } [\, (\re m , \re m + \re2
)_S + ( \re m + \re2 , \re m )_S + 4 ( \re m , \re m + \re 1 )_S + 4 (
\re m + \re 1 , \re m )_S \, ] 
\eea 

In our discussion so far we have ignored the short multiplets from the
free torus $T^4$ which forms a part of the Higgs branch of the \dd
system. We will see in section 6 that the short multiplets from
the free torus are not present in the supergravity. 
Thus for comparision
with supergravity it is sufficient for us to restrict our attention to
the shortmultiplets on ${\mathcal M}$.

\subsection{Stringy exclusion principle}

We see from the preceding discussion that the spin of short multiplets
in the SCFT is bounded by $(1+Q_1 Q_5)/2$. In the context of the
\ads/\cft\ correspondence (see Section 6) this is puzzling at first
since there is no corresponding bound on spin from
supergravity. However, since supergravity is only valid at $Q_1, Q_5
\to \infty$ these two facts are reconciled. The existence
of a maximum spin for finite $Q_1, Q_5$ has been called the
``stringy exclusion principle'' \cite{chp3:MalStr98}. Clearly
this bound cannot be understood in supergravity and should
be understood in terms an exact treatment of
strings in \ads.

\newpage
\section{Near horizon supergravity and SCFT \label{near-horizon-symmetry}}

In this section we will classify the supergravity fields according to
the symmetries of the near horizon geometry of the \dd system which
was derived in section. \ref{near-horizon-ads} and compare them with
the chiral multiplets of the SCFT on ${\mathcal M}$.

Let us examine the symmetries of the near horizon geometry
\eq{near-horizon}. The
bosonic symmetries arise from the isometries of $AdS_3$ and $S^3$.
The isometries of the $AdS_3$ space form the non-compact group
$SO(2,2)$, while the isometries of $S^3$ form the group $SO(4)_E=
SU(2)_E\times \widetilde{SU(2)}_E$. Though the compactification on
$T^4$ breaks the $SO(4)$ rotations of the coordinates $x_6, \ldots
,x_9$ we can still use this symmetry to classify supergravity
fields. We will call this symmetry $SO(4)_I$.  The \dd system
preserves eight out of the 32 supersymmetries of the type IIB
theory. In the near horizon limit the number of supersymmetries gets
enhanced from eight to sixteen \cite{Boonstra:1997dy,chp3:Town}. 
These symmetries fix
the form of the effective anti-de Sitter supergravity theory near the
horizon. The bosonic symmetries $SO(2,2)\times SO(4)_E = (SL(2,
R)\times SU(2)) \times {(SL(2, R) \times SU(2))}$ form the bosonic
symmetries of the anti-de Sitter supergravity in
three-dimensions. Simple anti-de Sitter supergroups in
three-dimensions were classified in \cite{chp3:towngun}. It can be
seen that the only simple supergroups whose bosonic part is
$SL(2,R)\times SU(2)$ are $Osp(3|2, R)$ and $SU(1,1|2)$. The former
contains the bosonic subgroup $O(3)\times SL(2,R)$. The supercharges
of the supergroup $Osp(3|2, R)$ transform as the vector representation
of the group $O(3)$, while the supercharges of the supergroup
$SU(1,1|2)$ transform as ${\bf 2}$ of the group $SU(2)$. The unbroken
supercharges of the \dd system transform in the spinor
representation of $SO(4)_E$ and therefore they transform as ${\bf 2}$
of $SU(2)$. This rules out $Osp(3|2, R)$.  Therefore the near horizon
anti-de Sitter supergravity is based on the supergroup
$SU(1,1|2)\times SU(1,1|2)$ with matter fields 
\footnote{The pure anti-de
Sitter supergravity based on the super group $SU(1,1|2)\times
SU(1,1|2)$ was constructed in \cite{chp3:Jus} using the fact that it
is a Chern-Simons theory.}.  

\subsection{Classification of the supergravity modes}
\label{chp3:classification}

In this section we analyze the spectrum of Type IIB supergravity
compactified on $AdS_3 \times S^3 \times T^4$. 
From \eq{s3-t4} we see the volume of $T^4$ is 
 $16\pi^4 \alpha^{' 2} Q_1/Q_5$. Therefore we ignore 
Kaluza-Klein modes on the $T^4$. The radius of the $S^3$ is
$\sqrt{\alpha^{\prime}}(g_6Q_1Q_5)^{1/4}$.
This 
is large when 
\be
g_sQ_1>>1 \;\;\mbox{and}\;\; g_sQ_5>>1
\ee
These inequalities imply that we are working in the regime where  
closed string perturbation theory
is valid and where all length scales are greater than the string
length.
Therefore we are justified in using supergravity. Kaluza-Klein
reduction of type IIB supergravity to six dimensions leads to six
dimensional $(2,2)$ supergravity.
We show that the Kaluza-Klein spectrum of the six dimensional
theory on $AdS_3\times S^3$ can be completely organized as short
multiplets of the supergroup $SU(1,1|2)\times SU(1,1|2)$.  We will
follow the method developed by \cite{chp1:Deboer}. 

The massless spectrum of $(2,2)$ six-dimensional supergravity
consists
of: 
a graviton, 8 gravitinos, 5 two-forms, 
16 gauge fields, 40 fermions and 25 scalars.
Since these are massless, 
the physical degrees of freedom 
fall into various representations $R_4$
of the little group $SO(4)_{L}$ of $R^{(5,1)}$. 
For example, the graviton transforms as a  $(\bf 3 ,\bf 3)$ under the
little group $SO(4)_{L}= SU(2)_L\times \widetilde{SU(2)}_L$.
On further compactifying
$R^{(5,1)}$ into $AdS_3 \times S^3$, each representation $R_4$
decomposes into various representations $R_3$ of $SO(3)$, the local
Lorentz group of the $S^3$. This $SO(3)\simeq SU(2)$ 
is the diagonal $SU(2)$ of $SU(2)_L\times \widetilde{SU(2)}_L$. 
For example, the graviton decomposes as $\bf{1} + \bf{3} +\bf{5}$ 
under the $SO(3)$, the local Lorentz group of $S^3$.
The dependence of each of these fields on
the angles of $S^3$ leads to decomposition in terms of Kaluza-Klein
modes on the $S^3$ which transforms according to some representation of
the isometry group $SO(4)$ of $S^3$. Only those representations of
$SO(4)$ occur in these decompositions which contain the representation
$R_3$ of $S^3$. To be more explicit, consider the field 
$\phi_{R_{SO(3)}} (x_0, x_5, r, \theta,\phi,\chi)$ 
which
transforms as some representation $R_{SO(3)}$ of the local Lorentz group
of $S^3$. The Kaluza-Klein expansion of this field on $S^3$ is given
by
\be
\phi_{R_{SO(3)}} (x_0, x_5, r, \theta,\phi,\chi ) =
\sum_{R_{SO(4)}} \tilde{\phi}_{R_{SO(4)}}(x_0, x_5, r)
Y^{R_{SO(4)}}_{R_{SO(3)}}(\theta, \phi, \chi).
\ee
Here
$Y^{R_{SO(4)}}_{R_{SO(3)}}(\theta, \phi, \chi)$ stands for the spherical
harmonics on $S^3$.  In the above expansion the only representation of
$R_{SO(4)}$ allowed are the ones which contain $R_{SO(3)}$. For
example, $\phi (x_0, x_5, r, \theta, \phi, \chi)$ which is a scalar under the 
local Lorentz group of $S^3$ can be expanded as
\be
 \phi (x_0, x_5, r, \theta, \phi, \chi) 
= \sum_{\bf{m}, \bf{m'};\; \bf{m}= \bf{m'}}
\tilde{\phi}_{\bf{m} \bf{m'}} (x_0, x_5, r) 
Y^{(\bf{m}, \bf{m'})} (\theta, \phi, \chi)
\ee
Once the complete set of Kaluza-Klein modes are obtained we will
organize them into short multiplets of the supergroup $SU(1,1|2)\times
SU(1,1|2)$.

Let us now consider all the massless field of $(2,2)$ supergravity in
six-dimensions individually.
The graviton transforms as $(\re{3},\re{3})$  of the little group
in 6 dimensions. The Kaluza-Klein harmonics of this field 
according to the rules discussed above are
\bea
\label{eq.gravi}
&(\re{1}, \re{1} ) + 2 (\re{2},\re{2}) + (\re{3},\re{1}) +
(\re{1},\re{3})  \\ \nonumber
&+ 3 \oplus_{\re{m}\geq \re{3}} (\re{m},\re{m}) + 2
\oplus_{\re{m}\geq \re{2}} [\, (\re{m} + \re 2,\re{m} ) + 
(\re{m} , \re{m} + \re2 ) \,]  \\ \nonumber
&+ \oplus_{\re{m}\geq \re 1} [\, (\re{m} + \re 4 ,\re{m} ) +
(\re{m}, \re{m} + \re 4 )  \, ]
\eea
The little group representations of the 8 gravitinos is 
$4(\re 2 , \re 3) + 4(\re 3 ,\re 2) $. Their Kaluza-Klein 
harmonics are
\bea
& 8 [\, (\re 1, \re 2 ) + (\re 2 , \re 1) \,] + 
16 \oplus_{\re m  \geq  \re 2 } 
[\, (\re m + \re 1 , \re m ) + ( \re m , \re m + \re 1 ) \, ]
\\  \nonumber
&+ 8 \oplus_{\re m \geq \re 1 }[ \, ( \re m + \re 3 , \re m ) +
(\re m , \re m + \re 3 ) \, ]
\eea
The Kaluza-Klein harmonics of the 5 two-forms transforming
in  $(\re 1 ,\re 3 ) + (\re 3 , \re 1 )$ of the little group are
\be
10 \oplus_{\re m \geq \re 2} (\re m, \re m ) + 10 \oplus_{\re m \geq
\re 1 } [\, (\re m +\re 2 , \re m ) + ( \re m , \re m + \re 2 ) \, ]
\ee
The Kaluza-Klein harmonics of the 16 gauge fields, $(\re 2 , \re 2)$
 are 
\be
\label{eq.gauge}
16 (\re 1 , \re 1 ) + 32  \oplus_{\re m \geq \re 2 } (\re m , \re m )
+ 16 \oplus_{\re m \geq \re 1 } [\, (\re m , \re m + \re 2 ) + ( \re
m + \re 2 , \re m ) \, ]
\ee
The 40 fermions $ 20 (\re 2 , \re 1) + 20 ( \re 1 , \re 2 )$ give
rise to the following harmonics
\be
40 \oplus_{ \re m \geq \re 1 } [\, ( \re m , \re m + \re 1 ) + (
\re m + \re 1 , \re m ) \, ]
\ee 
The 25 scalars $( \re 1 , \re 1 )$  give rise to the harmonics
\be
25 \oplus_{\re m \geq \re 1} (\re m , \re m )
\ee
Putting all this together the complete Kaluza-Klein 
spectrum of type IIB on
$AdS_3 \times S^3 \times T^4$ yields 
\bea
\label{eq.spec}
&42 ( \re 1 , \re 1 ) + 69 (\re 2 , \re 2 ) + 48 [\, (\re 1 , \re
2 ) + ( \re 2 , \re 1 ) \, ] + 27 [ \, (\re 1 , \re 3 ) + ( \re 3
, \re 1 ) \, ]  \\    \nonumber
&70 \oplus_{\re m \geq \re 3 } ( \re m , \re m ) + 56 \oplus_{\re
m \geq \re 2 } [\, (\re m , \re m + \re 1  ) + ( \re m + \re 1 ,
\re m ) \,] \\ \nonumber
&+ 28 \oplus_{\re m \geq \re 2 } [\, ( \re m , \re m + \re 2 ) +
( \re m + \re 2 , \re m ) \, ] + 8 \oplus_{\re m \geq \re 1 } [\,
(\re m , \re m + \re 3 ) + (\re m + \re 3 , \re m ) \, ] \\
\nonumber
&+ \oplus_{\re m \geq \re 1} [ \, ( \re m , \re m + \re 4 ) + ( \re m
+ \re 4 , \re m ) \, ]
\eea
We now organize  the above Kaluza-Klein modes into 
short representations of $SU (1,1 | 2) \times SU ( 1,1 |2 ) $
\cite{chp1:Deboer}. The short multiplet of $ SU(1,1 | 2)$ consists of
the following states 
\be
\begin{array}{ccc}
  j      & L_0   & \mbox{Degeneracy}     \\    
\hline 
\vspace{-.2ex}
h     & h       & 2h +1    \\    
h-1/2~~   & ~~h + 1/2 & 2(2h)     \\    
h-1   &h+1 & 2h -1  
\end{array}
\ee 
In the above table $j$ labels the representation of $SU(2)$ which
is identified as one of the $SU(2)$'s of the isometry
group of $S^3$. $L_0$ denotes the conformal weight of the state. 
We denote the short multiplet of $SU(1,1 |2) \times SU(1,1 |2)$ as 
$(\re{2h} + \re{1} , \re{2h'} + \re 1 )_S$. On organizing the
Kaluza-Klein spectrum into short multiplets we get the following
set
\bea
\label{eq.short}
&5 (\re 2 , \re 2 )_S + 6 \oplus_{\re m \geq \re 3 } (\re m , \re m
)_S \\ \nonumber &\oplus_{\re m \geq \re 2 } [\, (\re m , \re m + \re2
)_S + ( \re m + \re2 , \re m )_S + 4 ( \re m , \re m + \re 1 )_S + 4 (
\re m + \re 1 , \re m )_S \, ] 
\eea 
Equation \eq{eq.spec} shows that
there are $42 (\re 1 ,\re 1)$ $SO(4)$ representations in the
supergravity Kaluza Klein spectrum. We know that one of these arises from the
$s$-wave of $g_{55}$ from equation \eq{eq.gravi}.  This is one of the
fixed scalars. $16 (\re 1 ,\re 1)$ comes from the $s$-waves of the
$16$ gauge fields (the components along $x^5$) as seen in equation
\eq{eq.gauge}.  The remaining $25$ comes from the $25$ scalars of the
six dimensional theory. We would like to see where these $42 (\re 1
,\re 1)$ fit in the short multiplets of $SU(1,1|2)\times
SU(1,1|2)$. From equation \eq{eq.short} one can read that $20$ of them
are in the $ 5 (\re 2 , \re 2 )_S$ with $(j=0, L_0 =1 ;\, j=0, L_0 =
1)$. 6 of them are in in $6 (\re 3 ,\re 3 )_S$ with $(j=0, L_0=2;\,
j=0, L_0=2)$. These correspond to the fixed scalars. Finally, the
remaining $16$ of them belong to $4(\re 2 , \re 3 )_S + 4(\re 3 , \re
2 )_S $. $8$ of them have $(j=0, L_0=1 ;\, j=0, L_0=2 )$ and $8$ of
them have $(j=0, L_0=2 ;\, j=0, L_0=1 )$. These scalars can be
recognized as the intermediate scalars.

\mysec11{Comparison of supergravity short multiplets with SCFT}

In Section \ref{chp2:sec-shortmultiplets} we have listed the complete
set of short multiplets corresponding to single particle states of the
${\mathcal N}=(4,4)$ SCFT on the orbifold ${\mathcal M}$. Comparing equation
\eq{chp2:eq.short} and the list of short multiplets of single particle
states obtained from supergravity in \eq{eq.short} we find that they
are identical except for the presence of the following 
additional short multiplets  in the SCFT
\be
2(\re 1, \re 2)_S + 2(\re 2, \re 1)_S + (\re 1, \re 3)_S + (\re 3, \re
1)_S
\ee
These correspond to non-propagating degrees of freedom in the
supergravity \cite{chp1:Deboer}. 
Therefore they are not present in the list of
short multiplets obtained form supergravity \eq{eq.short}. 
Furthermore, note that 
in \eq{chp2:eq.short} 
we have ignored the contribution of short multiplets 
from the free $T^4$ which forms the part of the
Higgs branch of the \dd system. Thus the short multiplets in
supergravity also ignores the contribution from the free torus.

It is pertinent here to mention that the AdS/CFT duality for \dd
systems in theories with 16 supercharges were studied in
\cite{Gava:2000is,Gava:2001ne}.  These theories were obtained by
considering various orbifolds of type IIB.

\subsection{The supergravity moduli}
\label{sugramod}

In this section we will analyze in detail the massless
scalars in the near horizon geometry of the \dd system.
Type IIB supergravity
compactified on $T^4$ has 25 scalars. There are $10$ scalars $h_{ij}$
which arise from compactification of the metric.  $i, j, k \ldots $
stands for the directions of $T^4$. There are $6$ scalars $b_{ij}$
which arise from the Neveu-Schwarz $B$-field and similarly there are
$6$ scalars $b'_{ij}$ from the Ramond-Ramond $B'$-field. The remaining
3 scalars are the ten-dimensional dilaton 
$\phi_{10}$, the Ramond-Ramond scalar $\chi$ and
the Ramond-Ramond $4$-form $C_{6789}$. These scalars parameterize the
coset $SO(5,5)/(SO(5)\times SO(5))$.  The near horizon limit of the
\dd system is $AdS_3\times S^3\times T^4$ \eq{near-horizon}. 
In this geometry $5$ of
the $25$ scalars become massive \cite{chp3:MalStr98}.  
They are the $h_{ii}$ (the trace of
the metric of $T^4$ which is proportional to the volume of $T^4$), the
$3$ components of the anti-self dual part of the Neveu-Schwarz
$B$-field $b_{ij}^-$ and a linear combination of the Ramond-Ramond
scalar and the $4$-form \cite{Sei-Wit99}.  The massless scalars in the
near horizon geometry parameterize the coset $SO(5,4)/(SO(5)\times
SO(4))$ \cite{chp3:GivKutSei}.

As we have seen 
the near horizon symmetries form the supergroup $SU(1,1|2)\times
SU(1,1|2)$. 
We have classified  all
the massless supergravity fields of type IIB supergravity on
$AdS_3\times S^3\times T^4$ ignoring the Kaluza-Klein modes on $T^4$
according to the short multiplets of the supergroup $SU(1,1|2)\times
SU(1,1|2)$. The isometries of the anti-de Sitter space allow us to
relate the quantum number  $L_0 +\bar{L}_0$  to the mass of the scalar
field through the  relation \cite{chp3:MalStr98}.
\be
\label{chp3:massdim}
h+\bar{h}= 1+ \sqrt{1+m^2}
\ee
Here $m$ is the mass of the scalar in units of the radius of
$AdS_3$ and $(h,\bar h)$ is the eigenvalue of $L_0, \bar{L}_0$ under
the classification of the scalar in short multiplets of
$SU(1,1|2)\times SU(1,1|2)$. 
Thus the massless fields of the 
near horizon geometry of the \dd system fall into the top
component of the $5(\bf 2, \bf 2 )_S$ short multiplet.  We further
classify these fields according to the representations of the
$SO(4)_I$, the rotations of the $x_6,x_7,x_8,x_9$ directions. As we have
mentioned before this 
is not a symmetry of the supergravity as it is compactified on $T^4$,
but it can be used to classify states. The quantum number of the
massless supergravity fields are listed below.
\bea
\label{sugra-fields}
\begin{array}{lccc}
\mbox{Field} & SU(2)_I\times \widetilde{SU(2)_I} 
& SU(2)_E\times \widetilde{SU(2)}_E &
\mbox{Mass} \\
h_{ij } -\frac{1}{4}
\delta_{ij} h_{kk} & (\bf 3, \bf 3) & (\bf 1, \bf 1) & 0 \\
b'_{ij} & (\bf 3, \bf 1) + (\bf 1, \bf 3) &(\bf 1, \bf 1) &  0 \\
\phi_6 &  (\bf 1, \bf 1) &(\bf 1, \bf 1) &  0 \\
a_1 \chi + a_2 C_{6789}&  (\bf 1, \bf 1) &(\bf 1, \bf 1) &  0 \\
b^+_{ij} &  (\bf 1, \bf 3) &(\bf 1, \bf 1) & 0 
\end{array}
\eea
The linear combination appearing on the fourth line is the
one that remains massless in the near-horizon limit.
$\phi_6$ refers to the six-dimensional dilaton.
The $SU(2)_E \times \widetilde{SU(2)}_E$ stands for the $SO(4)$
isometries of the $S^3$.
All the above fields are s-waves of scalars in the near horizon
geometry. 

\subsection{AdS$_3$/CFT$_2$ correspondence\label{correspond}}

We have already seen in Section 6.3 that all the supergravity modes
can be organized as short multiplets of the SCFT on ${\mathcal M}$.
This is  evidence for Maldacena's $AdS/CFT$ correspondence.
Maldacena's conjecture \cite{Mal97, Wit98-ads, Gub-Kle-Pol98,
chp3:MalStr98}
for the case of the \dd system states
that string theory on $AdS_3\times S^3\times T^4$ is dual to the 
$1+1$ dimensional 
conformal field theory of the Higgs branch of gauge theory of the
\dd system. 
Here we briefly review the evidence for this conjecture from symmetries.
To describe the \dd system at a generic point in
the moduli space we can use the ${\mathcal N}=(4,4)$ SCFT on 
the orbifold
${\mathcal M} \times T^4 $ 
to describe  the Higgs branch of the gauge theory of the
\dd system as we have argued in Section 4. Here the dynamics on $T^4$
is decoupled from the symmetric product. However in 
string theory on $AdS_3 \times S^3\times T^4$, all fields couple to
gravity and no field is free. Thus in this case of the 
$AdS/CFT$ correspondence we have to ignore the free torus $T^4$.
This is the same reason that 
the $U(1)$ of the ${\mathcal
N=4}$ super Yang-Mills theory with gauge group $U(N)$ is ignored in
the correspondence with string theory on $AdS_5\times S^4$. The gauge
group used in the correspondence is in fact $SU(N)$
\cite{Witten:1998wy}. 
The volume of $T^4$ is of the order of string length and radius of
$S^3$ is large, therefore we can pass over from string theory on
$AdS_3\times S^3\times T^4$ to six-dimensional $(2,2)$ supergravity
on $AdS_3\times S^3$. We will compare symmetries in the supergravity
limit.  The identification of the isometries  of the near horizon
geometry with that of the symmetries of the SCFT are given in the
following table.

\begin{tabular}{l| l}
Symmetries of the Bulk & Symmetries of SCFT \\
\hline
\gap{1}
(a) Isometries of $AdS_3$ 
&The global part of the Virasoro group \\
 $SO(2,2)\simeq SL(2, R) \times \widetilde{SL(2,R)}$ &
$SL(2,R)\times \widetilde{SL(2,R)}$ \\
\hline
\gap{1}
(b) Isometries of $S^3$ &
R-symmetry of the SCFT \\
$SO(4)_E\simeq SU(2)\times SU(2)$ &
$SU(2)_R\times\widetilde{SU(2)}_R$ \\
\hline
\gap{1}
(c) Sixteen near horizon supersymmetries & 
\ Global supercharges  of 
${\mathcal N}= (4,4) $ SCFT \\
\hline
\gap{1}
(d) $SO(4)_I$ of $T^4$ & $SO(4)_I$ of $\tilde{T}^4$ \\
\hline
\end{tabular} 

To summarize the $SU(1,1|2)\times SU(1,1|2)$ symmetry of the near
horizon geometry is identified with the global part of the ${\mathcal
N}=(4,4)$ SCFT on the orbifold ${\mathcal M}$ together with the
identification of the $SO(4)_I$ algebra of $T^4$ and $\tilde{T}^4$.

\subsection{Supergravity moduli and the marginal operators}

We would like to match the twenty supergravity moduli appearing
in \eq{sugra-fields} with the twenty marginal operators 
appearing in \eq{untwist-operator} and \eq{twist-operator} 
by comparing their symmetry properties under the AdS/CFT correspondence
\cite{chp1:DavManWad2}. 

\noindent The symmetries, or equivalently quantum numbers, to be compared 
under the AdS/CFT correspondence are as follows:

(a) The isometries of the supergravity are identified with the global
symmetries of the superconformal field theory. For the $AdS_3$ case
the symmetries form the supergroup $SU(1,1|2)\times SU(1,1|2)$.  
The identification of this supergroup with the
global part of the ${\mathcal N}= (4,4)$ superalgebra leads
to the mass-dimension relation \eq{chp3:massdim}.
Since in 
our case the SCFT operators are marginal and the supergravity
fields are massless, the mass-dimension relation is obviously
satisfied. 

(b) The $SU(2)_E\times \widetilde{SU(2)}_E$
quantum number of the bulk supergravity field corresponds to the
$SU(2)_R\times \widetilde{SU(2)}_R$ 
quantum number of the boundary operator.
By an inspection of column three of the tables in
\eq{untwist-operator}, \eq{twist-operator} and \eq{sugra-fields}, 
we see that these quantum numbers
also match.
 
(c) The location of the bulk fields and the boundary operators as
components of the short multiplet can be found by the supersymmetry
properties of the bulk fields and the boundary operators.
Noting the fact that all the twenty bulk fields as
well as all the marginal operators mentioned above correspond to top
components of short multiplets, this property also matches.  

(d) The above symmetries alone do not distinguish between the twenty
operators or the twenty bulk fields. To further distinguish these
operators and the fields we identify the $SO(4)_I$ symmetry of the
directions $x_6,x_7,x_8,x_9$ 
with the $SO(4)_I$ of the SCFT. At the level
of classification of states this identification is reasonable though
these are not actual symmetries. Using the quantum numbers under this
group we obtain the following matching of the boundary operators and
the supergravity moduli.
\bea
\label{sugra-cft}
\begin{array}{llc}
\mbox{Operator} & \mbox{Field}  &  
SU(2)_I\times \widetilde{SU(2)}_I \\
\del x^{ \{ i }_A(z) \bar{\del}x^{ j\} }_A (\bar z) -
\frac14 \delta^{ij}
\del x^{k}_A \bar{\del}x^k_A \
& h_{ij} - \frac14 \delta_{ij} h_{kk} & (\bf 3, \bf 3 ) \\
\del x^{[i}_A(z) \bar{\del}x^{j]}_A (\bar z) 
& b'_{ij}
& (\bf 3, \bf 1 ) +
(\bf 1, \bf 3)  \\ 
\del x^i_A(z) \bar{\del}x^i_A (\bar z) 
& \phi
&( \bf 1, \bf 1 ) \\
{\mathcal T}^1 & b^+_{ij} & (\bf 1, \bf 3 ) \\
{\mathcal T}^0 & a_1\chi + a_2C_{6789} & (\bf 1, \bf 1 )
\end{array}
\eea
Note that both the representations ${\bf (1,3)}$ and ${\bf (1,1)}$ occur
twice in the above table. This could give rise to a two-fold ambiguity
in identifying either ${\bf (1,3)}$ or ${\bf (1,1)}$ operators with
their corresponding bulk fields. The way we have resolved it here is
as follows.  
The operators ${\mathcal T}^1$ and ${\mathcal T}^0$ correspond to
blow up modes of the orbifold, and 
as we will show in Section 7 that  these are related to the
Fayet-Iliopoulos terms and the $\theta$-term in the 
\dd gauge theory. 
Tuning these operators one can reach the singular SCFT
\cite{Sei-Wit99} that corresponds to fragmentation of the \dd
system. In supergravity, similarly, it is only the moduli $b^+_{ij}$
and $a_1\chi + a_2C_{6789}$ which affect the stability of the \dd
system 
\cite{Sei-Wit99, chp2:LarMar, chp2:Dijkgraaf}. 
As a result, it
is $b^+_{ij}$ (and not $b^{\prime +}_{ij}$) which
should correspond to the operator ${\mathcal T}^1$ and similarly
$a_1\chi + a_2C_{6789}$ should correspond to ${\mathcal T}^0$. Another
reason for this identification is as follows. $b^+_{ij}$ and $a_1\chi
+a_2C_{6789}$ are odd under world sheet parity while $b^{\prime
+}_{ij}$ and $\phi$ are even under world sheet parity. 
In a $Z_2$ orbifolded theory there is a $Z_2$ symmetry which can be
used to classify the states \cite{Vafa:1989ih}. 
Under this symmetry the $Z_2$ quantum number of the twisted sectors is
$-1$ and the $Z_2$ quantum number of the untwisted sectors is $+1$. 
If under the AdS/CFT correspondence
one can identify these $Z_2$ quantum numbers in the boundary SCFT and
the bulk then the correspondence we have made is further justified.

Thus, we arrive at a one-to-one correspondence between
operators of the SCFT and the supergravity moduli.

\newpage
\section{Location of the symmetric product}

In the previous section we 
have studied the moduli of the \dd system in detail. In
\eq{sugra-cft} we have listed all the supergravity moduli and their
corresponding operators in the SCFT on ${\mathcal M}$. 
The \dd system is unspecified until all its moduli are given. In this
section we will find the location of the free field orbifold 
${\mathcal N=(4,4)}$ SCFT theory on ${\mathcal M}$ in the the \dd moduli
space.

It is easy to see from the mass formula of the \dd system, that the
\dd system is marginally stable to decay. The mass per unit length 
of the \dd system
is given by
\be
\label{massnomod}
M = \frac{1}{g_s 2\pi \alpha'} ( Q_1 + vQ_5)
\ee
Here $v$ is defined in \eq{prescal}. Note that the above formula is
linear in $Q_1$ and $Q_5$, therefore it does not cost any energy for
the \dd system to decay to sub-systems with smaller values of $Q_1$
and $Q_5$. But, when any of the moduli in \eq{sugra-cft} is turned on
then the \dd system is stable. In fact there is a binding energy which
prevents its decay. To see this let us turn on one of the moduli in
\eq{sugra-cft}. Consider turning on 
the self-dual
Neveu-Schwarz $B$-field along the D5-brane direction. We can choose it
to be given by
\be
\label{bfieldef}
B_{ij} = \frac{1}{2\pi \alpha'} \left(
\begin{array}{cccc}
0 & b & 0 & 0 \\
-b & 0 & 0& 0 \\
0 & 0 & 0 & b \\
0 & 0 & -b & 0 
\end{array}
\right)
\ee
Here $i, j$ runs from $6, \ldots 9$.
For convenience let the metric of on the $T^4$ be $\delta_{ij}$ and
$v=1$. To demonstrate that there is a binding energy it is sufficient
to consider the case of $Q_1 =1$ and $Q_5 =1$. The mass 
\footnote{Here mass refers to mass per unit length}  of a single
D1 brane is given by 
\be
M_{\rm{D1}} = \frac{1}{g_s 2\pi \alpha'}
\ee
Similarly the mass of the D5 brane wrapped on $T^4$ 
with the B-field \eq{bfieldef}  is given by
\be
\label{massdb}
M_{\rm{D5}} = \frac{1}{g_s 2\pi \alpha'}(1+b^2)
\ee
It is easy to understand this mass formula from the Dirac-Born-Infeld
action
\be
S= \frac{1}{g_s 2\pi \alpha'} \sqrt{\mbox{Det} (G+ 2\pi B)}
\ee
where $G$ is the induced metric. Substituting the value of $B$ 
from \eq{bfieldef} and the metric of the $T^4$ and expanding the
Dirac-Born Infeld action in the static gauge we obtain the mass for
the D5 brane with the B-field as given by \eq{massdb}. The mass of the
\dd system with the B-field  is given by \cite{Obe-Pio98} 
\be
M = \frac{1}{g_s 2\pi \alpha'} \sqrt{ (Q_1 + (1-b^2)Q_5)^2 + 4b^2
Q_5^2)}
\ee
Here $Q_1$ and $Q_5$ stand for the number of D1 and D5 branes
respectively. Substituting $Q_1 =Q_5=1$ we get
\be
M_{\rm{D1-D5}} = \frac{1}{g_s 2\pi \alpha'} \sqrt{ 4+ b^4}
\ee
The the binding energy is given by
\be
\Delta M = (M_{\rm{D1}} + M_{\rm{D5}}) -M_{\rm{D1-D5}}
\ee
It is easy to see this binding energy is positive \footnote{If the NS
B-field was anti-self dual then the binding energy is zero, and the
\dd system is marginally stable. }.  One can repeat similar
calculations with the other moduli given in \eq{sugra-cft} and
demonstrate the existence of a positive binding energy. This issue of
the stability of the \dd system with the various moduli turned on
has been discussed in \cite{Sei-Wit99, chp2:Dijkgraaf, chp2:LarMar,
DMWY,Mikhailov:1999fd}.

It has been observed in \cite{Sei-Wit99} that the effective theory of
a single D1 brane separating off the \dd bound state is a
linear dilaton theory. This was derived by studying the dynamics of a
D1 brane close to the boundary of $AdS_3$. 
In section 7.1 we derive the effective theory of a set of $q_1$ D1
branes and $q_5$ D5 branes splitting of the \dd bound state.
The  AdS/CFT 
correspondence suggests  that 
this decay of the \dd system  should also be seen from the conformal
field theory of the Higgs branch of the \dd system.
It is convenient to extract the dynamics of the decay from the \dd
gauge theory. In fact 
such a decay signals a singularity in the world volume gauge
theory associated with the origin of the Higgs branch.  
The dynamics of the decay can be extracted from the \dd gauge theory
using the methods developed by \cite{Aharony:1999dw}. 
In section 7.3 we extract the
dynamics of the decay from the \dd gauge theory and show that it is
described by the same linear dilaton theory observed by
\cite{Sei-Wit99} in supergravity.

\begin{figure}
\centerline{
\epsffile{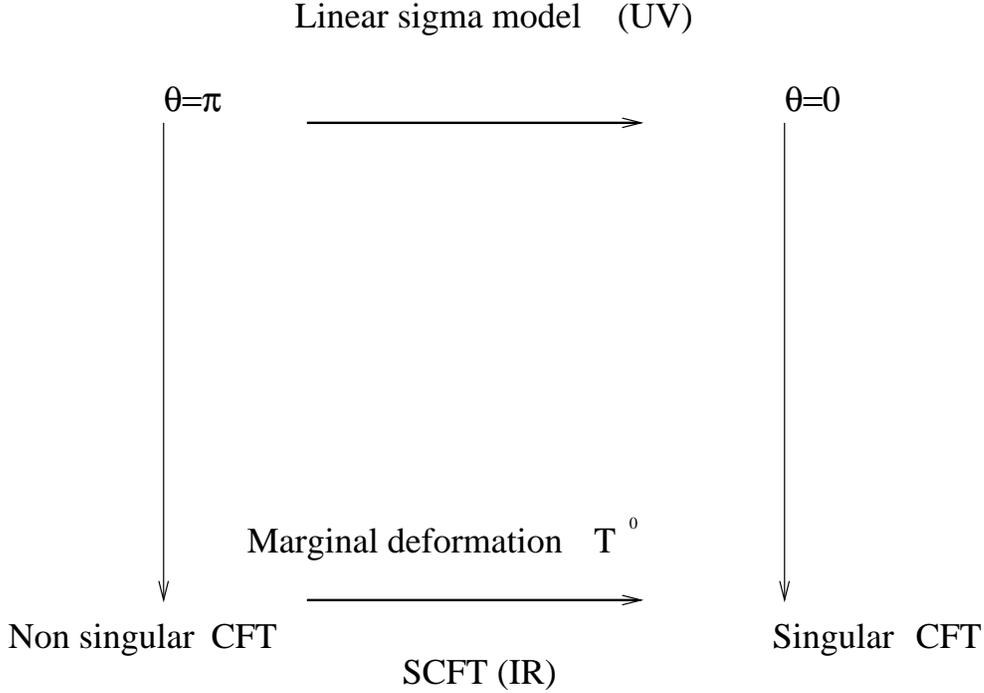}
}
\caption{Linear sigma model and CFT description of the $R^4/Z_2$
singularity}
\label{sigmodel}
\end{figure}

The singularity mentioned above leads to a singular conformal field
theory.  However, generic values of the supergravity moduli which do
not involve fragmentation into constituents are described by
well-defined conformal field theories and therefore string
perturbation theory makes sense.  
We have seen in section \ref{chp2:res} that the important 
singularity structure of
the ${\mathcal M}= 
(4,4)$ SCFT on the orbifold ${\mathcal M}$ is locally of the 
type $R^4/Z_2$. The resolution of this singularity gives rise to
marginal operators. An orbifold theory realized as a free 
field SCFT on $R^4/Z_2$
is nonsingular as all correlations functions are finite. 
The reason for this can
be understood from the linear sigma model description of the $R^4/Z_2$
singularity which will be discussed in section \ref{chp2:z_2sing}. 
We will see that the though
the $R^4/Z_2$ singularity is geometrically singular the 
SCFT is finite because it
corresponds to a non-zero theta term in the linear sigma model. The 
geometric resolution of
this singularity corresponds to adding Fayet-Iliopoulos 
terms to the D-term
equations of the linear sigma model. This deforms 
the $R^4/Z_2$ singularity to an
Eguchi-Hansen space. In the orbifold theory this deformation is caused
by the twist operator ${\mathcal T}^1$.
The Eguchi-Hansen space is 
asymptotically $R^4/Z_2$ but the
singularity at the origin is blown up to a 2-sphere. 
One can use the $SU(2)_R$ symmetry of the linear sigma model to rotate
the three 
Fayet-Iliopoulos terms to one term. This term 
corresponds to the radius of the blown up 2-sphere. 
The theta term of the linear
sigma model corresponds to $B$-flux through the 2-sphere. The change
of this $B$-flux is caused by deforming the orbifold SCFT by the twist
operator ${\mathcal T}^0$.
Thus SCFT realized as a free field theory on the orbifold $R^4/Z_2$ is 
regular even though the 2-sphere is squashed to zero size because of
the non-zero value of $B$-flux trapped in the squashed 2-sphere
\cite{Aspinwall:1995zi}. We summarize this discussion 
in (\fig{sigmodel}).

For most of our discussion we have assumed that the Higgs branch of
the \dd gauge theory is a resolution
${\mathcal N}=(4,4)$ theory on
${\mathcal
M}$. Furthermore we have realized this theory as a free field theory
with orbifold identification. This 
implies that we are at a  point in the moduli space of the
\dd  system at which the orbifold is geometrically singular but
because of the non-zero value of the theta term the SCFT is regular
and {\em not at the singularity corresponding to
fragmentation}. In other words, the orbifold SCFT corresponds to a
bound state of $Q_1$ D1 branes and $Q_5$ D5 branes (henceforth denoted
as the $(Q_1,Q_5)$ bound state). 
The supergravity solution \eq{near-horizon} has no moduli
turned on. This implies that the SCFT dual is singular and is far away
in moduli space from the regular conformal theory on ${\mathcal M}$. 
But, the fact
that we could show that the all the short multiplets of the
supergravity modes on $AdS_3\times S^3$ is in one-to-one
correspondence with the short multiplets of the SCFT on ${\mathcal M}$
implies that these multiplets are protected from non-renormalization
theorems. This will be discussed in detail in section 9.

\subsection{ Dynamics of the decay of the \dd system from gravity}

We consider a set of $(q_1, q_5)$ test 
D-branes with $q_1, q_5\ll Q_1, Q_5$
close to the boundary of $AdS_3$ but separated from the rest of the
branes of the \dd system.
When the test branes are close to
the boundary it is easy to see using the UV/IR correspondence that the
the gauge group is broken to $U(q_1)\times U(q_5)\times U(Q_1-q_1)\times
U(Q-q_5)$from $U(Q_1)\times U(Q_5)$ in the IR. 
Thus we can extract the infrared dynamics
of the decay of the \dd system from gravity.

Lets us first consider the case when $q_5=0$ \cite{Sei-Wit99}. 
The AdS/CFT
correspondence tells us that we need to consider $q_1$ D1 branes in
the background of $AdS_3\times S^3\times T^4$. 
(We will work in  the Euclidean $AdS_3$ coordinates.)
The radius of $S^3$ and
the anti-de Sitter space is given by $r_0=\sqrt{\alpha^{\prime}}
(g_6^2Q'_1Q'_5)^{1/4}$~\footnote{\label{ft:r0}We have called this quantity
$R =\sqrt{\alpha'}l$ in \eq{radius-ads},\eq{action-parameters};
the $r_0$ here is not be confused with the non-extremality
parameter, such as in \eq{non-extreme-2a}.},
where $Q'_1= Q_1-q_1$ and $Q'_5 = Q_5-q_1$. 
For the supergravity to be valid we need to
consider the limit $r_0\gg 0$. 
Let us focus on the distance between the
boundary of $AdS_3\times S^3\times T^4$ and the set of $q_1$
D1 branes. We are interested in the infrared description of the
splitting process. By the  
UV/IR correspondence the D1 branes should be close
to the boundary of the $AdS_3\times S^3\times T^4$ to obtain the
infrared description of the splitting process in the supergravity. We
assume that the D1 branes are fixed at a particular point on the $S^3$
and the $T^4$. The action of $q_1$ D1 branes in the background of
$AdS_3$ and the Ramond-Ramond two-form $B_{05}$ is given by the DBI
action. We can use the DBI action for multiple D1 branes as we are
interested only in the dynamics of the centre of mass of thee
collection of $q_1$ D1 branes. The DBI action of $q_1$ D1 branes is
given by
\be
\label{chp3:dbi}
S= \frac{q_1}{2\pi g_s\alpha^{\prime}} \int d^2 \sigma
e^{-\phi}\sqrt{\mbox{ det }( g_{\alpha\beta}^{{\rm ind}})} -
\frac{q_1}{2\pi
g_s\alpha^{\prime}} \int B
\ee
where $\sigma$ stands for the world volume coordinates and
$\alpha,\beta$ label these coordinates. $g_{\alpha\beta}^{{\rm ind}}$ 
is the
induced metric on the world volume. $B$ is the Ramond-Ramond 2-form
potential. We chose a gauge in which the world volume coordinates are
the coordinates of the boundary of the $AdS_3$. Let the metric on the
boundary be $g_{\alpha\beta}(\sigma)$. 
One can extend the metric $g_{\alpha\beta}(\sigma)$ on the boundary to
the interior of $AdS_3$ in the neighbourhood of the boundary
\cite{Sei-Wit99}. This is
given by
\be
\label{chp3:bcmetric}
ds^2= \frac{r_0^2}{t^2}\left( dt^2 + \hat{g}_{\alpha\beta}
(\sigma,t)d\sigma^{\alpha}d\sigma^{\beta} \right)
\ee
with
\be
\hat{g}_{\alpha\beta}(\sigma, 0) = g_{\alpha\beta}(\sigma), \;\;\;
\hat{g}_{\alpha\beta}(\sigma, t) = g_{\alpha\beta}(\sigma ) -
t^2P_{\alpha\beta} + O(t^3)+\ldots
\ee
Here $g_{\alpha\beta}P^{\alpha\beta}=R/2$. $R$ is the world sheet
curvature. The global coordinates of Euclidean $AdS_3$ is given by
\footnote{This can be obtained from \eq{eucl-hyp},\eq{metric-h}
by defining $Y_{-1}= l \cosh\phi,
\left(Y_0, Y_1, Y_2 \right) = l \sinh\phi \vec \Omega,\,
\vec \Omega \in S^2$, and then replacing the
notation $l$ by $r_0$, see the previous footnote
\eq{ft:r0}.}
\be
ds^2= r_0^2(d\phi^2 + \sinh^2\phi d\Omega^2)
\ee
where $d\Omega^2$ is the round metric on $S^2$. 
Near the boundary the metric is given by
\be
ds^2 = r_0^2(d\phi^2 +\frac{e^{2\phi}}{4} d\Omega^2)
\ee
Motivated by this we
use $\phi$ defined as 
$t=2e^{-\phi}$ to measure the distance from the boundary of
$AdS_3$. Substituting the metric in \eq{chp3:bcmetric} and the near
horizon value of the Ramond-Ramond 2-form and the
dilaton  in \eq{chp3:dbi} we obtain
the following effective action of the D1 branes near the boundary.
\bea
\label{chp3:d1eff}
S&=& \frac{q_1r_0^2 }{4\pi g_s \alpha^{\prime}}
\sqrt{\frac{Q'_5 v}{Q'_1}}
\int \sqrt{g} \left(
\partial_{\alpha}\phi\partial^{\alpha}\phi + \phi R -\frac{1}{2}R
 + O(e^{-2\phi}) \right) \\ \nonumber
&=& \frac{q_1Q_5}{4 \pi} \int \sqrt{g} \left( 
\partial_{\alpha}\phi\partial^{\alpha}\phi + \phi R -\frac{1}{2}R
+ O(e^{-2\phi})\right)
\eea

Now consider the case when $q_1=0$. The $q_5$ D5 branes are wrapped on
$T^4$. Therefore the world volume of the D5 branes is of the form
$M_2\times T^4$ where $M_2$ is any 2-manifold. The D5 branes are
located at a point on the $S^3$.
We ignore the fluctuations on $T^4$ as we are
interested in the dynamics on $AdS_3$. The DBI action of $q_5$
D5 branes is given by
\be
\frac{q_5}{32\pi^5g_s\alpha^{\prime 3} }
\left( \int d^6\sigma 
e^{-\phi} \sqrt{\mbox{det}(g_{\alpha\beta}^{{\rm ind}})}
-\int C^{6} \right)
\ee
where $C^6$ is the Ramond-Ramond 6-form potential coupling to 
the D5 brane. 
Performing a similar calculation for the D5 branes and substituting
the near horizon values of the 6-from Ramond-Ramond potential, the
dilaton and the volume of $T^4$ one obtains the following effective
actions for the D5 branes
\bea
\label{chp3:d5eff}
S&=& \frac{q_5r_0^2}{4\pi g_2\alpha^{\prime}}\sqrt{\frac{Q'_1 v}{Q'_5}}
\int \sqrt{g} \left(
\partial_{\alpha}\phi\partial^{\alpha}\phi + \phi R -\frac{1}{2}R
+ O(e^{-2\phi})\right) \\ \nonumber
&=& \frac{q_5Q_1}{4 \pi} \int \sqrt{g} \left( 
\partial_{\alpha}\phi\partial^{\alpha}\phi + \phi R -\frac{1}{2}R
+ O(e^{-2\phi})\right)
\eea

For the case when $q_1\neq 0$ and $q_5\neq 0$ and we just add the 
contribution from \eq{chp3:d1eff} and \eq{chp3:d5eff} to obtain the
effective action of the $(q_1, q_5)$ string in $AdS_3$. The reason we
can do this is because there is no force between the test D1 and
D5 branes. 
Thus to the leading order in
$\phi$ the total effective action
of the $(q_1, q_5)$ string near the boundary is given by
\be
\label{liouville}
S= \frac{(q_1Q_5 + q_5Q'_1)}{4\pi} \int 
\sqrt{g} \left( 
\partial_{\alpha}\phi\partial^{\alpha}\phi + \phi R -\frac{1}{2}R
\right)
\ee
Rescaling $\phi$ so that the the normalization of the kinetic energy
term is canonical one obtains a linear dilaton action with a back
ground charge given by
\be
\label{chargesugra}
Q_{{\rm SUGRA}} = \sqrt{2(q_1Q'_5 + q_5Q'_1)}
\ee
To summarize, the effective dynamics of the $(Q_1, Q_5)$ \dd system by
decaying into 
$(q_1, q_5)$ branes is governed by linear dilaton theory with
background charge given by \eq{chargesugra}. Note that the linear
dilaton theory in \eq{liouville} is strongly coupled at 
$\phi\rightarrow\infty$, the boundary of $AdS_3$.

If a similar analysis is performed for the 
supergravity solution with the self dual
NS $B$-field turned on \eq{full-solution},  one obtains a potential
for the linear dilaton which prevents the coupling to grow to infinity
at the boundary of $AdS_3$ \cite{DMWY}. Thus the effective theory is
non-singular in the presence of the NS $B$-field.


\subsection{The linear sigma model description of $R^4/Z_2$}
\label{chp2:z_2sing}

The linear sigma model is a $1+1$ 
dimensional $U(1)$ gauge theory with $(4,4)$ supersymmetry 
\cite{Witten:1995zh}. 
It has $2$
hypermultiplets charged under the $U(1)$. The scalar fields of the 
hypermultiplets can be
organized as doublets under the $SU(2)_R$ symmetry of the $(4,4)$
theory as
\bea
\chi_1=
\left(
\begin{array}{c}
A_1 \\ B_1^\dagger
\end{array}
\right)  \;\mbox{and} \;
\chi_2=
\left(
\begin{array}{c}
A_2 \\ B_2^\dagger
\end{array}
\right) 
\eea
The $A$'s have charge $+1$ and the $B$'s have charge $-1$ under the
$U(1)$. The vector multiplet has $4$ real 
scalars $\varphi_i$, $i=1,\ldots ,4$.
They do not transform under the $SU(2)_R$. One can 
include $4$ parameters
in this theory consistent with (4,4) supersymmetry. They are the $3$
Fayet-Iliopoulos terms and the theta term. 

Let us first investigate the hypermultiplet moduli space of this
theory with the $3$ Fayet-Iliopoulos terms and the theta term set to
zero. The Higgs phase of this theory is obtained by setting $\phi_i$
and the D-terms to zero. The D-term equations are
\bea
\label{dtermr4}
|A_1|^2 + |A_2|^2 -|B_1|^2 -|B_2|^2 &=&0 \\  \nonumber
A_1B_1 + A_2B_2 &=&0
\eea
The hypermultiplet moduli space is the space of solutions of the above
equations modded out by the $U(1)$ gauge symmetry. Counting the number
of degrees of freedom indicate that this space is $4$ dimensional. To
obtain the explicit form of this space it is convenient to introduce
the following gauge invariant variables
\bea
\label{dtermdef1}
M=A_1B_2 \; &\;& \; N= A_2B_1 \\ \nonumber
P= A_1B_1 &=& - A_2B_2 \\
\eea
These variables are not independent. Setting the D-terms equal to zero
and modding out the resulting space by $U(1)$ is equivalent to the
equation
\be
\label{surface}
P^2 +MN=0.
\ee
This homogeneous equation is an equation of the space $R^4/Z_2$. To see
this the solution of the above equation can be parameterized by $2$
complex numbers $(\zeta , \eta)$ such that
\be
\label{dtermdef2}
P= i\zeta \eta \;\; M=\zeta^2 \;\; N= \eta^2
\ee
Thus the point $(\zeta, \eta)$ and $(-\zeta, -\eta)$ are the same
point in the space of solutions of \eq{surface}. We have
shown that the hypermultiplet moduli space is $R^4/Z_2$. 

The above singularity at the origin of the moduli space 
is a geometric singularity in the hypermultiplet
moduli space. We now argue that this singularity is a genuine
singularity of the SCFT that the linear sigma model flows to in the
infrared. At the origin of the classical moduli space the Coulomb
branch meets the Higgs branch. In addition to the potential due to the
D-terms the linear sigma model contains the following term in the
superpotential\footnote{These terms can be understood from
the coupling $A_\mu A^\mu \chi^* \chi$ in six dimensions,
and recognizing that under dimensional reduction to two
dimensions $\varphi_i$'s appear from the components of $A_\mu$
in the compact directions.}
\be
V = 
(|A_1|^2 + |A_2|^2 + |B_1|^2 + |B_2|^2)(\varphi_1^2 + 
\varphi_2^2 +\varphi_3^2 +\varphi_4^2)
\ee
Thus at the origin of the hypermultiplet moduli space a flat direction
for the Coulomb branch opens up. The ground state at this point is not
normalizable due to the non-compactness of the Coulomb branch. This
renders the infrared SCFT singular.

This singularity can be avoided in two distinct ways. If one turns on
the Fayet-Iliopoulos D-terms, the D-term equations are modified to
\cite{Witten:1995zh}~\footnote{\label{ft:Fayet}
cf. Eqns. \eq{Tcpn1},\eq{Tcpn2}, whose parameters $\zeta,\zeta_c$
are related to the present parameters $r_1, r_2, r_3$ by
$ r_3 \equiv \zeta,\,  r_1 + i r_2 \equiv \zeta_c.$}  
\bea |A_1|^2 + |A_2|^2 -|B_1|^2 -|B_2|^2 &=&r_3 \\
\nonumber A_1B_1 + A_2B_2 &=&r_1+ir_2 
\eea 
Where $r_1,r_2,r_3$ are the
$3$ Fayet-Iliopoulos D-terms transforming as the adjoint of the
$SU(2)_R$. Now the origin is no more a solution of these equations and
the non-compactness of the Coulomb branch is avoided. In this case
wave-functions will have compact support on the Coulomb branch.
This ensures that the infrared SCFT is
non-singular. Turning on the Fayet-Iliopoulos D-terms thus correspond
to the geometric resolution of the singularity. The resolved
space is known  to be \cite{Witten:1995zh, Aspinwall:1995zi} 
described by 
an Eguchi-Hanson metric in which $r_{1,2,3}$ parameterize
a shrinking two-cycle.

The second way to avoid the singularity in the SCFT is to turn on the
theta angle $\theta$. This induces a constant electric field in the
vacuum. This electric field is screened at any other point than the
origin in the hypermultiplet moduli space as the $U(1)$ gauge field is
massive with a mass proportional to the vacuum expectation value of the
hypers. At the origin the $U(1)$ field is not screened and thus it
contributes to the energy density of the vacuum. This energy is
proportional to $\theta^2$. Thus turning on the theta term lifts the
flat directions of the Coulomb branch. This ensures that the
corresponding infrared SCFT is well defined though the hypermultiplet
moduli space remains geometrically singular. In terms of the
Eguchi-Hanson space, the $\theta$-term corresponds to a flux of the
antisymmetric tensor through the two-cycle mentioned above.

The $(4,4)$ SCFT on $R^4/Z_2$ at the orbifold point is well defined.
Since the orbifold has a geometric singularity but the SCFT is
non-singular it must correspond to the linear sigma model with a
finite value of $\theta$ and the Fayet-Iliopoulos D-terms set to
zero.  Deformations of the $R^4/Z_2$ orbifold by its $4$ blow up modes
correspond to changes in the Fayet-Iliopoulos D-terms and theta term
of the linear sigma model%
\footnote{If we identify the $SU(2)_R$ of the linear sigma-model
with $\widetilde{SU(2)}_I$ of the orbifold SCFT, then the
Fayet-Iliopoulos parameters will correspond to ${\mathcal T}^1$
and the $\theta$-term to ${\mathcal T}^0$. 
This is consistent with Witten's observation \cite{witt-higgs-br}
that $SO(4)_E$ symmetry  of the linear sigma-model (one
that rotates the $\phi_i$'s) corresponds to the $SU(2)_R$
of the orbifold SCFT.} 
The global description of the moduli of a
${\mathcal N} =(4,4)$ SCFT on a resolved $R^4/Z_2$ orbifold is provided by
the linear sigma model.  In conclusion let us describe this linear
sigma model in terms of the gauge theory of D-branes. The theory
described above arises on a single D1-brane in presence of $2$
D5-branes. The singularity at the point $r_1, r_2, r_3, \theta
=0$ is due to noncompactness of the flat direction of the Coulomb
branch.  Thus it corresponds to the physical situation of the D1-brane
leaving the D5-branes.

\subsection{The gauge theory relevant for the decay of the \dd  
system}
\label{chp2:linear_sigma}

As we have seen in Section 5.6 
the resolutions of the ${\mathcal N}=(4,4)$
SCFT on ${\mathcal M}$ is described by 4 marginal operators which were
identified in the previous subsection with the Fayet-Iliopoulos D-terms
and the theta term of the linear sigma model description of the
$R^4/Z_2$ singularity. We want to now indicate how these four
parameters would make their appearance in the gauge theory description
of the full \dd  system.

Motivated by the
D-brane description of the $R^4/Z_2$ singularity we look for the
degrees of freedom characterizing the break up of 
$(Q_1, Q_5)$ system
to $(q_1, q_5)$ and $(Q'_1, Q'_5)$ where $Q'_1= Q_1-q_1$ and  
$Q'_5 = Q_5-q_5$. 
Physically the relevant degree
of freedom describing this process is the relative coordinate between
the centre of mass of the $(q_1, q_5)$ system and the $(Q'_1,
Q'_5)$. We will describe the effective theory of this degree of
freedom below.

For the bound state $(Q_1, Q_5)$ the hypermultiplets,  $\chi$
are charged under the relative $U(1)$ of $U(Q_1)\times
U(Q_5)$, that is under the gauge field 
$A_\mu = \mbox{Tr}_{U(Q_1)}( A _\mu) - \mbox{Tr}_{U(Q_5)}(A_\mu)$.  
The relative $U(1)$ gauge multiplet corresponds to the degree of
freedom of the relative coordinate between the centre of mass of the
collection of $Q_1$ D1-branes and $Q_5$ D5-branes. At a generic point
of the Higgs phase, all the $\chi$'s have expectation
values, thus making this degree of freedom becomes massive. This is
consistent with the fact that we are looking at the bound state $(Q_1,
Q_5)$.

Consider the break up of the $(Q_1, Q_5)$ bound state to 
the bound states $(q_1, q_5)$ and $(Q'_1, Q'_5)$. To find out the
charges of the hypermultiplets under the various $U(1)$, we will
organize the hypers as
\bea
\chi = \left(
\begin{array}{cc}
\chi_{a\bar{b}} &  \chi_{a'\bar{b}'}    \\ 
\chi_{a'\bar{b}} & \chi_{a'\bar{b}'}
\end{array}
\right), \quad
Y^{(1)}_i= 
\left(
\begin{array}{cc}
Y_{i(a\bar{a})}^{(1)} & Y_{i(a\bar{a}')}^{(1)}  \\
Y_{i(a'\bar{a})}^{(1)} & Y_{i(a'\bar{a}')}^{(1)}
\end{array}
\right)
\quad \mbox{and} \quad
Y^{(5)}_i=
\left(
\begin{array}{cc}
Y_{i(b\bar{b})}^{(5)} & Y_{i(b\bar{b}')}^{(5)}  \\
Y_{i(b'\bar{b})}^{(5)} & Y_{i(b'\bar{b}')}^{(5)}
\end{array}
\right)
\eea
where $a,\bar{a}$ runs from $1,\ldots , q_1$, $b,\bar{b}$  
from $1,\dots ,q_5$,
$a'\bar{a}'$ from $1, \ldots Q'_1$ and 
$b',\bar{b}'$ from $1\ldots ,Q'_5$. 
We organize the scalars of the vector multiplet corresponding to the
gauge group $U(Q_1)$ and $U(Q_5)$ as
\bea
\phi_m^{(5)} =
\left(
\begin{array}{cc}
\phi_{m}^{(1)a\bar{a}} & \phi_{m}^{(1)a\bar{a}'} \\
\phi_{m}^{^(1)a'\bar{a}}& \phi_{m}^{(2)a'\bar{a}'}
\end{array}
\right)
\quad
\mbox{and}
\quad
\phi_m^{(5)} =
\left(
\begin{array}{cc}
\phi_{m}^{^(5)b\bar{b}} & \phi_{m}^{(5)b\bar{b}'} \\
\phi_{m}^{^(5)b'\bar{b}}& \phi_{m}^{(5)b'\bar{b}'}
\end{array}
\right)
\eea
where $m=1,2,3,4$. 

Let us call the the $U(1)$ gauge fields (traces) of 
$\, U(q_1),\, U(q_5),
U(Q'_1),\, U(Q'_5)\, $ as $\, A_1,\,  A_5, \, A'_1, \, A'_5\, $ 
respectively.  We
will also use the notation $A_\pm \equiv A_1 \pm A_5$ and $A'_\pm
\equiv A'_1 \pm A'_5$.

As we are interested in the bound states $(q_1, q_5)$ and $(Q'_1,
Q'_5)$, in what follows we will work with a specific classical
background in which we give {\em vev}'s to the block-diagonal hypers
$\chi_{ab},\chi_{a'b'}, Y^{(1)}_{i( a \bar a)},Y^{(5)}_{i(b
\bar b)},Y^{(1)}_{i( a' \bar a')}$ and $Y^{(5)}_{i(b' \bar b')}$.
These {\em vev}'s are chosen so that the classical background
satisfies the D-term equations \eq{D-terms}.

The {\em vev}'s of the $\chi$'s  render the fields
$A_-$ and $A'_-$ massive with a mass proportional to {\em vev}'s. In
the low energy effective Lagrangian these gauge fields can therefore
be neglected.  In the following we will focus on the $U(1)$ gauge
field $A_r=1/2(A_+ - A'_+)$ which does not get mass from the above
{\em vev}'s. The gauge multiplet corresponding to $A_r$ contains four
real scalars denoted below by $\varphi_m$. These represent the
relative coordinate between the centre of mass of the $(q_1, q_5)$
and the $(Q'_1, Q'_5)$ bound states. We will be interested in the
question whether the $\varphi_m$'s remain massless or otherwise. The
massless case would correspond to a non-compact Coulomb branch and
eventual singularity of the SCFT.

In order to address the above question we need to find the low energy
degrees of freedom which couple to the gauge multiplet corresponding
to $A_r$.  

The fields charged under $A_r$ are the hypermultiplets 
$\, \chi_{a\bar
b'},\;  \chi_{a'\bar b}, \; Y^{(1)}_{i(a\bar a')}, \;
Y^{(1)}_{i(a'\bar a)},$ \\
$Y^{(5)}_{i(b\bar b')},\,  Y^{(5)}_{i(b'\bar b)}\, $ 
and the vector multiplets
$\phi^{(1)a\bar a'}_{m}, \phi^{(1){a'\bar a}}_m, 
\phi^{(5)b\bar
b'}_{i}, \phi^{(5)b'\bar
b}_m$. In order to 
find out which of these are massless, we
look at the following terms in the Lagrangian of
 $U(Q_1)\times U(Q_5)$ gauge theory:
\bea
\label{lagrangian}
L &=& L_1 + L_2 + L_3 + L_4\\  \nonumber
L_1&=& 
\chi^\dagger\phi_m^{\dagger}
\phi_m\chi \\ \nonumber
L_2&=&
\chi^\dagger\phi_m^{\dagger}
\phi_m\chi_{a_1\bar{b}_3} \\  \nonumber
L_3&=&Tr([Y^{(1)}_i,Y^{(1)}_j][Y^{(1)}_i, Y^{(1)}_j]) \\  \nonumber
L_4&=&Tr([Y^{(5)}_i,Y^{(5)}_j][Y^{(5)}_i, Y^{(5)}_j]) \\   \nonumber
\eea
The terms $L_1$ and $L_2$ originate from terms of
the type $|A_M\chi|^2$ where $A_M \equiv (A_\mu, \phi_m)$
is the $(4,4)$ vector multiplet in two dimensions. The terms
$L_3$ and $L_4$ arise from commutators of gauge fields in
compactified directions.  

The fields $Y$ are in general
massive. The reason is that the traces $y^{ (1)}_{i}
\equiv \mbox{Tr}(Y_{i(a\bar{a})}^{(1)})$, representing
the centre-of-mass position in the $T^4$ of $q_1$
D1 branes, and  $y^{\prime (1)}_{i}
\equiv \mbox{Tr}(Y_{i(a'\bar{a}')}^{(1)})$, representing
the centre-of-mass position in the $T^4$ of $Q'_1$
D1 branes, are neutral and will have {\em vev}'s which
are generically separated (the centres of mass can be
separated in the torus even when they are on top
of each other in physical space). 
 The mass of $Y_{i(a\bar{a}')}^{(1)} , 
Y_{i(a'\bar{a})}^{(1)}$ can be read off from the term $L_3$ in 
\eq{lagrangian}, to be  proportional to
 $ (y^{ (1)} - y^{\prime (1)})^2$
Similarly the
mass of $Y_{i(b\bar{b}')}^{(5)} , Y_{i(b\bar{b}')}^{(5)} $ is
proportional to $(y^{(5)} -y^{\prime (5)})^2 $ 
(as can be read off from the term $L_4$ in \eq{lagrangian})
where $y^{(5)}$ and
$y^{\prime (5) }$ are the centers of mass of the 
$Q_5$ D5 branes and $Q'_5$
D5 branes along the direction of the dual four torus $\hat{T}^4$. 
(At special points when their centres of mass coincide, these fields
become massless. The analysis for these cases can also be carried out
by incorporating these fields in \eq{dterm}-\eq{resolved}, with no
change in the conclusion) The fields $\phi_{m}^{(1)a\bar{a}'},
\phi_{m}^{^(1)a'\bar{a}}$ are also massive. Their masses can be read
off from the $L_1$ in \eq{lagrangian}.
Specifically they arise from the following terms
\be
\chi^*_{a'_1\bar{b}'}\phi_m^{(1)a\bar{a}'_1*}
\phi_m^{(1)a\bar{a}'_2}
\chi_{a'_2\bar{b}'}+
\chi^*_{a_1\bar{b}}\phi_m^{(1)a'\bar{a}_1*}\phi_m^{(1)a'\bar{a}_2}
\chi_{a_2\bar{b}} 
\ee 
where $a_i$ run from $1,\dots q_1$ and
$a'_i$ run form $1,\ldots Q'_1$.  These terms show that
their masses are proportional to the expectation values of the hypers
$\chi_{a\bar{b}}$ and $\chi_{a'\bar{b}'}$. 
Similarly the terms of $L_2$ in
\eq{lagrangian} 
\be
\chi^*_{a'\bar{b}'_1}\phi_m^{(5)b'_1{b}*}\phi_m^{(5)b'_2\bar{b}}
\chi_{a'\bar{b}'_2}+
\chi^*_{a\bar{b}_1}\phi_m^{(5)b_1\bar{b}'*}\phi_m^{(5)b_2\bar{b}'}
\chi_{a\bar{b}_2} 
\ee 
show that the fields $\phi_{m}^{(5)b\bar{b}'}
\phi_{m}^{(5)b'\bar{b}}$ are massive with masses proportional to the
expectation values of the hypers $\chi_{a\bar{a}}$ and
$\chi_{a'\bar{b}'}$. In the above equation $b_i$ take values
from $1,\ldots , q_5$ and $b'_i$ take values from $1,\ldots
, Q'_5$. Note that these masses remain non-zero even in the limit
when the $(q_1,q_5)$ and $(Q'_1,Q'_5)$ are on the verge of
separating.

\emph{ 
Thus the relevant degrees of freedom describing the splitting process
is a $1+1$ dimensional $U(1)$ gauge theory of $A_r$ with $(4,4)$
supersymmetry.  The matter content of this theory consists of
hypermultiplets $\chi_{a\bar{b}'}$ with charge $+1$ and
$\chi_{a'\bar{b}}$ with charge $-1$. This theory consists of totally 
$ q_1Q'_5 +q_5Q'_1$ hypers.} We define the individual components of
hypers as the following doublets
\be
\chi_{a\bar{b'}} = \left( 
\begin{array}{c}
A_{a\bar{b}'} \\
B^\dagger_{a\bar{b}'}
\end{array}
\right)
\quad
\chi_{a'\bar{b}} = \left( 
\begin{array}{c}
A_{a'\bar{b}} \\
B^\dagger_{a'\bar{b}}
\end{array}
\right)
\ee
Let us now describe the dynamics of the splitting process. This is
given by analyzing the hypermultiplet moduli space of the effective
theory described above with the help of the D-term equations:
\bea
\label{dterm}
A_{a\bar{b}'}A^*_{a\bar{b}'}-A_{a'\bar{b}}A^*_{a'\bar{b}} - 
B_{b'\bar{a}}B^*_{b'\bar{a}}
+B_{b\bar{a'}}B^*_{b\bar{a}'} = 0 \\ \nonumber
A_{a\bar{b}'}B_{b'\bar{a}} -A_{a'\bar{b}}B_{b\bar{a}'} =0
\eea
In the above equations the sum over $a, b, a' , b'$ is understood.
These equations are generalized version of \eq{dtermr4} 
discussed for the
$R^4/Z_2$ singularity in Section 4.1. 
At the origin of the Higgs branch where the
classical moduli space meets the Coulomb branch this linear sigma model
would flow to an infrared conformal field theory which is singular. 
The reason for this is the same as for the
$R^4/Z_2$ case. The linear sigma model contains the following term in
the superpotential
\be
\label{potential}
V= (A_{a\bar{b}'}A^*_{a\bar{b}'} + 
A_{a'\bar{b}}A^*_{a'\bar{b}} + B_{b'\bar{a}}B^*_{b'\bar{a}} +
B_{b\bar{a}'}B^*_{b\bar{a}'})
(\varphi_1^2 + \varphi_2^2 + \varphi_3^2 +\varphi_4^2)
\ee
As in the discussion of the $R^4/Z_2$ case, at the origin of the
hypermultiplet moduli space the flat direction of the Coulomb branch
leads to a ground state which is not normalizable.  This singularity
can be avoided by deforming the D-term equations by
the Fayet-Iliopoulos terms (cf. Eqns. \eq{Tcpn1},\eq{Tcpn2}
and footnote \eq{ft:Fayet}):
\bea
\label{resolved}
A_{a\bar{b}'}A^*_{a\bar{b}'}-A_{a'\bar{b}}A^*_{a'\bar{b}} -
B_{b'\bar{a}}B^*_{b'\bar{a}} +B_{b\bar{a}'}B^*_{b\bar{a}'} =
r_3 \\ \nonumber A_{a\bar{b}'}B_{b'\bar{a}}
-A_{a'\bar{b}}B_{b\bar{a}'} = r_1 + ir_2 
\eea
We note here that
the Fayet-Iliopoulos terms break the relative $U(1)$ under discussion
and the gauge field becomes massive.  The reason is that the D-terms
with the Fayet-Iliopoulos do not permit all $A,B$'s in the above
equation to simultaneously vanish.  At least one of them must be
non-zero. As these $A, B$'s are charged under the $U(1)$, the non-zero
of value of $A,B$ gives mass to the vector multiplet. This can be seen
from the potential \eq{potential}. The scalars of the vector multiplet
becomes massive with the mass proportional to the {\em vev}'s of $A,B$.
Thus the relative $U(1)$ is broken.

The singularity associated with the non-compact Coulomb branch can also
be avoided by turning on the $\theta$ term, the mechanism being
similar to the one discussed in the previous subsection. If any of the
$3$ Fayet-Iliopoulos D-terms or the $\theta$ term is turned on, the
flat directions of the Coulomb branch are lifted, leading to
normalizable ground state is of the Higgs branch. This prevents the
breaking up of the $(Q_1,Q_5)$ system to subsystems. Thus we see that
the $4$ parameters which resolve the singularity of the ${\mathcal
N}=(4,4)$ SCFT on ${\mathcal M}$ make their appearance in the gauge theory
as the Fayet-Iliopoulos terms and the theta term.

It would be interesting to extract the singularity structure of the
gauge theory of the \dd  system through mappings similar to
\eq{dtermdef1}- \eq{dtermdef2}\footnote{The singularity structure for
a $U(1)$ theory coupled to $N$ hypermultiplets has been obtained in
\cite{DMWY}}.

\gap1
\noindent\underbar{\it The case $(Q_1,Q_5)\to (Q_1-1,Q_5)+
(1,0)$: splitting of a single  D1 brane}

It is illuminating to consider the special case
in which  1 D1 brane splits off from the bound state
$(Q_1,Q_5)$. The effective dynamics is again 
described in terms of a $U(1)$ gauge theory associated
with the relative separation between the single D1-brane
and the bound state $(Q_1-1,Q_5)$. The massless
hypermultiplets charged under this $U(1)$ correspond
to open strings joining the single D1-brane with
the D5-branes and are denoted by
\be
\chi_{b'} = \left( \begin{array}{c}
A_{b'} \\
B^\dagger_{b'}
\end{array}
\right)
\ee
The D-term equations, with the
Fayet-Iliopoulos terms, become in this case
 \be
\label{dterm1}
\sum_{b'=1}^{Q_5} \left( |A_{b'}|^2 - |B_{b'}|^2 \right)=
r_3, \quad  \sum_{b'=1}^{Q_5} A_{b'} B_{b'} = r_1 + ir_2
\ee
while the potential is
\be
\label{potential1}
V= \left[ \sum_{b'=1}^{Q_5} \left( |A_{b'}|^2 + |B_{b'}|^2 \right)
\right] (\varphi_1^2 + \varphi_2^2 +\varphi_3^2 +\varphi_4^2 )
\ee
In this simple case it is easy to see that the presences of the
Fayet-Iliopoulos terms in \eq{dterm1} ensures that all $A,B$'s do not
vanish simultaneously. The {\em vev}'s of $A,B$ gives mass to the
$\varphi$'s. Thus the relative $U(1)$ is broken when the
Fayet-Iliopoulos term is not zero.
The D-term equations above agree with those in
\cite{Sei-Wit99} which discusses the splitting
of a single D1-brane. It is important to emphasize
that the potential and the D-term equations describe
an {\em effective dynamics} in the classical background 
corresponding to the $(Q_1-1, Q_5)$ bound state. This
corresponds to the description in \cite{Sei-Wit99}
of the splitting process in an AdS$_3$ background
which represents a mean field of the above bound state. 

\subsection{Dynamics of the decay of the \dd system from 
gauge theory}
\label{chp2:sing}

We have seen in Section \ref{chp2:linear_sigma} 
that the effective theory
describing the dynamics of the splitting of the $(Q_1, Q_5)$ system to
subsystems $(q_1, q_5)$ and $(Q'_1, Q'_5$) is $(4,4)$, $U(1)$
super Yang-Mills coupled to $q_1Q'_5 + q_5 Q'_1)$ hypermultiplets.
The SCFT which this gauge theory flows in the infra-red is singular
if the Fayet-Iliopoulos terms and the theta term is set to zero. The
description of the superconformal theory of the
Higgs branch of a $U(1)$ gauge theory with $(4,4)$
supersymmetry and $N$ hypermultiplets 
near the singularity was found in
\cite{Aharony:1999dw}. The Higgs branch near the singularity  
was expressed in the Coulomb variables. The SCFT near the singularity
was derived using the R symmetry of the Higgs branch.
It consists of a bosonic $SU(2)$ 
Wess-Zumino-Witten model at level $N-2$, four free fermions and a
linear dilaton with background charge given by
\be
Q= \sqrt{\frac{2}{N}} (N-1)
\ee
The central charge of this SCFT is $6(N-1)$. Using this result for the
$U(1)$ theory describing the splitting we get a background charge for
the linear dilaton given by
\be
\label{backchargeg}
Q_{\rm Gauge Theory}=
\sqrt{\frac{2}{q_1Q'_5 + q_5Q'_1}}(q_1Q'_5 + q_5Q'_1 -1)
\ee
For large $Q'_1$ and $Q'_5$ we see 
from the above equation and \eq{chargesugra} 
that $Q_{\rm Gauge Theory} = Q_{\rm{SUGRA}}$

Consider the case of a single string splitting off the \dd bound
state, then the linear dilaton theory relevant for this decay has a
background charge of $Q= \sqrt{\frac{2}{Q_5}} (Q_5-1)$. This
effective theory is called the long string. On performing an S-duality
transformation this long D-string turns into a long fundamental
string. This argument demonstrates the existence of long fundamental
strings in the S-dual of the near horizon geometry of the \dd system.
We will discuss these solutions in detail in section 10.

\subsection{The symmetric product}

From the arguments of this section we see that the free field orbifold
conformal field theory on ${\mathcal M}$ does not correspond to the
\dd system given by the supergravity solution in \eq{near-horizon}.
This solution does not have any moduli turned on. We saw in this
section that in the absence of moduli the SCFT is singular. The
effective theory near the singularities does not just depend on the
product $Q_1Q_5$. For instance the theory near the singularity
corresponding to the decay of a single D1 brane is characterized by a
background charge of $\sqrt{\frac{2}{Q_5}}(Q_5-1)$. 
Thus it is not clear that
whether the symmetric product moduli space is connected to this
singular SCFT. 

In spite of this from the fact that the short multiplets of the SCFT on
${\mathcal M}$ agree with the supergravity modes on 
$AdS_3\times S^3$ we see
that at least for calculations involving 
correlations functions of the short multiplets we can
trust the SCFT on the symmetric product ${\mathcal M}$. 
The reason for this is that correlations functions involving
shortmultiplets are protected by non-renormalization theorems.

\newpage
\section{The microscopic derivation of Hawking radiation}
\def\om{\omega}
\def\tM{{\widetilde{\mathcal M}}}
\def\o{{\mathcal O}}
\def\z{{\vec z}}

From section 5 and section 6 we have seen that there is a one-to-one
correspondence of the supergavity modes and the short mulitplets of
the ${\mathcal N}=(4,4)$ SCFT on ${\mathcal M}$. 
In this section we
use this fact to obtain a precise understanding of Hawking radiation
from the \dd system starting from the microscopic SCFT.

As mentioned in the introduction after extracting out the low energy, 
degrees of freedom of the black hole
the next step towards
understanding Hawking radiation is to 
find the coupling of these degrees of freedom to the
supergravity modes. This is given by a specific SCFT operator
${\mathcal O}(z, \bar{z})$ which couples to the supergravity field
$\phi$ in the form of the interaction
\be
\label{bbcoupling}
S_{\rm{int}} = \mu \int d^2 z \phi(z, \bar{z}) {\mathcal O}(z,
\bar{z})
\ee
where $\mu$ is the strength of the coupling. 
We have no first principle method of determining the operator
corresponding ${\mathcal O}$  which couples with the supergravity mode
as the microsopic theory is an effective theory. Therefore we appeal
to symmetries to determine the operator. A coupling such as the one
given in \eq{bbcoupling} can exist only if
the operator ${\mathcal O}$ and the
field $\phi$ have the same symmetries. The identification of
the bulk and boundary symmetries  of the \dd system 
in section \ref{correspond} enables
the determination of  the operator which couples to a given supergravity
mode. Strictly speaking in the near horizon limit
\eq{near-horizon-btz} there is no coupling of the bulk to the
boundary. Here, we will assume $\alpha'$ small but strictly not zero,
so that we can discuss Hawking radiation.
From the near horizon limit of the \dd black hole whose metric in 5-d
is given in \eq{def-nl-nr}
we can infer that
the black hole is an excited state of the Ramond sector of the same
SCFT as that of the unexcited \dd system. Therefore the coupling
\eq{bbcoupling} should be the same as that of the \dd system.
The strength of the coupling $\mu$ is determined by comparing bulk and
boundary two point function using the AdS/CFT correspondence for the
\dd system. Once the interaction in \eq{bbcoupling} is determined the
calculation of Hawking radiation from the SCFT reduces to a purely
quantum mechanical evaluation of a scattering matrix in the SCFT.

In section 8.1 an 8.2 we identify the \dd black hole as an excited
state in the Ramond sector of the SCFT of the \dd system. 
We show how the entropy of the \dd black hole matches with this
excited state in the SCFT.
In section 8.3 we 
determine the coupling of the minimal scalar corresponding to the
metric fluctuation of the torus to the SCFT operator. In section 8.4 we
review the formulation of the absorption cross-section calculation
from SCFT as an evaluation of the thermal Green's function of
the operators ${\mathcal O}$
corresponding to the supergravity field $\phi$. We then evaluate the
absorption cross-section from SCFT and show that it agrees with the one
evaluated from supergravity including all the graybody factors.

In section 8.6 we address the Hawking radiation of fixed scalars from
the SCFT point of view. We show that fixing the SCFT operators using
symmetries resolves the disagreement observed in 
\cite{Krasnitz:1997gn} between  the
`effective string' calculation of the Hawking radiation and the
supergravity calculation. Finally in section 8.7 we outline how
Hawking radiation of the intermediate scalars also can be determined
from the SCFT.

\subsection{Near horizon limit and Fermion boundary conditions}

The near-horizon geometry of the \dd black hole 
is described in detail in
Sec. \ref{near-horizon-btz}.  We see from the remark (b) at the end of
that section that the boundary condition for fermions in the BTZ case
is periodic; this implies that the SCFT relevant for the \dd\ black
hole with Kaluza-Klein momentum $N=0$ is the Ramond vacuum of the
${\mathcal N}=(4,4)$ SCFT on the orbifold ${\mathcal M}$. The
microscopic states corresponding to the general \dd black hole are
states with $L_0\neq 0$ and $\bar{L}_0\neq 0$ excited over the Ramond
vaccum of the ${\mathcal N}= (4,4)$ SCFT on the orbifold ${\mathcal
M}$.  In the $AdS_3$ case, the fermion boundary condition is
antiperiodic; therefore the appropriate SCFT is that of the NS sector.

\subsection{The black hole state}
\label{chp4:state}

As we have seen, the general non-extremal 
black hole will have Kaluza-Klein 
excitations along both the directions on the $S^1$. In the SCFT
on ${\mathcal M}$, it is 
represented by states with 
$L_0 \neq 0$ and $\bar{L}_0 \neq 0$ over the Ramond 
vacuum. The black hole is represented by a density matrix 
(cf. Eq. \eq{density})
\be
\label{chp4:density-matrix}
\rho= \frac{1}{\Omega} \;\sum_{ \{i\} } |i\rangle\langle i|  
\ee
The states $|i\rangle$ 
belongs to the various twisted sectors of the orbifold theory.
They satisfy the constraint 
\be
L_0=\frac{N_L}{Q_1 Q_5} \;\;\;\; \bar{L}_0=\frac{N_R}{Q_1Q_5}
\ee
We have suppressed the index which labels the vacuum. $\Omega$ is the volume 
of the phase space in the microcanonical ensemble. It can be seen that the 
maximally twisted sector of the orbifold gives rise to the dominant
contribution to the sum in \eq{chp4:density-matrix} over the various
twisted sectors. The maximally twisted sector is obtained by the
action of the twist operator $\sum^{(Q_1 Q_5-1)/2}$ on the Ramond
vacuum. From the OPE's in \eq{chp2:OPEs} we see that the twist
operator $\sum^{(Q_1 Q_5-1)/2}$ introduces a cut in the complex plane
such that 
\be
\label{chp4:bc}
X_A (e^{2\pi i}z,e^{-2 \pi i} \bar{z})=X_{A+1} (z,\bar{z})
\ee
Thus this changes the boundary conditions of the bosons and the fermions.
Again from the OPEs in \eq{chp2:OPEs} one infers that  the excitations like 
$\partial \phi_1 |\sum^{(Q_1 Q_5-1)/2}\rangle$
over the maximally twisted sector have modes in units of $1/Q_1Q_5$. A
simple way of understanding that the maximally twisted sector 
has modes in units of $1/(Q_1Q_5)$ is to note that the boundary conditions 
in \eq{chp4:bc} imply that $X_A(z,\bar{z})$ is periodic with a period of $2\pi
Q_1Q_5$. 
This forces the modes to be quantized in units of $1/(Q_1Q_5)$.

We now show that the maximally twisted sector can account for the
entire entropy of the black hole. The entropy of the \dd black hole
can be written as 
\be
S_{SUGRA} = 2 \pi \sqrt{N_L} + 2 \pi \sqrt{N_R}
\ee
Using Cardy's formula, the degeneracy of the states in the maximally
twisted sector with $L_0=N_L/Q_1 Q_5$ and $\bar{L}_0=N_R/Q_1Q_5$
is given by
\be
\Omega =e^{2 \pi \sqrt{N_L} + 2 \pi \sqrt{N_R}}
\ee
By the Boltzmann formula,
\be
S(\mbox{maximally twisted})=2 \pi \sqrt{N_L} + 2 \pi \sqrt{N_R}
\ee
Thus the maximally twisted sector entirely accounts for the \dd
black hole entropy. $N_L$ and $N_R$ are multiples of $Q_1Q_5$ due to
the orbifold projection. Therefore, the entropy can be written as
\be
S = 2 \pi \sqrt{N_L Q_1 Q_5} + 2 \pi \sqrt{N_R Q_1 Q_5}
\ee

With this understanding, we restrict the calculations of Hawking
radiation and absorption cross-section only to the maximally twisted
sector. The probability amplitude for the Hawking process is given by
\be
\label{fermirule}
P=\frac{1}{\Omega} \sum_{f,i} |\langle f|S_{int}|i\rangle |^2
\ee
where $|f\rangle $ denotes the final states the black hole can decay into. We
have averaged over the initial states in the microcanonical ensemble.

It is more convenient to work with the canonical ensemble. We now
discuss the method of determining the temperature of the canonical
ensemble. Consider the generating function
\be
\label{chp4:generating}
Z=\mbox{Tr}_{R}(e^{-\beta_L E_0} e^{-\beta_R \bar{E}_0})
\ee
where the trace is evaluated over the Ramond states in the maximally
twisted sector. $E_0$ and $\bar{E}_0$ are energies of the left and the
right moving modes.
\be
E_0=\frac{L_0}{R_5}, \;\;\;\; \bar{E}_0=\frac{\bar{L}_0}{R_5}
\ee
From the generating function Z in \eq{chp4:generating} we see that the
coefficient of $e^{-(\beta_L N_L)/(Q_1Q_5R_5)}$ and $e^{-(\beta_R
N_R)/(Q_1Q_5R_5)}$ is the degeneracy of the states with
$L_0=N_L/Q_1Q_5$ and $\bar{L}_0=N_R/Q_1Q_5$
corresponding to the \dd black hole. A simple way to satisfy this
constraint is to choose $\beta_L$ and $\beta_R$ such that Z is peaked
at this value of $L_0$ and $\bar{L}_0$.

Evaluating the trace one obtains
\be
Z=\prod_{n=1}^{\infty} \left( \frac{1+e^{-(\beta_L
n)/(Q_1Q_5R_5)}}{1-e^{-(\beta_L n)/(Q_1Q_5R_5)}} \right)^4   \;\; 
\left(\frac{1+e^{-(\beta_R n)/(Q_1Q_5R_5)}}{1-e^{-(\beta_R
n)/(Q_1Q_5R_5)}} \right)^4
\ee
Then
\bea
\ln Z= 4 \left[\sum_{n=1}^{\infty} \ln(1+e^{-\beta_L n/Q_1Q_5R_5}) -
\sum_{n=1}^{\infty} \ln (1-e^{-\beta_L n/Q_1Q_5R_5}) \right] \\
+ 4 \left[\sum_{n=1}^{\infty} \ln(1+e^{-\beta_R n/Q_1Q_5R_5}) -
\sum_{n=1}^{\infty} \ln (1-e^{-\beta_R n/Q_1Q_5R_5}) \right] 
\eea
We can evaluate the sum by approximating it by an integral given by
\be
\label{chp4:lnZ}
\ln Z=4 Q_1 Q_5 R_5 \int_0^\infty dx\;\; 
\Big[\ln 
\left(\frac{ 1+e^{-\beta_L x}}
{ 1-e^{-\beta_L x}} \right)+
\ln\left(\frac{ 1+e^{-\beta_R x}}{ 1-e^{-\beta_R x}}\right) 
\Big]
\ee
From the partition function in \eq{chp4:generating} we see that 
\be
-\frac{\partial \ln Z}{\partial \beta_L} =\frac{
\langle N_L \rangle}{Q_1 Q_5 R_5}
\;\; \mbox{and} \;\;
-\frac{\partial \ln Z}{\partial \beta_R} =\frac{\langle N_R \rangle
}{Q_1 Q_5 R_5}
\ee
where $\langle \cdot
\rangle$ indicates the average value of $N_L$ and $N_R$. As the
distribution is peaked at $N_L$ and $N_R$ we assume that 
$\langle N_L \rangle=N_L$
and $\langle N_L \rangle=N_R$. Using \eq{chp4:lnZ} we obtain
\be
\frac{Q_1Q_5R_5 \pi^2}{\beta_L^2}= \frac{N_L}{Q_1Q_5R_5}
\;\;\; \mbox{and} \;\;\;
\frac{Q_1Q_5R_5 \pi^2}{\beta_R^2}= \frac{N_R}{Q_1Q_5R_5}
\ee
Thus
\be
\label{8:temp}
T_L=\frac{1}{\beta_L}=\frac{\sqrt{N_L}}{\pi R_5 Q_1Q_5}
\;\;\; \mbox{and} \;\;\;
T_R=\frac{1}{\beta_R}=\frac{\sqrt{N_R}}{\pi R_5 Q_1Q_5}
\ee
Above we have introduced a left temperature $T_L$ and a right
temperature $T_R$ corresponding to the left and the right moving
excitations of the SCFT to pass over to the canonical ensemble.
To see that the temperatures $T_L, T_R$ defined here are
the same as in \eq{4:temp}, note that the oscillator numbers
$N_L, N_R$ are $Q_1Q_5$ times the oscillator numbers that
enter \eq{4:temp}. The reason is that in \eq{4:temp} we defined
$N_L, N_R$ simply as the eigenvalues of $L_0, \bar {L_0}$;
in this secion we are working with the maximally twisted sectors
which have fractional oscillator numbers, thus to reach
the same energy we have to work with oscillator numbers
which are $Q_1Q_5$ times larger.

As mentioned before (cf. \eq{4:harmonic}) the temperature of the
combined system (conjugate to $E_L + E_R$), to be identified as the
Hawking temperature, is given by
\be
\label{8:hawking}
1/T_H= \frac12 \left(1/T_L + 1/T_R \right)
\ee 
\subsection{The coupling with the bulk fields for the \dd black hole}
\label{chp4:ramond-sector}

In section 6 we showed that there is a one to one map between the
supergravity fields on $AdS_3\times S^3$ and the short multiplets of
the ${\mathcal N}=(4,4)$  SCFT on ${\mathcal M}$. Therefore in
principle we can determine ${\mathcal O}$ for each bulk field $\phi$
in \eq{bbcoupling} just by matching the symmetries of the operator and
the field. But as we mentioned above that the black hole is
represented by an excited state in the Ramond sector and not for the
Neveu-Schwarz sector which corresponds to the $AdS_3$ boundary
conditions.
The couplings determined in the Neveu-Schwarz sector of the SCFT do
not change in the Ramond sector as interaction terms do not depend on
whether one is in the Ramond sector or the Neveu-Schwarz sector.
Therefore we can continue to use these couplings for the \dd black
hole.
The scaling dimension of an operator is given by
operator product expansions(OPEs) with the stress energy tensor. Since
OPEs are local relations , they do not change on going from the
Neveu-Schwarz sector to the Ramond sector, the same can be said of the
R-charge of the operator. 
In section 8. we will see
the calculation of Hawking radiation from the SCFT just depends on the
scaling dimension and the R-charge of the operator. Since this is
invariant whether one is in the Ramond sector or the 
Neveu-Schwarz sector
the operator ${\mathcal O}$ is the same as the one identified 
using  $AdS_3$ as the near horizon geometry.

\subsection{Determination of the strength of the coupling $\mu$}
\label{chp4:constant}

Before we perform the calculation of Hawking radiation/absorption
cross-section from the SCFT corresponding to the \dd black hole it
is important to determine the strength of the coupling $\mu$ in
\eq{bbcoupling}. 
In this section we will determine $\mu$ for the case
of minimal scalars $h_{ij}.$%
\footnote{From now on $h_{ij}$ will denote the 
traceless part of the metric
fluctuations of $T^4$.}
In \eq{sugra-cft} we have identified the SCFT operator
corresponding to these fields of the supergravity. 
The SCFT operator
is given by  
\be
{\mathcal O}^{ij}(z, \bar{z})=
\del x^{ \{ i }_A(z, \bar{z})
\bar{\del} x^{ j\} }_A (z, \bar{z} ) -\frac{1}{4} \delta^{ij} 
\del x^k_A \bar{\del} x^k_A (z, \bar{z})
\ee
Let us suppose the background metric of the torus $T^4$  is 
$g_{ij} = \delta_{ij} $. 
The interaction Lagrangian of the SCFT with the fluctuation $h_{ij}$
is given by
\be
\label{chp4:interaction} 
S_{\rm int} = \mu  T_{\rm eff} \int d^2 z\; 
\left[ h_{ij} \del x^i_A \bar{\del} x^j_A \right]
\ee
The effective string tension $T_{\rm eff}$ of the conformal
field theory , which also appears in the free part of the action 
\be
\label{chp4:free}
S_0 = T_{\rm eff} \int d^2 z\; \left[\del_z
x^i_A \del_{\bar z} x_{i,A} + {\rm fermions} \right]
\ee
has been discussed in \cite{Cal-Gub-Kle96,Mal96,Has-Wad97b}. The
specific value of $T_{\rm eff}$ is not important for the calculation
of the $S$-matrix for absorption or emission, since the factor just
determines the normalization of the two-point function of the operator
${\mathcal O}^{ij}(z, \bar{z})$. In this section we will argue that
the constant $\mu = 1$.

A direct string theory computation would of course provide the
constant $\mu$ as well (albeit at weak coupling).  This would be
analogous to fixing the normalization of the Dirac-Born-Infeld action
for a single D-brane by comparing with one-loop open string diagram
\cite{Pol-tasi96}. However, for a large number and more than one type
of D-branes it is a difficult proposition and we will not attempt to
pursue it here. Fortunately, the method of symmetries using the
AdS/CFT employed for determining the operator ${\mathcal O}$ helps us
determine the value of $\mu$ as well. For the latter, however, we need
to use the more quantitative version \cite{Wit98-ads, Gub-Kle-Pol98}
of the Maldacena conjecture.  We will see below that for this
quantitative conjecture to be true for the two-point function (which
can be calculated independently from the ${\mathcal N}=(4,4)$ SCFT and
from supergravity) we need $\mu=1$.

We will see  that the above normalization leads to precise
equality between the absorption cross-sections (and consequently
Hawking radiation rates) computed from the moduli space of the \dd
system and from semiclassical gravity. This method of fixing the
normalization can perhaps be criticized on the ground that it borrows
from supergravity and does not rely entirely on the SCFT.  However, we
would like to emphasize two things:\\ 
(a) We have fixed $\mu=1$ by
comparing with supergravity around AdS$_3$ background which does 
{\sl not} have a black hole.  On the other hand, 
the supergravity calculation of absorption
cross-section and Hawking flux is performed around a black hole
background represented in the near-horizon limit by the BTZ black
hole. From the viewpoint of semiclassical gravity  these two 
backgrounds are rather different. The fact that normalizing
$\mu$ with respect to the former background leads to the
correctly normalized absorption cross-section around the
black hole background is a rather remarkable prediction.\\
(b)  Similar issues are involved in fixing the coupling
constant between the electron and the electromagnetic field in the
semiclassical theory of radiation in terms of the physical electric
charge, and in similarly fixing the gravitational coupling of extended
objects in terms of Newton's constant.  These issues too are decided
by comparing two-point functions of currents with Coulomb's or
Newton's laws respectively. In the present case the quantitative
version of the AdS/CFT conjecture 
\cite{Wit98-ads, Gub-Kle-Pol98}
 provides the
counterpart of Newton's law or Coulomb's law at strong coupling.
Without this the best result one can achieve is that the Hawking
radiation rates computed from \dd branes and from semiclassical
gravity are {\sl proportional}.

We should remark that fixing the normalization by the use of
Dirac-Born-Infeld action, as has been done previously, is not
satisfactory since the DBI action is meant for single D-branes and
extending it to a system of multiple \dd branes does not always give
the right results as we have seen in Section \ref{chp3:sec-fixed}. The
method of equivalence principle to fix the normalization is not very
general and cannot be applied to the case of non-minimal scalars, for
example. 

Let us now compare the two-point function for the minimal scalar
$h_{ij}$ determined from the AdS/CFT correspondence and the SCFT to
determine the normalization constant $\mu$.
We will discuss the more quantitative version of the
AdS/CFT conjecture 
\cite{Wit98-ads, Gub-Kle-Pol98}
to compare the 2-point
correlation function of $\o_{ij}$ from supergravity and SCFT.

The relation between the correlators  are as
follows. Let the supergravity Lagrangian be
\bea 
\label{lag}
L &=& \int d^3 x_1 d^3 x_2 b_{ij,i'j'}(x_1,x_2) h_{ij}(x_1) 
h_{i'j'}(x_2)\nonumber\\
 &+& \int d^3 x_1 d^3 x_2 d^3
x_3 c_{ij,i'j',i''j''}(x_1, x_2, x_3) h_{ij}(x_1) h_{i'j'}(x_2)
h_{i''j''}(x_3) + \ldots\nonumber\\ 
\eea 
where we have only exhibited terms quadratic and cubic in the 
$h_{ij}$'s. The coefficient $b$ determines the propagator and
the coefficient $c$ is the tree-level 3-point vertex in supergravity. 
The coefficients $b$ and $c$ are local operators, $b$ is the kinetic
operator.

The 2-point function of the $\o_{ij}$'s (at large $g_sQ_1,g_sQ_5$)
is given by 
\cite{Wit98-ads, Gub-Kle-Pol98}
assuming $S_{\rm int}$
given by \eq{chp4:interaction}
\bea
\label{2pt}
& \langle \o_{ij}(z_1) \o_{i'j'}(z_2)\rangle
\\    \nonumber 
& = 2(\mu T_{\rm eff})^{-2}
\int d^3 x_1 d^3 x_2 \left[ b_{ij,i'j'}(x_1, x_2) 
K(x_1|z_1) K(x_2|z_2)\right],
\eea
where $K$ is the boundary-to-bulk Green's function for massless
scalars \cite{Wit98-ads}.
\be
\label{green}
K(x|z)= \frac{1}{\pi} \left[ \frac{x_0}{(x_0^2 + (|z_x - 
z|^2)}\right]^2
\ee
We use complex $z$ for coordinates of the SCFT, and
$x = (x_0, z_x)$ for the Poincar\'{e} coordinates of bulk theory.

\subsubsection{Evaluation of the tree-level vertices in supergravity}

We begin with the bosonic sector of Type IIB supergravity. The
Lagrangian is (we follow the conventions of \cite{chp4:BACHAS}) 
\bea  
I &=& I_{\rm NS} + I_{\rm RR}\nonumber\\
I_{\rm NS} &=& 
-\frac{1}{2k_{10}^2} \int d^{10} x \sqrt{-G}\left[ e^{-2\phi} \left( R -
4(d\phi)^2 + \frac{1}{12} (dB_{NS})^2 \right)  \right] \nonumber\\
I_{\rm RR} &=&  -\frac{1}{2k_{10}^2} \int d^{10} x \sqrt{-G}  
\left (\sum_{n=3, 7, ...}\frac{1}{2n!}
(H^{n})^2 \right)
\eea 
with $k_{10}^2= 64\pi^7 g^2_s \alpha^{\prime 4}$. 
We use $\hat{M}, \hat{N} \ldots $ to denote $10$ dimensional indices,
$i,j, \ldots$ to denote coordinates on the torus $T^4$, $M ,N
\ldots $ to denote the remaining $6$ dimensions and $\mu, \nu,
\ldots $ to denote coordinates on the AdS$_3$.  
We have separately indicated the terms
depending on Neveu-Schwarz Neveu-Schwarz and Ramond-Ramond backgrounds.

Our aim will  be to obtain the Lagrangian of the minimally coupled
scalars corresponding to the fluctuations of the metric of the $T^4$
in the \dd-brane system. We will find the Lagrangian up to
cubic order in the near horizon limit. Let us first focus
on $I_{\rm NS}$. 
We substitute the  values of the background fields of the \dd system
in the Type
IIB Lagrangian with the  following change in the metric
(cf. Eq. \eq{def-h})
\be
\label{changemetric}
f_1^{\frac{1}{2}} f_5^{-\frac{1}{2}}\delta_{ij} \rightarrow 
f_1^{\frac{1}{2}} f_5^{-\frac{1}{2}}(\delta_{ij} + h_{ij}).
\ee
where $h_{ij}$ are the minimally coupled scalars with trace
zero. These scalars are functions of the 6 dimensional
coordinates. Retaining the terms upto $O(h^3)$ and ignoring the
traces, the Lagrangian can be written as
\be
I_{\rm NS}= -\frac{V_4}{2k_{10}^2}\int d^6x \sqrt{-G}
\frac{G^{M N}}{4} \left[ \partial_M h_{ij} \partial_N h_{ij} +
\partial_M (h_{ik}h_{kj} ) \partial_N h_{ij} \right]
\ee
In the above equation we have used the near horizon limit and $V_4$ is
the volume of the $T^4$. 
It is easy to see that to $O(h^2)$, 
the minimally coupled scalars do not mix with 
any other scalars.
These minimally coupled scalars are all
massless (see the first line in \eq{sugra-fields}). 
Therefore to $O(h^2)$ 
they do not mix with any other scalars which are all
massive. For our purpose of computing the two point function using the
AdS/CFT correspondence it is sufficient to determine the tree-level
action correct to  $O(h^2)$.
The metric $G_{MN}$ near the horizon is
(writing $r$ for $U$ in Eq. \eq{2:ads}, and
including a factor of $\alpha'$)  
\be
ds^2 = \frac{r^2}{R^2} ( -dx_0^2 + dx_5
^2 ) + \frac{R^2}{r^2} dr^2 + R^2 d\Omega_3 ^2
\ee
We make a change of variables to the Poincar\'{e} coordinates 
by substituting
\be
z_0 = \frac{R}{r}, \ z_1 = \frac{x_0}{R}, 
\ z_2 = \frac{x_5}{R}  
\ee
The metric becomes 
\be
ds^2 = R^2 \frac{1}{z_0^2} ( dz_0^2 - dz_1^2 + dz_2^2) +
R^2 d\Omega_3 ^2 .
\ee
Here $R= \sqrt{\alpha^{\prime}}(g_6^2 Q_1Q_5)^{1/4}$ 
is the radius of curvature
of $AdS_3$ (also of the $S^3$) (see \eq{radius-ads}).
For $s$-waves the minimal scalars do not
depend on the coordinates of the $S^3$. Finally, in Poincar\'{e} coordinates  
$I_{\rm NS}$ (correct to cubic order in $h$) can be written as
\be
I_{\rm NS}= -\frac{V_4}{8k_{10}^2} R^3 
V_{S^3} \int d^3z \sqrt{-g} g^{\mu\nu} \left[ \partial_\mu
h_{ij} \partial_\nu h_{ij} + \partial_\mu (h_{ik} h_{kj})
\partial_\nu h_{ij} \right],
\ee
where $V_{S^3}=2\pi^2$, the volume of a three-sphere of unit radius.

Now we would like to show that to all orders in h, $I_{RR}=0$ in the near 
horizon geometry.
The relevant terms in our case are 
\be
I_{RR}=-\frac{1}{4\times 3! k_{10}^2} \int d^{10} x \sqrt{-G} 
H_{\hat{M} \hat{N} \hat{O} } H^{\hat{M} \hat{N} \hat{O}}.
\ee
We substitute the values of $B$ due to the magnetic and electric
components of the Ramond-Ramond charges and the value of $G$. The
contribution from the electric part of $B'$, after
going to the  near-horizon limit and performing the integral over $S^3$ and
$T^4$ is 
\be
\frac{V_4}{4 k_{10}^2} R V_{S^3} \int d^3 z \sqrt{-g}
\sqrt{{\mbox{det}}(\delta_{ij} + h_{ij})}
\ee
The contribution of the magnetic part of $B'$ in the
same limit is 
\be
-\frac{V_4}{4 k_{10}^2} R V_{S^3} \int d^3 z \sqrt{-g}
\sqrt{{\mbox{det}}(\delta_{ij} + h_{ij})}
\ee
We note that the contribution of the electric and the magnetic
parts cancel giving no couplings for the minimal scalars
to the Ramond-Ramond background. Therefore the tree-level supergravity action 
correct to cubic order in $h$ is given by \footnote{The cubic
couplings of all fields in type IIB supergravity on $AdS_3\times
S^3\times T^4$ were 
determined in \cite{Arutyunov:2000by}.}
\be
\label{chp4:effective}
I= -\frac{Q_1Q_5}{16\pi} \int d^3 z \left[ 
\del_\mu h_{ij} \del_\mu h_{ij}
+ \del_{\mu} (h_{ik} h_{kj}) \del_{\mu} h_{ij}\right]
\ee

The coefficient $Q_1Q_5/(16\pi)$ is U-duality invariant.
This is because it is a function of only the integers $Q_1$ and
$Q_5$.
This can be tested by computing the same coefficient from the
Neveu-Schwarz/fundamental string background which is related to the
\dd system by S-duality. The Neveu-Schwarz/fundamental string
back ground also gives the same coefficient. U-duality transformations
which generate $B_{NS}$ backgrounds \cite{DMWY} 
also give rise to
the same coefficient.

\subsubsection{Two-point function}

The two-point function of the operator ${\mathcal O}_{ij}$ can be evaluated
by substituting the value of $b_{ij, i'j'}(x_1,x_2)$ obtained from 
\eq{chp4:effective} into \eq{2pt} and using the boundary-to-bulk
Green's function given in \eq{green}. On evaluating the integral in
\eq{2pt} using formulae given in \cite{chp4:FreMatMatRas}, we find that
\be
\label{chp4:coefficient}
\langle \o_{ij}(z) \o_{i'j'}(w) \rangle =
(\mu T_{\rm eff})^{-2}\delta_{ii'}\delta_{jj'} 
\frac{Q_1 Q_5}{16 \pi^2}\frac{1}
{|z - w|^4}
\ee  
This is exactly the value of the two-point function obtained
from the SCFT described by the free Lagrangian \eq{free} provided
we put $\mu=1$. 

We have compared the two-point function  obtained from the supergravity
corresponding to the near horizon geometry of the \dd system with no
moduli to the orbifold SCFT. As we have argued before the orbifold
SCFT corresponds to the \dd system with moduli. Thus naively this
comparison seems to be meaningless. On further examination we note that
the coefficient $b_{ij, i'j'}$ in \eq{2pt} was U-duality invariant.
Since the \dd system with moduli can be obtained through U-duality
transformations we know that this coefficient will not change for the
\dd system with moduli. It is only the value 
of this  coefficient which fixes $\mu$ to be $1$. Thus the comparison
we have made is valid.
It is  remarkable that even at strong coupling
the two-point function of $\o_{ij}$ can be computed 
from the free Lagrangian
\eq{chp4:free}.  This is consistent with the non-renormalization 
theorems involving the ${\mathcal N}=(4,4)$ SCFT which will be
discussed in Section 9.

The choice $\mu=1$ ensures that the perturbation \eq{chp4:interaction}
of \eq{chp4:free} is consistent with the perturbation implied in  
\eq{changemetric}. 
We will see in the next section  
that this choice leads to precise equality
between 
absorption
cross-sections (consequently Hawking radiation rates) calculated from
semiclassical gravity and from the \dd branes. 
The overall
multiplicative constant $T_{\rm eff}$ will not be important for the
absorption cross-section calculation. This factor finally cancels off
in the calculation as we will see in Section \ref{chp4:abscft}.

Higher point correlations functions in the orbifold conformal
field theory were
determined in \cite{Lunin:2000yv,Lunin:2001pw} using general methods
of computing correlations functions of twist fields on symmetric
product orbifolds developed by 
\cite{Arutyunov:1998gt,Arutyunov:1998gi} 

\subsection{Absorption cross-section as thermal Green's function}
\label{chp4:abs-green}

Let us now relate the 
absorption cross-section of a supergravity
fluctuation $\delta \phi$ to the thermal Green's function of the
corresponding operator of the ${\mathcal N} =(4,4)$  SCFT on the orbifold
${\mathcal M}$ \cite{chp4:Gubser}.
The notation $\delta\phi$
implies that we are considering the supergravity field to be of the
form
\be
\phi = \phi_0 + \mu \delta \bar{\phi}
\ee 
where $\phi_0$ represents the background value and $\mu$ is the
strength of the coupling. 
\bea
S &=& S_0 +  \int d^2 z [\phi_0 + \mu  \delta \bar{\phi}] 
{\mathcal O}(z,\bar z)
\nonumber \\
&=& S_{\phi_0} + S_{int}
\nonumber \\
\eea
where
\be
S_{\phi_0} = S_0 + \int d^2 z \;\;  \phi_0 {\mathcal O}(z, \bar z)
\ee
\be
S_{int} = \mu \int d^2 z \;\; \delta  \bar{\phi} {\mathcal O}(z, \bar z)
\ee
${\mathcal O}$ is the operator corresponding to supergravity field $\phi$.
$S_0$ is the Lagrangian of the SCFT which includes the deformations
due to various backgrounds in the supergravity. For example, the free
Lagrangian in \eq{chp4:free} corresponds to the case when 
the field $a_1\chi + a_2 C_{6789}$ in \eq{sugra-fields} is turned on.

We  calculate the absorption of  a quanta $\delta\bar\phi =
\kappa_5 e^{-ipx}$  corresponding to the operator ${\mathcal O}$ 
using the Fermi's Golden Rule. $\kappa_5$ is related to the
five-dimensional Newton's  constant $G_5$ and the ten-dimensional
Newton's constant $G_{10}$ as 
\be
\label{chp4:kappa}
\kappa_5^2 = 8\pi G_5 = \frac{8\pi G_{10}}{V_4 2\pi R_5} = 
\frac{64 \pi^7 g_s^2 \alpha^{\prime 4}} {V_4 2\pi R_5}
\ee
We see that $\kappa_5$ is proportional to $\alpha^{'2}$. In the
Maldacena limit \eq{chp4:Maldalim} the coupling of the bulk
fluctuation to the SCFT drops out. 
We retain this term for our
absorption cross-section calculation.
In fact we will see below that the absorption cross-section
turns out to be proportional $g_6^2 \sqrt{Q_1 Q_5} \alpha^{'2}$. 

In this computation of the absorption cross-section
the black hole is represented by a canonical
ensemble at a given temperature.
The above interaction gives
the thermally averaged transition probability ${\mathcal P}$ as
\be
\label{chp4:green}
{\mathcal P} = \sum_{i,f} \frac{e^{-\beta\cdotp p_i}}{Z} P_{i\rightarrow f} 
= \mu^2\kappa_5^2 Lt
\sum_{i,f} \frac{e^{-\beta\cdotp p_i}}{Z} (2\pi)^2 \delta^2(p+p_i
-p_f) |\langle f| {\mathcal O} (0,0) |i\rangle |^2
\ee
Here $i$ and $f$ refer to initial and final states respectively.
$p_i, p_f$ refers to the initial and final momenta of these states. 
$L = 2\pi R_5$ denotes
the length of the string and $t$ is the time of interaction. As we have seen 
in Section \ref{chp4:state},
the inverse temperature $\beta$ has two components $\beta_L$ and
$\beta_R$. The relation of  these temperatures to the 
parameters of the \dd black hole is
\be
\beta_L= \frac{1}{T_L} \;\;\mbox{and} \;\; \beta_R= \frac{1}{T_R}
\ee
The left moving momenta $p_+$ and the right moving momenta $p_-$ are
in a thermal bath with inverse temperatures $\beta_L$ and 
$\beta_R$ respectively. $\beta\cdotp p$ is defined as
$
\beta\cdotp p = \beta_L p_+ + \beta_R p_-
$. $Z$ stands for the partition function of the thermal ensemble. 

The Green's function in Euclidean time is given by
\be
{\mathcal G} (-i\tau, x) = \langle{\mathcal O}^{\dagger} (-i\tau, x)
{\mathcal
O}(0,0) \rangle =
\mbox{Tr} (\rho T_{\tau} \{ {\mathcal O}^{\dagger} (-i\tau , x) {\mathcal
O}(0,0) \} )
\ee
where $\rho=e^{-\beta\cdotp \hat{p}}/Z$.  Time ordering is defined as
$T_{\tau}$ with respect to $-\mbox{Imaginary} (t)$. This definition
coincides with radial ordering on mapping the coordinate 
$(\tau , x)$ from the cylinder to the plane.
The advantage of doing this is that the integral
\be
\label{chp4:green1}
\int dt\;dx\; e^{ip \cdotp x} {\mathcal G}(t-i\epsilon , x) =
\sum_{i,f} \frac{e^{-\beta\cdotp p_i}}{Z} (2\pi)^2 \delta^2(p+p_i
-p_f) |\langle f| {\mathcal O} (0,0) |i\rangle |^2
\ee
The Green's function 
${\mathcal G}$ is determined by the two-point function of the operator
${\mathcal O}$. This is in turn determined by 
conformal dimension $(h, \bar{h})$  of the operator ${\mathcal O}$ 
and the normalization of the two-point function.

As we have to subtract out the emission probability we get the
cross-section as
\be
\sigma_{abs} {\mathcal F} t = {\mathcal P} (1-e^{-\beta \cdotp p} )
\ee
where ${\mathcal F}$ is the flux and ${\mathcal P}$ is given by
\eq{chp4:green}. Substituting the value of ${\mathcal P}$ from
\eq{chp4:green1} we get
\be
\sigma_{abs} =\frac{\mu^2\kappa_5^2 L}{{\mathcal F}} \int dt\;dx
({\mathcal G} (t-i\epsilon , x) - {\mathcal G} (t+i\epsilon, x) )
\ee
In the above equation we have related the evaluation of the
absorption cross-section to the evaluation of the thermal Green's
function. Evaluating the integral one obtains
\bea
\label{chp4:cross-section}
\sigma_{abs}&=& \frac{\mu^2\kappa_5^2 L {\mathcal C_O}}{{\mathcal F}}
\frac{ (2\pi T_L)^{2h -1}  (2\pi T_R)^{2\bar{h} -1} }
{ \Gamma(2h) \Gamma(2\bar{h}) }
\frac{ e^{\beta\cdotp p/2} - (-1)^{2h + 2\bar{h}} e^{-\beta\cdotp p/2} 
}{2} \\ \nonumber
&\;&\left| 
\Gamma (h + i\frac{p_{+}}{2\pi T_L} )
\Gamma (\bar{h} + i\frac{p_{-}}{2\pi T_R} )
\right|^2
\eea
where ${\mathcal C_O}$ is the coefficient of the leading order
term in the OPE of the two-point  function of operator ${\mathcal O}$. 

\subsection{Absorption cross-section of minimal scalars from the \dd
SCFT}
\label{chp4:abscft}

In the previous section we related the thermal Green's function of the
SCFT operator to the absorption cross-section. We will apply the
results of the previous section for the case of the minimal scalars.
We will consider the case of the minimal scalars corresponding to the
fluctuation of the metric of $T^4$. 
Let the background metric of the torus be $\delta_{ij}$.
Consider the minimal scalar $h_{67}$. 
We know the SCFT operator corresponding to this  has
conformal dimension $(1,1)$. From Section \ref{chp4:constant} we know
that $\mu =1$. The interaction Lagrangian is given by
\be
S_{int} = 2 T_{\rm eff} \int d^2 z h_{67} \del x^6_A (z, \bar{z}) 
\bar{\del}
x^7_A (z, \bar{z}) 
\ee
where we have set $\mu=1$. The factor of $2$ arises because of the
symmetric property of $h_{67}$. $S_0$ is given by 
\be
S_{0} =T_{\rm eff} \int d^2 z \del x^i_A (z, \bar{z}) \bar{\del}
x^j_A(z, \bar{z})
\ee
Comparing with the previous section the operator ${\mathcal O} = 2T_{\rm
eff} \del x^6_A (z, \bar{z}) \bar{\del} x^7_A(z, \bar{z})$. 
For the absorption of a quanta of energy $\omega$ using
\eq{chp4:cross-section} we obtain
\be
\label{chp4:scftabs}
\sigma_{abs}=
2 \pi^2 r_1^2 r_5^2 \frac{\pi \om}{2} 
\frac{\exp(\om/T_H)-1}{(\exp(\om/2T_R)-1) (\exp(\om/2T_L)-1)}
\ee
where we have $L=2\pi R_5$, ${\mathcal F}= \omega$, \eq{chp4:kappa} for $\kappa_5$ 
and \eq{chp4:coefficient} for ${\mathcal C_O}$. Comparing the
 absorption cross-section of the minimal scalars obtained from
 supergravity in \eq{class-abs} with \eq{chp4:scftabs} we find that
\be
\sigma_{abs}(\mbox{ SCFT}) = 
\sigma_{abs}({\mbox{ Supergravity}})
\label{8.agreement} 
\ee
Thus the SCFT calculation and the supergravity calculation of the
absorption cross-section agree exactly.  As we have mentioned earlier,
this implies an exact agreement of decay rates between SCFT and the
semiclassical calculation.

It is important to note that we have used the ${\mathcal N}=(4,4)$
SCFT realized as a free SCFT on the orbifold ${\mathcal M}$ as the
background Lagrangian $S_0$. As we have said before, this SCFT is
non-singular and therefore cannot correspond to the case of the \dd
system with no moduli. In Section \ref{chp4:sugra-abs-moduli} we have
argued that the supergravity calculation of the absorption
cross-section is independent of moduli.  Therefore it makes sense to
compare it with the SCFT result for the case with moduli turned on. In
the next section we will show that the SCFT calculation is also
independent of the moduli.

Before closing this section, we should mention the semiclassical
calculation of decay rates from BTZ black holes and its comparison to
CFT \cite{Birmingham:1997rj,Muller-Kirsten:1998mt, Ohta:1998xh}.  The
greybody factor for BTZ black hole is connected with that for the \dd\
black hole in \cite{Lee:1998pd,Lee:1999ua}.

\subsubsection{Absorption cross-section for the blow up modes
\label{sec:blow-up}}

Another point worth mentioning is that the method followed above 
for the calculation of the absorption cross-section from the SCFT can be
easily extended for the case of minimal scalars corresponding to the
four blow up modes. These minimal scalars are listed in the last two
lines of \eq{sugra-cft}. They are the self-dual NS B-field and a
linear combination of the Ramond-Ramond four form and the zero form.
The operators for these scalars are the $Z_2$
twists in the SCFT. Absorption cross-section  calculations for these
scalars cannot be performed on the `effective string' model based on
the DBI action.
The simple reason being that these operators are not present in the 
`effective' string model. Thus the `effective' string model does not
capture all the degrees of freedom of the \dd black hole.

\subsection{Fixed scalars}
\label{chp3:sec-fixed}

Out of the 25 scalars mentioned earlier which form part of the
spectrum of IIB supergravity on $T^4$, five become massive when
further compactified on $AdS_3 \times S^3$. There is an important
additional scalar field which appears after this compactification:
$h_{55}$. Let us remind ourselves the 
notation used for the coordinates: $AdS_3:
(x_0, x_5, r), S^3:(\chi, \theta, \phi); T^4:(x_6, x_7, x_8,
x_9)$. $r, \chi,\theta,\phi$ are spherical polar coordinates for the
directions $x_1,x_2,x_3,x_4$. In terms of the D-brane wrappings, the
D5 branes are wrapped along the directions 
$x_5, x_6, x_7, x_8, x_9$ 
and D1 branes are aligned 
along $x_5$. The field $h_{55}$ is scalar in the sense
that it is a scalar under the local Lorentz group $SO(3)$ of $S^3$.

In what follows we will specifically consider the three scalars
$\phi_{10}, h_{ii}$ and $h_{55}$. The equations of motion of these
fields in supergravity are coupled and have been discussed in detail
in the literature 
\cite{Cal-Gub-Kle96,Klebanov:1996gy,Krasnitz:1997gn,
Taylor-Robinson:1997kx,Lee:1998xz}. 
It turns out that the six-dimensional dilaton 
$\phi_6 = \phi_{10} - h_{ii}/4$
which is a linear combination of $h_{ii}$ and $\phi_{10}$ remains massless; 
it is
part of the twenty massless (minimal) scalars previously discussed.
The two other linear combinations $\lambda$ and $\nu$ 
 are defined as (see case (2) after Eq. \eq{def-h})
\bea
\lambda &=& \frac{h_{55}}{2} -\frac{\phi_{10}}{2} + \frac{h_{ii}}{8} \\
\nonumber
\nu &=& \frac{h_{ii}}{8}
\eea
$\lambda$ and $\nu$ satisfy coupled
differential equations. They can be decoupled by the following
position independent linear transformation
\cite{Krasnitz:1997gn}\footnote{We have set $r_0=0$ in the equations
in \cite{Krasnitz:1997gn} as we are looking at the background
corresponding to the \dd system. }.
\bea
\lambda &=& (\cos \alpha) \phi_+  + (\sin \alpha) \phi_- \\ \nonumber
\nu &=& -(\sin \alpha) \phi_+ + (\cos \alpha) \phi_-
\eea
where $\alpha$ can be found by solving the equation
\be
\tan \alpha - \frac{1}{\tan\alpha} = \frac{2}{\sqrt 3} 
\frac{Q_1 + v 
Q_5}{Q_1 -vQ_5}
\ee
Then $\phi_\pm$ obey the following equations
\be
\label{fixscalar}
\left[\frac{1}{r^3} \partial r^3 \partial_r + \omega^2 f_1f_5 -
8\frac{Q_\pm^2}{r^2(r^2 + Q_\pm^2)^2} \right]\phi_\pm =0
\ee
where $\omega$ is the frequency of wave and
\be
Q_\pm = \frac{\alpha'g_s}{3}\left(Q_5 + \frac{Q_1}{v} \mp \sqrt{Q_5^2 +
(\frac{Q_1}{v})^2 - \frac{Q_5 Q_1}{v}} \right)
\ee
These are examples of fixed
scalars. The pick up masses in the 
background geometry of the \dd system. To see this take the near
horizon limit defined in \eq{prescal} 
in the equation \eq{fixscalar}, we get (see also
\cite{Teo:1998dw})
\be
\left[
\frac{1}{l^2U} \frac{\partial}{\partial U}
\left( U^3\frac{\partial}{\partial U}\right) + \frac{\omega^2 l^2}{U^2}
- \frac{8}{l^2} \right] \phi_\pm =0
\ee
Note that this is the Klein-Gordan equation of a massive scalar in
$AdS_3$ with (mass)$^2= 8$ is units of the $AdS_3$ units. Further more
since the equation for both $\phi_+$ and $\phi_-$ is identical in the
near horizon limit the equation of motions of $\lambda$ and  $\nu$
become decoupled. They obey the massive Klein-Gordon equation in
$AdS_3$.  The near horizon mass of $\lambda$ and $\nu$ is $m^2= 8$ in
units of the radius of $AdS_3$.

Understanding the absorption and emission properties of fixed scalars
is an important problem, because the D-brane computation and
semiclassical black hole calculation of these properties 
\emph{are at variance} \cite{Klebanov:1996gy, Krasnitz:1997gn}.  The discrepancy
essentially originates from the `expected' couplings of $\lambda$ and
$\nu$ to SCFT operators with $(h, \bar h)=(1,3)$ and $(3,1)$ (see also
\cite{Cal-Gub-Kle96}).  These SCFT operators lead to qualitatively
different graybody factors from what the fixed scalars exhibit
semiclassically.  The semiclassical graybody factors are in agreement
with D-brane computations if the couplings were only to (2,2)
operators.

The coupling to $(1,3)$ and $(3,1)$ operators is guessed from
qualitative reasoning based on the Dirac-Born-Infeld action. Since we
now have a method of deducing the couplings to the bulk fieds based on
near-horizon symmetries, let us use it in the case of the fixed
scalars.

(a) By the mass dimension relation \eq{chp3:massdim} 
we see that the fixed
scalars $\lambda$ and $\nu$ correspond to operators with weights
$h +\bar{h}=4$.

(b) The fixed scalars have $SU(2)_E\times \widetilde{SU(2)}_E$ quantum
numbers $(\bf{1}, \bf{1})$. 

As all the supergravity fields are
classified according to the short multiplets of $SU(1,1|2)\times
SU(1,1|2)$ we can find the field corresponding  to these quantum
numbers among the short multiplets.
Searching through the short multiplets (see below \eq{eq.short}),
we find that the fixed scalars belong to the
short multiplet $( \re 3 , \re 3 )_S$ of
$SU(1,1|2) \times SU(1,1|2)$. They occur as top component of 
$( \re 3 , \re 3 )_S$.  There are six fixed scalars in all.
We conclude that the operators with
$(h, \bar{h}) = (1,3)$ or
$(h, \bar{h}) = (3,1)$  (which were inferred by the DBI method) are ruled out 
by the analysis of symmetries.

In summary, since the $(1,3)$ and $(3,1)$ operators are ruled out by
our analysis, the discrepancy between the D-brane calculation and the
semiclassical calculation of absorption and emission rates disappears.
Using the coupling to $(2,2)$ operators as we derived above, we can
compute $\sigma_{abs}$ for fixed scalars using
\eq{chp4:cross-section}.  This agrees exactly with the result
\eq{3:abs-fixed}.  

\subsection{Intermediate Scalars}

We only make the remark that the classification presented in Section
\ref{chp3:classification} correctly accounts for all sixteen
intermediate scalars, and predict that they should couple to SCFT
operators with $(h,\bar h) =(1,2)$ belonging to the short multiplet
$({\bf 2}, {\bf 3})_S$ or operators with $(h,\bar h) =(2,1)$ belonging
to the short multiplet $({\bf 3}, {\bf 2})_S$ (see below equation
\eq{eq.short}). This agrees with the `phenomenological' prediction
made earlier in the literature
\cite{chp3:KleRajTse}.

\newpage

\section{Non-renormalization theorems}

Let us review the three major agreements between the ${\mathcal N}=
(4,4)$ SCFT on ${\mathcal M}$ and supergravity. We showed in section 6
that the spectrum of short multiplets of ${\mathcal N}= (4,4)$ SCFT on
${\mathcal M}$ was in one to one correspondence with the supergravity
modes in the near horizon geometry of the \dd system. In sections 5
and 8 we saw that the entropy calculated from the microscopic SCFT
agreed with that of the \dd black hole. Finally in section 8 we have
seen that the calculation of Hawking radiation from the SCFT agreed
precisely with the semiclassical calculation including the gray body
factors.

In this section we will discuss the validity of these calculations
both in the boundary SCFT and the bulk supergravity. The 
validity of the calculations 
performed on the conformal field theory side in
general do not overlap with that on the supergravity. 
Conformal field theory calculations are performed using the 
${\mathcal N}=(4,4)$ free
orbifold theory on ${\mathcal M}$. In section 7 we saw that the
SCFT corresponding to supergravity is singular and presumably involves 
a large deformation in the moduli space
from the free orbifold conformal field theory. 
Therefore the calculations performed in 
supergravity are valid in the region of
moduli space when the conformal field theory is singular.

Let us compare the moduli space of deformations on
the SCFT and the supergravity. The supergravity and the SCFT moduli
are listed in \eq{sugra-cft}. They are in one to one correspondence.
There are $20$ moduli in all. On the supergravity side the moduli
parameterize the homogeneous space 
$\widetilde{\mathcal M}=SO(4,5)/SO(4)\times SO(5))$
\cite{chp3:GivKutSei, Sei-Wit99, chp2:Dijkgraaf}. On the SCFT
${\mathcal N}=(4,4)$ supersymmetry highly constraints the metric on
the moduli space to also 
be $\widetilde{\mathcal M}$ \cite{chp4:Cecotti}. 
Therefore for every
point on the moduli space of the SCFT there is a corresponding point
on the SCFT. As we have seen in section 7 the 
\dd supergravity solution  and the free orbifold SCFT are at different
points in the moduli space.
Therefore it is
natural to conclude that there are non-renormalization theorems 
that allow us to interpolate between the  calculations done using the
free orbifold theory and supergravity calculations.
In this section we will detail these non-renormalization theorems for
the three calculations, the spectrum of short multiplets, the entropy
and Hawking radiation. 

\subsection{The spectrum of short multiplets}

Short multiplets 
in the SCFT are built on chiral primaries both for the left and right
movers (see section \ref{chp2:short-multiplets}).
The chiral primaries satisfy a BPS bound. Their R-charge is
the same as the conformal dimension. Thus their spectrum in
independent of any perturbation of the conformal field theory which
preserves the ${\mathcal N}= (4,4)$ supersymmetric structure
\cite{chp2:Wit-susy,Witten:1981nf}
The entire structure of short multiplets is then dictated by the
${\mathcal N}=(4,4)$ SCFT algebra. 
Therefore the spectrum of short multiplets is
invariant under deformations of the SCFT. In section
\ref{chp2:sec-shortmultiplets} the entire set of short multiplets was
evaluated using the free orbifold theory on ${\mathcal M}$. This will
remain invariant under deformations of this SCFT from the free
orbifold point. A further piece of evidence that the shormultiplets
structure does not change under derformations is that the number of
chiral primaries which form the bottom component of a short multiplet
can be counted in the case of SCFT on ${\mathcal M}$ using a
topological partition function \cite{Maldacena:1999bp}. 
This partition function is
invariant under deformation of the SCFT.

The short multiplet structure of the supergravity modes obtained in
section \ref{chp3:classification} ignored the 
winding and momentum modes on the torus.
In fact the mass spectrum would not change if there were any metric
deformation or Neveu-Schwarz $B$-field through the torus. These affect
only winding and momentum modes on the torus.  
Thus the short multiplet structure of the supergravity modes presented
in \eq{eq.short} are invariant under deformations of the moduli that
involve the traceless components of the metric and the self-dual NS
$B$-field \eq{sugra-cft}. 

We have just demonstrated that
calculation of the spectrum of short multiplets in the
supergravity and the SCFT is independent of the moduli 
This allows us
to compare the short multiplet spectrum on both sides and obtain
agreement.

\subsection{Entropy and area}

As we have seen in section 8 
the black hole is represented as a state with $L_0\neq
0, \bar{L}_0 \neq 0$ over the Ramond sector of the SCFT. The entropy
in the SCFT is calculated by evaluating the asymptotic density of
states of with these values of $L_0$ and $\bar{L}_0$. 
The asymptotic density of states  in a conformal
field theory is given by Cardy's formula \cite{Cardy:ie}. 
This depends on the
level and the central charge of the conformal field theory. The
central charge of a conformal field theory is independent of moduli.
The Entropy is given by the logarithm for the asymptotic density of
states. 
Therefore the entropy calculated from the SCFT
remains invariant for various values of moduli.

The Bekenstein-Hawking entropy in supergravity is evaluated in the
five-dimensional Einstein metric and is equal to the area of the
Horizon. 
From the equations of motion of
type IIB supergravity \cite{Cal-Gub-Kle96}, 
we can explicitly see that the five
dimensional Einstein metric is not changed by turning on the sixteen
moduli listed on the top three lines of \eq{sugra-cft}. 
These are the traceless components of the metric, the self-dual
NS $B$-field along the torus and the six-dimensional dilaton.
This can also be seen explicitly form the supergravity solution with
moduli \eq{full-solution} constructed in \cite{DMWY}.
Thus the area of the \dd black hole does not change with moduli in
supergravity. This ensures that the evaluation of the entropy from
SCFT will agree with that in supergravity even though they are
evaluated at different points in the moduli space.

\subsection{Hawking radiation}

Before we discuss the dependence of Hawking radiation on the moduli 
let us examine the various approximations made in the derivation of
Hawking radiation both in supergravity and in SCFT.
Recall that the semi-classical calculations of Hawking radiation was
done in the dilute gas limit \eq{dilute-gas} $r_0, r_n\ll r_1, r_5$.
We also made the approximation of low energies compared to the horizon
radius 
$\omega r_5 \ll 1$. As the SCFT calculation relies very much on the
near horizon limit and the enhanced symmetries near the horizon, 
one can ask the question whether in the near horizon limit the dilute
gas approximation is obeyed. 
Let us convert $r_0, r_n, r_1, r_5$  to near horizon variables.
\bea
r_0 = U_0\alpha'  \quad r_n = U_0\sinh \sigma \alpha' \\ \nonumber
r_1 = \sqrt{\frac{g_sQ}{v} \alpha'} \quad
r_5 = \sqrt{g_sQ_5 \alpha'}
\eea
Now it is easy to see in the near horizon limit $\alpha' \rightarrow
0$ the dilute gas approximation always holds.
 
We also use the low energy approximation in the SCFT calculation.
In the microscopic calculations of Hawking radiation from the SCFT we
restricted our attention to first order in perturbation theory. In
fact we used the Fermi-golden rule to obtain the absorption
cross-section in \eq{fermirule}. It is easy to see that for the metric
fluctuation the higher order terms 
in perturbation theory go in powers of
$w^2r_5^2$ \cite{Das:1997kt}. Thus higher order terms in the SCFT
calculation are suppressed due to the low energy approximation.

In spite of the fact that the dilute gas approximation and the low
energy approximation are made on both the supergravity and SCFT side
the Hawking radiation calculation from the \dd black hole 
in supergravity 
and the SCFT are done at different point in the
moduli space. We now discuss why in spite of this they both agree.

\subsubsection{
Independence of Hawking radiation calculation on moduli:
Supergravity}
\label{chp4:sugra-abs-moduli}

We recall that the \dd black hole solution in the absence
of moduli is \cite{Str-Vaf96,Cal-Mal96} 
obtained from the \dd system by
further compactifying $x^5$ on a circle of radius $R_5$ and adding left(right)
moving Kaluza-Klein momenta along $x^5$.  
The corresponding supergravity solution is given 

The absorption cross-section of minimal scalars in the absence of
moduli is given by \eq{class-abs} \cite{Dha-Man-Wad96, Mal-Str96}.
We will now show that the absorption cross-section remains unchanged
even when the moduli are turned on.

{}From the equations of motion of type IIB supergravity
\cite{Cal-Gub-Kle96}, we can explicitly see that the five-dimensional
Einstein metric $ds_{5,Ein}^2$ is not changed by turning on the
sixteen moduli corresponding to the metric $G_{ij}$ on $T^4$ and the
Ramond-Ramond 2-form potential $B$. As regards the four blowing up
moduli, the invariance of $ds_{5,Ein}^2$ can be seen from the fact
that turning on these moduli corresponds to $SO(4,5)$ transformation
(which is a part of a U-duality transformation) and from the fact that the
Einstein metric does not change under U-duality.  
Now we know that the minimal scalars $\phi^i$ all
satisfy the wave-equation
\be 
D_\mu \del^\mu \phi^i =0 
\ee 
where the Laplacian is with respect to the Einstein metric in five
dimensions. Since it is  only this  wave equation that
determines the absorption cross-section completely, we see that
$\sigma_{abs}$ is the same as before.

It is straightforward to see that the Hawking rate, given by
\eq{class-decay} is also not changed when moduli are turned on.

\subsubsection{Independence of Hawking radiation on  moduli:SCFT }
\label{chp4:marginal}

In this section we will study the 
independence of the Hawking radiation on \dd moduli.
In Section 2 we have listed the twenty (1,1) operators $ {\mathcal O}_i(z,\bar z) $
in the SCFT based on the symmetric product orbifold ${\mathcal M}$ which
is dual to the \dd system.  
Turning on  various moduli $\phi^i$ of supergravity
corresponds to perturbing the SCFT
\be
\label{CFTperturbation}
S= S_0 +  \sum_i \int d^2 z \;\;\bar \phi^i {\mathcal O}_i(z,\bar z) 
\ee
where $\bar \phi^i$ denote the near-horizon limits of the
various moduli fields $\phi^i$. 
We note here that $S_0$ corresponds to the free SCFT based on the
symmetric product orbifold ${\mathcal M}$. 
As we have seen in section 7 that this SCFT is non-singular
(all correlation functions are finite), it does not correspond to the
marginally stable BPS solution originally found in
\cite{Str-Vaf96, Cal-Mal96}
Instead, it corresponds to a five-dimensional
black hole solution in supergravity with suitable ``blow-up'' moduli
turned on.

Let us now  calculate the
absorption cross-section of a supergravity
fluctuation $\delta \phi_i$ to the thermal Green's function of the
corresponding operator of the SCFT.
The notation $\delta\phi_i$
implies that we are considering the supergravity field to be of the
form
\be
\phi^i = \phi^i_0 + \mu \delta \phi^i
\ee 
where $\phi^i_0$ represents the background value and $\mu$ is the
strength of the coupling. 
\bea
\label{CFTperturbation1}
S &=& S_0 +  \int d^2 z [\bar \phi^i_0 + \mu  \delta \bar 
\phi^i] {\mathcal O}_i(z,\bar z)
\nonumber \\
&=& S_{\phi_0} + S_{int}
\nonumber \\
\eea
where
\be
\label{S-phi-0}
S_{\phi_0} = S_0 + \int d^2 z \;\; \bar \phi_0^i {\mathcal O}_i(z, \bar z)
\ee
\be
\label{S-int}
S_{int} = \mu \int d^2 z \;\; \delta \bar \phi^i {\mathcal O}_i(z, \bar z)
\ee
As we have seen in the Section \ref{chp4:abs-green}
the absorption cross-section of the 
supergravity fluctuation
$\delta\phi^i$ involves 
essentially
the two-point function of the operator ${\mathcal O}_i$ calculated with respect
to the SCFT action $S_{\phi_0}$. Since ${\mathcal O}_i$ is a marginal operator,
its two-point function is completely determined apart from a constant.
Regarding the marginality of the operators ${\mathcal O}_i$, it is easy to
establish it upto one-loop order by direct computation
($c_{ijk}=0$). The fact that these operators are exactly marginal can
be argued as follows. The twenty operators ${\mathcal O}_i$ arise as top
components of five chiral primaries. It is known that the number of
chiral primaries with $(j_R, \tilde j_R)=(m,n)$ is the Hodge number
$h_{2m,2n}$ of the target space ${\mathcal M}$ of the SCFT. Since this
number is a topological invariant, it should be the same at all points
of the moduli space of deformations.

We showed in Section \ref{chp4:constant} that if the
operator ${\mathcal O}_i$ corresponding to $h_{ij}$ 
is canonically normalized (OPE has residue 1) and if
$\delta \phi_i$ is canonically normalized in supergravity, then the
normalization of $S_{int}$ as in \eq{S-int} ensures that
$\sigma_{abs}$ from SCFT agrees with the supergravity result.  The
crucial point now is the following: once we fix the normalization of
$S_{int}$ at a given point in moduli space, at some other point it may
acquire a constant ($\not=1$) in front of the integral when 
${\mathcal O}_i$ and
$\delta\phi_i$ are canonically normalized at the new point. This would
imply that $\sigma_{abs}$ will get multiplied by this constant, in
turn implying disagreement with supergravity.  We need to show that
this does not happen.
 
To start with a simple example, let us first restrict to the moduli
$g_{ij}$ of the torus $\widetilde{T^4}$. We have
\be
\label{full-action}
S = \int d^2 z \, \del x^i \bar \del x^j g_{ij}
\ee
The factor of string tension has been absorbed in
the definition of $x^i$.

In Section \ref{chp4:constant} we had $g_{ij}
= \delta_{ij} +   h_{ij}$, leading to
\bea
S &=& S_0 + S_{int} \nonumber\\
S_0 &=&  \int d^2 z \;\;\del x^i \bar \del x^j \delta_{ij}
\nonumber\\
S_{int} &=&   
\int d^2 z \;\;\del x^i \bar \del x^j h_{ij}
\nonumber \\
\eea
In the above equation we have set $\mu =1$.
As we have remarked above, this $S_{int}$ gives rise to the
correctly normalized $\sigma_{abs}$. 

Now, if we expand around some other metric 
\be
g_{ij} = g_{0ij} +  h_{ij}
\ee
then the above action \eq{full-action} implies 
\bea
S &=& S_{g_0} + S_{int} \nonumber\\
S_{g_0} &=&  \int d^2 z \;\; \del x^i \bar \del x^j g_{0ij}
\nonumber\\
S_{int} &=& 
\int d^2 z \;\; \del x^i \bar \del x^j h_{ij}
\nonumber \\
\eea
Now the point is that neither $h_{ij}$ nor the operator $ {\mathcal O}^{ij} =
\del X^i \bar \del X^j $ in $S_{int}$ is canonically normalized at $g_{ij}=
g_{0ij}$. When we do use the canonically normalized operators, do we
pick up an additional constant in front?

Note that 
\be
\label{zamol-example}
\langle {\mathcal O}^{ij} {\mathcal O}^{kl} \rangle_{g_0}=g_0^{ik} g_0^{jl}|z-w|^{-4}
\ee
and
\be
\label{sugra-propagator-example}
\langle h_{ij} (x) h_{kl} (y) \rangle_{g_0} =
g_{0,ik} g_{0,jl} {\mathcal D}(x,y)
\ee
where $ {\mathcal D} (x,y)$ is the massless scalar propagator 

This shows that

Statement (1): 
{\em The two-point functions of $ {\mathcal O}^{ij} $ and $h_{ij}$
pick up inverse factors }.

As a result, $S_{int}$ remains correctly normalized when re-written in
terms of the canonically normalized $h$ and $ {\mathcal O} $ and no
additional constant is picked up.

The above result is in fact valid in the full twenty dimensional
moduli space $\tM$ because Statement (1) above remains true
generally. 

To see this, let us first rephrase our result for the special case of
the metric moduli \eq{full-action} in a more geometric way.  The
$g_{ij}$'s can be regarded as some of the coordinates of the moduli
space $\tM$ (known to be a coset $SO(4,5)/(SO(4)\times SO(5))$).  The
infinitesimal perturbations $h_{ij}, h_{kl}$ can be thought of as
defining tangent vectors at the point $g_{0,ij}$ (namely the vectors
$\del/\del g_{ij}, \del/\del g_{kl}$). The (residue of the) two-point
function given by \eq{zamol-example}\ defines the inner product
between these two tangent vectors according to the Zamolodchikov
metric \cite{chp4:Zamolodchikov,chp4:Cecotti}.
 
The fact that the moduli space $\tM$ of the ${\mathcal N}=(4,4)$
SCFT on ${\mathcal M}$ is the coset $SO(4,5)/(SO(4) \times SO(5))$ is
argued in \cite{chp4:Cecotti}. If the superconformal theory has 
${\mathcal N}= (4,4)$ supersymmetry and if the dimension of the moduli space is $d$
then it is shown in \cite{chp4:Cecotti} that the moduli space of the 
symmetric the product SCFT is given by
\be
\label{chp4:m-space}
\frac{SO(4,d/4)}{SO(4)\times SO(d/4)}
\ee
As a simple check note that the dimension of the space is
\eq{chp4:m-space} is $d$.
The outline of the argument is a follows.
An ${\mathcal N}=(4,4)$ SCFT
has superconformal $SU(2)_R \times \widetilde{SU(2)_R}$ 
symmetry. We have
seen that the bottom component of the short multiplet which contains
the marginal operator $\bf{(2,2)_S}$ transforms as a $\bf{(2,2)}$
under $SU(2)_R \times \widetilde{SU(2)_R}$. The top component which
corresponds to the moduli transforms as a $\bf{(1,1)}$ under the
R-symmetry. 
The holonomy group of the 
Zamolodchikov metric 
should leave invariant the action of $SU(2)_R \times
\widetilde{SU(2)_R}$. Then the holonomy group should have a form 
\be
\label{chp4:moduli}
K \subset SU(2) \times SU(2) \times \tilde{K} \subset SO(d)
\ee
Then \eq{chp4:moduli}
together with ${\mathcal N}=(4,4)$ supersymmetry and the left-right
symmetry of the two $SU(2)_R$'s of the SCFT fixes the moduli space
to be uniquely that given in \eq{chp4:m-space}.
We have found in Section  2 
that there are 20 marginal operators for the ${\mathcal
N}=(4,4)$ SCFT on the orbifold ${\mathcal M}$. 
Therefore the dimension of the moduli
space is $20$. 
Thus $\tM$ is given by
\be
\tM=\frac{SO(4,5)}{SO(4)\times SO(5)}
\ee

Consider, on the other hand, the propagator (inverse two-point
function) of $h_{ij}, h_{kl}$ in supergravity. The moduli space action
of low energy fluctuations is nothing but the supergravity action
evaluated around the classical solutions $g_{0,ij}$.  The kinetic term
of such a moduli space action defines the metric of moduli space. The
statement (1) above is a simple reflection of the fact that the
Zamolodchikov metric defines the metric on moduli space, and hence 

Statement (2):  {\em The propagator of supergravity fluctuations, viewed
as a matrix, is the inverse of the two-point functions in the SCFT.}

The last statement is of course not specific to the moduli $g_{ij}$
and is true of all the moduli.  We find, therefore, that fixing the
normalization of $S_{int}$ \eq{S-int} at any one point $\phi_0$
ensures that the normalization remains correct at any other point
$\phi'_0$ by virtue of Statement (2). We should note in passing that
Statement (2) is consistent with, and could have been derived from
AdS/CFT correspondence as applied to the two-point function.

Thus, we find that $\sigma_{abs}$ is independent of the
moduli, in agreement with the result from supergravity.

\newpage
\section{Strings in $AdS_3$} 
The AdS/CFT correspondence  
\cite{Mal97,Wit98-ads,Gub-Kle-Pol98,chp3:MalStr98,Aharony:1999ti}  
for the case of the D1-D5 system states that
type IIB string
theory on $AdS_3\times S^3\times T^4$ 
is dual to the $1+1$ dimensional
conformal field theory of the Higgs branch of the gauge theory of the
\dd system. In our study of the bulk geometry we have worked only in
the supergravity approximation of string theory on $AdS_3\times S^3$.
To fully explore the AdS/CFT correspondence for the \dd system we need
to understand string propagation in the bulk geometry.
String theory on $AdS_3\times S^3\times T^4$ involves 
Ramond-Ramond fluxes
through the $S^3$. Progress in formulation of string propagation in
Ramond-Ramond
backgrounds for the case of $AdS_3$ have been made in 
\cite{Pesando:1998wm,Rahmfeld:1998zn,Park:1998un,Yu:1998qw,
Berkovits:1999im,Dolan:1999dc}
Unfortunately, these models have been difficult to quantize 
and results have been hard to obtain.
It is more convenient to study string theory on the background which
is S-dual to the D1-D5 system. The near horizon geometry of the S-dual
is also $AdS_3\times S^3\times T^4$ but with Neveu-Schwarz H-fluxes 
through the $S^3$ and $AdS_3$. String propagation on this dual
background can be quantized. 
It also provides an exact string background in which the
metric $g_{00}$ is non-trivial. 
An important result from the study of
strings in $AdS_3$ has been in understanding the role of long strings
in the spectrum. We saw in section 7 that there exists long D1 brane
solutions near the boundary of $AdS_3$. Therefore, in the S-dual
geometry this would mean that there exists long fundamental strings.
These strings have been constructed as classical solutions and also
have been identified in the full quantum spectrum
\cite{Maldacena:2000hw,Maldacena:2000kv}. They play an
important role in constructing a consistent spectrum of strings in
$AdS_3$. 

In this section 10.1 we will introduce the S-dual of the \dd system. 
We then
formulate string propagation on $AdS_3$ and study its spectrum in
section 10.2. 
Our discussion will be
based on \cite{Maldacena:2000hw}. To understand the essential aspects
of the spectrum it is enough to focus on the $AdS_3$ part of the
geometry, we also  restrict our discussion to bosonic strings in
$AdS_3$.
In section 10.3 briefly review 
string propagation in Euclidean $AdS_3$ which was 
initiated in \cite{chp3:GivKutSei}. We then
write down the long string solution in Euclidean $AdS_3$ and discuss
its symmetries \cite{Sei-Wit99}.  Finally in section 10.4 
we discuss string theory on 
thermal $AdS_3$ backgrounds. The one loop free energy of a gas of
strings in $AdS_3$ was evaluated in \cite{Maldacena:2000kv}. It was
shown to be modular invariant and the space-time spectrum read off
from it matched with the proposal for the spectrum in
\cite{Maldacena:2000hw}.

We just mention two topics which we will not have the time to review. 
For a discussion of
correlation functions for string theory on $AdS_3$ and their role in
the AdS/CFT correspondence see
\cite{Maldacena:2001km} and references there in. Branes in $AdS_3$ has
been extensively studied for a recent work 
see \cite{Lee:2001gh,Ponsot:2001gt} and references there
in.

\subsection{The S-dual of the \dd system}

Let us consider the S-dual of the \dd whose metric is given in
\eq{6d-black-string}. S-duality takes $\phi\rightarrow -\phi$,
$C^{2}\rightarrow B_{NS}$ and $ds^2 \rightarrow e^{-\phi}ds^2$.
Performing these operations on the supergravity solution given in
\eq{6d-black-string} along with a rescaling of the coordinates by
$\sqrt{g_s}$ (coordinates are rescaled to keep the 10-dimensional
Newton's constant invariant) we obtain
\bea
\label{sdual}
ds^{ 2} &=& f_1^{-1} (-dt^2 + dx_5^2) + 
f_5( dx_1^2 + \cdots + dx_4^2) \\ \nonumber
&+& (dx_6^2 + \cdots + dx_9^2),  \\ \nonumber
e^{-2\phi} &=& \frac{1}{g_s^{\prime 2}} f_1f_5^{-1},  \\ \nonumber
B_{05} &=& \frac{1}{2}(f_1^{-1} -1), \\ \nonumber
H_{abc} &=& \frac{1}{2} \epsilon_{abcd}\partial_{d} f_5, \;\;\;\;\; a,
b, c, d = 1, 2, 3, 4 
\eea
where $g_s^{\prime} = 1/g_s$ and 
\be
\label{dualharm}
f_1= 1+ \frac{16\pi^4 g_s^{\prime 2} \alpha^{\prime 3}
Q_1}{V_4^{\prime} r^2}, 
\;\;\;\;\;
f_5 = 1+ \frac{\alpha' Q_5}{r^2}.
\ee
Here $V_4^{\prime}$ refers to the volume of the $T^4$ measured in the
scaled coordinates. From the Neveu-Schwarz fluxes it 
is now easy to see that this system
is the supergravity solution of  $Q_5$
Neveu-Schwarz branes with $Q_1$ fundamental strings smeared over the
four torus $T^4$. 

Let us now take the near horizon limit of this solution. This is given
by
\bea
\label{sdualimit}
\alpha' \rightarrow 0, &\quad& \frac{r}{\alpha'} \equiv U =
\mbox{fixed} \\ \nonumber
v' \equiv \frac{V_4'}{16\pi^4 \alpha^{\prime 2}} = \mbox{fixed},
&\quad&
g_6^{\prime} \equiv \frac{g'_s}{\sqrt v'} = \mbox{fixed}
\eea
Under this scaling limit the metric given in \eq{sdual} reduces to
\bea
\label{sdualnearh}
ds^2 &=& \alpha' U^2 Q_5( -dt^{2} + dx_5^{2} )
+ \alpha' Q_5 \frac{dU^2}{U^2} + \alpha' Q_5 d\Omega^2 \\ \nonumber
&+& (dx_6^2 + \cdots dx_9^2), \\ \nonumber
e^{-2\phi} &=& \frac{Q_1}{v'Q_5}, \\ \nonumber
H_{05U} &=& \alpha' Q_5 U,  \\ \nonumber
H_{\theta\phi\chi} &=& \alpha' Q_5, 
\eea
Here we have rescaled coordinates $t$ and $x_5$ by $\sqrt{Q_1Q_5
g_6^{\prime 2}}$.  Thus the near horizon geometry of $Q_1$ fundamental
strings and $Q_5$ NS branes is $AdS_3\times S^3\times T^4$ with
Neveu-Schwarz fluxes. The radius of $S^3$ is $\sqrt{Q_5\alpha'}$. Note
that the near horizon geometry depends on $Q_1$ only through the
string coupling constant which is proportional to the ratio  $Q_1/Q_5$.

It is convenient to study string propagation on this geometry as
it  consists  only of Neveu-Schwarz fluxes. 
We will restrict out attention to string propagation only on the
$AdS_3$ part of the geometry. Strings on $S^3$ with $H$ flux through
the sphere is an $SU(2)$ WZW model at level $Q_5$, while strings on
$T^4$ is a free field conformal theory. 
We refer to these conformal field theories as the internal
conformal field theory. To simplify the discussion we will study only
bosonic strings on $AdS_3$.

\subsection{String propagation on $AdS_3$}

String propagation on $AdS_3$ with $H$ flux 
is an exact conformal field theory
$AdS_3$ is a $SL(2, R)$ group manifold. To see this consider the
$SL(2,R)$ group element parameterized by
\be
\label{group}
g= \exp\left( i\frac{t+\phi}{2}\sigma_2 \right) 
\exp(\rho \sigma_3 )
\exp\left(i\frac{t-\phi}{2} \sigma_2\right),
\ee
where $\sigma_i$ are the Pauli matrices. We use the following
generators for the $SL(2, R)$ Lie algebra
\be
\label{generators}
T^{3} = -\frac{i}{2} \sigma^2, \;\;\;\;\; T^{\pm} = 
\frac{1}{2} (\sigma^3 \pm i\sigma_1).
\ee
Then the metric on the $SL(2,
R)$ group manifold is given by 
\be
\label{gmetric}
g_{\mu\nu} = \frac{1}{2} \mbox{Tr}( g^{-1} \partial_\mu g g^{-1}
\partial_\nu g), 
\ee
where $\mu,\nu$ are indices referring to $\rho, t, \phi$.
Evaluating the metric using  the parameterization given in \eq{group}
we get
\be
\label{metricg}
ds^2 = -\cosh^2 \rho dt^2 + d\rho^2 + \sinh^2 \rho d\phi^2,
\ee
which is the metric on $AdS_3$ expressed in the global coordinates 
$(t, \phi,\rho)$ \eq{globcord}. 
Thus string propagation on $AdS_3$ can be expressed
in terms of the WZW action given below
\be
\label{wzwact}
S = \frac{Q_5}{4\pi \alpha'} \int_M d x^{+} dx^{-}
\mbox{Tr} (g^{-1} \partial^+ g g^{-1} \partial^- g) + \frac{Q_5}{12\pi
\alpha'} \int_N \mbox{Tr} ( \omega^3).
\ee
Here $M$ is the embedding of the world sheet into the group manifold
$AdS_3$ and $N$ is any $3$ dimensional manifold whose boundary is $M$
and $\omega = g^- dg$ the Maurer-Cartan 1-form.
The second term is called the Wess-Zumino term.
We have used Minkowski signature on the world sheet and 
$x^{\pm} = \tau \pm \sigma$, where $\tau$ and $\sigma$ are  
the world sheet time and position coordinate.
As a check on the action compare the $H$ field in 
\eq{sdualnearh} and the one induced by the Wess-Zumino term. It is
easily seen that both are proportional to the volume form on $AdS_3$.

\ni\underbar{\it Classical solutions}

Now that we have the action of strings in $AdS_3$ we obtain
the form of the classical solutions. 
The equations of motion derived form the action in \eq{wzwact} is
given by
\be
\label{eomwzw}
\partial^- (\partial^+ g g^{-1}) =0.
\ee
Thus a general solution of this action is given by
\be
\label{solwzw}
g = g_{+}(x^+) g_{-} (x^{-})
\ee
It is thus easy to construct classical solutions. Consider the
following solution
\be
g_+(x^+) = e^{\frac{i\alpha\sigma^{2}}{2} x^{+}} \quad
g_-(x^-) = e^{\frac{i\alpha\sigma^{2}}{2} x^{-}}
\ee
For this solution we have $g=e^{-\alpha \tau \sigma^2}$. From the
parametrization of the group element in \eq{group} we see that the
solution is a timelike geodesic with $\rho=0, \phi=0$ and $t= \alpha
\tau$. Note that the solution does not have $\sigma$ dependence,
therefore it represents a particle trajectory. Space like geodesics
can also be constucted. The solution
\be
g_+(x^+) = e^{\frac{\alpha\sigma^3}{2} x^{+}} \quad
g_-(x^-) = e^{\frac{\alpha\sigma^3}{2} x^{-}} \quad
\ee
represents a spacelike geodesic with $g= e^{\alpha \tau \sigma^3}$.
From \eq{group} we see that the trajectory is given by $\rho = \alpha
\tau$ which is space like.

It is interesting to note that there is a symmetry which allows the
generation of new solutions given one solution. The transformation
\be
\label{spec-new-sol}
g_+ = e^{i\frac{1}{2} \omega x^+ \sigma_2} \tilde{g}_+ \quad
g_- = e^{i\frac{1}{2} \omega x^- \sigma_2} \tilde{g}_+ \quad
\ee
where $\tilde{g}_+$ and $\tilde{g}_-$ are the old solution is also a
solution. From \eq{group} we see that this acts on
$t$ and $\phi$ as
\be
t\rightarrow t + \omega \tau  \quad
\phi\rightarrow \phi + \omega \sigma 
\ee
The periodicity of the string worldsheet under $\sigma \rightarrow
\sigma + 2\pi$ is obeyed. This transformation 
is called spectral flow. 
It stretches the geodesic in the $t$-direction 
gives $\sigma$ dependence to the geodesics. 
In fact now the solution represents a string winding $w$ times around
the centre $\rho=0$ of $AdS_3$. The spectral flow of timelike
geodesics are called short strings, their energy is bounded from above.
While the spectral flow of spacelike
geodescis are called long strings, their energy is bounded from below.
Thus long strings are like scattering states while short strings are
like bound states in $AdS_3$ \cite{Maldacena:2000hw}. 
These long strings as we will see  in section 10.4 
are the duals of the
long D-strings discussed in section 7.

\ni\underbar{\it Symmetries of the $SL(2,R)$ WZW action}

The WZW action has an infinite set of conserved charges given by
\be
\label{currents}
J^{a}_n = Q_5 \int_0^{2\pi} 
\frac{d x^+}{2\pi } e^{inx^+} \mbox{Tr} ( T^a \partial_+ g
g^{-1}) \;\;\;\;\;\;
\bar{J}^{a}_n = 
Q_5 \int_0^{2\pi} \frac{d x^- }{2\pi } e^{inx^-} \mbox{Tr} ( T^a 
\bar{\partial_-} g g^{-1})
\ee
They obey the commutation relations
\bea
\label{currentalgeb}
[J_n^3, J_m^3] &=& - \frac{Q_5}{2} n \delta_{n+m , 0}, \\ \nonumber
[J_n^3, J_m ^{\pm} ] &=& \pm J_{n+m}^{\pm}, \\ \nonumber
[J_n^+, J_m^-] &=& -2J^3_{n+m} + Q_5n \delta_{n+m ,0}.
\eea
There is a similar set of commutation relations for the right movers
$\bar{J}^a_n$. Using the Sugawara construction one can define the
Virasoro generators, they are given by
\bea
\label{sugawara}
L_0 &=& \frac{1}{Q_5 -2} \left[
\frac{1}{2} (J_0^+ J_0^- + J_0^- J_0^+) - (J_0^3)^2  \right. \\
\nonumber
&& \quad\quad +\left. 
\sum_{m=1}^\infty ( J_{-m}^+ J_m^- + J_{-m}^- J_m^+ - 2J_{-m}^3J_m^3)
\right] \\ 
L_{n\neq0} &=& \frac{1}{Q_5-2} \sum_{m=1}^\infty (J^+_{n-m} J^-_m +
J^-_{n-m} J^+_m - 2 J^3_{n-m} J^3_m ) \nonumber
\eea
These generators obey the Virasoro algebra with the central charge
given by
\be
\label{centrcharge}
c= \frac{3Q_5}{Q_5-2}
\ee

\subsection{Spectrum of strings on $AdS_3$}

From the fact that WZW action admits the $SL(2, R)$ current algebra we
see  that 
the physical spectrum of a string in $AdS_3$ must be in unitary
representations of the current 
We construct the unitary 
representation of the $SL(2, R)$ current algebra by
first constructing unitary representation of the global part of the
$SL(2, R)$.  This is given by
\be
\label{globalgeb}
[J^3_0, J^{\pm}_0] = \pm J^{\pm}_0 \;\;\;\;\;
[J^+_0, J^-_0] = -2J^3_0
\ee
From this algebra we see states are classified by the eigen
values of $J_0^3$ which we denote by $m$
and $j$  which is related to the Casimir $c_2 = \frac{1}{2} (J_0^+J_0^-
+ J_0^-J_0^+) - (J_0^3)^2$ by $c_2 = -j(j-1)$.

Unitary representations of $SL(2, R)$ fall in five classes
\footnote{See \cite{Dixon:1989cg} for a review.}:
\begin{enumerate}
\item
Identity: \\
The trivial representation $|0\rangle$. This representation has $j=0,
m=0$ and $J_0^\pm |0\rangle=0$.
\item
Principal discrete representations (lowest weight): \\
These are representation of the form
\be
\label{lowwgt}
{\mathcal D}_j^+ = \{ |j;m\rangle : m = j, j+1, j+2\, \cdots \}, 
\ee
Here $|j;j\rangle$ is annihilated by $J_0^-$. The tower of states over
$|j;j\rangle$ is built by the repeated 
action of $J_0^+$. The norm of these
states is positive and the representation is unitary 
if $j$ is real and $j>0$. 
$j$ is restricted to be half integer if we are considering
representation of the group $SL(2,R)$, however for the universal cover
of $SL(2,R)$ which is our interest, $j$ can be any positive integer.
\item
Principal discrete representations (highest weight): \\
These are representation of the form
\be
\label{highwgt}
{\mathcal D}_j^- = \{|j;m\rangle: m = -j, -j-1, -j-2, \cdots\},
\ee
where $|j;j\rangle$ is annihilated by $J_0^+\rangle$. The 
representation has positive norm and is unitary if $j$ is real and
$j>0$. This representation is the charge conjugate of ${\mathcal
D}_j^+$.
\item
Principal continuous representations: \\
A representation is of the form 
\be
\label{cont}
{\mathcal C_j^\alpha} = \{ |j, \alpha ;m\rangle: m = \alpha, \alpha\pm
1, \alpha\pm 2, \cdots \}
\ee
Without loss of generality, we can restrict $0\leq \alpha <1$. The
representation has positive norm and is unitary if $j= 1/2+ is$ where
$s$ is real.
\item
Complementary representations: \\
These are of the form
\be
\label{comp}
{\mathcal E}^\alpha_j = \{ | j, \alpha;m \rangle : m =\alpha,
\alpha\pm 1, \alpha\pm 2, \cdots \},
\ee
Again without loss of generality we can restrict $0\leq\alpha <1$
The representation has positive norm and is unitary  if $j$ is real
and $j(1-j)>\alpha(1-\alpha)$.
\end{enumerate}

Among these representations, we restrict to those which admit square
integrable wave functions 
in the point particle limit.  
As $AdS_3$  is non-compact,
square-integrability refers to delta function normalizable wave
functions.
This imposes the restriction $j>1/2$ 
\footnote{This condition is also the
condition for the Breitenlohner-Freedman bound 
\cite{Breitenlohner:1982bm} on ${\mathcal D}_j^\pm$
which states the mass of a scalar in $AdS_3$ is given by $m^2 =
j(j-1)\geq -\frac{1}{4}$. }.
It is known that 
${\mathcal C}_{j=1/2+is}^\alpha \otimes {\mathcal C}_{j=1/2+is}^\alpha$
and ${\mathcal D}_j^\pm \otimes {\mathcal D}_j^\pm$ with $j>1/2$ form
the complete basis of square integrable wave functions on $AdS_3$.
So it is sufficient to work with these representations only.
Let us call a unitary
representation of $SL(2,R)$ with this restriction as 
${\mathcal H}$.  

Now that we have the unitary
representation of the global part of the current
algebra we can obtain the unitary representation of the $SL(2,R)$
current algebra by considering ${\mathcal H}$ as its primary states
which are annihilated by $J_n^3, J_n^{\pm}$ where $n>0$. 
Then the full representation
is obtained by the action of $J_n^3, J_n^{\pm}$ with $n<0$ on
${\mathcal H}$. We denote this full representation 
by 
$\hat{{\mathcal D}}_j^\pm$ and 
$\hat{{\mathcal C}}_{j=1/2+is}^\alpha$.
In general, representation of the 
$SL(2, R)$ current algebra contains negative norm states. String
theory  on $AdS_3$ is consistent if one can remove these negative
states by imposing the Virasoro constraint on the Hilbert space for
a single string state. If we consider the bosonic string theory on
$AdS_3$ then the Virasoro constraint is given by
\be
\label{virasoro}
(L_n^{\rm{total}} -\delta_{n, 0})|\mbox{Physical}\rangle =0, \;\;\;
n\geq 0,
\ee
Here $L_n^{\rm{total}}$ refers to the Virasoro generator of the $c=26$
conformal field theory including the $SL(2,R)$ WZW model. It has been
shown that there are no negative norm states for 
$\hat{{\mathcal D}}_j^\pm$ with $0<j<k/2$ and  
$\hat{{\mathcal C}}_{j=1/2+is}^\alpha$ (See \cite{Evans:1998qu,
Satoh:1999jc,Maldacena:2000hw} for a list of references).

In \cite{Maldacena:2000hw} it was seen that the $SL(2, R)$ WZW model
admits a symmetry given by
\be
\label{specflow}
J_n^3 = \tilde{J}_n^3 +\frac{k}{2} w\delta_{n,0}, \;\;\;\;
J_n^{+} = \tilde{J}^+_{n-w}, \;\;\;\;
J_n^{-} = \tilde{J}^-_{n+w}, 
\ee
where $w$ is any integer. The map of $J$'s to $\tilde{J}$'s preserves
the commutation relations \eq{currentalgeb}. The Virasoro generators
$\tilde{L}_n$ can be found using the Sugawara construction and they
are related to $L_n$'s by the map
\be
\label{specflowv}
L_n = \tilde{L}_n - w\tilde{J}_n^3 - \frac{k}{4} w^2 \delta{_n, 0}
\ee
This symmetry of the $SL(2, R)$ WZW model is called spectral flow. 
It is the same symmetry as the one which allowed the generation of new
classical solutions from old ones in section 10.2. This can be seen
easily by computing the change in the stress energy tensor and the
$SL(2,R)$ generators under the map \eq{spec-new-sol}. They are
identical to \eq{specflowv} and \eq{specflow} respectively.
The spectral flow  maps
one representation to another. For the case of 
a compact group like $SU(2)$ it does not
generate a new representation, but for the non-compact group $SL(2,R)$
it generates new representations. Let us call the resulting
representations 
$\hat{{\mathcal D}}^{\pm, w}_{\tilde{j}}$ 
and
$\hat{{\mathcal C}}_{1/2+is}^{\alpha, w}$, where $\tilde{j}$ denotes
the $SL(2,R)$ spin before the flow. 
The representations obtained by
the spectral flow also have negative norms states. 
It has been shown in 
\cite{Maldacena:2000hw}, that there are no negative norm states for
representation obtained from spectral flow for
$1/2<\tilde{j}<(k-1)/2$.

Now we have the ingredients to state the proposal for the spectrum of
strings on $AdS_3$ \cite{Maldacena:2000hw}. The spectrum consists of
two kinds of representations, the spectral flow of the continuous
representation with the same amount of spectral flow on the left and
right $\hat{{\mathcal C}}_{1/2+is, L}^{\alpha, w}\otimes
\hat{{\mathcal C}}_{1/2+is, R}^{\alpha, w}$ along with the spectral
flow of the discrete representations
$\hat{{\mathcal D}}^{\pm, w}_{\tilde{j},L }\otimes
\hat{{\mathcal D}}^{\pm, w}_{\tilde{j}, R}$. The value of $\tilde{j}$
is restricted to be $1/2 <\tilde{j} < (k-1)/2$. These representations
should be tensored with the representations of the internal CFT which
contributes to the net central charge. We then have to impose the
Virasoro constraints. 
The expressions for the energy and the virasoro constraints for both
the discrete and the continuous representation are given in
\cite{Maldacena:2000hw}.
This proposal was verified in \cite{Maldacena:2000kv}
by reading out the spectrum from 
the modular invariant one-loop partition function in
thermal $AdS_3$ background. The discrete and continuous states
obtained form the one-loop partition function was in agreement with
the above proposal. We will review \cite{Maldacena:2000kv} in section
10.5.

\subsection{Strings on Euclidean $AdS_3$}

In this section we formulate string theory on Euclidean $AdS_3$, we
denote Euclidean $AdS_3$  by $\H$ (see Appendix C).
Again we will restrict our attention to a single Poincare patch in  
$\H$.
Consider the following coordinate
redefinition of the coordinates in 
the Euclidean version of \eq{sdualnearh}
\be
\label{coredef}
U = e^{\phi} \;\;\;\;\; \gamma = it + x_5  \;\;\;\;
\bar{\gamma} = -it + x_5 
\ee
Then the metric on $\H$  becomes (cf. \eq{eucl-ads-poincare}
with $h= e^{-\phi}$)
\be
\label{eucads}
ds^2 = l^2 ( d\phi^2 + e^{2\phi} d\gamma d\bar{\gamma})
\ee
Here $l^2 = \alpha' Q_5$. The value of the B-field can be read out
again from \eq{sdualnearh} and is given by 
\be
\label{eucbfield}
B = l^2 e^{2\phi} d\gamma \wedge d \bar{\gamma}
\ee
The B-field is necessary for worldsheet conformal invariance. The
B-field in Euclidean $AdS_3$ is imaginary. We work with a Euclidean
worldsheet theory. This makes the contribution of B-field to the world
sheet Lagrangian real. The world sheet action is given by
\be
\label{eucws}
S= \frac{l^2}{2\pi \alpha'}\int d^2z (\partial \phi \bar{\partial}
\phi + e^{2\phi} \bar{\partial} \gamma \partial \gamma )
\ee
Let us write this action in a more convenient form. Introducing
auxiliary fields $\beta$ and $\bar{\beta}$ of weights $(1,0)$ and
$(0,1)$ we can write the action as
\be
\label{euaux}
S= \frac{l^2}{2\pi \alpha'} \int d^2z (\partial\phi \bar{\partial}\phi
+ \beta\bar{\partial} \gamma + \bar{\beta} \partial \bar{\gamma} -
e^{-2\phi} \beta\bar{\beta})
\ee
Integrating out the auxiliary fields $\beta$ and $\bar{\beta}$ in the
above equation we obtain \eq{eucws}. 
Scaling $\phi$ so that it has the canonical normalization and taking
into account of the measure we obtain.
\be
\label{euwsren}
S=\frac{1}{4\pi} \int d^2z \left( \partial \phi \bar{\partial} {\phi} -
\frac{2}{\alpha_+} \hat{R} \phi + \beta\bar{\partial} \gamma +
\bar{\beta} \partial \bar{\gamma} - \beta\bar{\beta} \exp(
-\frac{2}{\alpha_+} \phi )\right)
\ee
Here $\alpha_+ = \sqrt{2Q_5 -4}$ and $\hat{R}$ is the world sheet
curvature.
Notice that the coefficient of the exponent has been renormalized.
The action becomes free at $\phi \rightarrow \infty$ which is near  
the boundary of $AdS_3$. The world sheet propagators of the fields in
\eq{euwsren} are
\be
\label{wsprop}
\langle\phi(z) \phi(0)\rangle = -\log |z|^2 \;\;\;\;\; 
\langle \beta(z) \gamma(0) \rangle = \frac{1}{z}
\ee
This action \eq{euwsren} 
admits a $SL(2, R) \times SL(2, R)$ current algebra. 
This representation of the $SL(2,R)$ current algebra is called the
Wakimoto representation \cite{Wakimoto:1986gf}.
The holomorphic
currents are given by
\bea
\label{eucural}
J^3 &=& \beta\gamma + \frac{\alpha_+}{2} \partial \phi \\ \nonumber
J^+ &=& \beta\gamma^2 + \alpha_+ \gamma \partial \phi + Q_5 \partial
\gamma \\   \nonumber
J^- &=& \beta
\eea
Similar definitions for the antiholomorphic currents exist. The modes
of these currents generate the $SL(2,R)$ current algebra given in
\eq{currentalgeb} with the same central charge as given in
\eq{centrcharge}. 


\subsubsection{The long string worldsheet algebra}

We now derive the worldsheet degrees of freedom of the long string
solution in $\H$. 
Using supersymmetry one can derive
the worldsheet theory of the long string exactly, unlike our classical
analysis in section 7.1. We will follow the discussion given in
\cite{Sei-Wit99}.
The long string solution in the static gauge is given by
\cite{Sei-Wit99, chp3:GivKutSei, Maldacena:2000hw}
\be
\label{longsol}
\phi = \phi_0,\;\;\;\; \gamma(z, \bar{z}) = z ,\;\;\;\; \bar{\gamma}
(\bar{z}, \bar{z}) = \bar{z}
\ee
The ghosts corresponding to $\gamma$ and $\bar{\gamma}$ decouple
\cite{Sei-Wit99}.
Thus the bosonic worldsheet degrees of freedom of the long string will
be the coordinate $\phi$ characterizing its radial  position  in
$\H$, the coordinates on the sphere $S^3$  which forms an $SU(2)$
current algebra $j^a$, $a = 1, \ldots ,3$ and the coordinates of $T^4$. 
The 4 fermionic partners of the coordinates on $T^4$ with the bosons
form a ${\mathcal N}=(4,4)$ superconformal algebra with central charge
6. 
The superpartners of
the coordinate $\phi$ and the $SU(2)$ current
algebra $j^a$ 
of the sphere $S^3$ 
4 free fermions $S^\mu$ with $\mu = 1, \ldots, 4$. 
From the fact that the radius of $S^3$ is
$Q_5$ the we know that the $SU(2)$ current algebra is at level
$Q_5-2$ \footnote{The shift in the level is because we have used
decoupled fermions (see for instance in \cite{Callan:1991at}).}.
These arguments lead us to the following operator product expansions
among the fields
\bea
\label{longope}
S^{\mu}(z) S^\nu(w) &=& -\frac{\delta^{\mu\nu}}{z-w} \\ \nonumber
\partial\phi(z) \partial \phi (w) &=& - \frac{1}{(z-w)^2} \\ \nonumber
j^{a}(z) j^b(w) &=& -\frac{\delta^{ab}(Q_5 -2)}{2(z-w)^2} +
\frac{\epsilon^{abc}j^c}{(z-w)}
\eea
To keep our discussion less cumbersome we have ignored the
anti-holomorphic fields.
As the space time preserves 16 supersymmetries we should construct out
of these field an ${\mathcal N}=(4,4)$ superconformal algebra. It is
known \cite{Callan:1991at, Sei-Wit99} that the following construction
has the required properties of the long sting.
\bea 
\label{longalgeb}
T&=& -\frac{1}{2} \partial S^\mu S^\mu - \frac{j^a j^a}{Q_5} -
\frac{1}{2} \partial \phi \partial \phi + \frac{\sqrt{2} (Q_5
-1)}{2\sqrt{Q_5}} \partial^2 \phi \\ \nonumber
J^a &=& j^a + \frac{1}{2} \eta^{a}_{\mu\nu} S^{\mu\nu} \\ \nonumber
G^\mu &=& \frac{\sqrt{2}}{2} \partial\phi S^\mu - \frac{2}{\sqrt{Q_5}}
\eta^a_{\mu\nu} j^a S^\nu + \frac{1}{6\sqrt{Q_5}}
\epsilon_{\mu\nu\rho\sigma} S^\nu S^\rho S^\sigma - \frac{Q_5
-1}{\sqrt{Q_5}} \partial S^\mu
\eea
These generators form an ${\mathcal N}= (4,4)$ superconformal algebra 
given in
\eq{chp2:scft-algebra} with the following definition for the two
component super charge in \eq{chp2:scft-algebra}
\be
\label{longsupcha}
G^{a} = 
(
G^{1} + i G^{2},  G^{3} -iG^{4}
)
\ee
The following points are worth noting
\begin{enumerate}
\item 
Note from the definition of the stress energy tensor 
that the field $\phi$ is a linear dilaton with background charge
$Q= \sqrt{\frac{2}{Q_5}} (Q_5-1)$. This is precisely the background
charge of the linear dilaton theory 
for the case of a single D1 brane
splitting off the \dd bound state (see below \eq{backchargeg}).
\item
The central charge of this algebra is $6(Q_5-1)$ and not 6, as one
would have expected if this algebra was describing the 
spacetime geometry
\item
Note also that the R-symmetry generators in \eq{longalgeb} involve the
bosonic fields $j^a$ which corresponds to the symmetry of $S^3$. 
This is a characteristic of the Higgs branch \cite{witt-higgs-br}.
\end{enumerate}

\subsection {Strings on the thermal $AdS_3$}

To discuss thermal boundary conditions on $AdS_3$ it is convenient to
parameterize Euclidean $AdS_3$ using the following coordinates 
\be
\label{newcoord}
\gamma= ve^\phi \quad \bar{\gamma} =\bar{v} e^\phi
\ee
In terms of these new coordinate the metric on $\H$ becomes
(see \eq{eucads},\eq{eucl-ads-poincare})
\be
\label{newmetric}
ds^2 = l^2 (d\phi^2 + (dv + vd\phi) (d\bar{v} + \bar{v} d\phi))
\ee
The worldsheet action with the B-field given by \eq{eucbfield} is
\be
\label{neweucws}
S= \frac{Q_5}{2\pi} \int d^2z (\partial \phi \bar{\partial}\phi +
(\partial \bar{v} + \partial \phi \bar{v}) (\bar{\partial} v +
\bar{\partial} \phi v) ) 
\ee
Thermal $AdS_3$ is then defined by the following identifications
(see Appendix C.4)
\bea
\label{thermbc}
v \sim ve^{i\mu\beta} \\ \nonumber
\bar{v} \sim \bar{v} e^{-i\mu\beta} \\ \nonumber
\phi \sim \phi + \beta
\eea
Here $\beta$ is the inverse temperature and $\mu$ is the chemical
potential.

We now set up the one-loop evaluation of the partition function on
$\H$ \cite{Maldacena:2000kv}. 
From this it is easy to evaluate the space time free 
energy and thus determine the spectrum of strings in $AdS_3$. We have
to evaluate the path integral on a torus with modular parameter $\tau$.
The conformal field theory consists of the worldsheet Lagrangian
\eq{neweucws} with the identifications \eq{thermbc}, the $b$ $c$
ghosts and an internal conformal field theory.
Let the partition function of the internal conformal field theory  be
given by
\be
\label{intercft}
{\mathcal Z_M} = (q\bar{q})^{-\frac{c_{\rm int} }{24} } 
\sum_{h, \bar{h} } D(h, \bar{h}) q^h \bar{q}^{\bar{h}}
\ee
where $q= e^{2\pi i \tau}$, 
$D(h,\bar{h})$ is the degeneracy  of the state with weight 
$(h, \bar{h})$. Putting this partition function and that of the $b,
c$ ghosts together we get \cite{Maldacena:2000kv},
\bea
\label{one-loop-part}
Z(\beta, \mu) &= \frac{\beta(k-2)^{\frac{1}{2}} }{8\pi}
\int_0^\infty \frac{d\tau_2}{\tau_2^{3/2}}
\int_{-1/2}^{1/2} d\tau_1
e^{4\pi \tau_2(1-\frac{1}{4(k-2)} ) }
\sum_{h, \bar{h} } D(h, \bar{h}) q^h \bar{q}^{\bar{h} }\\ \nonumber
&\times
\sum_{m=1}^\infty
\frac{e^{-(k-2)m^2 \beta^2/4\pi \tau_2} }{|\sinh(m \hat{\beta}/2)|^2}
\left|
\prod_{n=1}^\infty 
\frac{1-e^{2\pi i n \tau} }{(1-e^{m\hat{\beta} + 2\pi i n\tau})
(1-e^{-m\hat{\beta} + 2\pi i n\tau})}
\right|^2
\eea
where $\hat{\beta} = \beta + i\mu \beta$.

From the one-loop partition function it is easy to extract the 
spacetime free energy $F$, 
which is given by $Z(\beta, \mu) = -\beta F$. One can rewrite the free
energy as a sum over states in the single particle string Hilbert
space ${\mathcal H}$
\be
\label{one-loop-free}
F(\beta, \mu) = \frac{1}{\beta} \sum_{\rm{string} \in {\mathcal H} }
\log 
\left(
1- e^{-\beta (E_{\rm{string}} + i\mu l_{\rm{string}} ) }
\right)
\ee
where $E_{\rm{string}}$ and $l_{\rm{string}}$ are the energy and
angular momentum of the string state. One can compare
\eq{one-loop-part} and \eq{one-loop-free} and show that the spectrum is
precisely the one proposed by \cite{Maldacena:2000hw} which was
discussed in section 10.3.

\newpage
\section{Applications of \ads-\cft\  duality}

The microscopic derivation (Section 8) of Hawking radiation from the
D1-D5 black hole shows that the microstates of the black hole are to
be identified with states of the ``boundary CFT''. One way to
understand this is to note that \cite{Mal97,Aharony:1999ti} in the
black hole description (large 'tHooft coupling $g_s Q$) the
propagation of closed string quanta in the curved geometry (see
Section 3) consists of (a) free propagation in the asymptotically flat
region and (b) propagation in \ads\ geometry (throat region); in the
weak coupling description the closed string quanta (a$'$) propagate
freely in flat space and (b$'$) occasionally interact with the D1-D5
system which, for the low energy scales associated with Hawking
quanta, is described by a CFT. Since we are describing the same
physical process as (a)+ (b) at strong coupling and as (a$'$)+(b$'$)
at weak coupling and since (a) is clearly equivalent to (a$'$), we say
that (b) is equivalent (dual) to (b$'$). That is, as seen by a closed
string probe, supergravity in \ads\  is equivalent to the CFT of the
D1-D5 system. This was indeed the reasoning behind the discovery of
the AdS/CFT duality \cite{Mal97}.

While the above equivalence gives us important insight in the context
of the full geometry including the asymptotically flat part, the
equivalence between asymptotically \ads\ spaces and the boundary \cft\ 
has a number of important applications. This subject has been
discussed in fair amount of detail in \cite{Aharony:1999ti}. Our
discussion here should be regarded as complementary to it; we will
focus here mainly on (a) thermodynamics and phase transitions in \ads\ 
and (b) black hole formation by particle collision.  We will begin
with (a).

\subsection{Hawking-Page transition in AdS$_3$}

In the context of flat space, it is well-known
\cite{Hawking:de,Gross:cv} that a thermal state in flat Minkowski
space, no matter how low the temperature, is unstable to formation of
a black hole:
\begin{itemize}
\item
For a flat space with infinite volume, the mass of a
thermal state is infinite; therefore the state will gravitationally
collapse to a black hole. The conclusion remains true even if one
restricts to a large but finite volume. 
\item
Furthermore, because of the negative specific heat implied by
$T= 1/(8\pi\,GM)$ (Eq.\eq{sch-bh}), a black hole can only be in an
\underbar{unstable} equilibrium with radiation at the same
temperature (in a sufficiently large volume to keep the temperature
constant). Any fluctuation resulting in an increase of the black hole
mass would reduce its temperature below that of its surroundings,
causing more absorption than emission so that the black hole continues
to grow. This implies a breakdown of the canonical ensemble.
\end{itemize}

As found in \cite{Hawking:1982dh}, the situation is different in an
AdS space. Let us write the metric of AdS$_3$  as
(see \eq{2:ads}, \eq{a3:ads})
\be
\label{11:ads}
ds^2= -dt^2 (1 + \frac{r^2}{l^2}) + dr^2 (1 + 
\frac{r^2}{l^2})^{-1} + r^2 d\phi^2
\ee
and that of BTZ  as (see \eq{2:btz},\eq{a3:btz},\eq{a3:lapse})
\be
\label{11:btz}
ds^2 = -\left[ \frac{r^2}{l^2} - M + (\frac{J}{2r})^2 \right] dt^2
+ \left[ \frac{r^2}{l^2} - M + (\frac{J}{2r})^2 \right]^{-1} dr^2
+ r^2(\frac{-J}{2r^2} dt +d\phi )^2 
\ee
where $M$ and $J$ refer to the mass and angular momentum of the BTZ
black hole ($J \le M$). 
The arguments of \cite{Hawking:1982dh}, applied to the present case,
state that a thermal state describing radiation in AdS$_3$ at a finite
temperature has a finite mass. For $l\gg 1$ the energy density of
radiation is given approximately by the formula $\rho \propto T^3$
where $T$ is the locally measured temperature $T = T_0(1 +
\frac{r^2}{l^2})^{-1/2}$ which, for $r\gg l$, goes as $T \propto
1/r$. The total energy is given by the integral $M
\approx \int l\ dr\ d\phi \rho$ which is clearly finite since 
$\rho \propto r^{-3}$ at large distances $r$. Thus, unlike in flat
space, thermal radiation in AdS$_3$ can be a stable configuration, if
its (free) energy is less  than that of a black hole at the
same temperature (see below).

We will also see below that the BTZ black hole has a positive specific
heat (see, e.g. the formula for the temperature \eq{a3:btz-temp}),
unlike the Schwarzschild black hole. Thus a stable equilibrium between
a BTZ black hole and radiation in AdS$_3$ is possible; a fluctuation
resulting in an increase of mass of the black hole {\em increases} its
temperature above that of the radiation, causing more emission than
absorption by the black hole, thus restoring its energy back to the
equilibrium value.

\gap2

\ni{\it Euclidean Free energy from supergravity}

\gap2

As we argued in Section 2, the AdS$_3$ and BTZ spacetimes
\eq{11:ads},\eq{11:btz} (times $S^3 \times K$)
are near-horizon limits of solutions of type IIB supergravity (namely
the D1-D5 string and the D1-D5 black hole). In fact they are exact
solutions of type IIB supergravity in their own right. Stated
differently, AdS$_3$ and BTZ are solutions of  three-dimensional
supergravity obtained by a Kaluza-Klein reduction of type IIB
theory on $S^3 \times K$. 

Let us elaborate this a bit more. The action for pure anti-de Sitter
supergravity, based on the super group $SU(1,1|2)\times SU(1,1|2)$
(see Section 6), in a three-dimensional spacetime with cosmological
constant $\Lambda= -1/l^2 < 0$, is given by (see
\cite{chp3:Jus,Nishimura:1999gg}) \footnote{Recently the
three-dimensional $SO(4)$ gauged supergravity which provides the
coupling of the pure anti-de Sitter supergravity based on the
supergroup $SU(1,1|2)\times SU(1,1|2)$ to the lowest matter multiplets
including the massless scalar fields has been constructed in
\cite{Nicolai:2001ac}.}
\bea
S &&= \frac{1}{16 \pi G^{(3)}_N} \int d^3 x
\left[\right. eR+\frac{2}{l^2}e
-\epsilon^{\mu\nu\rho} \bar{\psi}_{\mu} {\mathcal D}_{ \nu}
\psi_{\rho} -  8l\epsilon^{\mu\nu\rho} (
A_{\mu}^i
\del_{\nu} A_{\rho}^i - 
\nn \\
&& \frac{ 4i\epsilon_{ijk} }{3}
A_{\mu}^i
A_{\nu}^j A_{\rho}^k )
-\epsilon^{\mu\nu\rho} \bar{\psi'}_{\mu} {\mathcal D'}_{ \nu}
\psi'_{\rho} +  8l\epsilon^{\mu\nu\rho}(
A'^i_{\mu}
\del_{\nu} A'^i_{\rho} - \frac{ 4i\epsilon_{ijk} }{3}
A'^i_{\mu}
A'^j_{\nu} A'^k_{\rho})  \left. \right]
\label{11:action}
\eea
Here  $G^{(3)}_N$ is the three-dimensional Newton's constant,
and ${\mathcal D}_{\nu } = \del_{\nu} +
\omega_{ab \nu }\gamma^{ab}/{4} - e_{a
\nu}\gamma^{a}/{(2l)} -
2 A_{ \nu}^i\sigma^i$ and ${\mathcal D'}_{\nu } =
\del_{\nu} + \omega_{ab \nu }\gamma^{ab} /{4} + e_{a
\nu}\gamma^{a}/{(2l)} -2 A'^i_{ \nu}\sigma^i$.  The basic fields
appearing in the Lagrangian are the vielbein $e^a_\mu, \psi_\rho,
A^i_\mu, \psi'_\mu$ and $A'^i_\mu$. The $\omega$'s
are spin connections.

As we remarked before \eq{11:action}, the same three-dimensional
supergravity can be obtained from type IIB string theory compactified
on $K \times S^3$ (with constant flux on $S^3$). This identifies the
cosmological constant and Newton's constant in terms of type IIB
parameters and the parameters of compactification
(cf. Eqn. \eq{radius-ads}):
\bea
\label{action-parameters}
G^{(3)}_N &=& \frac{4 \pi^4 g^2_s}{V_4 l^3} \\
\nn
l^4 &=& \frac{16 \pi^4 g^2_s Q_1 Q_5}{V_4}
\\
\nn
\eea 
where $V_4$ is the volume of $T^4$ and $g_s$ is the string coupling
(we are working in the units $\alpha'=1$).

From our remarks above, it is obvious that \ads\ and BTZ are solutions
of \eq{11:action}.  We will now use this supergravity Lagrangian to
calculate the Euclidean free energy. Our method will be similar in
spirit to that of
\cite{Gibbons:1976ue,Hawking:1982dh,Witten:1998zw}. The presentation
will more closely follow
\cite{David:1999zb,chp3:MalStr98,Dijkgraaf:2000fq}.

Let us recall that the free energy of \ads\ supergravity is given by
\be
\label{11:free-energy}
Z_{\rm sugra}  = \int 
{\mathcal D}_\alpha[\hbox{fields}] \exp[-S_E]
\ee
where $S_E$ is the Euclidean version of the action written in
\eq{11:action}.

\gap2

\ni{\it Boundary Conditions:}

\gap2

In the above equation, $\alpha$ denotes boundary conditions that
define the functional integral. The semiclassical evaluation of the
functional integral is performed by summing over the saddle point
configurations which satisfy the given boundary conditions.

The boundary conditions are specified as follows:
\begin{enumerate}
\item  Asymptotically \ads\  Euclidean metrics satisfy
the boundary condition: 
\be
ds^2 \stackrel{r\to\infty}{\rightarrow}  
dr^2/r^2 + r^2  | d\phi + i d\ttt |^2
\label{metric-bc}
\ee
where $\phi,\ttt$ satisfy the periodicity
conditions:
\be
\phi + i\ttt \equiv \phi + i\ttt + 2\pi(n + m \tau)
\label{a-b-cycles}
\ee
This defines the (conformal) boundary of the space to 
be a torus with modular parameter $\tau$.

For Euclidean \ads, Eqns. \eq{phi-period},\eq{a3:twist} imply
that the  modular parameter for the boundary torus
is 
\be
\label{a3:tau-ads}
\tau = i\beta/(2\pi), \quad \beta = 1/T - i\Phi
\ee 
where the complex temperature $\beta$ (see \eq{a3:ads-temp} 
relates to a partition function of the form ${\rm Tr}
\exp[- H/T + i \Phi J]$.

For Euclidean BTZ, on the other hand, Eqns. \eq{phi-period-btz},
\eq{btz-beta} imply that the modular parameter for the boundary torus
(cf. \eq{metric-bc}) is now 
\be
\tilde \tau = i\beta/2\pi=-1/\tau
\label{ads-vs-btz}
\ee
where $\tau$ is given by \eq{def-tau}.

The inverse relation \eq{ads-vs-btz} between the modular parameters
$\tau, \tilde \tau$ arises because of the difference in our
identifications of space and Euclidean time in
\eq{u-phi}, \eq{u-phi-btz}. Indeed from this viewpoint, Euclidean BTZ
is a special case of an SL(2,Z) family of instanton configurations
\cite{chp3:MalStr98,Dijkgraaf:2000fq}.  In the more general case we
replace the parameter $\tau$ in \eq{a3:btz-quotient} by
\be
\tilde \tau = \frac{a\tau +b}{c\tau + d}, \qquad
\left( \begin{array}{cc} a & b \\ c & d \end{array}
\right) \in SL(2,Z)
\ee 
and \eq{u-phi-btz} is replaced by
\be
2u = \frac{i}{c\tau + d}(\phi + i\ttt)
\label{u-phi-general}
\ee
In \cite{Dijkgraaf:2000fq} an elliptic genus calculated
from the boundary CFT is interpreted as a sum over this
entire SL(2,Z) family of gravitational instantons.
In our simplified treatment below, only \ads\ and BTZ
configurations will dominate the path integral.
As an upshot of the above discussion is the
relation (cf. \eq{a3:tau-ads}, \eq{btz-beta})
\be
Z_{AdS}(\tau=i\beta/2\pi) = Z_{BTZ}(\tau=i\beta_0/2\pi)
\quad \beta_0= (4\pi^2/\beta)
\label{ads-vs-btz1}
\ee
We should note  that for Euclidean BTZ black hole, the
complex temperature gets fixed by the geometry
(see \eq{def-tau}), whereas for Euclidean \ads\ the
complex temperature is for us to specify.

\item  Gauge field: the boundary condition
on the gauge fields will not play an important role in the following
discussion (see Eq. (5.7) of \cite{Dijkgraaf:2000fq} for details).

\item Fermions: the above discussion,
especially \eq{metric-bc},\eq{a-b-cycles} points to two non-trivial
cycles of the boundary torus. The fermion fields can have either
periodic (P) or antiperiodic (AP) boundary conditions along either
cycle. Thus we can have any of the four boundary conditions (P, P),
(P, A), (A, P), (A, A), where by convention the entry denotes boundary
condition along the cycle $\phi + i\ttt \to\phi + i\ttt + 2n\pi $ and
the second entry denotes boundary condition along $\phi + i\ttt
\to\phi + i\ttt + i\beta, \tau=i\beta/2\pi$.
In the language of 2D SCFT, a periodic boundary condition correspond
to Ramond (R) fermions and an antiperiodic b.c.  corresponds to
Neveu-Schwarz (NS) fermions. By the remarks at the end of  Section
\ref{near-horizon-btz} (see also Section 6) we find that the BTZ
fermions correspond to Ramond boundary conditions along the ``space''
cycle; thus a standard Euclidean partition function in BTZ should
correspond to the boundary condition (P,AP) (thermal fermions
represented by Tr $(e^{-\beta H})$ happen to be antiperiodic along the
``time'' cycle). We will compute below such a partition function. By
the duality \eq{ads-vs-btz1} this should be related to (AP,P) for the
fermions in \ads. This is consistent with the remarks at the end of
Section \ref{near-horizon-btz} (see also Section 6) that \ads\
fermions are represented by NS boundary conditions at the boundary.

\end{enumerate}

\gap2

\ni{\it Finding the  Saddle points}

\gap2

We will evaluate \eq{11:free-energy} by finding saddle points of the
action subject to the specific boundary conditions
mentioned above. By virtue of the
equation of motion $R=-6/l^2$, the Euclidean action $S$ of a classical
spacetime $X$ is simply its volume times a constant. To be precise,
\bea
\label{euclideanaction}
S(X) &=& \frac{1}{4 \pi l^2 G^{(3)}_N} \hbox{Vol}(X)
\nn\\
\hbox{Vol}(X) &=& \int_0^\beta d\ttt
\int_{r_0}^R dr \int_0^{2\pi} d\phi \sqrt{g}
\eea
The ranges of $\phi, \tau$ follow from the identifications mentioned
above. The lower limit $r_0$ of the $r$-integral is identically zero
for \ads\ and the conical spaces, whereas for BTZ it denotes the
location of the horizon (the Euclidean section is defined only upto
the horizon). The upper limit $R$ is kept as an infrared regulator to
make the volume finite. We will in practice only be interested in free
energies relative to \ads\ and the $R$-dependent divergent term will
disappear from that calculation.

\gap2

\ni{\it Free energy of BTZ:}

\gap2

We will now compute the free energy of BTZ relative to \ads\ (in a
manner similar to \cite{chp1:Btz,Witten:1998zw}). As detailed in
Appendix C, the \ads\ solution can be at any temperature $\beta $
\eq{a3:ads-temp} while the temperature of the black hole is fixed to
be $\beta_0$ \eq{btz-beta},\eq{def-tau} which is given by the
geometry. To compare with the $AdS_3$ background one must adjust
$\beta$ so that the geometries of the two manifolds match at the
hypersurface of radius $R$ (in other words, we must use the {\em same}
infrared regulator on all saddle points of the functional
integral). This gives the following relation
\be
\label{betabeta0}
\beta_0 = \beta\sqrt{\frac{1+l^2/R^2}{1-M^2l^2/R^2}}
\ee
\bea
\label{freebtz}
S(BTZ)-S(AdS_3) 
&=& \frac{1}{4 \pi l^2 G^{(3)}_N}\left[ \int_{BTZ} d^3 x
\sqrt{g} - \int_{AdS_3} d^3 x \sqrt{g}\right]  \\ \nonumber
&=& \frac{1}{4 \pi l^2 G^{(3)}_N}\left[\pi \beta_0 (R^2 - r_+^2) -
\pi\beta R^2\right]   
\eea
Substituting the value of $\beta_0$ in terms of $\beta$
\footnote{We should make a remark here to avoid any potential 
confusion between equations like \eq{ads-vs-btz1} and \eq{betabeta0};
the former equation is a statement that the Euclidean partition
function of \ads, corresponds to a certain complex temperature,
matches with that of BTZ at a different temperature. In
calculating the supergravity partition function,
however, we want to regard both as having the same temperature ($R\to
\infty$ limit of \eq{betabeta0}); the difference 
between which bulk geometry corresponds to \ads\ and which to BTZ
arises here by the choice of ``space'' and ``time'' (see Appendix
C.4).}  and taking the limit $R\rightarrow\infty$ we obtain
\be
S(BTZ) -S(AdS_3) = \frac{1}{4 \pi l^2 G^{(3)}_N}\left[\pi\beta_0 l^2 -
\pi^2 r_+ l^2\right] 
\ee
where the complex  temperature $\beta_0$ is given by Eqns.
\eq{def-tau},\eq{a3:btz-temp}.
Using these variables the difference in the action becomes
(identifying $\beta_0$ with $\beta$)\cite{David:1999zb}
\be
\label{freebtzvalue}
S(BTZ) -S(AdS_3) = \frac{1}{4 \pi l^2 G^{(3)}_N}
\left[\frac{\pi}{2}(\beta + \beta^* ) l^2  
- \pi^3 l^4 \left(
\frac{1}{\beta }+ \frac{1}{\beta^* }
\right) \right] 
\ee
By using \eq{ads-vs-btz1} we find
\be
S(\ads) = \frac{1}{4 \pi l^2 G^{(3)}_N}
\left[\frac{\pi}{2}(\beta + \beta^* ) l^2
\right]
\label{free-ads}
\ee
\be
S(BTZ) = \frac{1}{4 \pi l^2 G^{(3)}_N}
\pi^3 l^4 \left[
\frac{1}{\beta }+ \frac{1}{\beta^* }
\right]
\label{freebtzhigh}
\ee
It is clear that at low temperatures the \ads\ saddle point dominates
the path integral, where at high temperatures the BTZ dominates.  The
transition from \ads\ at low temperature to BTZ at high temperature is
the 3-dimensional analogue of the Hawking-Page transition.

\gap2

\ni\underbar{\it Euclidean Free energy from CFT}

\gap2

The aim of this section is to calculate the partition function of the
$(4,4)$ CFT on the orbifold $T^{4Q_1 Q_5}/S(Q_1 Q_5)$. The partition
function depends on the boundary conditions of the fermions of the
CFT.  We will first calculate the partition function when the bulk
geometry is that of the BTZ black hole.

\gap1

\ni\underbar{\it CFT partition function corresponding to BTZ}

\gap1

The fermions of the CFT are periodic along the angular coordinate of
the cylinder if the bulk geometry is that of the BTZ black hole. This
can be seen by observing that the zero mass BTZ black hole admits
killing vectors which are periodic along the angular coordinate
\cite{chp4:CouHen}. Therefore the zero mass BTZ black hole correspond to
the Ramond sector of the CFT. The general case of the BTZ black hole
with mass and angular momentum correspond to excited states of the CFT
over the Ramond vacuum with
\bea
L_{0} + \bar{L}_0 &=& M l \\  \nonumber
L_0 - \bar{L}_0 &=& J_E
\eea
where $M$ and $J_E$ are the mass and the (Euclidean) angular momentum
of the BTZ black hole. Therefore the partition function of the BTZ
black hole should correspond to
\be
Z= \mbox{Tr}_R (e^{2\pi i \tau L_0} e^{2\pi i
\bar{\tau} \bar{L}_0} )
\ee
The Hilbert space of the CFT on the orbifold $T^{4Q_1 Q_5}/S(Q_1 Q_5)$
can be decomposed into twisted sectors labeled by the conjugacy
classes of the permutation group $S(Q_1 Q_5)$.  The conjugacy classes
of the permutation group consists of cyclic groups of various
lengths. The various conjugacy classes and the multiplicity in which
they occur in $S(Q_1 Q_5)$ can be found from the solutions of the
equation
\be
\sum_{n=0}^{Q_1Q_5} n N_n = Q_1 Q_5
\ee
where $n$ is the length of the cycle and $N_n$ is the multiplicity
of the cycle. The Hilbert space is given by
\be
{\mathcal H} = \bigoplus_{\sum n N_n = Q_1 Q_5} \bigotimes_{n>0}
 S^{N_n} {\mathcal H}_{(n)}^{P_n}
\ee
$S^N {\mathcal H}$ denotes the symmetrized product of the Hilbert space
${\mathcal H}$, $N$ times. By the symbol ${\mathcal H}_{(n)}^{P_n}$ we mean
the Hilbert space of the twisted sector with a cycle of length $n$ in
which only states which are invariant under the  projection
operator 
\be
P_n = \frac{1}{n} \sum _{k=1}^{n} e^{2\pi i k (L_0 - \bar{L}_0)}
\ee
are retained.
The values of $L_0$ or $\bar{L}_0$ in the twisted sector of length $n$
is of the form $p/n$ where $p$ is positive integer. This projection
forces the value of $L_0 -\bar{L}_0$ to be an integer on the twisted
sector. It arises because the black hole can exchange only integer
valued Kaluza-Klein momentum with the bulk \cite{chp1:DavManWad1}.

The dominant contribution to the partition function arises from the
maximally twisted sector. That is, from the longest single cycle of
length $Q_1 Q_5$. It is given by
\be
Z= \sum_{m,n} d(Q_1 Q_5 n +m) d(m) e^{2 \pi i n \tau} 
e^{2\pi i m \tau /Q_1 Q_5} e^{-2\pi i m \bar{\tau}/Q_1 Q_5}
\ee
Where $d$'s are the coefficients defined by the expansion
\be
Z_{T^4} = \left[ \frac{\Theta_2 (0|\tau }{\eta ^3 (\tau )}\right] ^2
     = \sum_{n\geq 0}  d(n) e^{2 \pi i \tau n}
\ee
In the above equation $Z_{T^4}$ is the partition function of the
holomorphic sector of the CFT on $T^4$. We will first evaluate the sum
\be
P(m, \tau) = \sum _{n=0} ^{\infty} d(Q_1 Q_5 n + m) e^{2 \pi i n \tau}
\ee
For large values of $Q_1 Q_5 $ we can use the asymptotic form of
$d(Q_1 Q_5 n +m)$
\be
d(Q_1 Q_5 n + m) \sim \exp\left(2 \pi \sqrt{Q_1 Q_5 n +m}\right)
\ee
Substituting the above value of $d(Q_1 Q_5 n +m )$ in $P(m, \tau )$ we
obtain a sum which has an integral representation as shown below.
\bea
P(m, \tau) &=& \sum_{n=1}^{\infty} e^{2 \pi \sqrt{Q_1 Q_5 n + m} + 
2\pi i n \tau } + d(m) \\     \nonumber
&=& {\mathcal P}\frac{i}{2} \int_{-\infty}^{\infty} dw \coth \pi\omega 
e^{2\pi \sqrt{i Q_1 Q_5 \omega + m } -2 \pi \omega \tau}  + 
d (m) -\frac{e^{2\pi \sqrt m}}{2}
\eea
where ${\mathcal P}$ denotes ``principal value'' of the integral.
 
We are interested in the high temperature limit of the partition
function. The leading contribution to the integral in the limit
$\tau\rightarrow 0$ is
\be
P(m, \tau) \sim \sqrt{i\pi Q_1 Q_5/ \tau}
e^{i \pi Q_1 Q_5/2\tau -i 2 \pi m \tau/Q_1 Q_5}
\ee
Substituting the above value of $P(m, \tau)$ the partition function
becomes
\be
Z= \sqrt{i\pi Q_1 Q_5/ \tau}  
\sum _{m=0}^{\infty} d(m) e^{-2
\pi i m \bar{\tau}/Q_1 Q_5} 
\sim \exp\left(i \pi Q_1 Q_5 (1/2\tau -1/2\bar{\tau}) \right)
\ee
Thus the free energy at high temperatures is given by
\be
-\ln Z = \frac{-i \pi Q_1 Q_5}{2} 
\left(\frac{1}{\tau} -\frac{1}{\bar{\tau}}
\right)
\ee
This exactly agrees with \eq{freebtzhigh}\ with the 
identification $\tau = i\beta_+/(2 \pi l)$. 

We will not attempt to calculate the \ads\ partition function here
(with fermion b.c. (AP,P)), except to note that by using spectral flow
arguments \cite{Dijkgraaf:2000fq} it is possible to show that the CFT
result agrees with \eq{free-ads}. Indeed in \cite{Dijkgraaf:2000fq} a
more general agreement between the CFT elliptic genus and the
corresponding \ads\ quantity is demonstrated for the entire SL(2,Z)
family of solutions of which \ads\ and BTZ are special cases.

\subsection{Conical defects and particles in $AdS_3$}

In this section we will mention some of the 
continuing developments
regarding point particle dynamics and black hole formation in \ads.

It has long been noted \cite{Deser:tn,Deser:dr} that there is a
one-parameter family of spacetimes, given by
\be
ds^2 = l^2 \Big[ 
-\Big(\gamma + \frac{r^2}{l^2} \Big)\ dt^2 + 
\frac{dr^2}{\gamma + \frac{r^2}{l^2}}+ r^2 d\phi^2 \Big]
\label{11:conic}
\ee
which represent the so-called point masses in \ads.  These spacetimes
are all asymptotically \ads, and they interpolate between \ads\  and BTZ
($\gamma=-M=1$ is \ads, $\gamma=-M\le 0$ is BTZ ($J=0$), and
$\gamma=-M \in (0,1)$ are the conical spaces).  The conical spaces are
called so because they have a defect angle $\Delta= 2\pi(1 -
\sqrt\gamma)$ (see Appendix C).  The parameter $\gamma$ is related to
the ``mass'' of the point particle that causes the conical defect. 
To say it in more detail, consider a free  particle
(a geodesic) ${\sf g}$ in \ads, given by
\be 
{\bf Y}(s) = e^{s p^a {\bf \gamma}_a}.
\label{11:geodesic}
\ee 
where we have used coordinates \eq{a3:sl2}. The quantities
$p^a$ play the role of ``momenta'' (see Section \ref{btz-creation}
for more detail). For a static particle
\be
p^0= m, p^1 =  p^2=0. 
\label{11:momenta}
\label{11:momenta-static}
\ee
The geodesic \eq{11:geodesic} satisfies, in the coordinates
\eq{a3:matschull},
\be
\phi(t)=0, \, \rho(t)= 0. 
\label{11:geodesic-static}
\ee
We state without proof here (for more details see Section
\ref{btz-creation}) that the gravitational back
reaction of the point particle \eq{11:geodesic} 
amounts to cutting a wedge out of \ads\  and identifying the edges.
Such an identification is achieved by
\eq{a3:conic-similarity}-\eq{a3:gamma-m}.

\gap2

\ni\underbar{\it String theory embedding and CFT duals}

\begin{enumerate}

\item 
It has been shown in \cite{David:1999zb} that the conical space ${\mathcal
C}$ \eq{11:conic} can be embedded as a solution of three-dimensional
supergravity based on the supergroup $SU(1,1|2)\times SU(1,1|2)$
described by \eq{11:action}. The latter appears as the low energy
description of Type IIB string theory on $S^3 \times T^4$. At a first
sight supersymmetrization of conical spaces seems to be an
impossibility since the candidate Killing spinors \cite{chp4:CouHen}
typically pick up phases when transported around the conical
singularity, and therefore are not single valued. It was shown in
\cite{Izquierdo:1994jz} that this problem could be avoided by
employing an extended supersymmetry ${\mathcal N}= (2,0)$ (see
\cite{Achucarro:gm} for the notation ${\mathcal N}= (p,q)$ supergravity)
and assigning a background value to the gauge field (which occurs in
the supergravity multiplet). By choosing the background value
appropriately the gravitational holonomy picked up by the Killing
spinors can be cancelled by the gauge holonomy. Since the extended
supergravity ${\mathcal N}= (2,0)$ can be embedded, further, in the
$SU(1,1|2) \times SU(1,1|2)$ supergravity which is ${\mathcal N}= (4,4)$,
the conical space ${\mathcal C}$ can be embedded as solutions of this
latter supergravity. We therefore land up with a solution ${\mathcal
C}\times S^3 \times T^4$ of type IIB supergravity \cite{David:1999zb}.  
The background value of the $SU(2)$ gauge field is given by $ A =
\frac{l}2 \gamma d\phi {\bf \tau}_3 $; from the point
of IIB theory this is one of the ``Kaluza-Klein'' gauge
fields that appear on reduction on $S^3$.

The CFT dual of the conical spaces \eq{11:conic}, from this viewpoint,
is identified \cite{David:1999zb} as the spectrally flowed CFT Hilbert
space with spectral flow parameter
\be
\label{eta-gamma}
\eta = \sqrt\gamma. 
\ee
The spectral flow provides a one-parameter interpolation
between the NS ($\eta=0$) and R
($\eta=1$) sectors of the CFT, just as the
conical spaces (parameterized by $\gamma$)
interpolate between \ads\ ($\gamma=0$)
and zero-mass BTZ ($\gamma=1$).
Note that the spectrally flowed energy formula of the CFT
ground state:
\[
L_0 | 0 \rangle_\eta = \bar  L_0 | 0 \rangle_\eta
= - \frac{c}{24} \eta^2 | 0 \rangle_\eta.
\]
This precisely agrees with the ADM mass of the conical spaces provided
one uses \eq{eta-gamma}. Indeed, the free energy at a finite
temperature also agrees with the CFT calculation \cite{David:1999zb}.

\item 

In the above embedding, the identification \eq{a3:conic-similarity}
acts apparently only on the \ads, and not on the $S^3
\times T^4$, although the {\em vev} of the SU(2) gauge field
indirectly does affect the $S^3$. In
\cite{Balasubramanian:2000rt,Maldacena:2000dr}  a different,
though perhaps not entirely unrelated, embedding of \eq{11:conic} into
type IIB theory is used where the holonomy acts simultaneously on
\ads\ as well as on $S^3$ (as we go around the conical singularity in
\ads, we also go around a circle in $S^3$), leading to an embedding in
type IIB theory as $ \left( \ads
\times S^3 \right)/Z \times T^4$. This description naturally 
arise as near-horizon limit of spinning black holes.
 
The CFT dual for the conical spaces in this approach are (a
microcanonical ensemble of) RR states which depend on the parameter
$\gamma$ [see \cite{Balasubramanian:2000rt}, Eqns. (118)-(121)].

\item 

For $\gamma=1/N^2$ the conical space \eq{11:conic} becomes the
orbifold $\ads/Z_N$ which has an exact world sheet CFT description and
hence can be embedded as an exact solution in string theory
\cite{Martinec:2001cf}. This can be done in two ways:
\begin{itemize}
\item $\ads /Z_N \times S^3 \times K$. This theory is
tachyonic, and forms a model of closed string tachyons.
\item $(\ads \times S^3)/Z_N \times K$. This theory is 
supersymmetric and has no tachyons. 

The \ads/CFT$_2$ dual of the conical defect from this viewpoint is
given in terms of a fractionally moded ${\mathcal N}=(4,4)$
superconformal algebra, the fractional moding being given by
$1/N$ spectral flow \cite{Martinec:2001cf} from the Ramond sector.

\end{itemize}

\end{enumerate}

\subsubsection{Black hole creation by particle
collision}\label{btz-creation} 

In 3D gravity ($\Lambda <0$) there are explicit solutions
\cite{Matschull:1998rv} where a BTZ black hole is created by point particle
collision. We will very briefly mention some salient
points here:
\begin{itemize}

\item The conical spaces \eq{11:conic} above are a special case of
the conical spacetimes for a {\em moving} point particle. 
We will describe here the simplest case of a moving particle
which turns out be that of a massless particle. In this
case, the geodesic ${\sf g}$ \eq{11:geodesic} is specified by
the ``meomenta''
\be
p^0= p^1 = \tan \epsilon, p^2=0. 
\label{11:massless}
\ee
In coordinates \eq{a3:matschull}  ${\sf g}$ now satisfies
\be
\phi(t)=0, \, \rho(t)= \tan(t/2). 
\label{11:geodesic-massless}
\ee
The gravitational back-reaction of such a particle can be
exactly computed and the resulting spacetime is obtained
by cutting out a wedge $W$ from \ads, with edges
$\del W= w_+ \cup w_-$,
\[
w_\pm:  \frac{2\rho}{1+ \rho^2} \sin(\epsilon \pm \phi)
= \sin t \ \sin \epsilon
\]
where $w_+$ are $w_-$ are {\em identified}. The identification is
achieved by quotienting \ads\ by an isometry as in
Eqns. \eq{a3:conic-similarity} and \eq{a3:holonomy} with the momenta now given
by \eq{11:massless}.  It is clear that the geodesic ${\sf g}$
\eq{11:geodesic-massless} is a fixed point set of the quotienting
operation \eq{a3:conic-similarity}. Note that under this identification $w_+
\equiv w_- = {\sf g}$. Thus a moving particle is described by a wedge
$W$ as constructed above. The spacetime constructed this way is
\ads/$Z$, where the $Z$ is a discrete subgroup of the isometry group
as just described. Hence the resulting spacetime remains an exact
solution of three-dimensional gravity with $\Lambda<0$. The energy of
the particle is related to the parameter $\epsilon$ which determines
the holonomy of the conical spacetime.

\item

It is easy to generalize the above procedure to construct spacetimes
representing two particles moving towards each other. 
Each geodesic, ${\sf g}^{(1)}$ or ${\sf g}^{(2)}$,
represents a wedge, $W^{(1)}$ or $W^{(2)}$, with edges given by
\beas
w^{(1)}_\pm:&& \frac{2\rho}{1 + \rho^2} \sin(\epsilon \pm \phi)
= \sin t \ \sin \epsilon 
\nn\\
w^{(2)}_\pm:&& \frac{2\rho}{1 + \rho^2} \sin(\epsilon \pm \phi)
= - \sin t \ \sin \epsilon 
\eeas
Like before $w^{(1)}_+ \equiv w^{(1)}_- \equiv {\sf g}^{(1)}$ under
the holonomy matrix ${\bf u}^{(1)}={\bf 1} + \tan \epsilon \left({\bf
\gamma}_0 + {\bf \gamma}_1 \right)$ and similarly for the second
particle. The full spacetime is represented in terms of two charts,
one obtained by quotienting with the holonomy matrix ${\bf u}^{(1)}
{\bf u}^{(2)}$, and the other one by quotienting with ${\bf u}^{(2)}
{\bf u}^{(1)}$. Once again, the full spacetime is 
AdS$_3/Z$, where the quotienting is by a discrete 
isometry subgroup; hence the resulting spacetime is an
exact solution of \ads-gravity.

\item

Since the above construction gives the full spacetime, the
time-development of the collision process is computed by looking at
the Poincare discs (see definition near \eq{a3:matschull}) at various
times $t$. Thus, on the Poincare disc corresponding to $t=0$, the two
wedges $W^{(1)}$ and $W^{(2)}$ meet; this, therefore represents the
time when the two massless particles collide . At later times $t>0$,
the Poincare discs exhibit a single wedge
(the two wedges get identified!) which corresponds to the
worldline
\be
\frac{2\rho}{1 + \rho^2} = \sin(t) \tan(\epsilon)
\label{11:combined}
\ee
For energies $\epsilon$ low enough so that $\pi/4 > \epsilon > 0$,
$\tan(\epsilon) < 1$. It is easy to verify that \eq{11:combined} then
represents a timelike worldline. In this energy range, therefore, the
collision of two massless particles results in a single massive
particle (see Figure 5, \cite{Matschull:1998rv}). For higher energies $\pi/2 >
\epsilon > \pi/4$, {\it i.e.}  $\tan(\epsilon) > 1$, hence the
geodesic \eq{11:combined} is spacelike. As shown in  \cite{Matschull:1998rv},
this spacelike worldline  is identified with the {\it future
singularity} of a BTZ black hole 
(see Figure 6,  \cite{Matschull:1998rv}). Indeed
the holonomy matrix is identified as exactly the one appropriate for a
BTZ black hole. 

The spacetime constructed this way therefore corresponds to formation
of a BTZ black hole by a collision of two particles. Once again, since
the identifications used correspond to discrete isometries, we
have an exact solution of \ads\ gravity.

\end{itemize}

The embedding of the above solutions in string theory or
\ads\ supergravity remains an open problem, although it is
likely that the constructions described here will admit
straightforward generalizations. A more interesting problem is to
understand the CFT dual of these solutions. If we succeed in applying
any one of the candidate CFT duals of conical spaces as described
above, we will have a unitary quantum mechanical description of black
hole formation.

We should mention that black hole formation in three dimensions
can also be described as a collapsing scalar field
confiuration  \cite{Choptuik:1993jv} or as collpsing
dust shells  \cite{Peleg:1994wx}. The former
process, which exhibits a critical scaling behaviour,
has been discussed in the context of \ads/\cft\ correspondence
in \cite{Birmingham:2001hc}. CFT duals of collapsing
dust shell solutions in AdS spaces are discussed in detail
in \cite{Danielsson:1999zt,Danielsson:1999fa,Giddings:2001ii}.

\newpage
\section{Concluding remarks and open problems}

In this Report (see Section 
\ref{sec:plan} for a more detailed summary) we have
presented a detailed calculable formalism of the near extremal black
hole (Sections 2,3) of type IIB string theory in terms of the D1-D5
system of branes (Sections 4,5). We discussed (Section 8) how for this
black hole the thermodynamics and also the rates of Hawking radiation
of all the massless particles can be reproduced from string theory to
match the results derived in supergravity. The facility of
extrapolating weak coupling calculations to the strong coupling regime
owes to the high degree of supersymmetry that exists in the effective
Lagrangian of the low energy degrees of freedom in the string theory
(Sections 7,9). As we emphasized in this Report, a crucial input
(Section 6) in the calculation of Hawking rates comes from the AdS/CFT
correspondence of Maldacena \cite{Mal97} without which the Hawking
radiation from some massless scalars is impossible to calculate from
CFT (in particular, we saw in Sections \ref{chp3:sec-fixed} and
\ref{sec:blow-up} that we get \emph{incorrect results} 
if we use the earlier DBI approach to derive the interaction of the
D1-D5 bound state with the Hawking quanta). We also presented a review
of \ads/\cft\ correspondence beyond the supergravity approximation
(using the NS5 version, see Section 10), and some applications of this
correspondence for black hole formation in three dimensions (Section
11) by thermal transition and by collision of point particles.

In the light of what is achieved the following open 
problems naturally suggest themselves:
\begin{enumerate}

\item 
The emergence of $AdS_3(\times S^3)$ spacetime from the the ${\mathcal
N}=(4,4)$ SCFT is an open important open problem. In the context of
$AdS_5\times S^5$ Dorey et al \cite{Dorey:1998qh} (see also
\cite{Blau:2001gj}) have argued that this spacetime emerges from the
moduli and the fermion zero modes associated with the large N saddle
point of the dual ${\mathcal N}=4$ SUSY gauge theory. Another approach
to the question of how the radial direction of the AdS space arises
from the viewpoint of the boundary theory is via Liouville theory
where the Liouville or conformal degree of freedom is interpreted as a
space dimension \cite{Das:1989ds,Dhar:1990km,Das:1990da,Dhar:2000ai}.
For a recent discussion of formulation of holography in $AdS_3$ using
Liouville theory see \cite{Krasnov:2000ia,Krasnov:2001ui}.
Also see \cite{Krasnov:2002rn} for a discussion of black hole creation
by point particles using Liouville theory.

\item
Although the D1-D5 system provides a correct derivation of the black
hole entropy, there is no precise understanding of \emph{why} the
entropy is given by the area of the event horizon. The reason why we
lack this understanding is that the counting of microstates (see
Section 8) is peformed in the dual theory.  An important question to
answer is to understand these states in the language of
supergravity/string theory. Related questions have been discussed in
\cite{Banks:1998dd,Balasubramanian:1999ri}.  See also
\cite{Brown:1986nw} where a conformal algebra is constructed in
AdS$_3$ supergravity with a central charge that coincides with
that of the boundary CFT obtained from the D1-D5 system. 

\item
A matrix string description of the D1-D5 black hole was proposed in
\cite{Dijkgraaf:1997nb}. It will be interesting to make a detailed
comparison between this work and the microscopic formulation presented
in this Report.

\item
Most of the discussion (especially where it involves the
micro-description) in the present Report involves BPS or near-BPS
black holes.  There is a vast amount of literature on
nonsupersymmetric black holes in supergravity, but we are quite far
from a microscopic understanding of them. Works which address this
problem include

\begin{itemize}

\item 
D0 brane blackholes (see \cite{Kabat:2001ve} and references therein):
Quantum mechanics of $N$ D0 branes at a finite temperature is analysed
at large 'tHooft coupling $g_s N$, using a mean-field theory
approximation. This leads to an entropy in good agreement (over a
cetain range of temperatures) with the Bekenstein-Hawking entropy of a
ten dimensional type IIA black hole carrying 0-brane charge.

\item Correspondence principle and string-black hole 
phase transitions (see also Section 11.3): It has been shown
\cite{susskind,polhoro} that the entropy formula for a large class of
nonsupersymmetric black holes agrees with that for a highly excited
string state (with the same mass and charge) at a correspondence point
where the curvature at the horizon becomes of the order of the string
scale. This suggests a phase transition \cite{Horowitz:1998jc,venezia}
from a string state to a black hole (see also
\cite{Abel:1999dy,Barbon:2001di}). A precise understanding of such a
transition is obviously important. 

\item Closed string tachyons and black hole 
entropy \cite{Dabholkar:2001if}: A string theoretic version is
suggested of the Gibbons-Hawking derivation
\cite{Gibbons:1976ue} of entropy of a  Schwarzschild 
black hole. The Schwarzschild black hole is in fact replaced by a cone
for computing the entropy. The corresponding orbifold string theory
has a closed string tachyon whose dynamics is dealt with using the
techniques of \cite{Adams:2001sv}. It is obvious interest to
understand this calculation in terms of miscrostates.
\end{itemize}

\item
One of the potential applications of the AdS/CFT correspondence
(especially in the conext of \ads) is black hole \emph{dynamics}.  In
Section 11.2 we briefly discussed the problem of black hole formation
by particle collisions in \ads (see also
\cite{Birmingham:2001hc} which discusses Choptuik scaling
\cite{Choptuik:1993jv} in the context of
\ads/\cft\ correspondence). Clearly, the problem of evaporation of
black holes too maps into interesting time dependent phenomena in the
dual gauge theory/CFT. The Euclidean Ads/CFT correspondence
describes equilibrium physics; the approach to this 
equilibrium is an important unsolved problem.

\item Issues of singularities and the D1-D5 system:
The D1-D5 system (with K3 compactification) has been used to resolve
naked singularities \cite{Johnson:1999qt}. The detailed understanding
of this mechanism is an important problem, especially in what precise
sense the gauge theory acts as a source for the geometry, and how
precisely the ``matching'' of the gauge theory and the geometry can be
understood.

\item
Recently \cite{Berenstein:2002jq} a particular scaling limit of AdS
spaces (``pp wave'') is found where string theory in the Ramond-Ramond
background is solved exactly, leading to new insights into the AdS/CFT
correspondence. It is important to explore the consequences for the
D1-D5 system, in particular to understand from the CFT the IIB
\emph{string spectrum} in the pp-wave limit of $AdS_3\times S^3$
\cite{Berenstein:2002jq,Zayas:2002rx}.

\item 
We found in Section \ref{sec:2dbh} that the two dimensional black hole
\cite{Mandal:1991tz,Witten:1991yr,Elitzur:1991cb} arises as a limit of
the non-extremal 5-brane. In \cite{Kazakov:2000pm} a holographic
description of the two dimensional black hole is proposed in terms of
a quantum mechanical matrix model. It will be interesting to see how
this matrix model fits into (a holographic description of) the
five-brane theory, e.g. whether it can perhaps lead to the
phenomenological model of \cite{Maldacena:1996ya} consisting of a gas
of strings on the five brane (with tension $1/(2\pi Q_5 g_s \alpha')$ and
central charge $= 6$), that gives the Bekenstein-Hawking entropy of this
black hole $S= U_0^2 g_{\rm{YM}} \sqrt{Q_5}$. It will be interesting
to compare both these approaches with the recent calculation
\cite{Hanany:2002ev} of the partition function of the two-dimensional
black hole.

\end{enumerate}

\gap5
\acknowledgments

We would like to thank L.~Alvarez-Gaume, S.~Das, T.~Damour, A.~Dhar,
M.R.~Douglas, D.~Gross, S.F.~ Hassan, K.~Krasnov, K.~Maeda,
J.~Maldacena, G.~Moore, K.S.~Narain, A.~Sen, L.~Susskind, G.~Veneziano and
E.~Witten for useful discussions. The work of J.R.D is supported by
NSF grant PHY00-98396.

\newpage
\appendix
\section{Euclidean derivation of Hawking temperature}\label{gibbons-hawking}

We will present a derivation (the purist may regard this as a
``mnemonic for a derivation'', for more detailed accounts see,
e.g. \cite{chp1:Wald,Hawking:ig,Gibbons:1976ue}) of Hawking
temperature for a class of black holes represented by the metric
\be
ds^2 = - F(r) C(r) dt^2 + dr^2/ C(r) + H(r) r^2 d\Omega^2
\ee
which of course include \eq{sch-bh} and \eq{rn-bh}. The black hole could be
three, four or higher dimensional, the number of angles represented by
$d\Omega^2$ varying accordingly. 

We will assume that $C(r)$ vanishes at the real value $r= r_h$ and is
non-vanishing for $r> r_h$ ($C(r)=0$ can have smaller roots than
$r_h$; they are irrelevant for the present discussion). We will also
assume that $F(r)$ and $H(r)$ are smooth and positive for $r \ge
r_h$. For asymptotically flat black holes, $C(r), F(r), H(r) \to 1$ as
$r\to \infty$. $r=r_h$ will correspond to the location of the event
horizon.

We will focus our attention near $r=r_h$, where
\be
C(r) = C'(r_h) (r - r_h) + O(r-r_h)^2
\label{a1:taylor}
\ee
It is useful to define a new radial coordinate $\rho$:
\be
d\rho^2 =dr^2/C(r), r \in ]r_h, \infty], \rho \in ]0, \infty]
\ee
In the near-horizon region \eq{a1:taylor}
\be
\rho = 2 \sqrt{\frac{r- r_h}{C'(r_h)}} [ 1 + O(r- r_h)]
\ee 
The near-horizon metric is given by
\be 
ds^2 = -dt^2 F(r_h)\left[C'(r_h)\right]^2\, 
\frac{\rho^2}4 + d\rho^2 + 
H(r_h)r_h^2 d\Omega^2
\ee
The Euclidean continuation $t = -i \tau$ is given by
\be
ds_E^2 = d\tau^2 F(r_h)\left[C'(r_h)\right]^2\, 
\frac{\rho^2}4 + d\rho^2 + r_h^2 d\Omega^2
\ee
If we require the $\tau,\rho$ plane not to have a conical singularity
\cite{Gibbons:1976ue}, we must assign the following periodicity 
\be
\tau \equiv \tau + \beta,\; \beta= \frac{4\pi}{C'(r_h)\sqrt{F(r_h)}}
\label{a1:beta}
\ee
To see this, write $\tau= (\beta/2\pi) \varphi$, so that
\be
ds_E^2 = d\varphi^2 {\rho^2} \left[\frac{\beta C'(r_h)\sqrt{F(r_h)
}}{4\pi}\right]^2 + d\rho^2 + H(r_h) r_h^2 d\Omega^2
\ee
Absence of conical singularity implies that the quantity
inside the square brackets must be equal to one.

A periodic Euclidean time with period $\beta$ implies a 
temperature
\be
T_H= \hbar/\beta = \frac{\hbar C'(r_h)\sqrt{F(r_h)}}{4\pi}
\label{a1:temp}
\ee
\begin{itemize}

\item
Example 1: for the RN black hole \eq{rn-bh}, $ r_h= r_+,C(r)= (1-
r_-/r)(1- r_+/r), F(r)=1$, hence $ C'(r_h) = (r_+ - r_-)/r_+^2, $
leading to \eq{rn-temp}.

\item
Example 2: for the non-extremal five-dimensional black hole
\eq{non-extremal-5d}, $r_h= r_0, C(r)= h f^{-1/3}, F(r)= f^{-1/3},
H(r)= f^{1/3},$ hence 
\[
C'(r_h)\sqrt{F(r_h)} = h'(r_0) f(r_0)^{-1/2}
= 2r_0^{-1} \left(\cosh(\alpha_1)
\cosh(\alpha_5)\cosh(\alpha_n) \right)^{-1},
\] 
leading to \eq{2:temp}.

\end{itemize}

\gap{1} 

\noindent\underbar{\em Entropy and the first law} 

\gap{1}

We note that the entropy \eq{bek-hawk}, namely
\be
S= \frac{\pi r_h^2}{G_N \hbar}, 
\label{a1:entropy}
\ee
together with \eq{a1:temp}, satisfies the first law of
thermodynamics \eq{1:rn-first-law}. E.g. for the RN black hole
\eq{rn-bh}, consider two neighbouring states, differing in
the value of $r_+$  (but having the same
value of $r_-$). Using $\Phi= A_0(r_h)= Q/r_+,$ we get
\bea
\Phi d Q && = \frac{1}{r_+} Q dQ =  \frac{r_-}{2G_N r_+}\ dr_+
\nn \\
T_H dS &&= \frac{r_+ - r_-}{4\pi r_+^2}\ d\frac{\pi r_+^2}{G_N} 
= \frac{r_+ - r_-}{2G_N r_+}
\nn \\
d M && = d \frac{r_+ + r_-}{2 G_N} = \frac{1}{2 G_N} dr_+ = 
T_H dS + \Phi dQ
\label{a1:first-law}
\eea
In the first line we have used $Q^2 = r_+ r_-/G_N \Rightarrow
Q dQ = \half d (Q^2) = \frac{r_-}{2G_N} dr_+$.

We thus verify Eqn. \eq{1:rn-first-law}.

\newpage
\section{A heuristic motivation for \underbar{Rules} 
1 and 2 in Section (2.4)}\label{algorithm}


We present a brief, heuristic, motivation for the algorithm of section
\ref{non-extreme}. Suppose we view a static Schwarzschild black hole,
of ADM mass $m$, from the viewpoint of the five-dimensional ($R^4
\times S^1$) Kaluza-Klein theory (we use coordinates
$x^{0,1,2,3}$ for $R^4$ and $x^5$ for the
$S^1$). The 5-momentum $(p_0, \vec p, p_5)$ would be given by
\be
p_0 = m, \quad  p_5 \propto \hbox{charge} =0, 
\quad  \vec p=0
\ee
The $p_5$ equation follows because the Schwarzschild 
black hole is neutral. 

A way of generating charged solutions (see, e.g.,
\cite{Sen:1992ua,Horowitz:1992jp,Russo:1997if} for details)
is to unwrap the above solution in five non-compact dimensions,
perform a boost in the 0-5 plane (which is a symmetry of the
non-compact theory), and compactify the new $x'_5$ direction to get a
charged (RN) black hole in four non-compact dimensions. The momenta
should transform as
\be
p'_0 \equiv M = m \cosh \delta, 
\quad p'_5 \equiv Q 
= m \sinh \delta, \quad  {\vec p}'=0
\ee 
In the above, we have absorbed the factor of radius in the
definition of the charge so that the extremality condition
reads $M=Q$. The extremal limit  can be attained by
\be
\delta\to \infty, m\to 0, m e^\delta \to {\rm constant}
\ee
so that  
\be
\label{approach-extremal}
Q \to M = me^\delta/2, \quad  p'_R \equiv p'_0 - p'_5\to 0 
\ee
\gap1
\noindent\underbar{\sl Near-extremal limit}\nl
\gap1
The near-extremal limit is obtained by 
keeping the leading corrections in
the expansion parameter $e^{-\delta}$. Thus,
\be
\tilde Q/M = \tanh\,\delta \simeq 1 - m^2/(2 
{\tilde Q}^2),\qquad  p'_R \ll p'_L, \; p'_L \equiv p'_0 + p'_5
\ee
In terms of these parameters, the four-dimensional metric for a 
near-extremal charged (RN) black hole is given by
\bea
ds_4^2 &=& - f dt^2 + f^{-1} dr^2
+ r^2 d\Omega_2^2 \nn\\
f &\equiv& 1 - 2M/r + {\tilde Q}^2/r^2 = f_{\rm ext} h(r)
\nn\\
f_{\rm ext} &\equiv &  (1 - {\tilde Q}/r)^2,  
\quad  h(r)= (1 - \mu/r) 
\nn\\
\mu &=& m^2/\tilde Q  
\eea
The last equality implies
\be
\tilde Q= \mu \sinh^2 \delta,
\ee
same as \eq{boost} above. Also, the second equality
agrees with Rule 1 for relating the non-extremal $g_{tt},
g_{rr}$ to their extremal counterparts. 

Of course, we have considered here only the near-extremal case. The
remarkable thing about the algorithm mentioned in Section
\ref{non-extreme} is that it works for arbitrary deviations from
extremality.

\newpage
\section{Coordinate systems for \ads\ and related spaces}

For a more detailed discussion and derivations, see, e.g.
\cite{hawking-ellis,Banados:1993gq}. This
Appendix is meant to serve as a compendium of some definitions and
results about the geometry of \ads\ and related spaces. We discuss
both Lorentzian and Euclidean signatures.

\subsection{AdS$_3$}

AdS$_3$ is defined as a hyperboloid
\be
-Y_0^2 - Y_{-1}^2 + Y_1^2 + Y_2^2 \equiv Y_+ Y_- + \sum_{\mu=0,1}
Y_\mu Y^\mu = -l^2
\label{a3:hyperboloid}
\ee
in $R^{2,2}$ with metric 
\be
ds^2 =  -dY_0^2 - dY_{-1}^2 + dY_1^2 + dY_2^2 = dY_\mu dY^\mu + 
dY_+dY_-
\label{a3:r22}
\ee
Here $Y_\pm = Y_2 \pm Y_{-1}$.
The condition \eq{a3:hyperboloid} can be equivalently
stated by saying that a point in \ads\ is represented by
an SL(2) matrix $(1/l){\bf Y}$ where
\be
{\bf Y} = Y_{-1} {\bf 1} + Y^a {\bf \gamma}_a,\, a=0,1,2 
\label{a3:sl2}
\ee
where 
\be
\label{a3:gamma}
{\bf \gamma}_0= \left( \begin{array}{cc}
0 & 1\\ -1 &0 \end{array} \right), \;
{\bf \gamma}_1= \left( \begin{array}{cc}
0 & 1\\ 1 &0 \end{array} \right), \;
{\bf \gamma}_2= \left( \begin{array}{cc}
1 & 0\\ 0 &-1 \end{array} \right).
\ee

\ni\underbar{\it Global coordinates:}
\bea
\label{hyp-global}
Y_{-1} &&= l \cosh\mu \sin t; \quad  Y_0= l \cosh\mu \cos t
\nn\\
Y_1 &&= l \sinh \mu \cos \phi; \quad Y_2 = l \sinh \mu \sin \phi
\eea
The metric is
\be
\label{globcord}
ds^2 = l^2 \Big( -\cosh^2\mu\ dt^2 + d\mu^2 + \sinh^2 \mu\ d\phi^2 \Big)
\label{a3:global}
\ee
By redefining $\sinh\mu = r/l$ we get
\be
ds^2 = l^2 \Big[ 
-\Big(1 + \frac{r^2}{l^2} \Big)\ dt^2 + 
\frac{dr^2}{1 + \frac{r^2}{l^2}}+ \frac{r^2}{l^2} d\phi^2 \Big]
\label{a3:ads}
\ee
\be
\hbox{Range~ of~ coordinates:} \phi \in[0,2\pi],\ 
\mu ({\rm equivalently}~~ r) \in (0, \infty),\
t \in [0, 2\pi]
\label{range-ads}
\ee
If we unwrap $t$ to the range 
$t\in (-\infty, \infty)$ we get the so-called C\ads
(covering space) which is geodesically complete. 

There is another popular form, given by the coordinate 
transformation $\rho= \tanh(\mu/2)$, leading to
\be
ds^2 = - \left(\frac{1 + \rho^2}{1- \rho^2}\right)^2 dt^2
+ \left( \frac{2}{1 - \rho^2}\right)^2 \left(d\rho^2 + \rho^2d\phi^2
\right)
\label{a3:matschull}
\ee
The section at any given $t$ has the metric of a disc, called
the Poincare disc.

\ni\underbar{\it Poincare coordinates}:
\[
Y_+ = l^2/u, \quad Y_\mu= l x_\mu/u
\]
($Y_-$ determined by \eq{a3:hyperboloid}).
The metric is
\be
\label{ads-poincare}
ds^2= \frac{l^2}{u^2}\Big( du^2 + dx^\mu dx_\mu \Big)
\ee
Range: $u\in (0,\infty), x^0=t\in(-\infty, \infty),
x^1\in(-\infty, \infty)$: covers a half of \ads \eq{a3:hyperboloid}.

\subsection{BTZ black hole}
The three-dimensional black hole \cite{chp1:Btz,Banados:1993gq}
in an asymptotically
AdS spacetime ($\Lambda:=-1/l^2$), of mass $M$ and
angular momentum $J$, is given by the metric
\bea
\label{a3:btz}
ds^2 &&= - N^2(r) dt^2 + dr^2/N^2(r) + r^2 
\left( N_\phi(r)  dt + d\phi \right)^2
\nn\\
N^2(r) &&=  \frac{r^2}{l^2} - M + (\frac{J}{2r})^2,
\quad 
N_\phi(r) = \frac{J}{2r^2}
\nn\\
\hbox{Range}:&& t \in (-\infty, \infty), r\in (0, \infty),
\phi \in [0, 2\pi].
\label{range-btz}
\eea

Properties: 
\begin{itemize}
\item 
The lapse function $N(r)$ vanishes at $r=r_\pm$
where
\be
\label{a3:rpm}
r_\pm = l \Big[\frac{M}2 \left( 1 \pm \sqrt{1 - (\frac{J}{Ml})^2} \right)
\Big]^{1/2}
\ee
Thus,
\be
\label{a3:lapse}
N^2(r) = \frac{1}{l^2r^2}
\left(r^2 - r_+^2 \right)\left(r^2 - r_-^2 \right),
\quad
N_\phi(r) = \frac{r_+ r_-}{r^2 l}.
\ee
$r_+$ corresponds to the event horizon.
\item $g_{00} = -N(r)^2 + r^2 N_\phi(r)^2=\frac{r^2}{l^2} - M$ vanishes at 
\[
r_{\rm erg} = l \sqrt{M}
\]
which represents the surface of infinite red-shift.
\item 
These three special values of $r$ satisfy $r_-\le r_+ \le r_{\rm erg}$.
The region between $r_+$ and $ r_{\rm erg}$ is called the ``ergosphere''.
\end{itemize}
Furthermore, the above black hole \eq{a3:btz} can be
obtained as a quotient of the \ads space \eq{a3:hyperboloid}. We show this
for the $J=0, M >0$ BTZ,  given by
\be
\label{a3:j0btz}
ds^2 = -\left[ \frac{r^2}{l^2} - M  \right] dt^2
+ \left[ \frac{r^2}{l^2} - M  \right]^{-1} dr^2
+ r^2 d\phi^2 
\ee
In the patch $Y_{-1}^2 \ge Y_1^2, Y_0^2 \le Y_2^2$ we define
\bea
\label{a3:hyp-btz}
Y_{-1} &&= \pm \frac{r}{\sqrt M} \cosh \phi\sqrt M, 
\quad Y_0 = \left( \frac{r^2}{M} -
l^2\right)^{1/2} \sinh \frac{t\sqrt M}{l}
\nn\\
Y_1 &&= \frac{r}{\sqrt M} \sinh \phi\sqrt M,
\quad Y_2 = \pm \left( \frac{r^2}{M} -
l^2\right)^{1/2}  \cosh \frac{t\sqrt M}{l}
\eea
With this, the metric \eq{a3:r22} coincides with \eq{a3:j0btz}.
In \eq{a3:j0btz} the angle $\phi \equiv \phi + 2\pi$. This
implies, through \eq{a3:hyp-btz} a discrete quotienting of
the original hyperboloid. We thus find that the BTZ metric
is equivalent to \ads$/Z$. (We have shown this here
in a certain coordinate patch, but it can be proved
more generally \cite{Banados:1993gq}.) 


\subsection{Conical spaces}


\be
ds^2 = l^2 \Big[ 
-\Big(\gamma + \frac{r^2}{l^2} \Big)\ dt^2 + 
\frac{dr^2}{\gamma + \frac{r^2}{l^2}}+ r^2 d\phi^2 \Big]
\label{a3:conic}
\ee
Range: $r\in (0, \infty), t\in (-\infty, \infty),
\phi \in [0, 2\pi].$

Define $r= \sqrt \gamma \tilde r, t= \tilde t/\sqrt \gamma,
\phi = \tilde \phi/\sqrt \gamma$, get
\be
ds^2 = l^2 \Big[ 
-\Big(1 + \frac{\tilde r^2}{l^2} \Big)\ d\tilde t^2 + 
\frac{d\tilde r^2}{1 + \frac{\tilde r^2}{l^2}}+ 
\tilde r^2 d\tilde\phi^2 \Big]
\label{a3:conic2}
\ee
\be
\label{a3:conic-range}
\hbox{Range:}~~~ 
\tilde \phi \in [0, 2\pi \sqrt\gamma].
\ee
Defect angle $\Delta= 2\pi(1 - \sqrt\gamma)$.

\mysec11{Conic as \ads/Z}

We will show that \eq{a3:conic2} can be 
obtained from \eq{a3:hyperboloid},\eq{a3:r22} modulo
the following identification (using the
notation of \eq{a3:sl2})
\be
\label{a3:conic-similarity}
{\bf Y} \equiv {\bf u}^{-1} {\bf Y} {\bf u}
\ee
The holonomy matrix ${\bf u}$ is given by
the momentum $p^a, a=0,1,2$ of the particle:
\be
{\bf u} = u {\bf 1} + p^a {\bf \gamma}_a
\label{a3:holonomy}
\ee
In case of a static particle, of mass $m$ 
\be
p^0= m, p^1 =  p^2=0. 
\label{a3:momenta}
\ee
Hence ${\bf u}= u {\bf 1} + m \gamma_0 $,
where $u= \sqrt{1 - m^2}$ by the SL(2) 
condition. The identification \eq{a3:conic-similarity},
affects only the components $Y_1, Y_2$ and reads
\[
\left( \begin{array}{c} Y_1 \\ Y_2 
	\end{array}
\right)
\equiv \left( \begin{array}{cc} 1-2m^2 & -2m\sqrt{1-m^2}
\\ 2m\sqrt{1-m^2} & 1-2m^2 	\end{array}
\right)
\left( \begin{array}{c} Y_1 \\ Y_2 
	\end{array}
\right)
\]
Note that the matrix is an $SO(2)$ rotation matrix
with $\cos(\alpha)= 1 - 2m^2$.
In terms of the coordinates \eq{hyp-global},\eq{a3:conic2} this 
implies an identification
\[
\tilde\phi \equiv \tilde \phi + \cos^{-1}(1 - 2m^2)
\]
Comparing with \eq{a3:conic-range} we get a relation
between the parameter $\gamma$ and the mass $m$
\be
1- 2m^2 = \cos(2\pi \sqrt\gamma)
\label{a3:gamma-m}
\ee
 

\subsection{Euclidean sections and Thermal Physics}


Euclidean \ads can be defined by the
global coordinates of \eq{a3:ads} with the
replacement ${\tt t}= -it$, leading to
the metric 
\be
ds^2 = l^2 \Big[ 
\Big(1 + \frac{r^2}{l^2} \Big)\ d\ttt^2 + 
\frac{dr^2}{1 + \frac{r^2}{l^2}}+ r^2 d\phi^2 \Big]
\label{a3:eads}
\ee
where (cf. \eq{range-ads}
\be
\phi \equiv \phi + 2\pi
\label{phi-period}
\ee
Like \eq{a3:sl2} in the Lorentzian case, a point in Euclidean \ads\ can
be alternatively defined as a hermitian matrix $(1/l) {\bf Y}$
of unit determinant (the space of such matrices is
called $\H$)  where
\be
\label{a3:h}
{\bf Y} = Y_{-1} {\bf 1} + Y^a \tilde{\bf \gamma}_a,\, a=0,1,2 
\ee
and the new $\tilde \gamma$ matrices are obtained by
replacing $\gamma_0$ in \eq{a3:gamma} by
$\tilde \gamma_0=i \gamma_0$. The determinant condition now reads
\be
- Y_{-1}^2 + Y_0^2 + Y_1^2 + Y_2^2 = - l^2
\label{eucl-hyp}
\ee
The metric in this parameterization is 
\be 
ds^2 = {\rm Tr} d{\bf Y}^2 
= - dY_{-1}^2 + dY_0^2 + dY_1^2 + dY_2^2 
\label{metric-h}
\ee 
To make contact with the global coordinates of \eq{a3:eads} we first
introduce the by the following parameterization of $\H$
\be
{\bf Y}/l= \left(\begin{array}{cc}e^u & 0 \\ 0 & e^{-u}\end{array} 
\right)\ \left(\begin{array}{cc} \sqrt{1+r^2/l^2} & 
r/l \\ r/l & \sqrt{1+r^2/l^2} \end{array} 
\right)\ \left(\begin{array}{cc}e^{\bar u} & 0 \\ 0 & e^{-
\bar u}\end{array} 
\right)
\label{h-global}
\ee
In order to cover $\H$ only once, we
must identify
\be
\label{u-period}
2u \equiv 2u + i 2n \pi
\ee
The metric \eq{metric-h} then  becomes 
\be
ds^2 =l^2\Big[ \frac{(du + d\bar u)^2}{1 + r^2/l^2} 
	+ \frac{dr^2}{1 + r^2/l^2} - \frac{r^2}{l^2}(du-d\bar u)^2
\Big]
\label{metric-u}
\ee
A comparison of the peroidicities  \eq{u-period}
and \eq{phi-period} suggests that we define
\be
\label{u-phi}
2u= \ttt + i\phi.
\ee
With this definition we recover \eq{a3:eads} from \eq{metric-u}. 

For Euclidean \ads, the periodicity in the $\ttt$ direction is to be
supplied as an input from physics. A thermal ensemble implies in the
usual fashion $ \ttt \equiv \ttt + 1/T $.  In addition to a
temperature one may want to introduce an angular potential (conjugate
to angular momentum $J \sim i\del/\del\phi$); such an ensemble,
described by ${\rm Tr} \exp[- H/T + i \Phi J]$ implies a more general
(twisted) identification
\be
\label{a3:twist}
\phi + i\ttt \equiv  \phi + i\ttt + \Phi + i/T :=
\phi + i\ttt + i\beta
\ee
where the second step is a definition of the ``complex
temperature'' (cf. \eq{a1:beta})
\be
\label{a3:ads-temp}
\beta = 1/T - i \Phi
\ee
\mysec11{Euclidean Poincare coordinates}

The Euclidean verison of \eq{ads-poincare} is defined by
\be
{\bf Y}/l= \left(\begin{array}{cc}e^u & 0 \\ 0 & e^{-u}\end{array} 
\right)\ \left(\begin{array}{cc} h + w\bar w/h & 
w/h \\ \bar w/h & 1/h \end{array} 
\right)
\label{eucl-ads-poincare-def}
\ee
The metric \eq{metric-h} becomes
\be
ds^2 = \frac1{h^2}\left( dw d\bar w + dh^2 \right)
\label{eucl-ads-poincare}
\ee
\mysec11{Euclidean BTZ}

Euclidean BTZ is usually defined by defining ${\ttt}= -it, J_E=iJ$ in
\eq{a3:btz}:
\be
\label{a3:ebtz}
ds^2 = \left[ \frac{r^2}{l^2} - M - (\frac{J_E}{2r})^2 \right] d\ttt^2
+ \left[ \frac{r^2}{l^2} - M - (\frac{J_E}{2r})^2 \right]^{-1} dr^2
+ r^2 \left[\frac{iJ_E}{2r^2} d\ttt +d\phi \right]^2 
\ee
where (cf. \eq{range-btz})
\be
\phi \equiv \phi + 2\pi
\label{phi-period-btz}
\ee
Like in the Lorentzian case, we can obtain the metric \eq{a3:ebtz} as
EBTZ= E\ads/Z \cite{chp3:MalStr98,Dijkgraaf:2000fq}, as follows.
Define a quotient of \eq{a3:h}, by \cite{chp1:Btz}
\be
\label{a3:btz-quotient}
{\bf Y} \equiv \left( \begin{array}{cc} e^{-i\pi \tau}& 0\\
0 & e^{i\pi \tau} \end{array} \right) \ {\bf Y}
\ \left( \begin{array}{cc} e^{i\pi\bar \tau}& 0\\
0 & e^{-i\pi \bar\tau} \end{array} \right) 
\ee
In terms of \eq{h-global} the above identification
\eq{a3:btz-quotient} reads
\be
\label{btz-period}
2u \equiv 2u + 2\pi i n \tau
\ee
Note that this identification is in addition to \eq{u-period}.
The two identifications defines for us two independent cycles
in the $u$-plane. We will show below that the description
\eq{a3:ebtz} follows by identifying the cycle \eq{btz-period}
with the ``space'' cycle, namely \eq{phi-period-btz}. This
can be easily done by  defining
\be
2u = -i\tau (\phi + i \ttt)
\label{u-phi-btz}
\ee
The cycle \eq{u-period} now becomes the ``time'', unlike
in the case of thermal \ads, where it was ``space''.
If we define
\be
\tau = |r_-| - ir_+
\label{def-tau}
\ee
and a new  radial coordinate $\tilde r$ in
terms of the $r$ of \eq{h-global}, as follows
\be
r^2/l^2 = \frac{\tilde r^2/l^2 - \tau_2^2 }{|\tau|^2}
\ee
where
\be
r_\pm^2 =  \frac{l^2 M}{2} 
\left[1 \pm \sqrt{1 + \frac{J_E^2}{M^2
l^2}} \right]
\label{rpm-euclid}
\ee
(note that $r_- = i | r_-|$ is purely imaginary), then the metric on
$\H$, for $\tilde r \ge r_+$, becomes
\eq{a3:ebtz} once we drop the tilde from $\tilde r$.
\mysec11{Temperature}

Note that the periodicity along the
new ``time'' cycle \eq{u-period}, through \eq{u-phi-btz}
implies the following complex periodicity
(complex temperature) 
\bea
\phi + i\ttt && \equiv \phi + i\ttt + i\beta_0,
\nn\\
\beta_0 &&= 2\pi i(1/\tau) = \frac{2\pi r_+}{r_+^2-r_-^2}
+ i \frac{2\pi |r_-|}{r_+^2-r_-^2} 
\label{btz-beta}
\eea
The real temperature $T_0\equiv \left(\Re{\beta_0}\right)^{-1}$ 
is given by
\be
T_0= \frac{r_+^2-r_-^2}{2\pi r_+}
\label{a3:btz-temp}
\ee
which agrees with the expression in \cite{chp1:Btz}.

\mysec11{Entropy}

As can be derived from the partition function calculation
in Section 11.1, the entropy of the BTZ black hole
is given by \cite{chp1:Btz}
\be
S= \frac{2\pi r_+}{4 G^{(3)}_N }
\label{a3:entropy}
\ee
which of course agrees with the Bekenstein-Hawking formula
\eq{bek-hawk} as well.

\newpage
\bibliographystyle{utphys}
\bibliography{repr}

\end{document}